# Active Noise Control Portable Device Design

## (***Proposal Plan***)

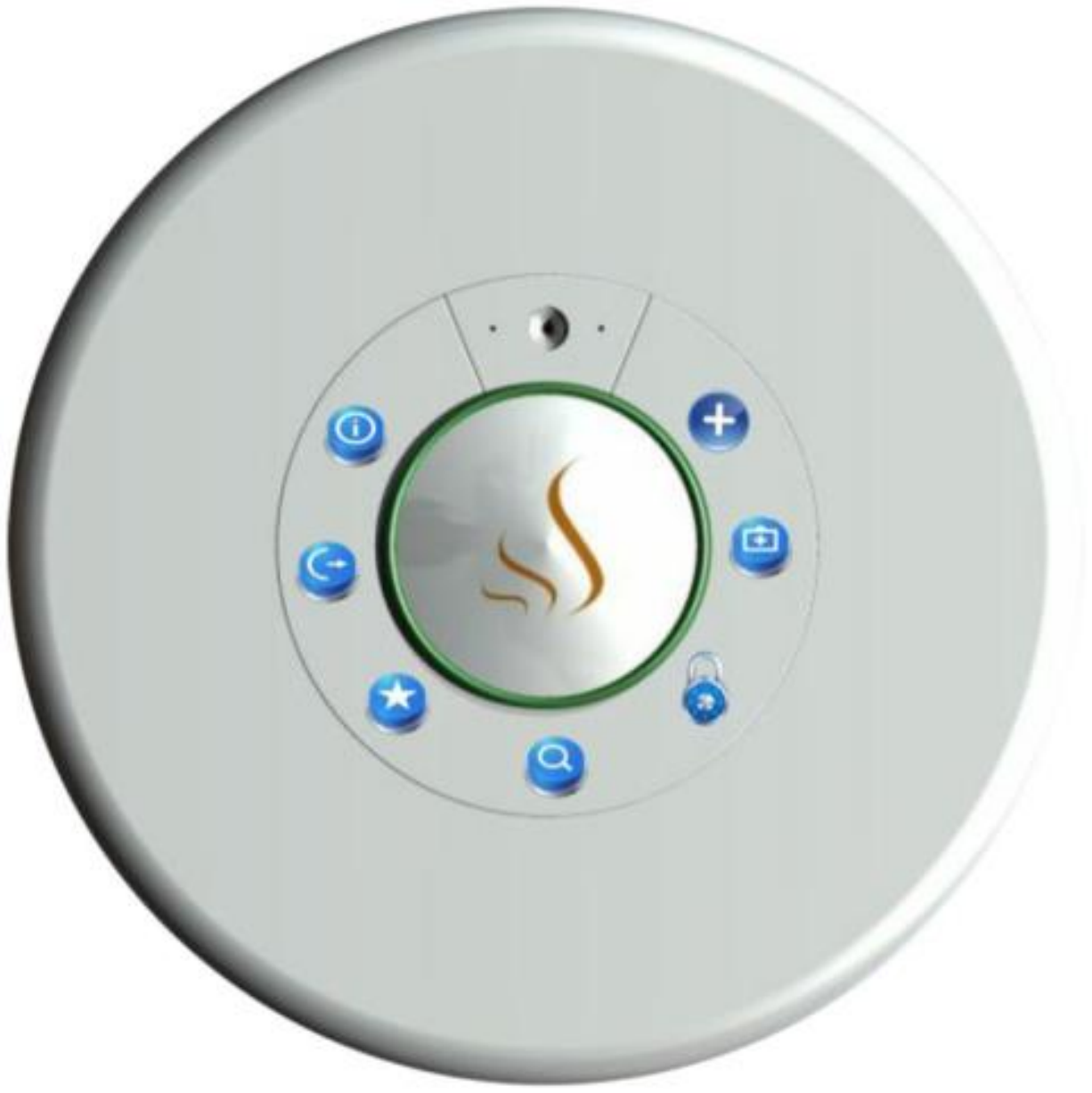

***Wu kai and Chen Yuanyuan***

# Contents

# 1. Introduction

While our world is filled with its own natural sounds that we can’t resist enjoying, it is also chock-full of other sounds that can be irritating, this is noise. Noise not only influences the working efficiency but also the human’s health. The problem of reducing noise is one of great importance and great difficulty. The problem has been addressed in many ways over the years. The current methods for noise reducing mostly rely on the materials and transmission medium, which are only effective to some extent for the high frequency noise. However, the effective reduction noise method especially for low frequency noise is very limited [1].

Here we come up with a noise reduction system consist of a sensor to detect the noise in the environment. Then the noise will be sent to an electronic control system to process the noise, which will generate a reverse phase frequency signal to counteract the disturbance. Finally, the processed smaller noise will be broadcasted by the speaker. Through this smart noise reduction system, even the noise with low-frequency can be eliminated.

The system is also integrated with sleep tracking and music player applications. It can also remember and store settings for the same environment, sense temperature, and smart control of home furniture, fire alarm, etc. This smart system can transfer data easily by Wi-Fi or Bluetooth and controlled by its APP.

In this project, we will present a model of the above technology which can be used in various environments to prevent noise pollution and provide a solution to the people who have difficulties finding a peaceful and quiet environment for sleep, work or study.

# 2. Product Idea Evaluation

## 2.1 Product Idea Description

As we all know, good and quiet environment is very important, which not only influences the working efficiency but also the human's health. Based on my personal experience, many people including me are very sensitive to noise and we are easily to be disturbed during sleep by annoying noise in the life, such as motorcycle, plane, etc. Some scientific evidence also confirmed the bad effects of noise. Per an early report from the World Health Organization, noise exposure in long term can result in illness and even death to people in busy cities [2]. Among various types of noise, the part only from traffic can lead to 3% of deaths due to heart attack and strokes all over the Europe. If we estimate based on the percentage, 210,000 people died of traffic noise in the world. In addition, above 600,000 potential years of healthy life were deprived by noise-related death and disabilities in Europe [3].

The importance of noise control has also been realized in Singapore. The National Environment Agency (NEA) of Singapore set the maximum permissible noise levels for construction and industrial is 60 dB(A) from 7am to 7pm and 50 dB(A) from 7pm to 7am [4]. However, the most popular noise in Singapore is coming from the traffic noise, typically ranging from 70 to 80 dB(A) at 15 meters from the highway [4], which is hard to be controlled. And since many residential buildings are closed to road or MRT, which affects many people by disturbing sleep, increasing risk of heart diseases, reducing working efficiency, it's very meaningful to reduce those noises.

The environmental research coming from US also reported that noise pollution can lead to health hazard, and reducing noise levels will help save economic expenses on medical issues. According to the analysis, 5-decibel noise reduction will decrease 1.4% of high blood pressure occurrence and 1.8% of coronary heart, thus saving $3.9 billion for economic cost [6].

Currently, the main method for noise reducing mostly relies on the materials and transmission medium. For example, segregation board is often used to shield the noise in the construction site. These traditional ways are only effective to some extent for the high frequency noise. However, the effective reduction noise method especially for low frequency noise is very limited.

To solve this problem, we want to develop a system to reduce the noise by offsetting the signal frequency of noise. This is a new technology named active noise control.

## 2.2 What Is Active Noise Control (ANC) System?

Active noise control is a method to reduce the noise by adding the second sound specifically designed to cancel the initial noise.

Based on Fourier theory, all the voices have different frequencies, phases and amplitude values in the frequency domain, and the voice can be regard as a sinusoidal curve with peak amplitude. Based on this important theory, we can use the electronic circuit to generate the same frequency and amplitude value, but with $-180^{\circ}$ phase to combine to cancel out the environment noise[7-8]. The details are shown as below:

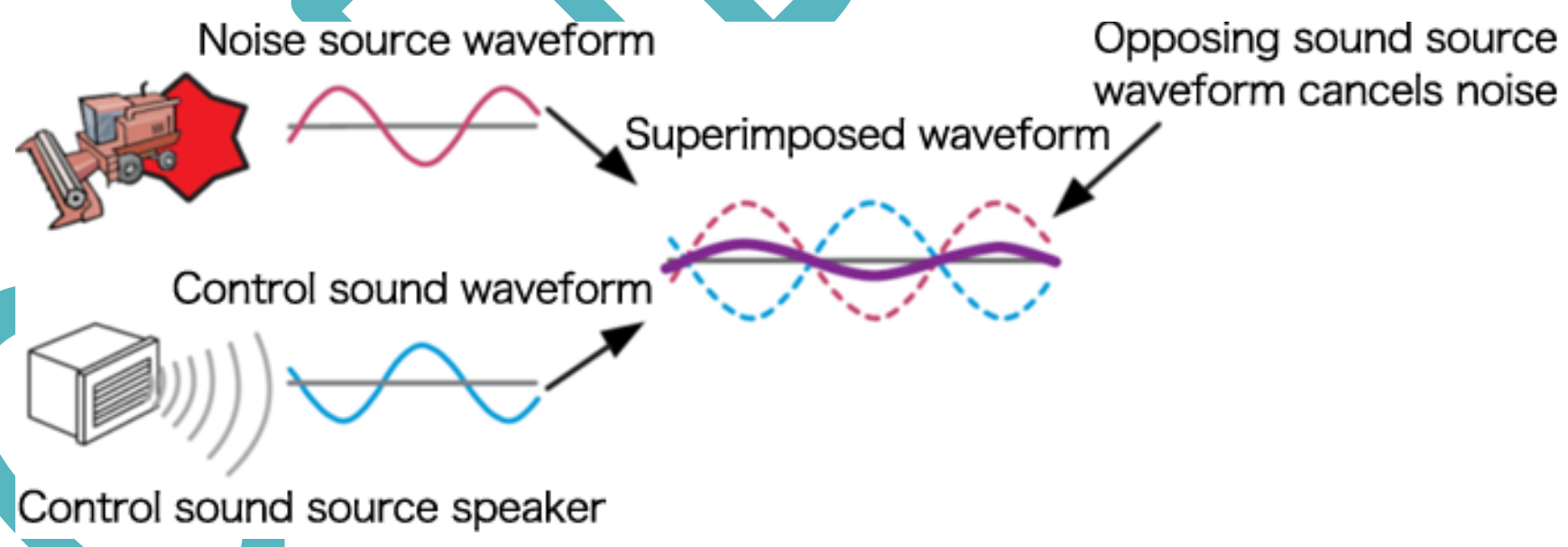

Active noise control technology has gained significant popularity in various domains, including the headphone industry [7-15], due to its effective performance in reducing low-frequency noise. It is noteworthy that the research group in Singapore has employed this technology to mitigate traffic noise in an open window setting [16-36]. Nevertheless, it is imperative to acknowledge that certain imperfections persist in practice [7,37], which can impact the efficacy of noise reduction and even compromise the resilience of the active noise control (ANC) system. In order to tackle these concerns, several innovative algorithms or implemented structures have been suggested [38-50], operating on the high-performance DSP processor or FPGA [51]. However, it is undeniable that these high-performance processors significantly contribute to the overall expenses associated with the active noise cancellation (ANC) system. Therefore, the proposed device will leverage novel algorithms [52-55] that enable the entire algorithm to be executed on a more cost-effective processor. By integrating novel Deep-learning-based Active Noise Control (ANC) strategies [56-62], this portable device exhibits enhanced convergence behavior and stability compared to conventional adaptive ANC systems.

**Figure 1 Noise Reduction Mechanism [7]**

## 2.3 How Does the Noise Reduction System Work?

According to the principle of active noise control, we can design the noise reduction system as below:

Firstly, the noise detector will sense noise, and send to the noise reducing system. Secondly, the noise reducing system will calculate the noise frequency and generate a reverse frequency.
Thirdly, noise reducing system speaker will broadcast the inverse frequency and cancel out the noise. Finally, the system will also detect the reduction efficiency and give the system feedback to do the adjustment.

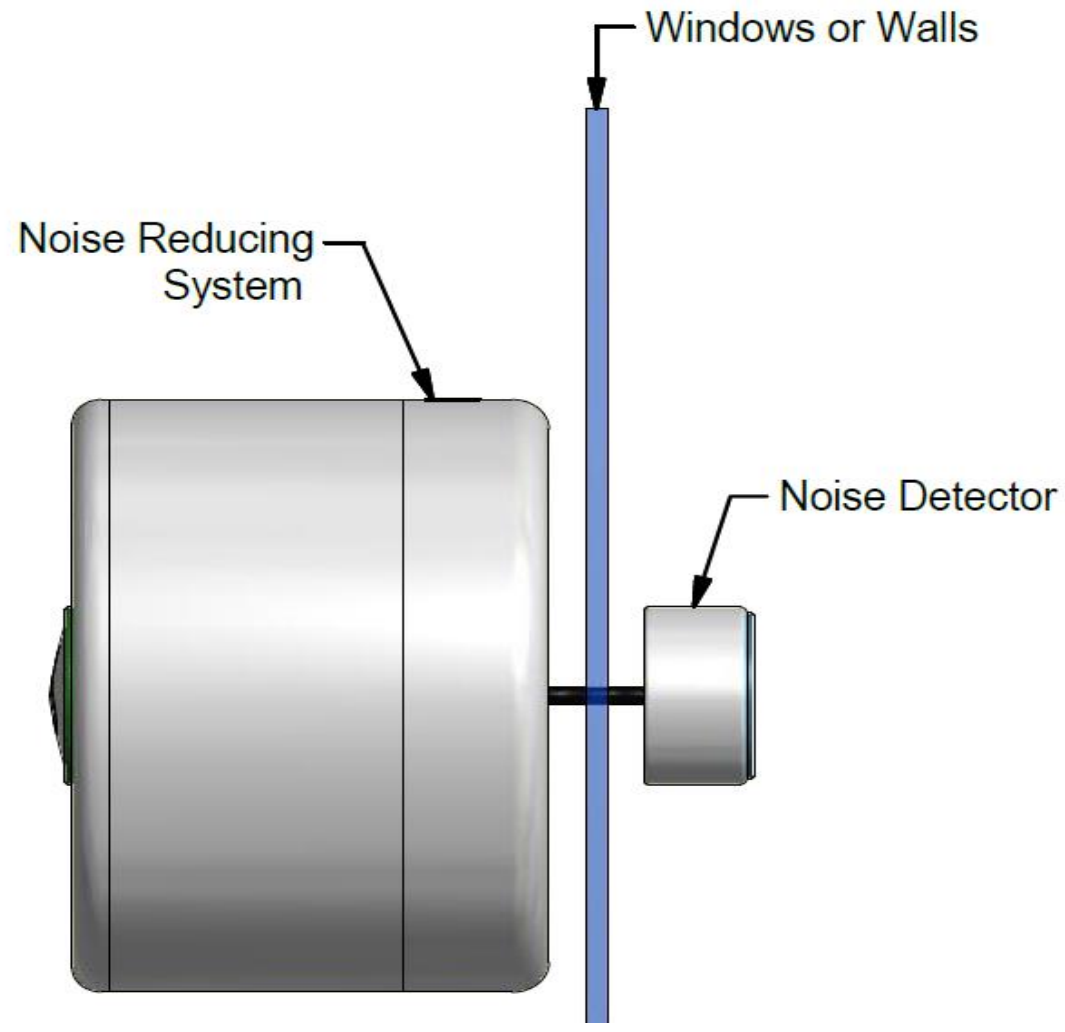

**Figure 2 Noise Reduction System**

After reduce the noise, we also use the system to detect the sleep quality and record the data. The data also can be transferred by Bluetooth or WIFI technology for review the health record and take the action to improve the sleep quality. Below picture are main application functions to control or review the data:

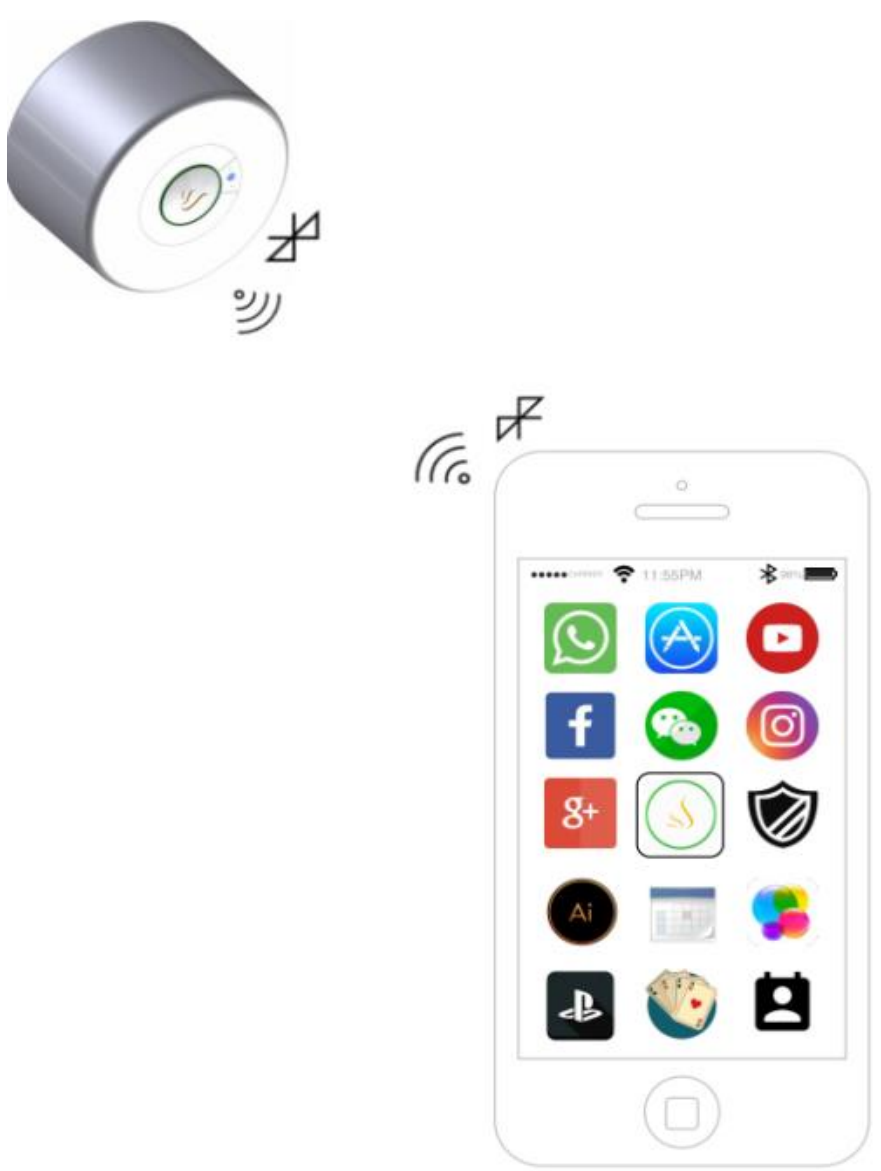

**Figure 3 Data transfer method**

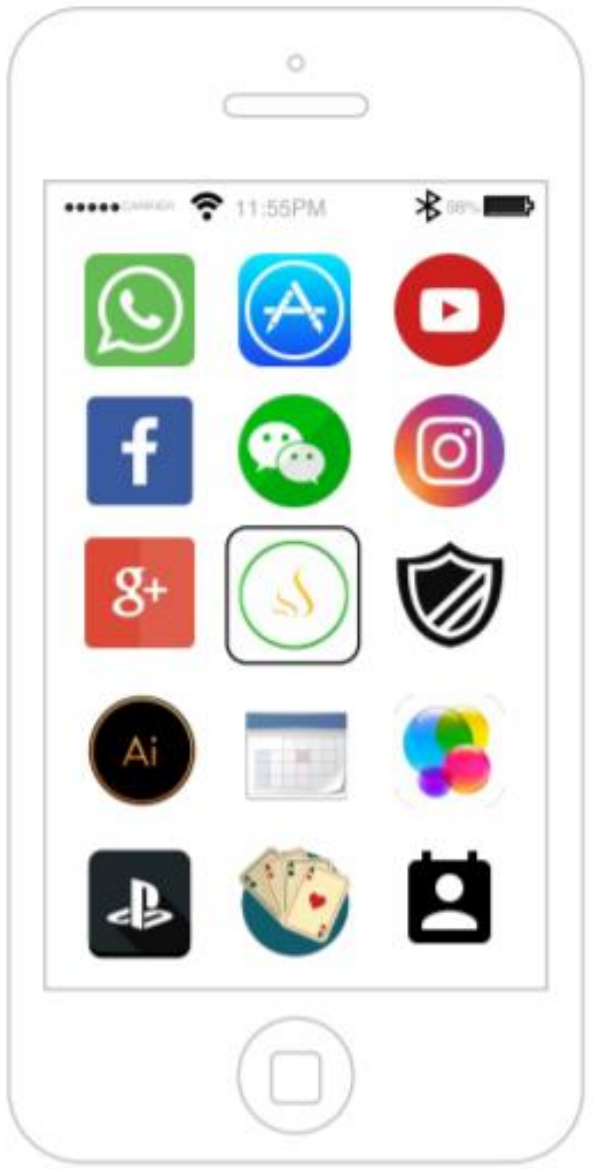

**Figure 4 Application user Interface**

To make our product more useful and more competitive in the market, we design to add some new features in our system. The system has the different noise reducing levels based on the users' requirement. After processing the noise frequency, the intelligent system also can store the data for future process. Once the same situation happens again, the system will automaticity recognizes and generates the corresponding action.

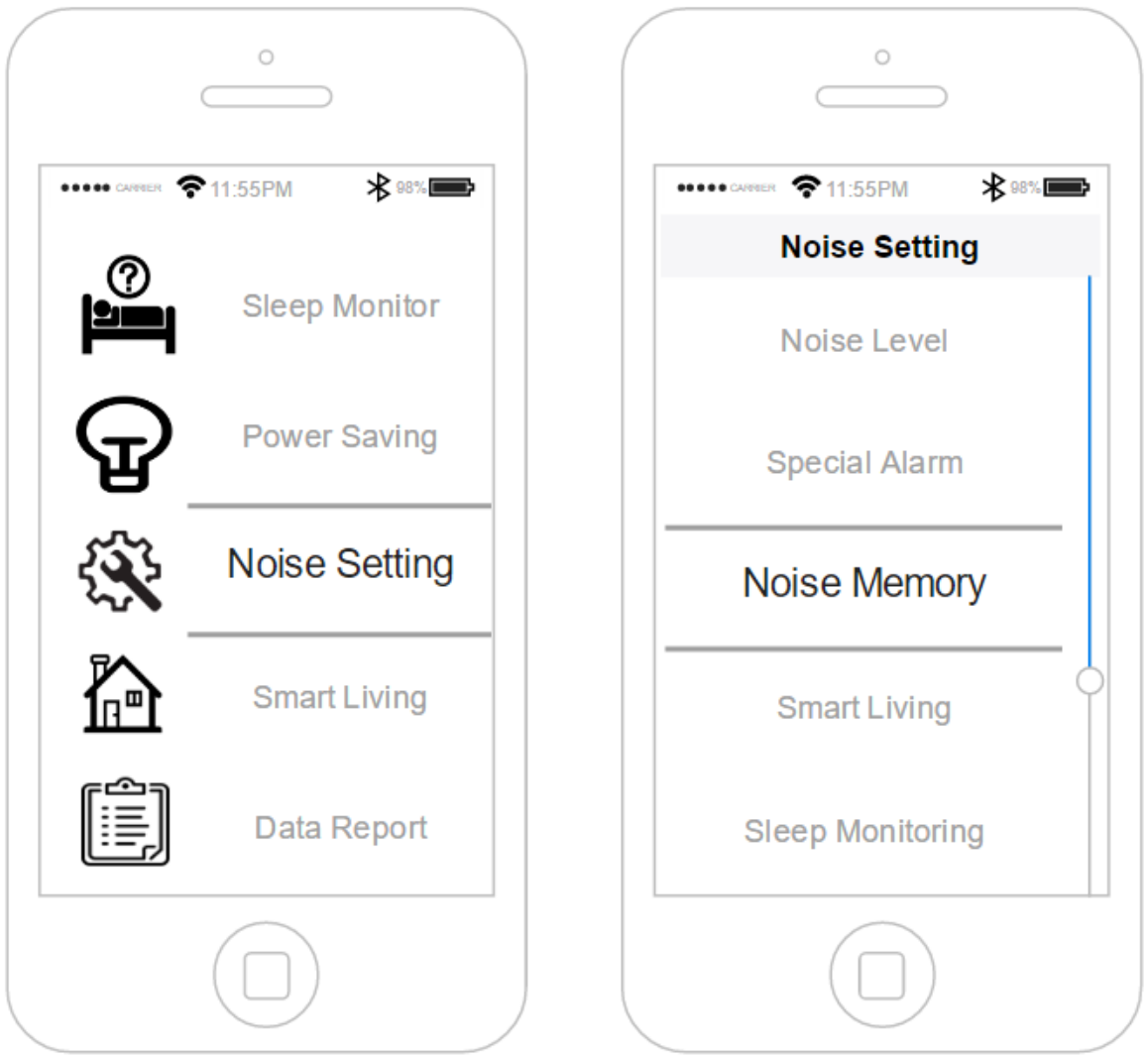

**Figure 5 Noise setting interface**

For the user convenience, we can also use the Bluetooth and Wi-Fi connection through the mobile device. In addition, our product can also be used as a speaker, and people can amplify their music and control the sound volume by their mobile phone. They can also choose the white noise or other voice mode in order to get a good quality of sleeping.

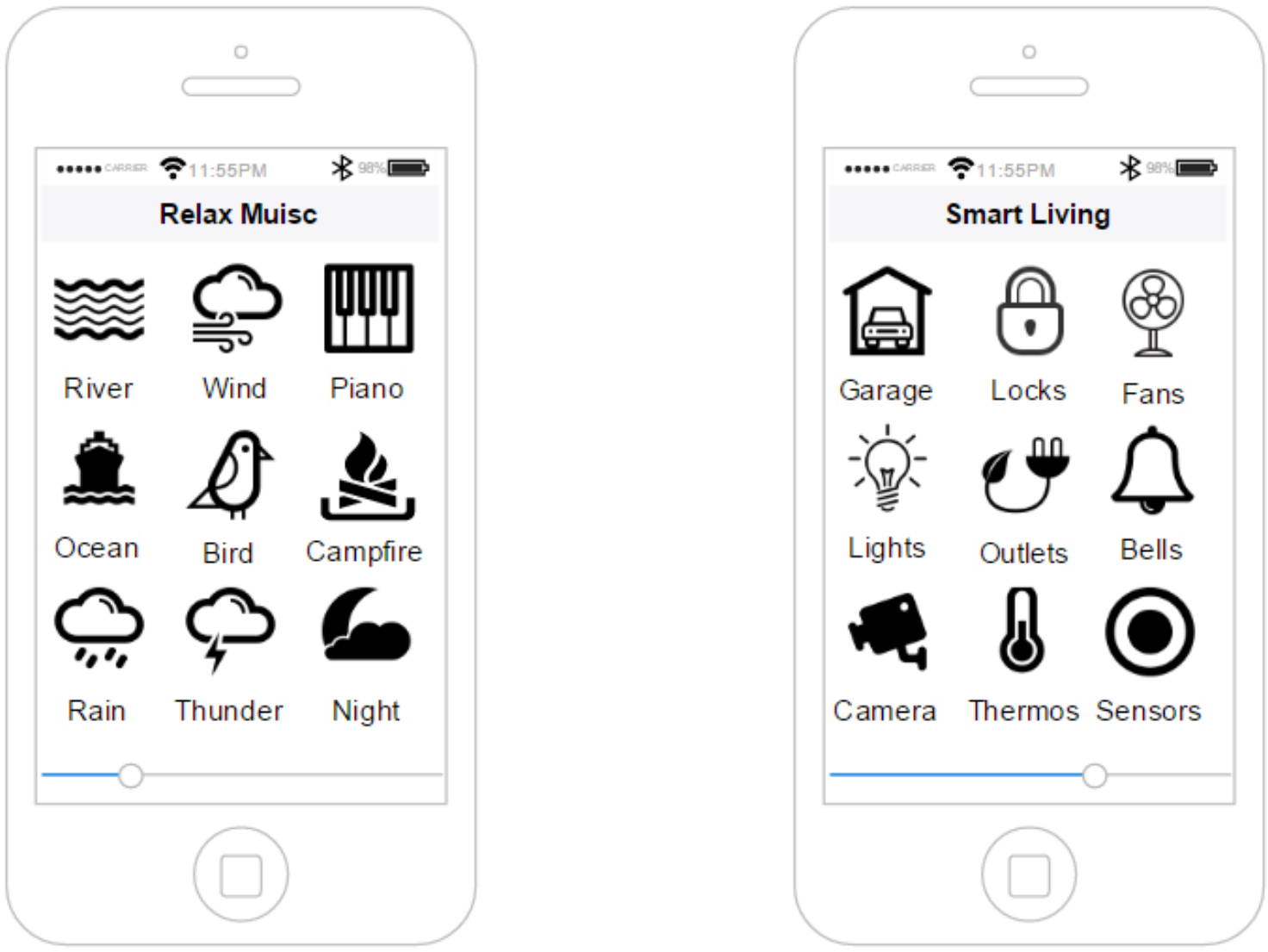

**Figure 6 Apps function interface**

The smart home system is making our life easier and safer, which will continue to play its important role in the future. We also designed the intelligent system which can be matched with the current marketing product. Our device can be used as a server to control the smart home device, for example, you can use our mobile app to control the smart lighting. It also can connect with your home security camera or fire alarm. If there's anything happening, the system will send the alarm to your mobile device and inform you to take some measures. The related drawing is shown as below:

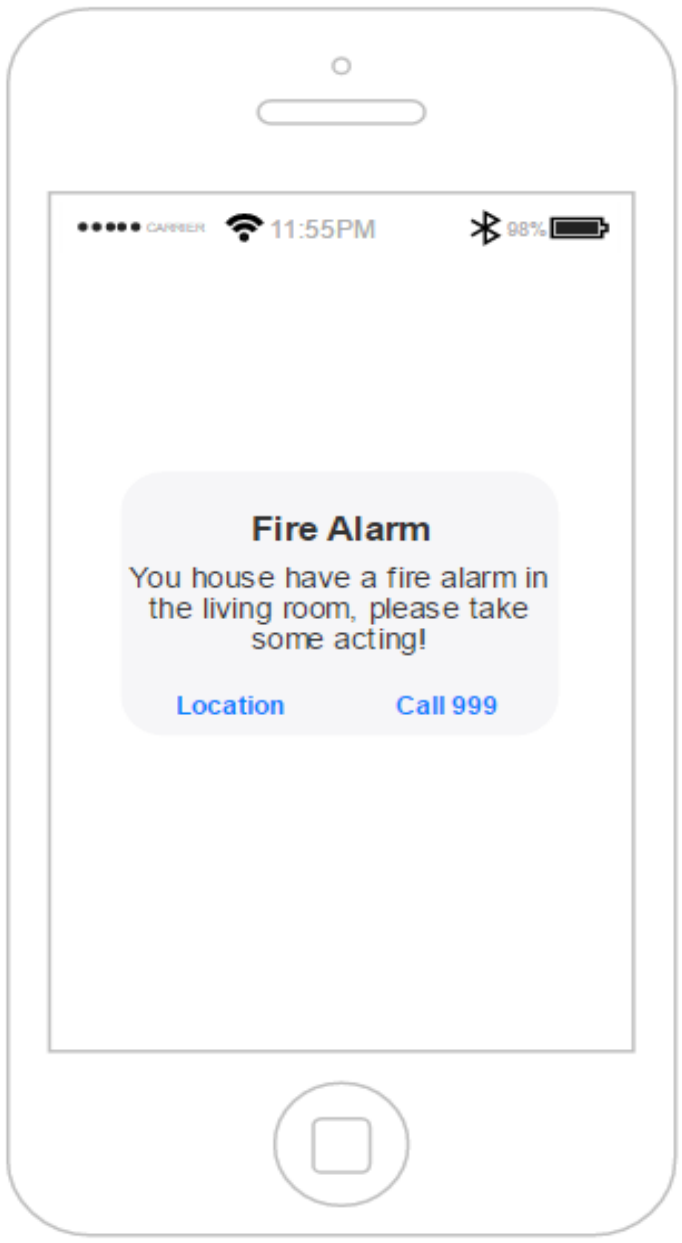

**Figure 7 Apps alarm interface**

In summary, the main functions and features of our product can achieve are below.

(1) Reduce the noise effectively even for low-frequency noise;
(2) Customers can adjust the reduced noise levels based on different requirements, which can be adjusted easily by mobile APP Eliminoise$^{TM}$;
(3) It has low energy consumption;
(4) The product can remember the stored values and noise environment so that it will automatically control and adjust accordingly;
(5) The product can monitor the sleeping quality, store data, and transfer to the smart phones by Bluetooth or WIFI conveniently, as a part of healthy data;
(6) This product can sense the actual temperature and adjust wisely;
(7) The product can also be used as speakers, to amplify the music and easily controlled by smart phones.
(8) The product can combine with the intelligent system at home, to control the electrical furniture and fire alarm, etc.

## 2.4 Opportunity Evaluation

| | |
|---|---|
| **SIZE** | This system targets buildings or houses which needs to reduce noise as much as possible. My noise reduction system aims to reduce 15 dB off from original noise source.<br><br>According to the research data [63], in USA with a population of 318.9 million, the number of people with heart disease due to traffic noise is<br>15 million×3.2% = 480,000.<br><br>We can roughly estimate the number of people with heart disease in the world (total population is 7 billion) due to traffic noise with 20% tolerance is<br>7 billion/318.9million×480000×20%=2.1 million.<br><br>We estimate 55% of the affected people in the whole market will buy our product, so the total marketing size is below:<br><br>Cost per unit: $150/unit (mainly coming from research fee, manufacture fee, etc.)<br><br>Selling price per unit: $300/unit<br><br>Profit: $(300-150)×2.1 million×0.55=$173.8 million |
| **SUSTAINABILITY** | Currently, there are not effective systems commercially to reduce the noise, including such as traffic and construction noise. Although the workers in the construction site try to build a barrier to shield noise, it's still not effective especially for low frequency noise.<br><br>My noise reduction system is a novel invention in this market. And its estimated price is $300 per unit, but it can be reduced even further with the technology development. In addition, we |
| | can also design some different kinds of models to satisfy different groups of people.<br><br>At present, there are no similar competitors, providing a good opportunity to develop this new product, which is very effective for reducing noise. At the same time, we can also get a lot of feedback and improve our product before other competitors appear. |
| **SCALABILITY** | The noise reduction system desired in this report is very useful for various applications in the life, however it has not been applied widely. Besides, not only in Singapore, every country in the world will need this to achieve good environment for business and education.<br><br>With the development of technology, this system will continue being improved and the cost will be also reduced. We can satisfy the different group of people. Nowadays with the fast-economic development all over the world, people requires more enjoyable environment to focus on the jobs. The noise reduction system will help improve working efficiency greatly and it can be expanded to other countries easily. |

## 2.5 Mission Statement

| | |
|---|---|
| **Product Description** | A noise reduction system to reduce the noise especially caused by traffic noise or construction noise. To improve sleep quality, working efficiency for people by reducing noise to provide an enjoyable rest, working and study environment. |
| **Key Business Goals** | • Development budget: S$1000,000<br>• Development timeframe:<br>• Design period: 01-Feb-2017 to 20-May-2017<br>• Testing and calibration: 01-Jun-2017 to 30-Aug-2017<br>• Launch to market: 1-Jan-2018<br>• Market Share: 55% of population by noise polluted<br>• Selling Price: $300 USD<br>• Gross Margins: 69.2%<br>• Cost: $92.5 USD<br>• Intellectual properties: apply a patent to protect this invention and register a trademark |
| **Primary Market** | • Name of market: residents in the city, especially near MRT, bus stops, etc.; office workers, schools, construction site;<br>• Market size (rough estimate): 2.1 Billion×0.55=1.2 Billion<br>• Business model:<br>You need to provide figures. |
| **Secondary Markets** | • Museums, libraries, hospitals; |
| **Assumptions & Constraints** | • Platforms<br>• Technologies<br>• Certification<br>• Quality |
| **Stakeholders** | • Suppliers of PCB boards, speakers, LCD screen and so on.<br>• Manufacturers of noise reduction device<br>• Shareholders<br>• Service operators<br>• Distributors and resellers<br>• Regulatory agencies (NEA)<br>• Certification bodies<br>• Customers<br>• Buildings developers |

# 3. Customer Needs

## 3.1 Customer Needs Analysis

### 3.1.1 Identification of Stake Holder

The following Stake Holders are identified for our product.

a) Suppliers

Suppliers provide the hardware or software to the project. They are the resource to our projects. They play a great important part in produce quality. As they have been in industry for a long time, we can get advice from them for the details on the parts selection or app design to make a high-quality product.

b) Manufacturer

Manufacturer is another important part for the project. They will help to make the special parts that are unique to the project. They also help us to assemble different parts together to make it a final product. They will be the key to ensure the durability of the product.

c) Shareholders

Shareholders are important stake holder as they invest our company. Their contribution is the resource that we can start running our business. The successfulness of the product also matters the revenue that the shareholders can get. That is why shareholders are key stake holders.

d) Service operator

Service operators can be divided into two parts, one is installation and maintenance service provider and the other is the server service provider.

Installation and maintenance service provider will be the main part that controls the subsequent sales service to the customer. They will also be the main part that gives the impression of our product to the customers. In other words, this is also a part of our brand or our product. As our product has some functions that need a server to store the data from the device and have the remote notification function, a good and stable server will also be a key to make the product a success. Server service operators will help to maintain and manage the server we will be used for our product.

e) Distributors and resellers

Distributers and resellers are important to us. They are the channels that our product can be sold to customer. The way they do business will affect our sales volume. Besides sell products for our company, they can also provide some marketing information like usually what the functions customers like and what kind of products already exist and their sales situation. They are one of the information channels for our product development.

f) Government agency

As our product, can gather the data of the noise happened around users' neighborhood, the data will be very useful for the government. They can monitor the noise level and take the necessary action. They may directly consider to purchase and install the device to some residential or business building to reduce the noise. When a user complains about the noise situation, the data will be a clear and direct proof to them. That is why we think the government agency is another stake holder to our project.

g) Certification bodies

As we are building an electronic device, it will be necessary for us to get our product to get certified before we can launch it. Depending on the country we are going to sell and launch the products, the certification bodies may also be deferent like FCC in US and CCC in China. These certification bodies include quality and EMC certifications etc. Without this certification, we cannot even export the products to the country that we want to sell.

h) Customers

Our customer is the main stake holder of our product. They are our target market that will purchase our product. Our product function and features will be decided based on the survey resulting from our target customers. They will give us feedback on our product so that we can make improvements on our products. They will also be the reason we need to have our service provider to carry out the subsequent sales services. They are also the main reasons that we can service and exist in the market.

i) Building developer

If the products have a good result in reducing the noise and can integrate into the smart home system, building developers will be a direct purchaser of our products and integrate the item into their buildings. This will also help them to have a new selling point.

j) Health care professionals

Since we have sleep monitor function and can record the noise data for the user, when the user wants to consult a health care professional, the data will give a great reference for the health care professional to give more precise advice to the user for greater health.

### 3.1.2 Gather data from Customers

We have invited 300 customers to take part in our survey and we have face to face interviewed 10 investors and 10 enterprise customers to get the ideas of our project. Full survey form and interview questions can be found in the appendix.

From the following charts, we can see the responses we got from customer.

## Situation investigation

1 Do you think noise pollution related problems are very serious?

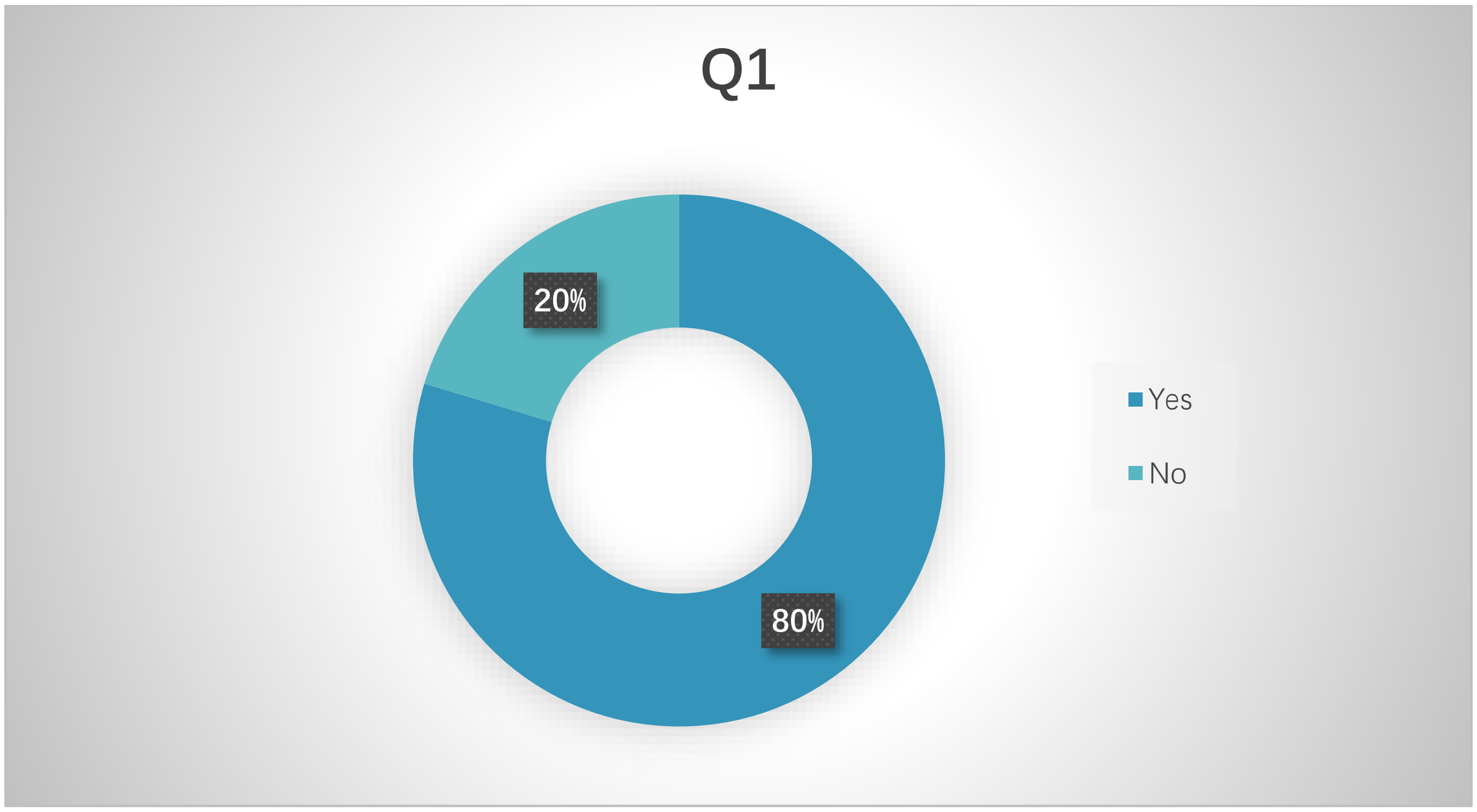

2 Do you have any trouble caused by noise?

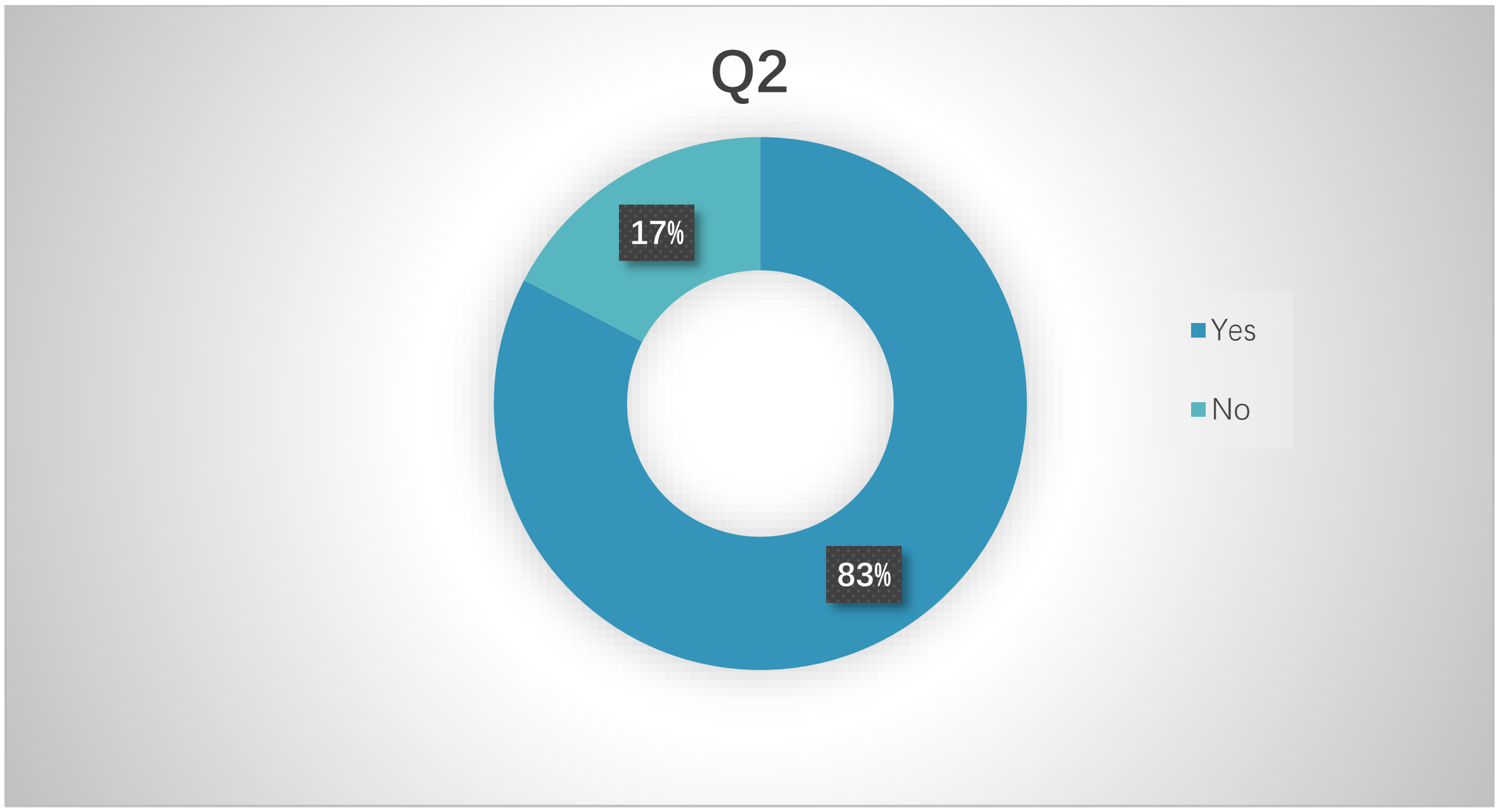

3 Normally, when will the noise appear? (Multiple choice)

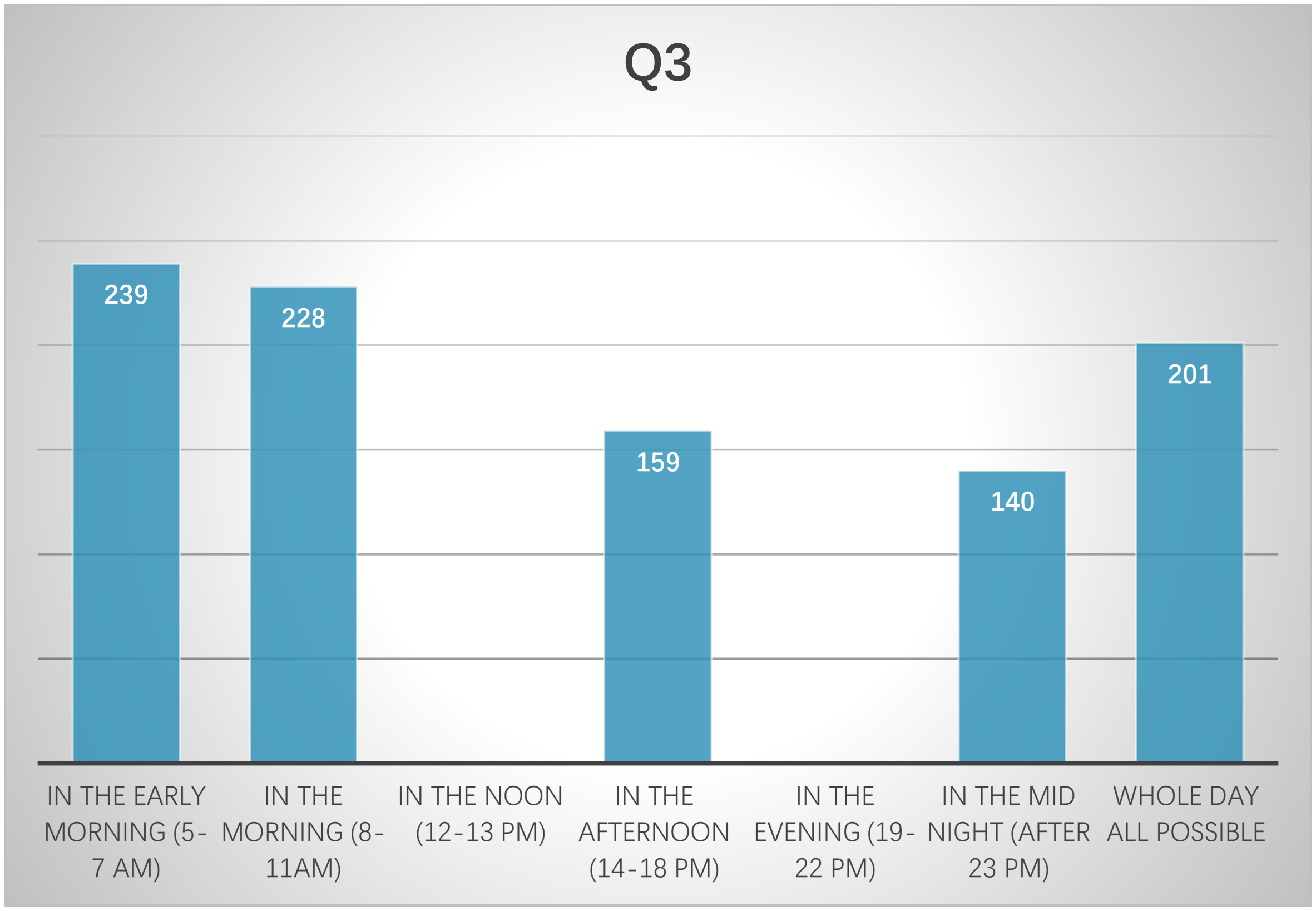

4 Usually, how often will you be affected by the noise appeared?

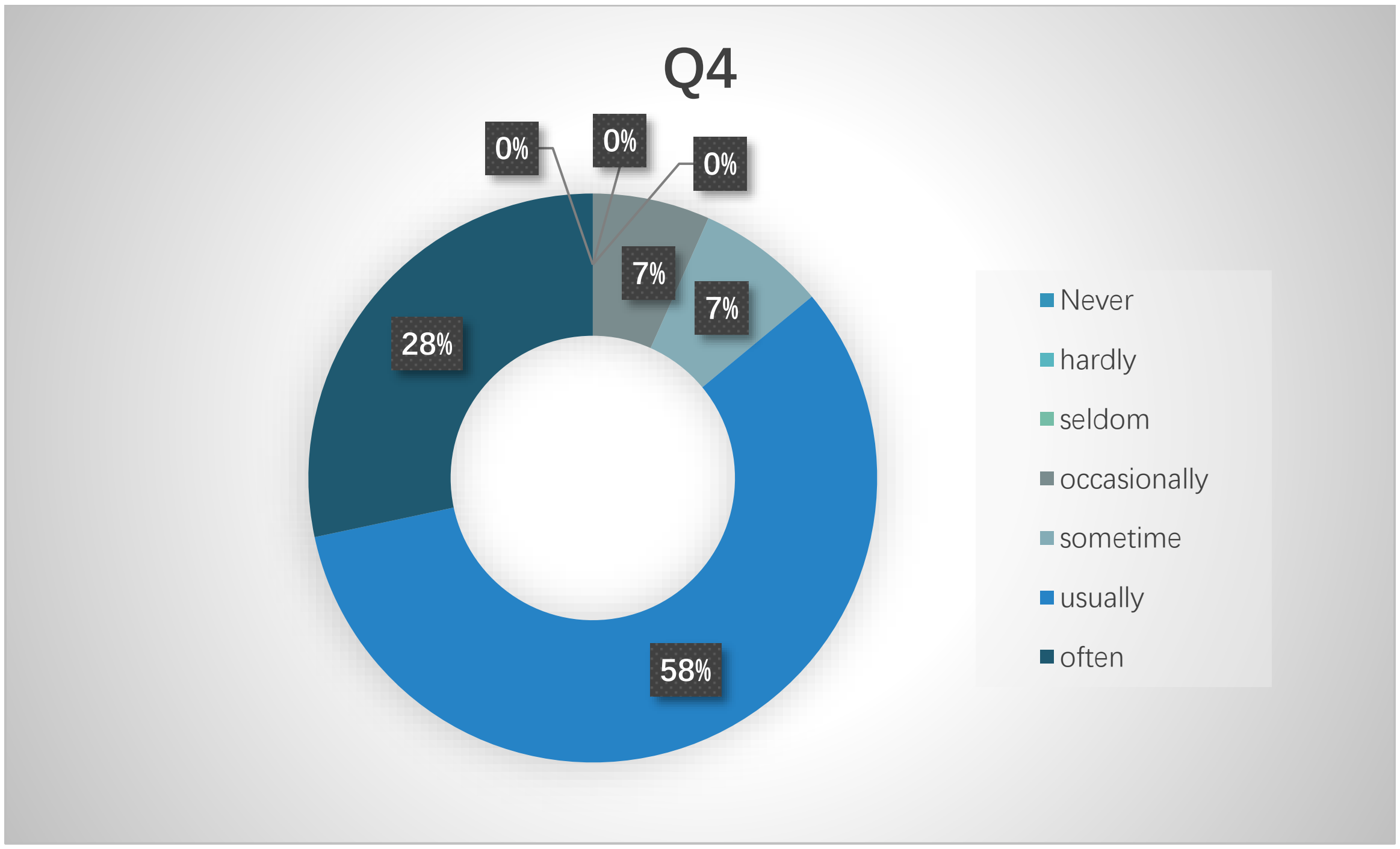

5 Normally, where does the noise happen? (Multiple choice)

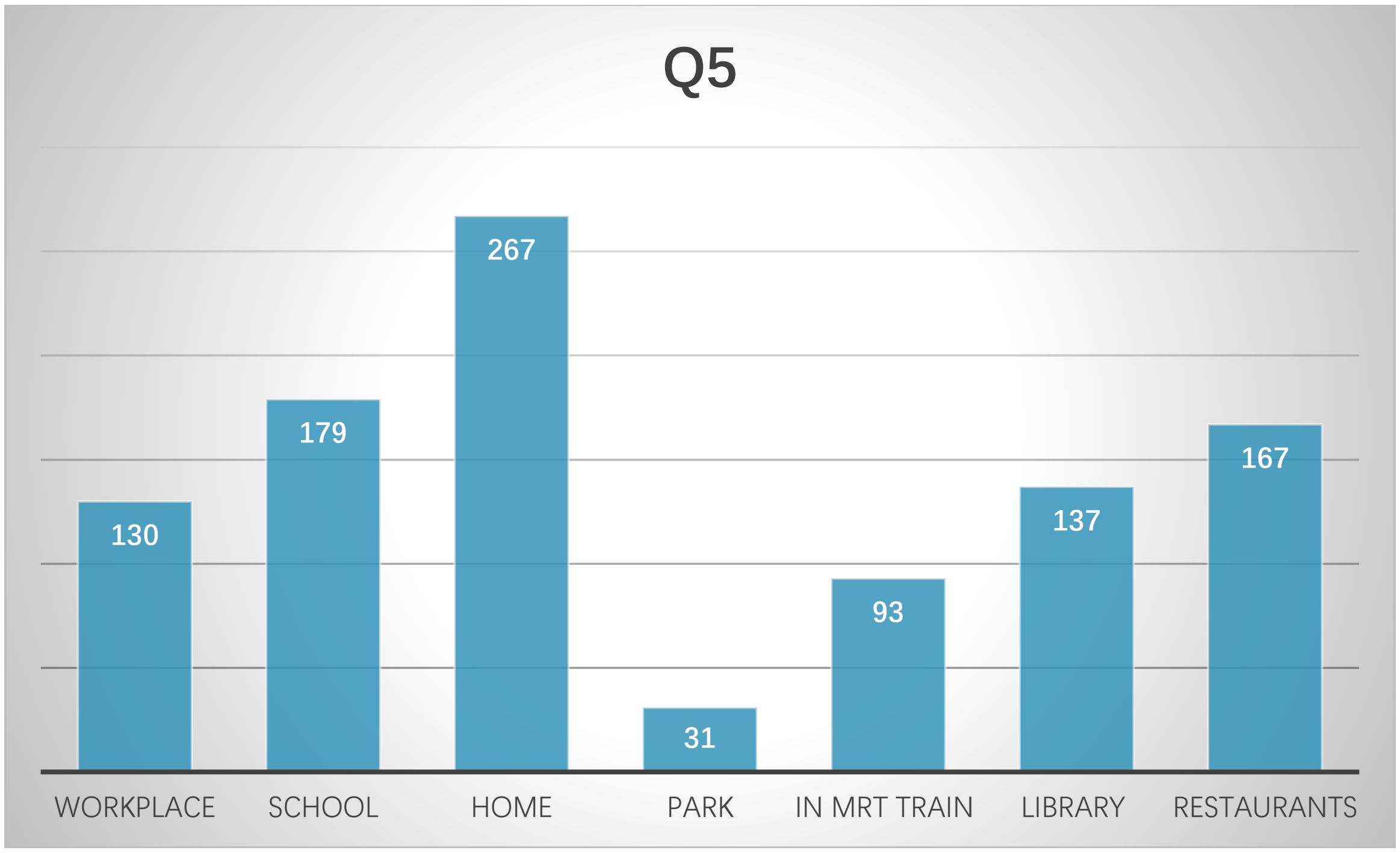

6 Usually, what is the source of the noise?

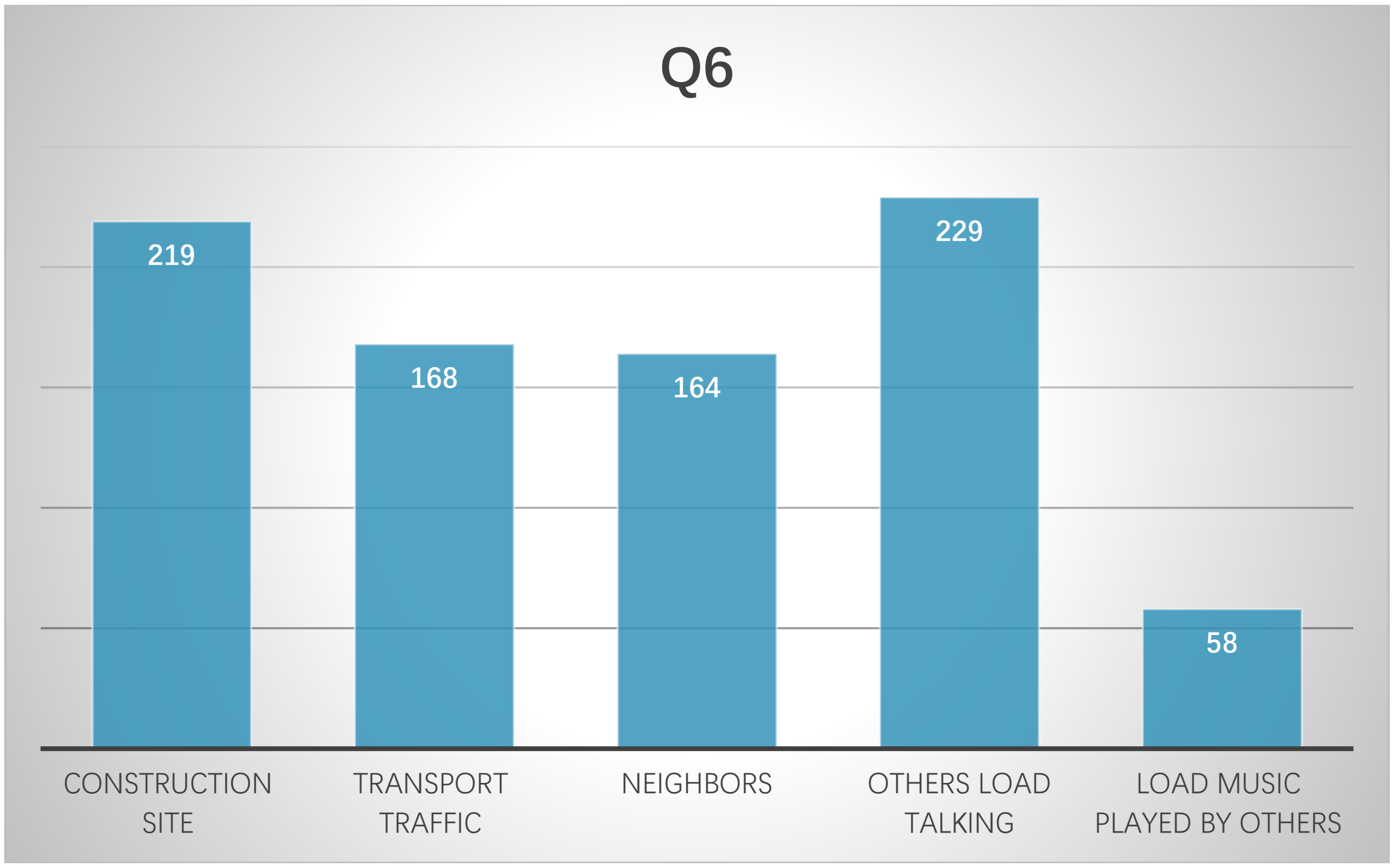

7 Is the source you identified from Q6 a permanent or temporary source?

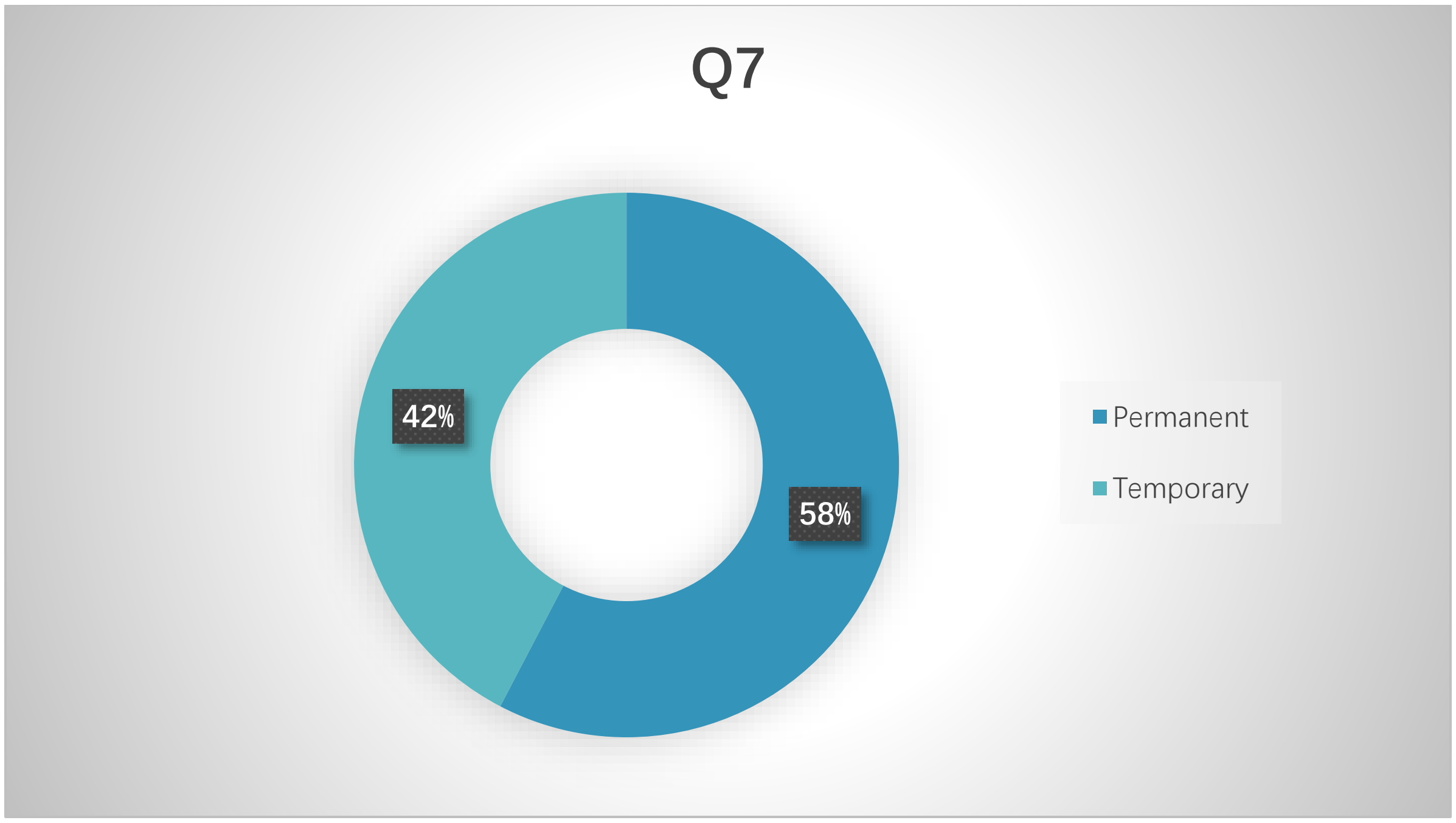

8 Are here any effects caused by the noise that affect your life?

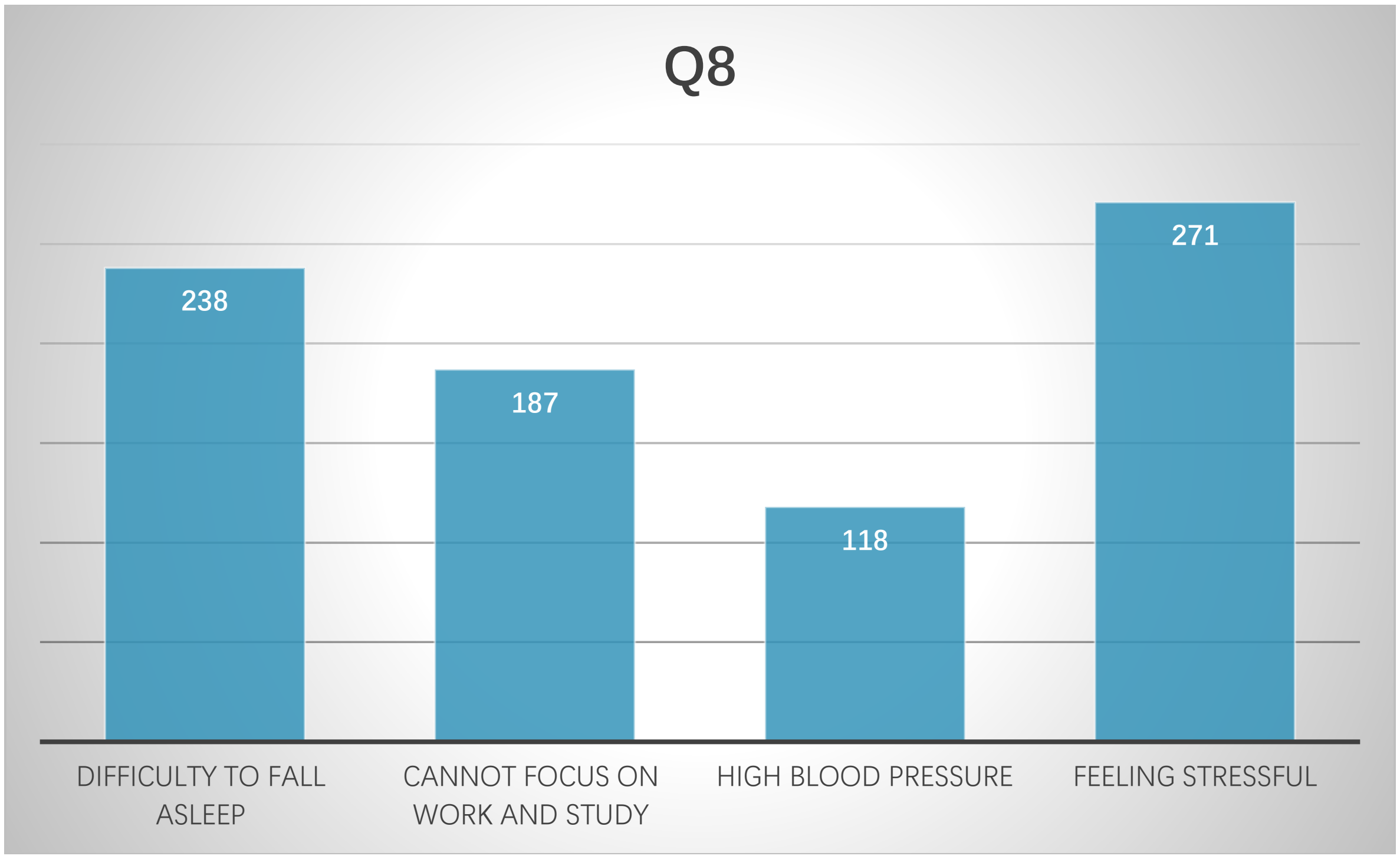

9 Currently, what is your solution to deal with the noise?

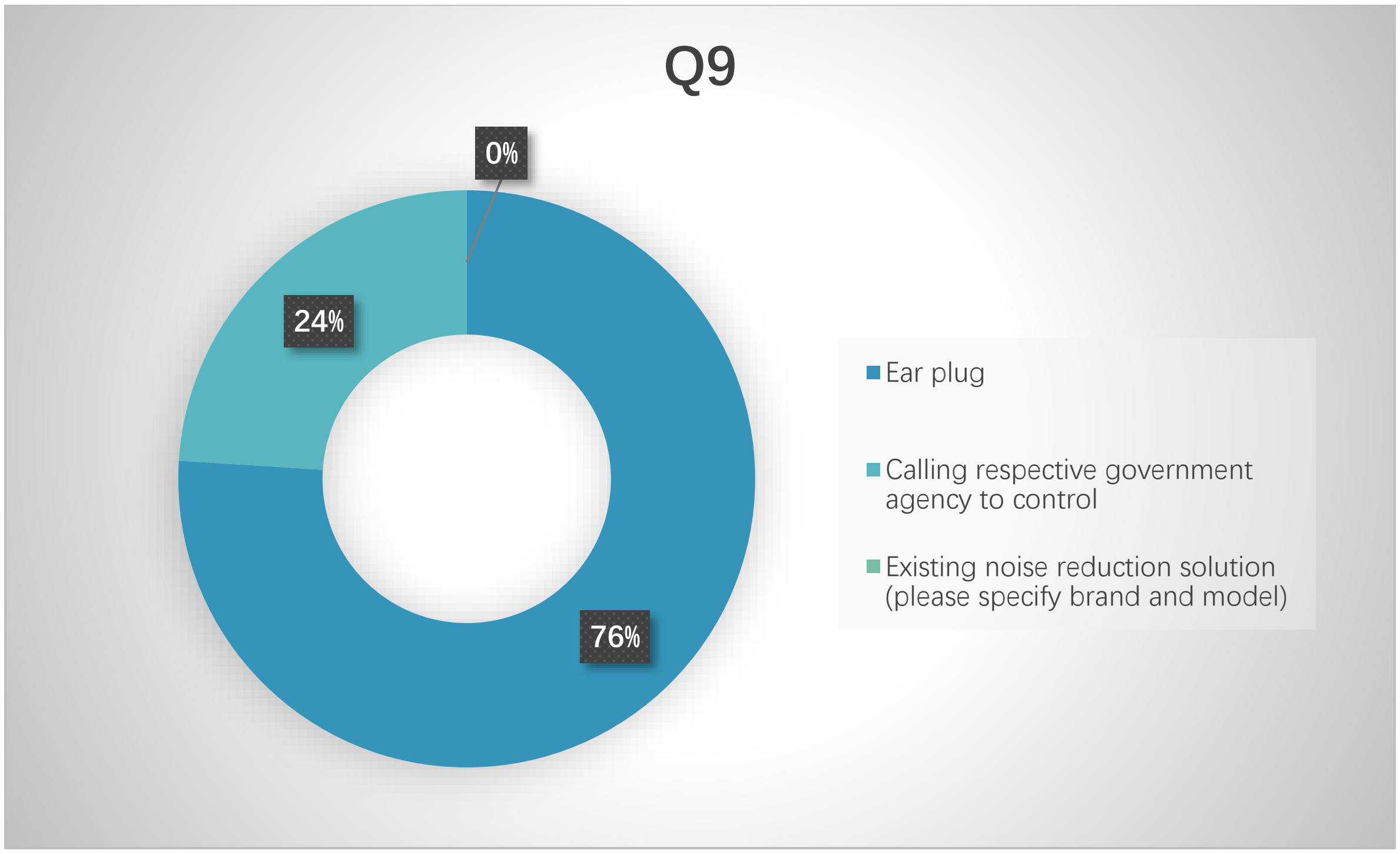

10 Do you know the affects that can be caused by long-term noise?

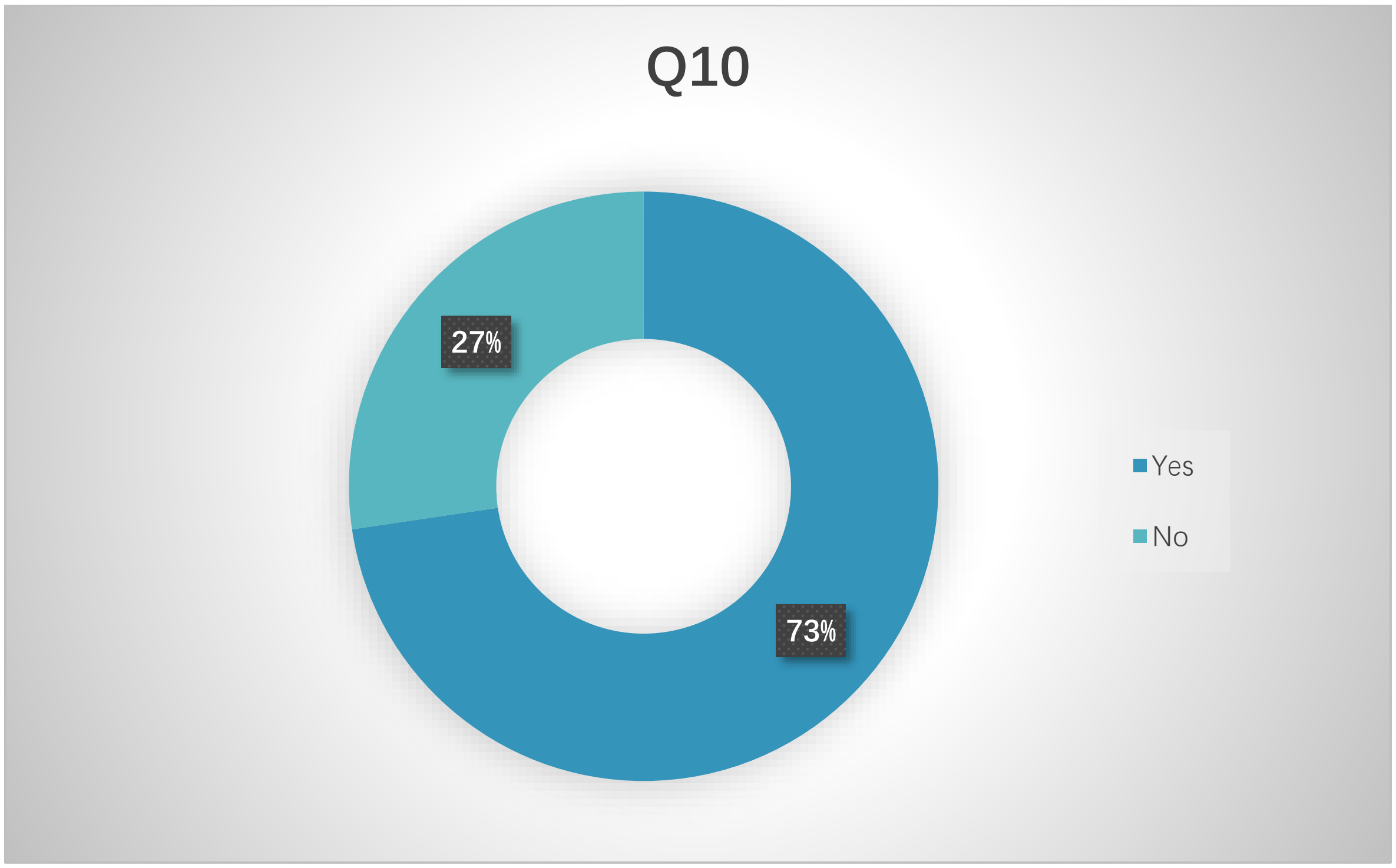

11 Does the current solution applied by the government meet your requirement to reduce the noise caused by construction sites, MRT or airport?

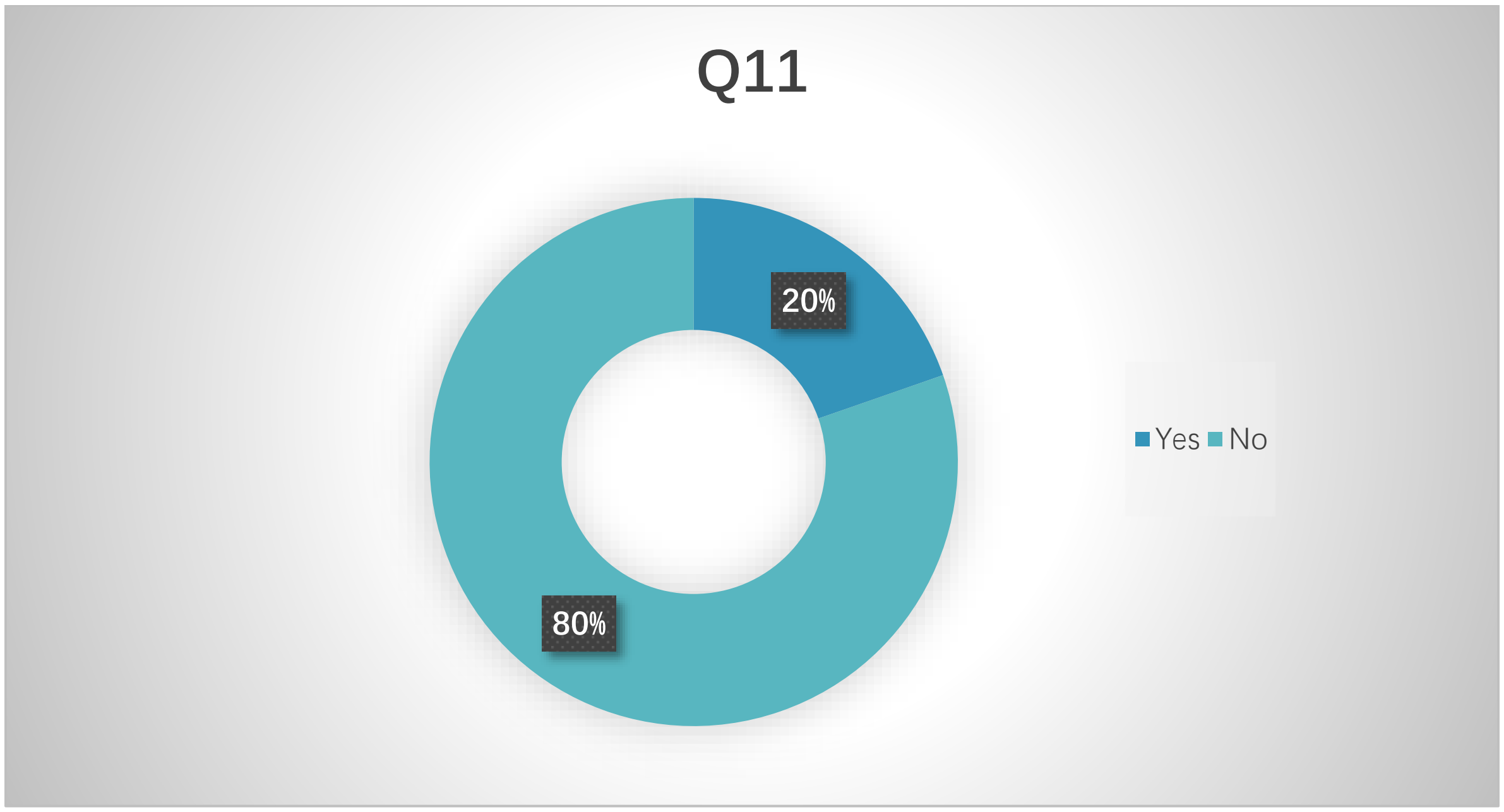

12 What kind of the following sound do you think is noise?

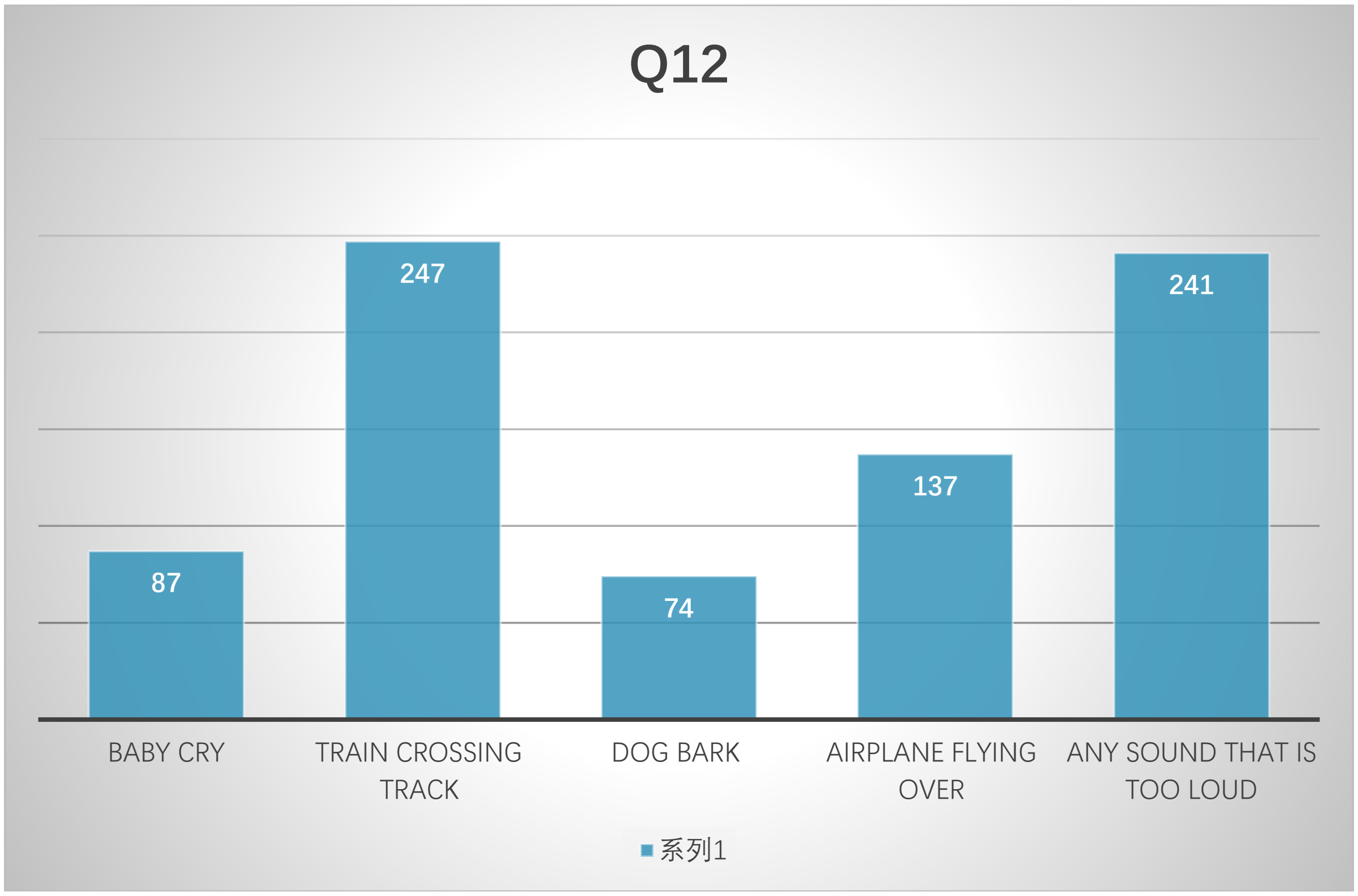

## Product Functionality

1. If there is a product that can help you to reduce noise, what are the functions that you think is good to have on this product?

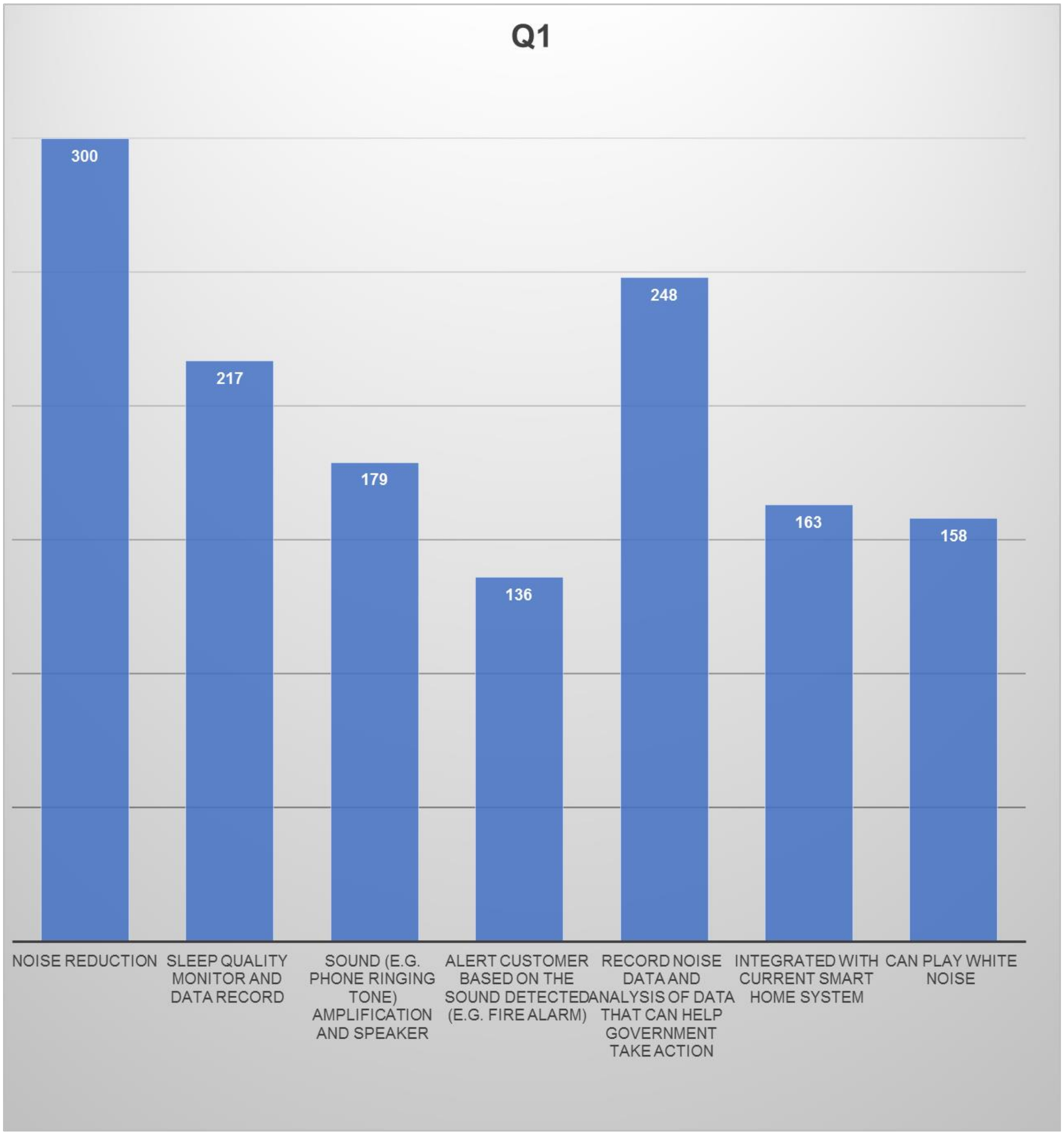

2. What are the features that you think this device should have?

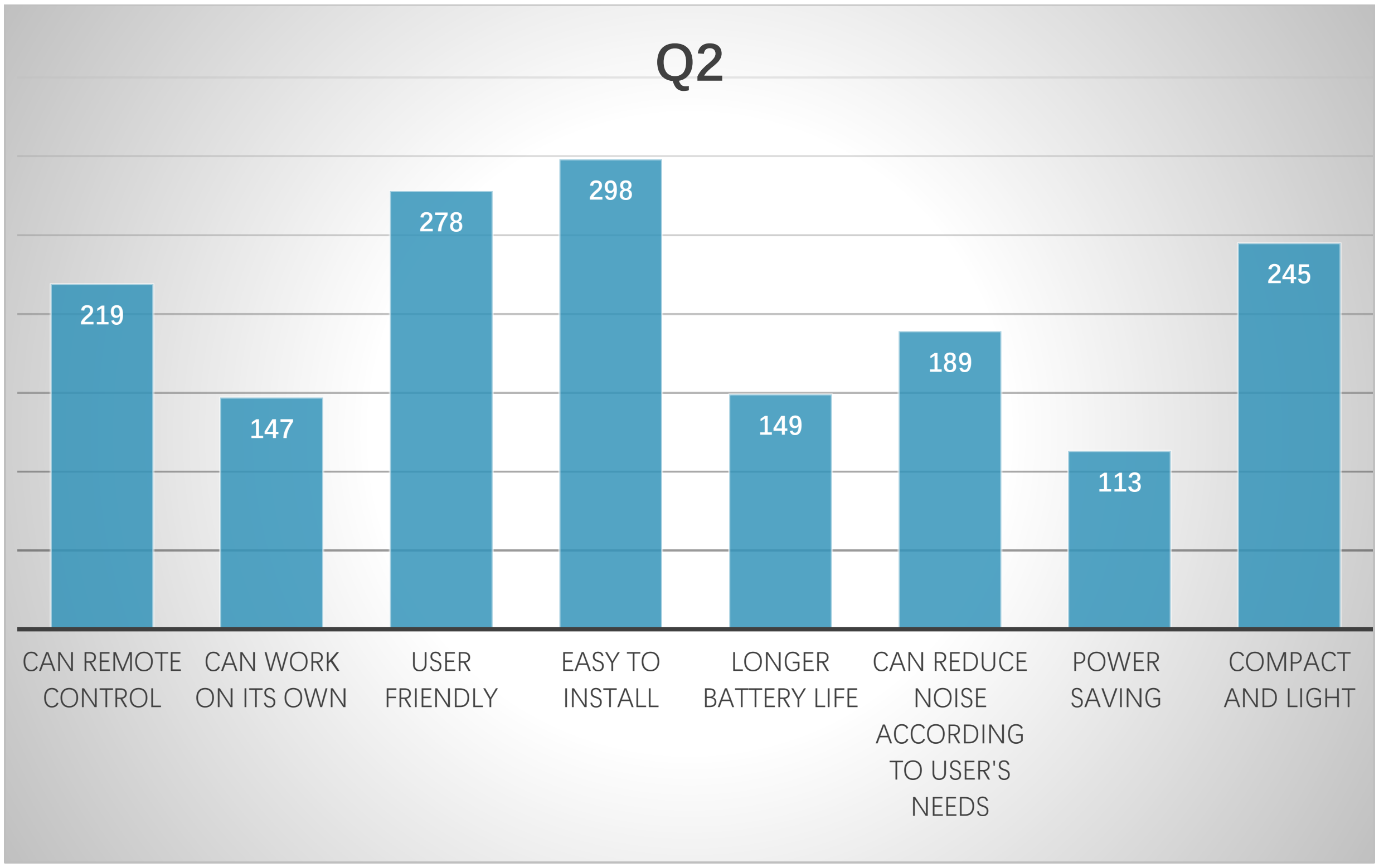

3. If it is possible for government to monitor and control noise source based on the noise reduction product, do you think it is necessary and useful?

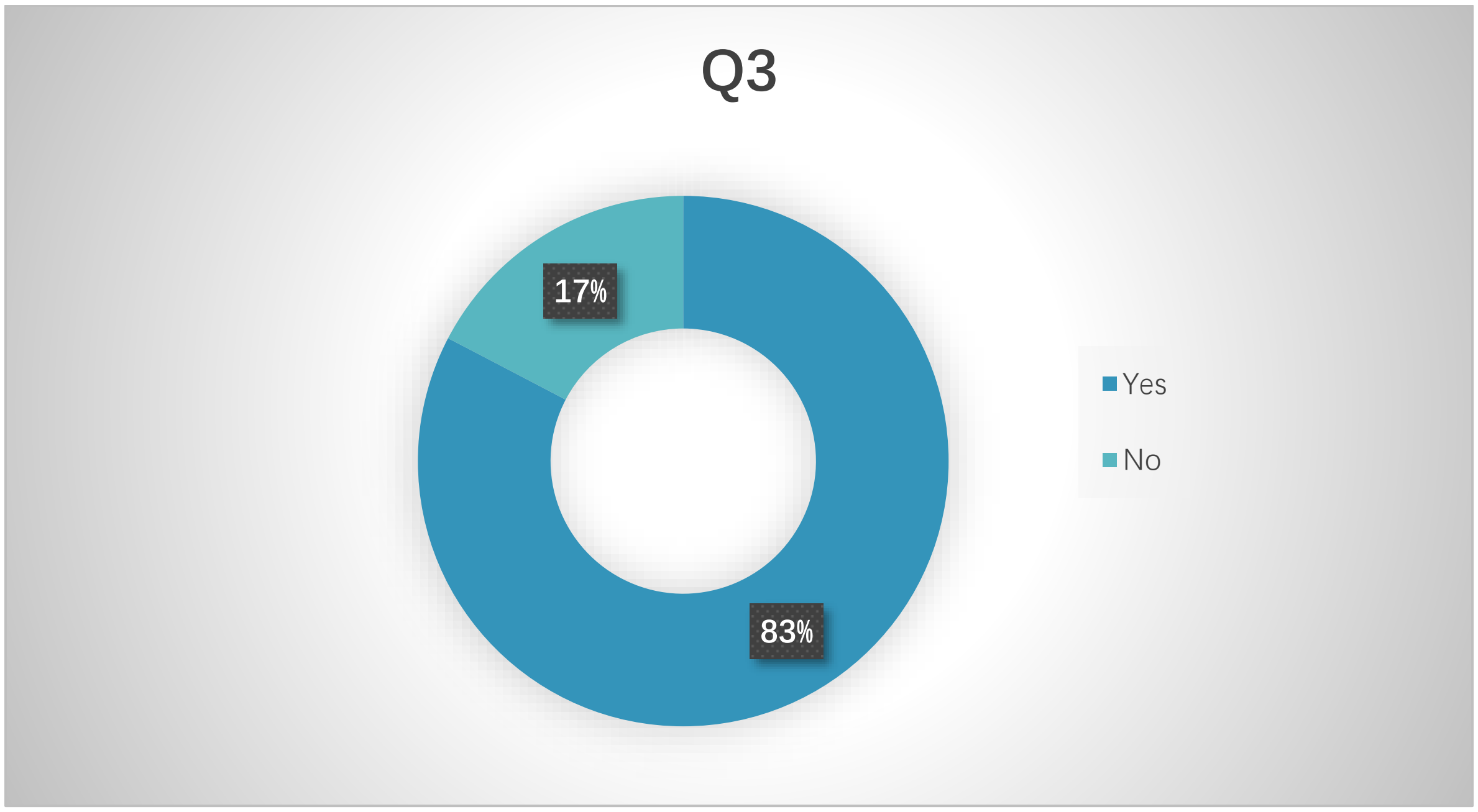

4. If this product can also help you to monitor your sleep quality, do you think the function is necessary?

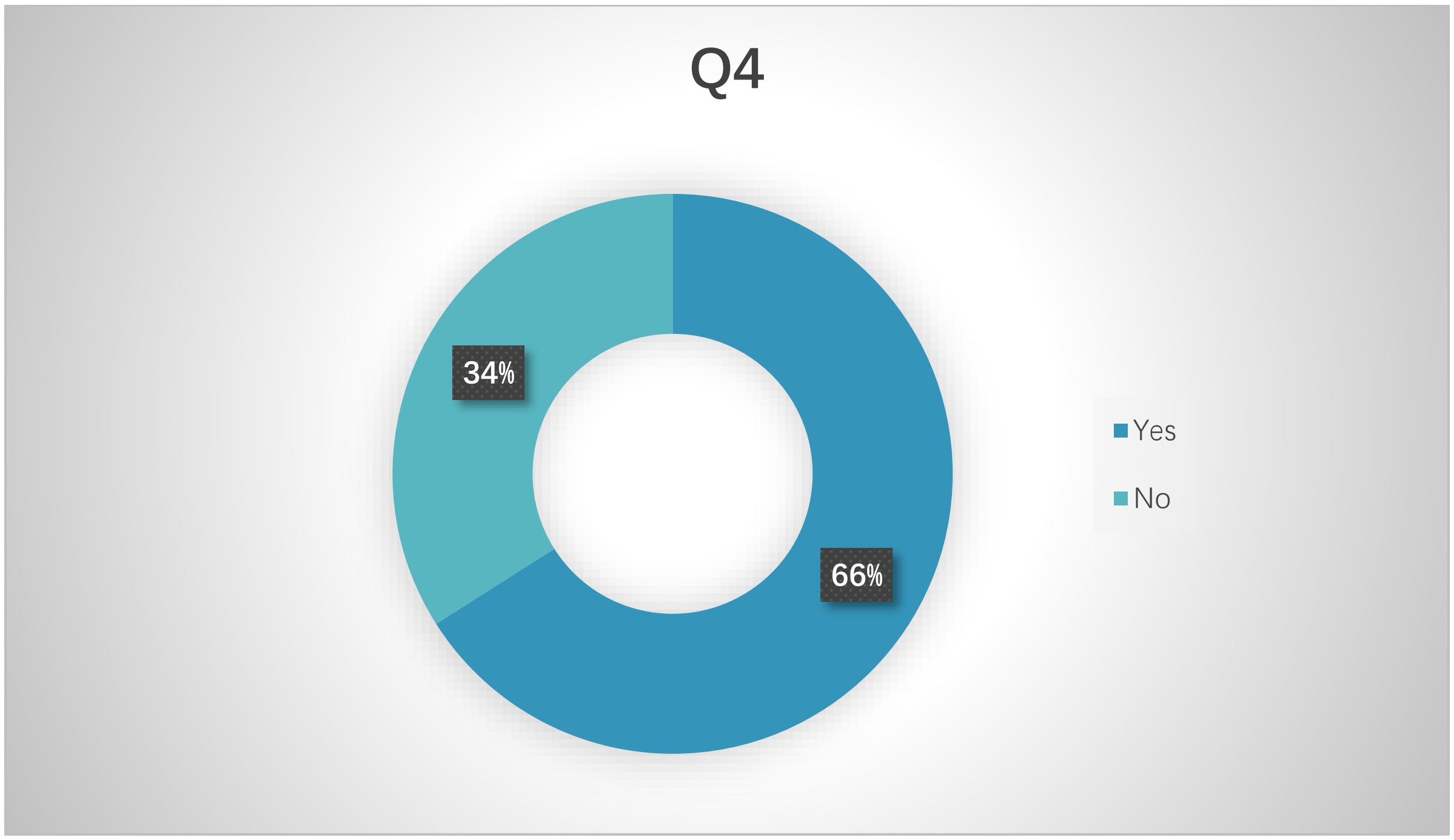

5. If the device will be using green energy as part of the power supply, which do you think is more suitable?

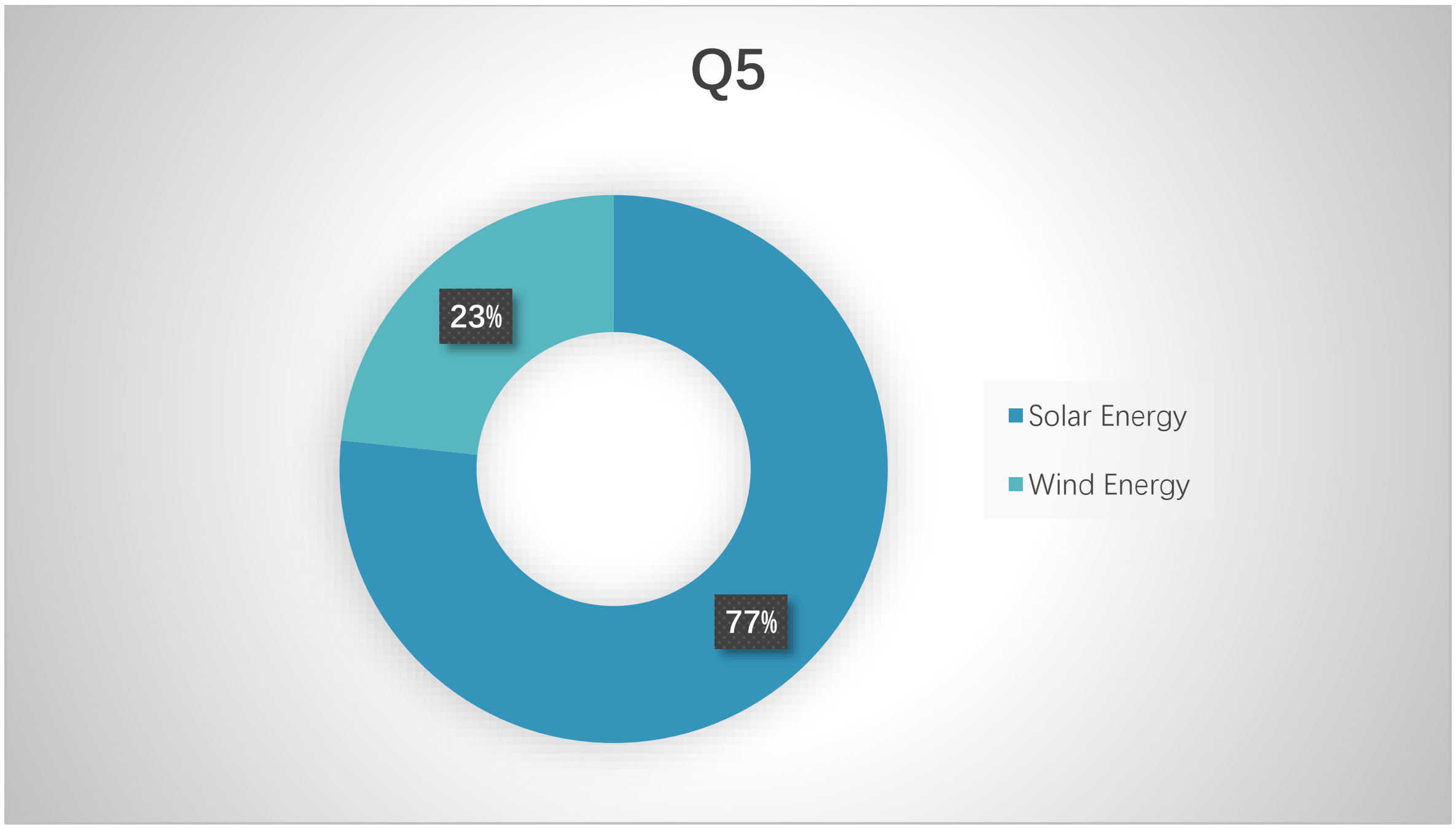

6. Do you prefer the product has the wireless remote control function? Which do you prefer?

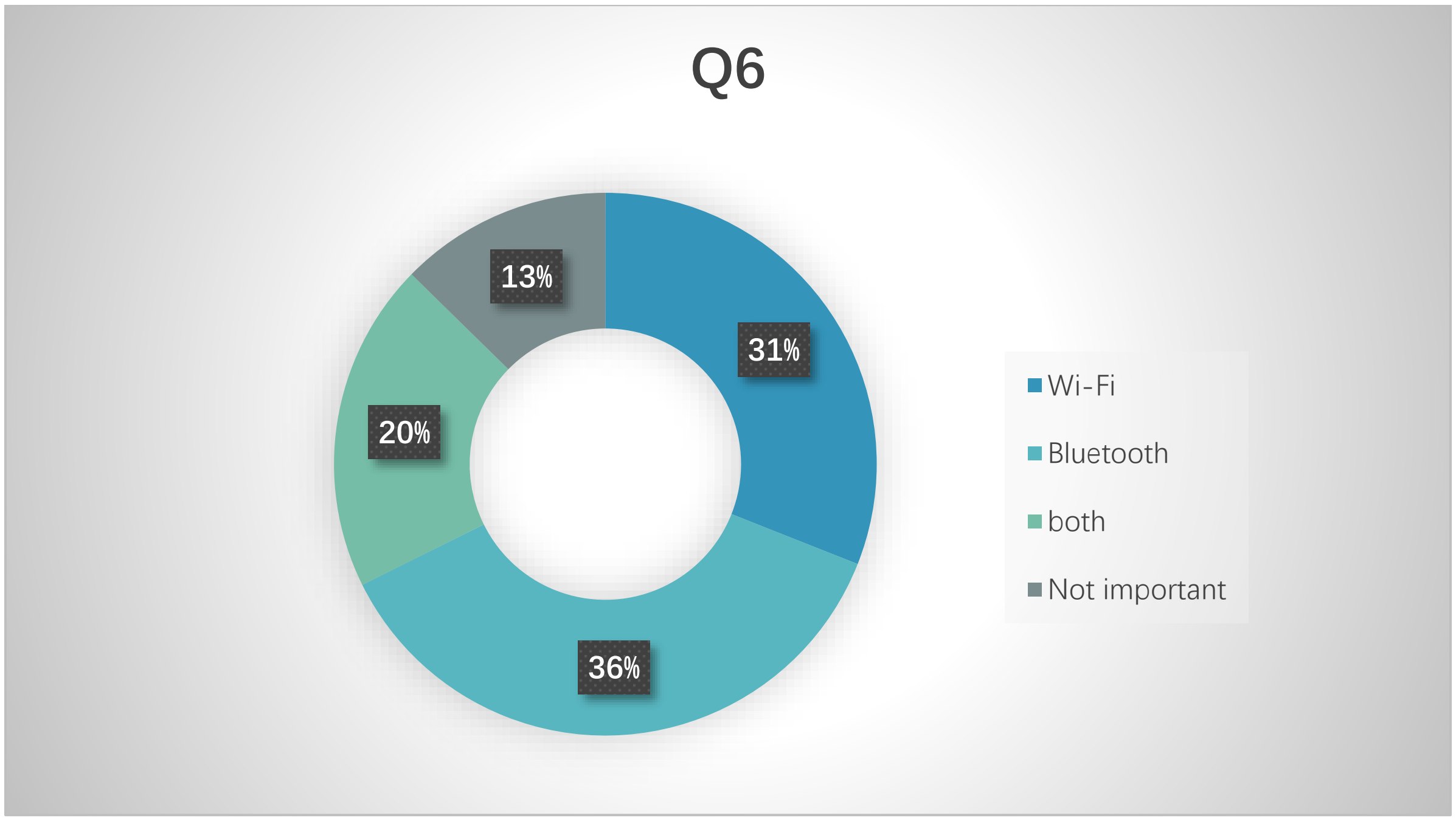

7. If can do remote control, do you want it be controlled on your mobile device like your smart phone using an app?

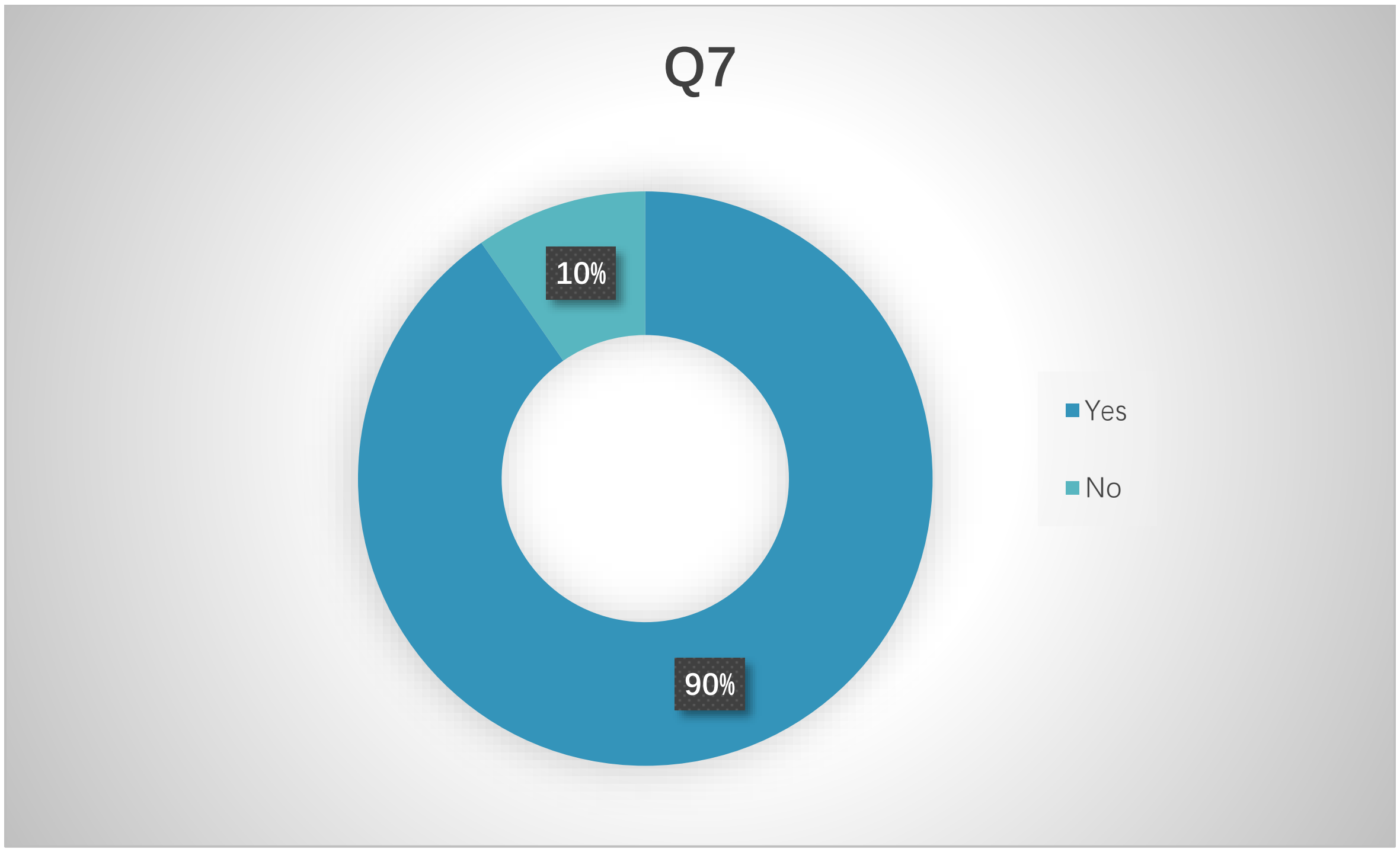

## Other information

1. If the product with these functions is priced at S$ 300, will you purchase it? If No, please specify the percentage the price should be reduced.

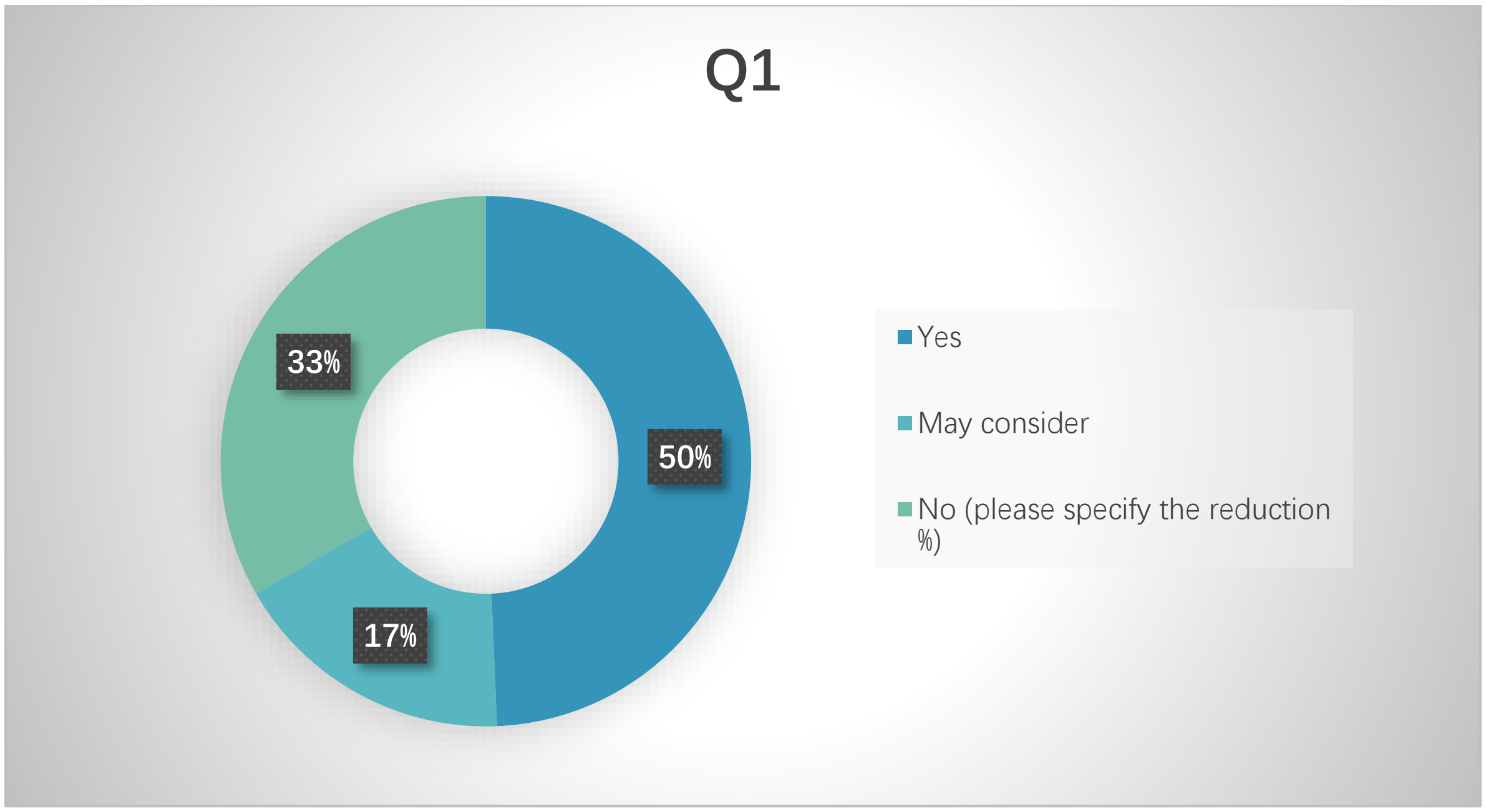

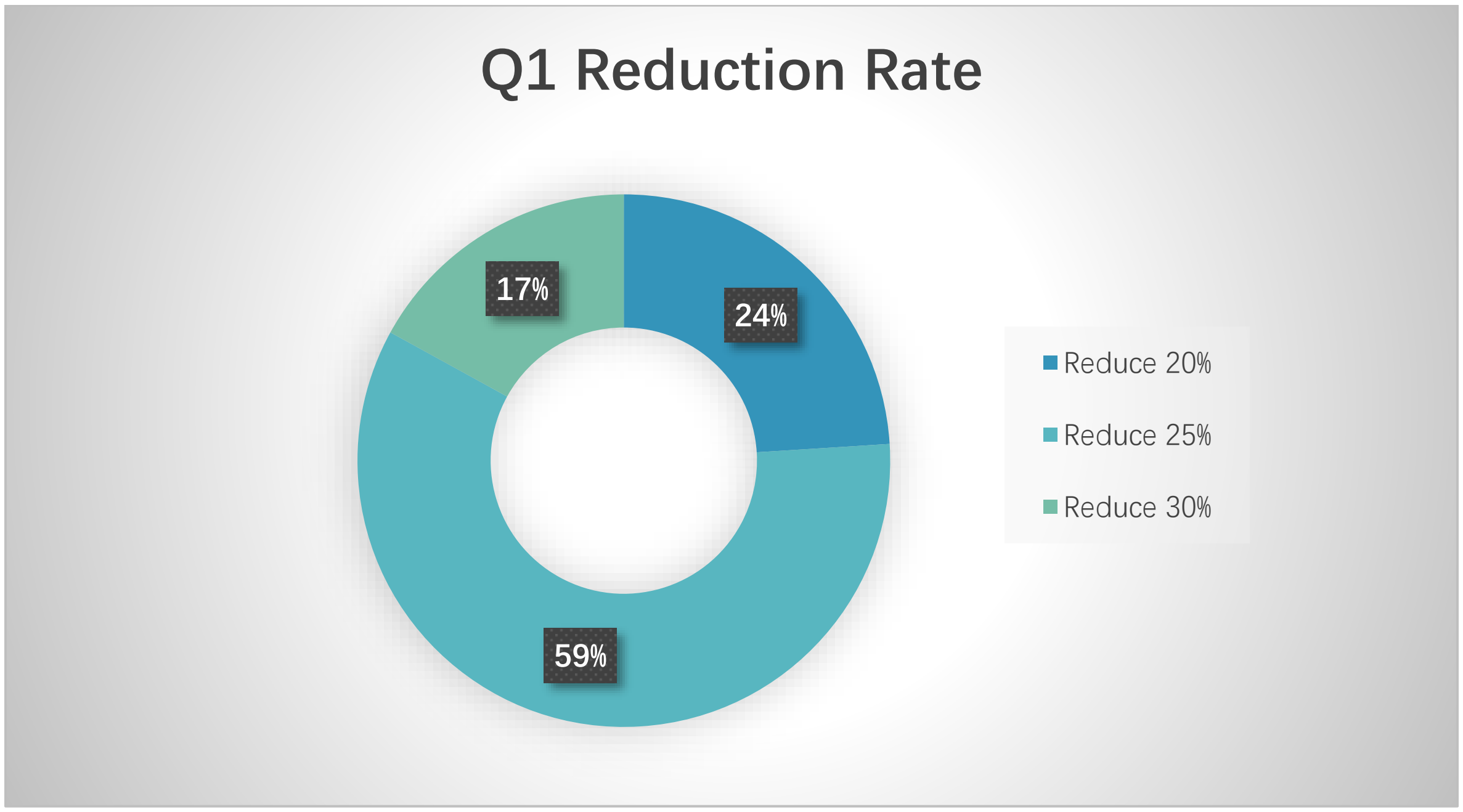

2. Will you make the noise into consideration when you making the following decision?

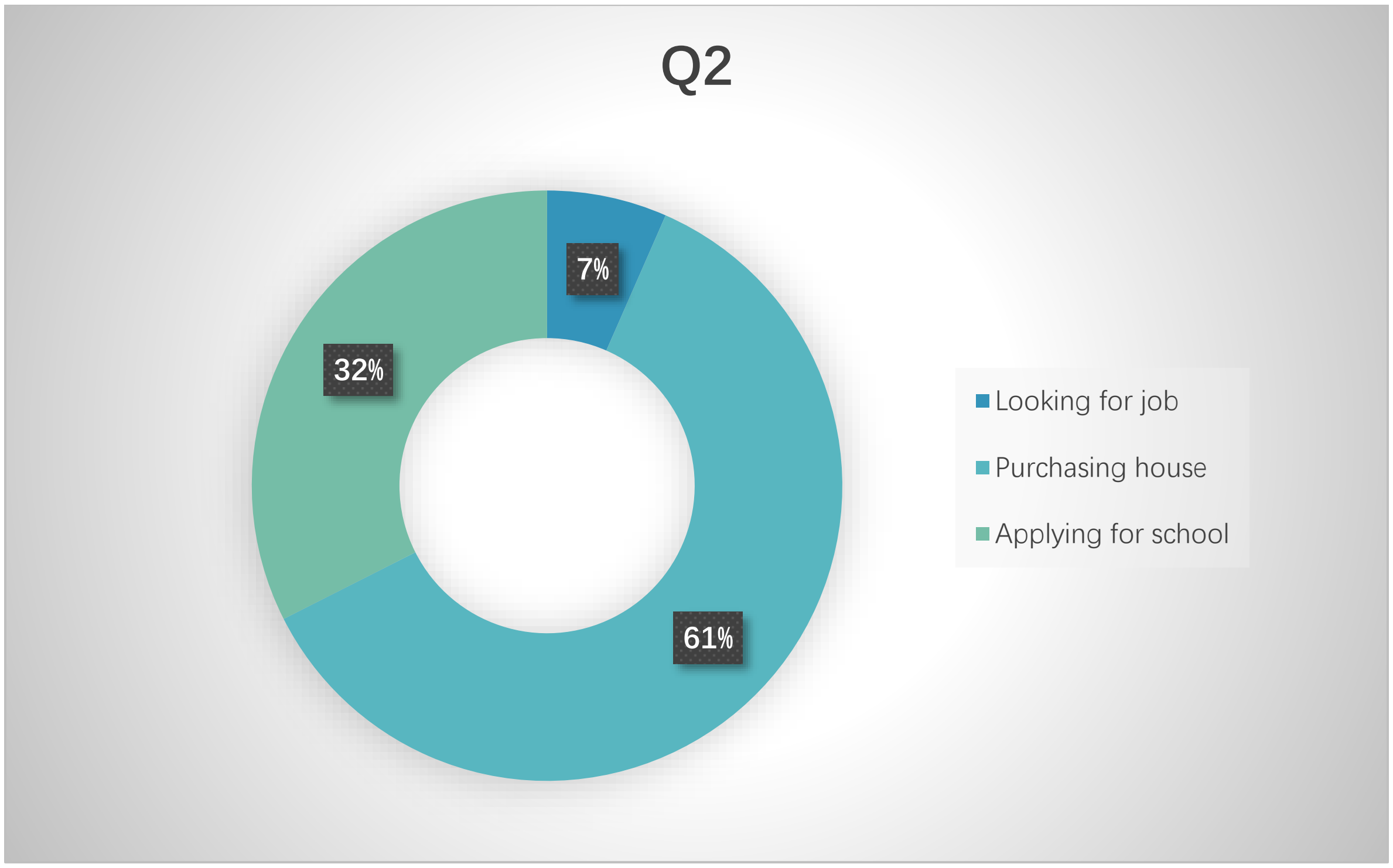

3. Can we have your age range?

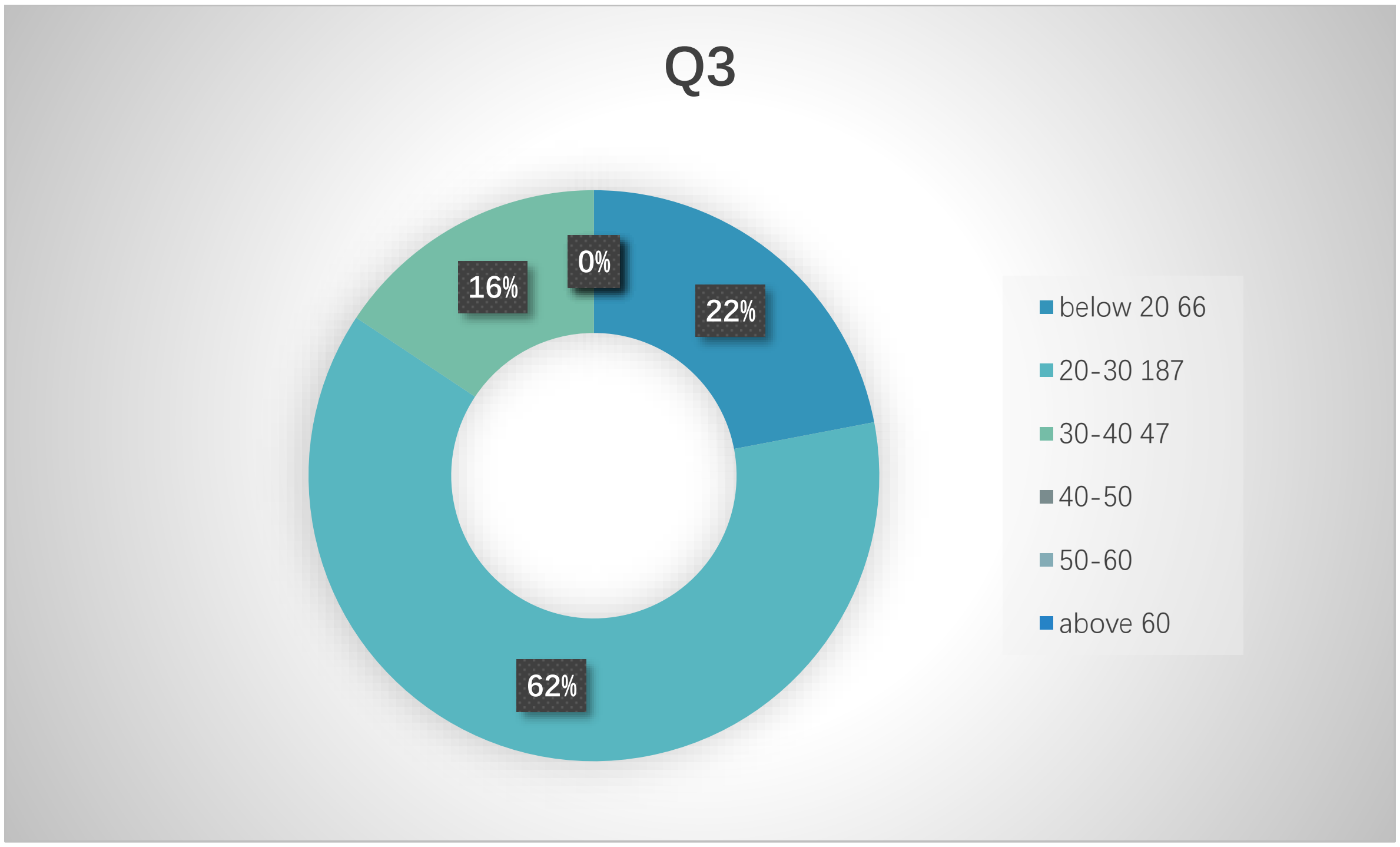

4. Gender?

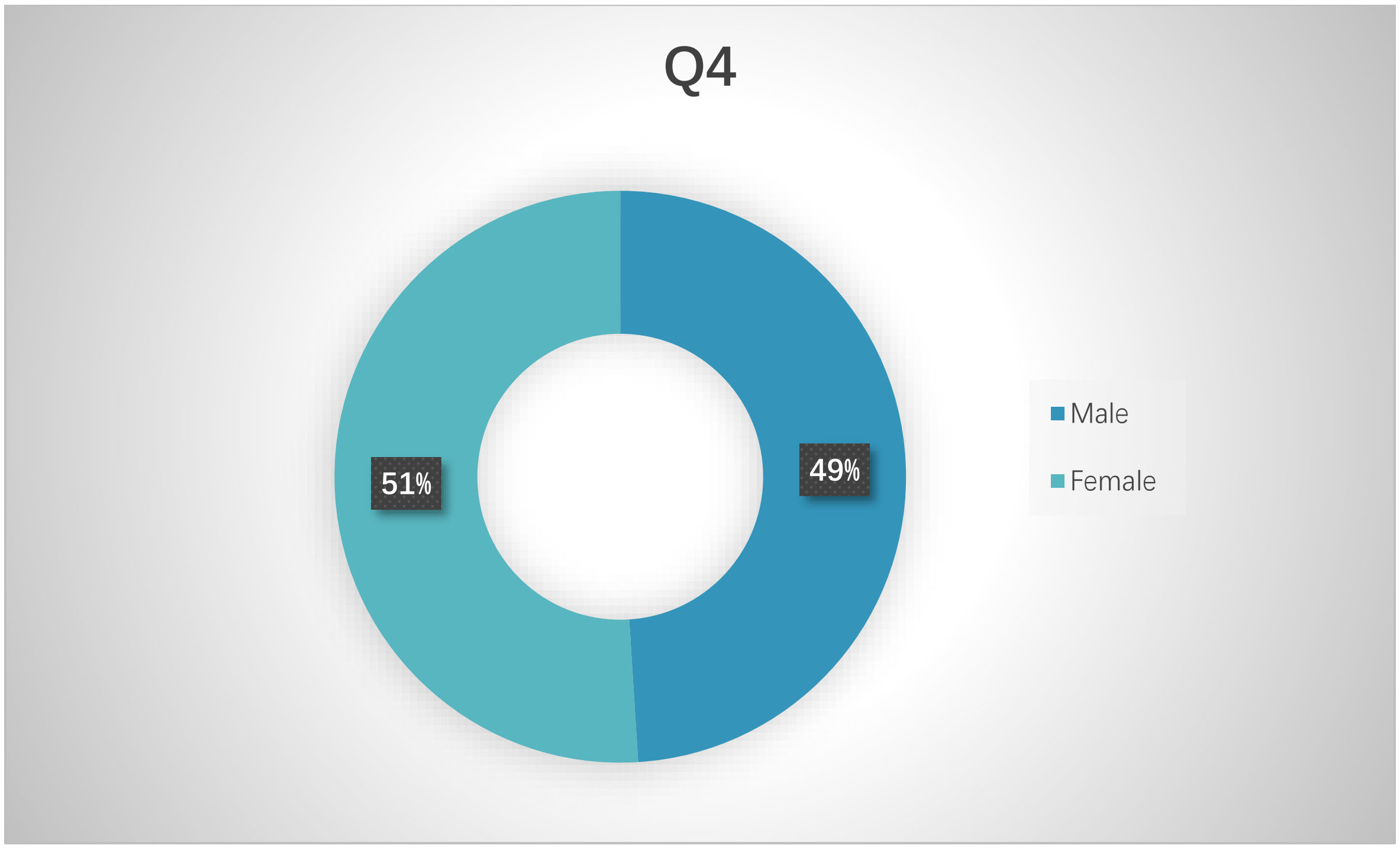

5. Salary range?

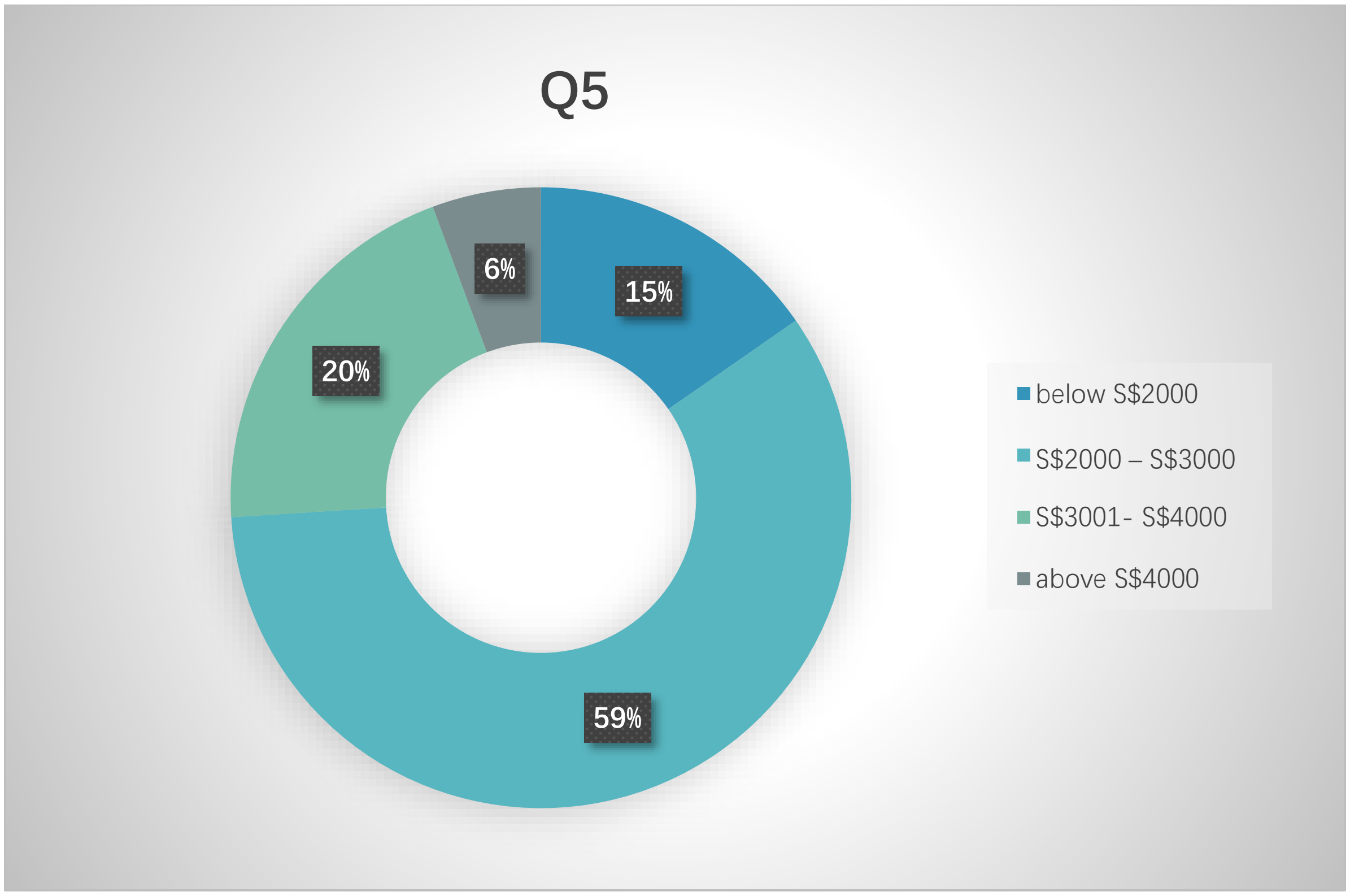

## Interview questionnaire (sample size 10)

## For investors

1. Are you willing to invest in a product that can do the noise reduction for domestic and business usage?

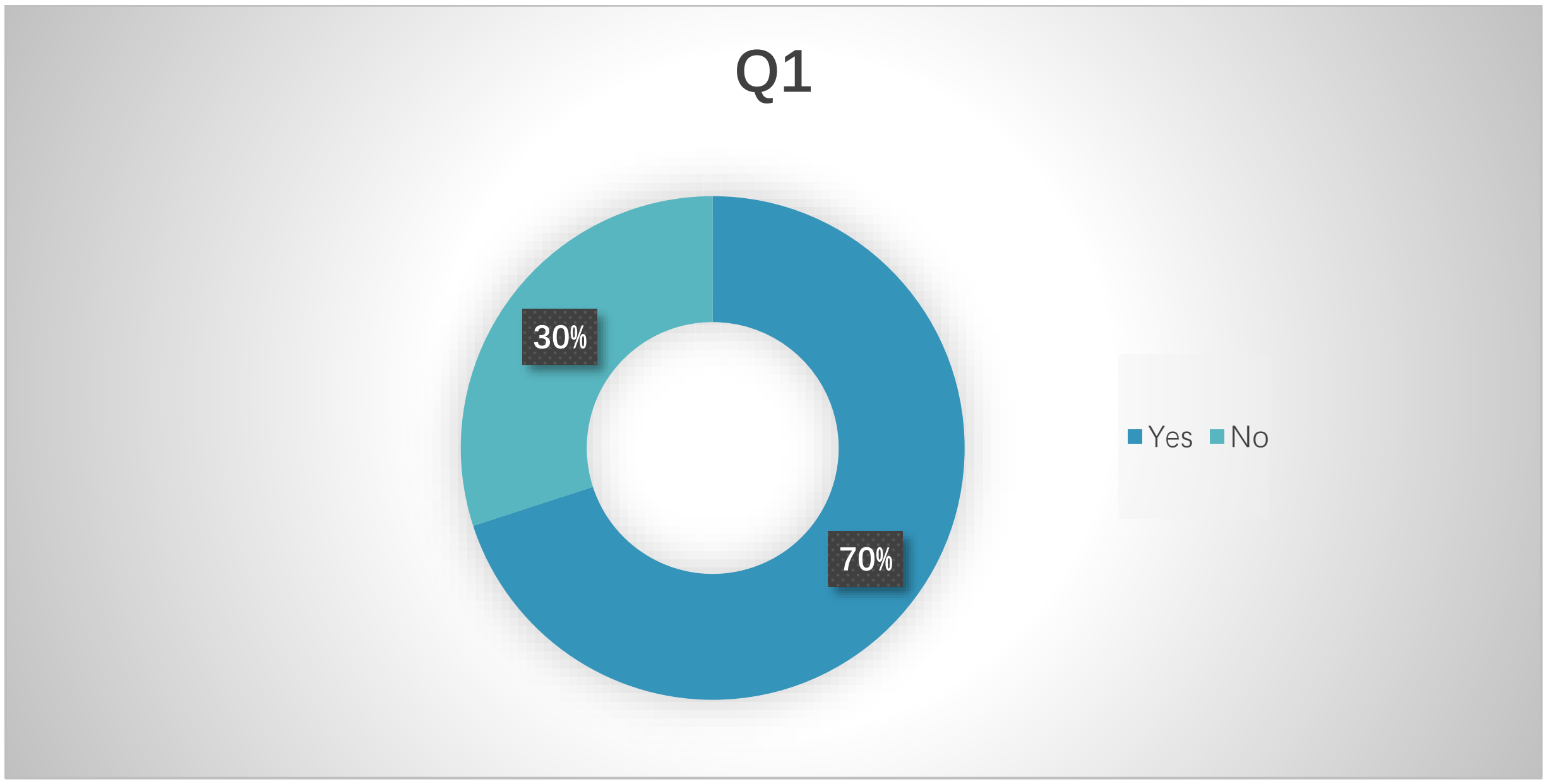

2. If the predicted market share is 1 billion, return rate for the project is 10%. We need S$ 500,000 to start up the business. Are you willing to be an investor? If no, please specify the suitable amount that you are willing to invest in.

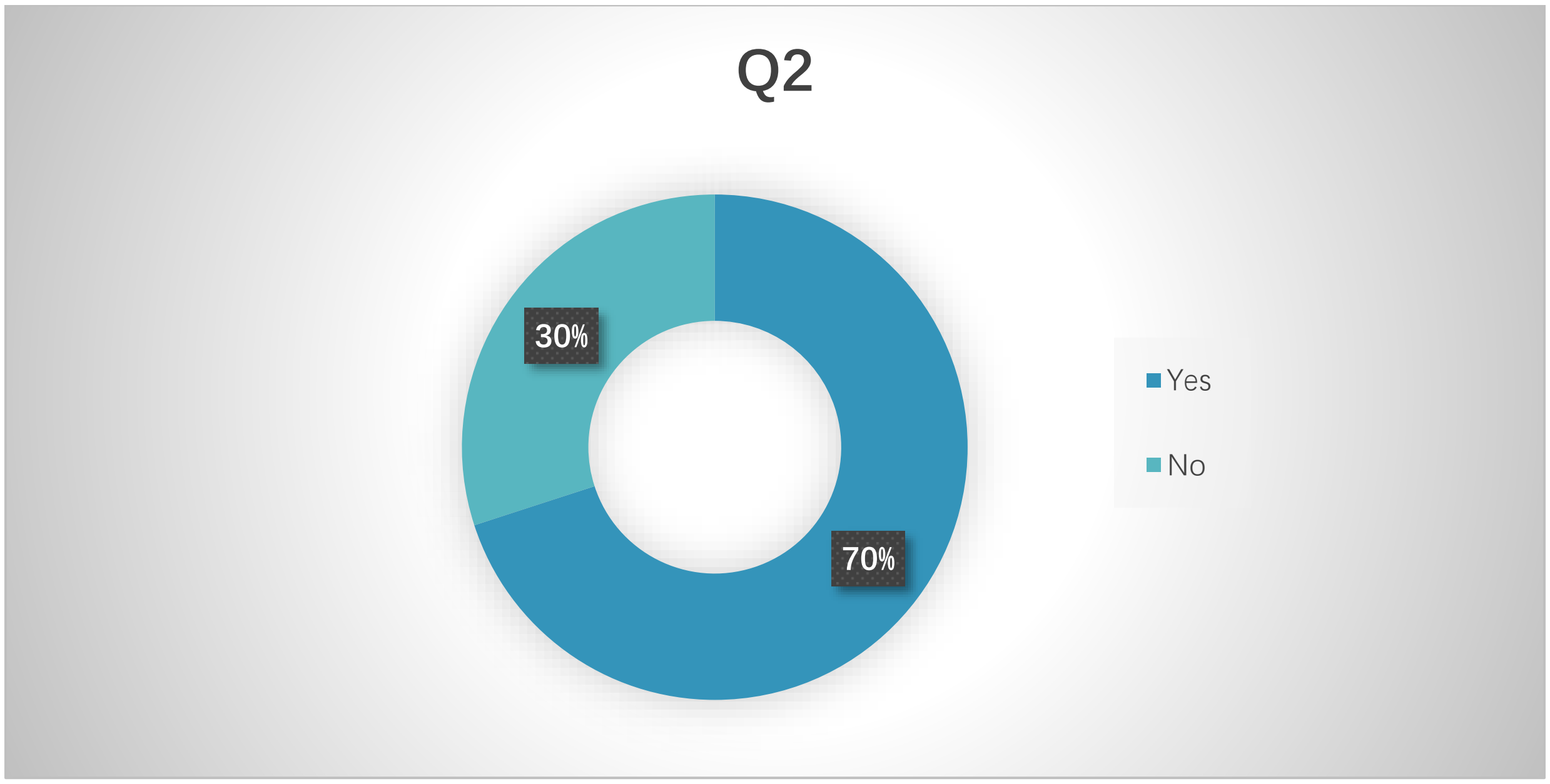

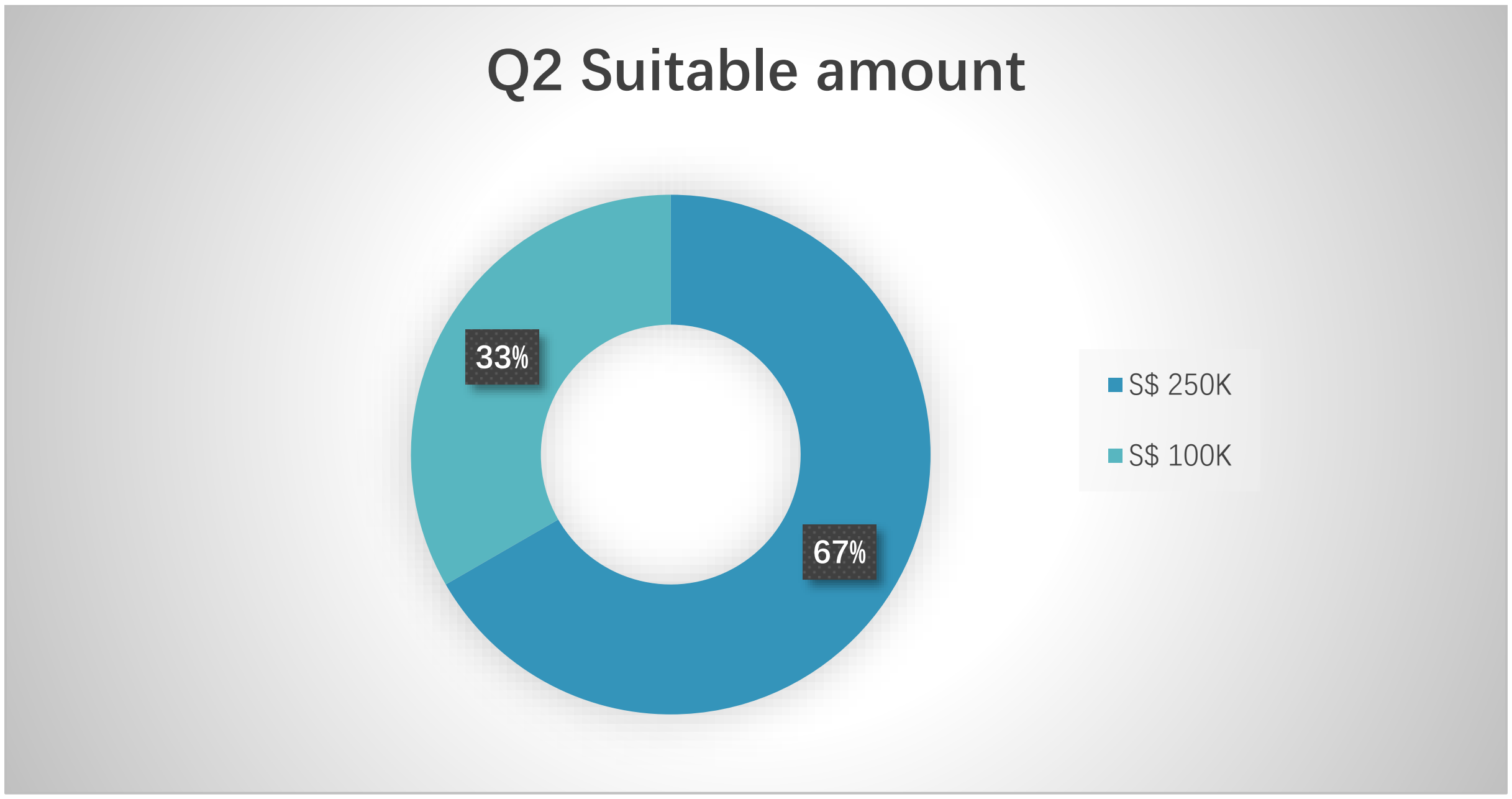

3. In which form are you willing to invest?

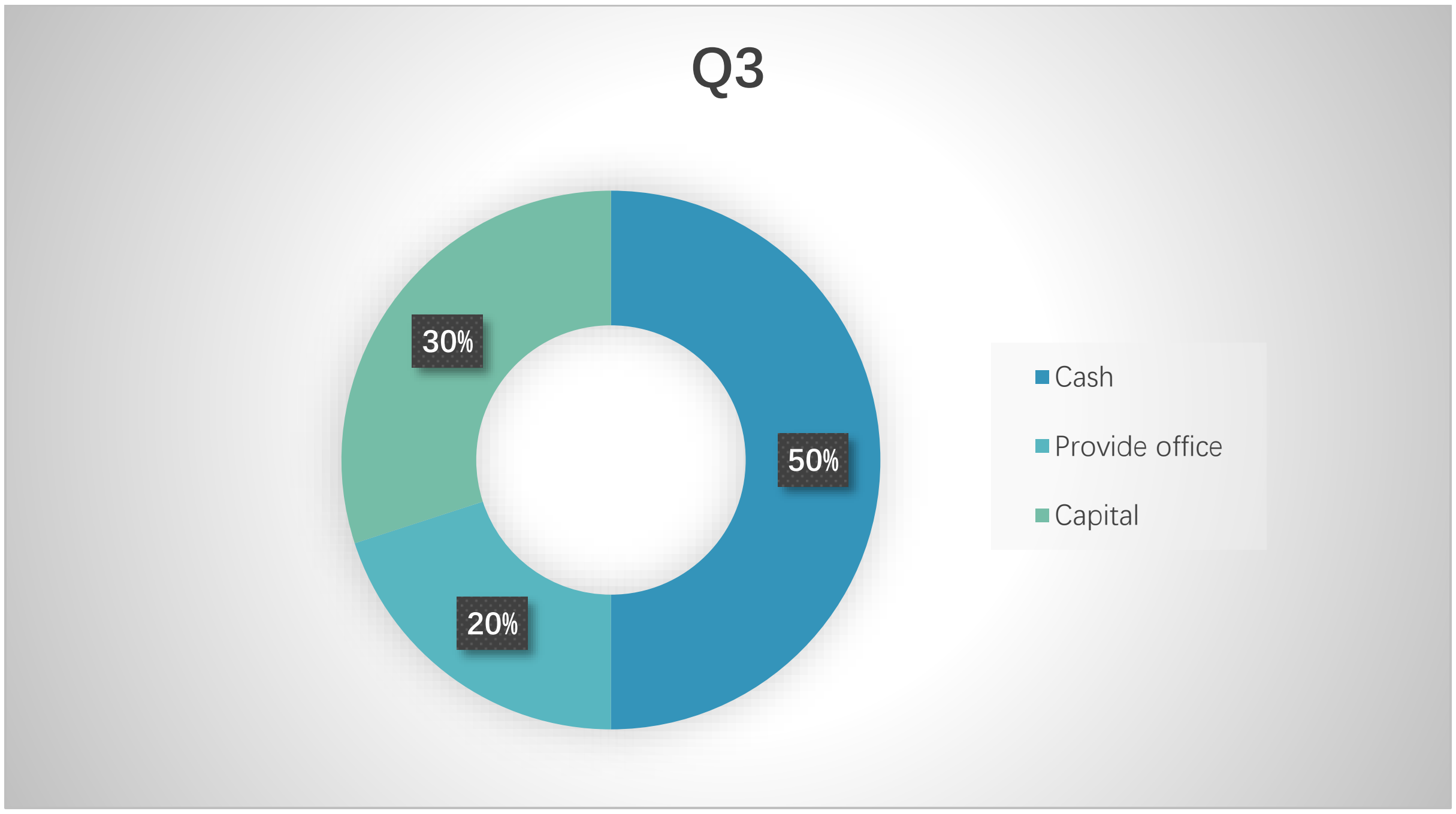

4. Which type of return do you prefer?

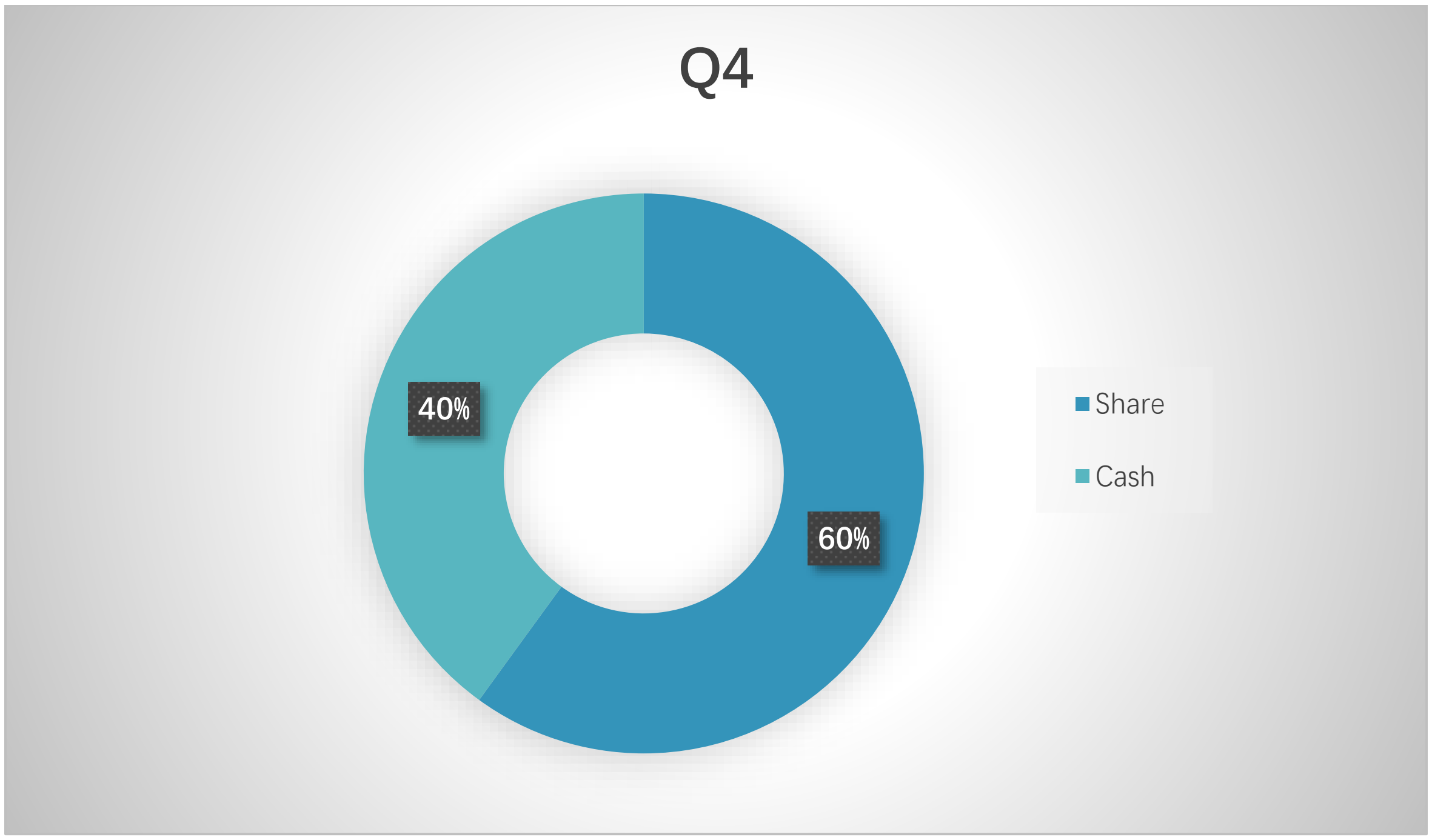

## **<u>For enterprise customer (sample size 10)</u>**

1. If the product can help you to reduce the noise in your working environment, do you want to purchase?

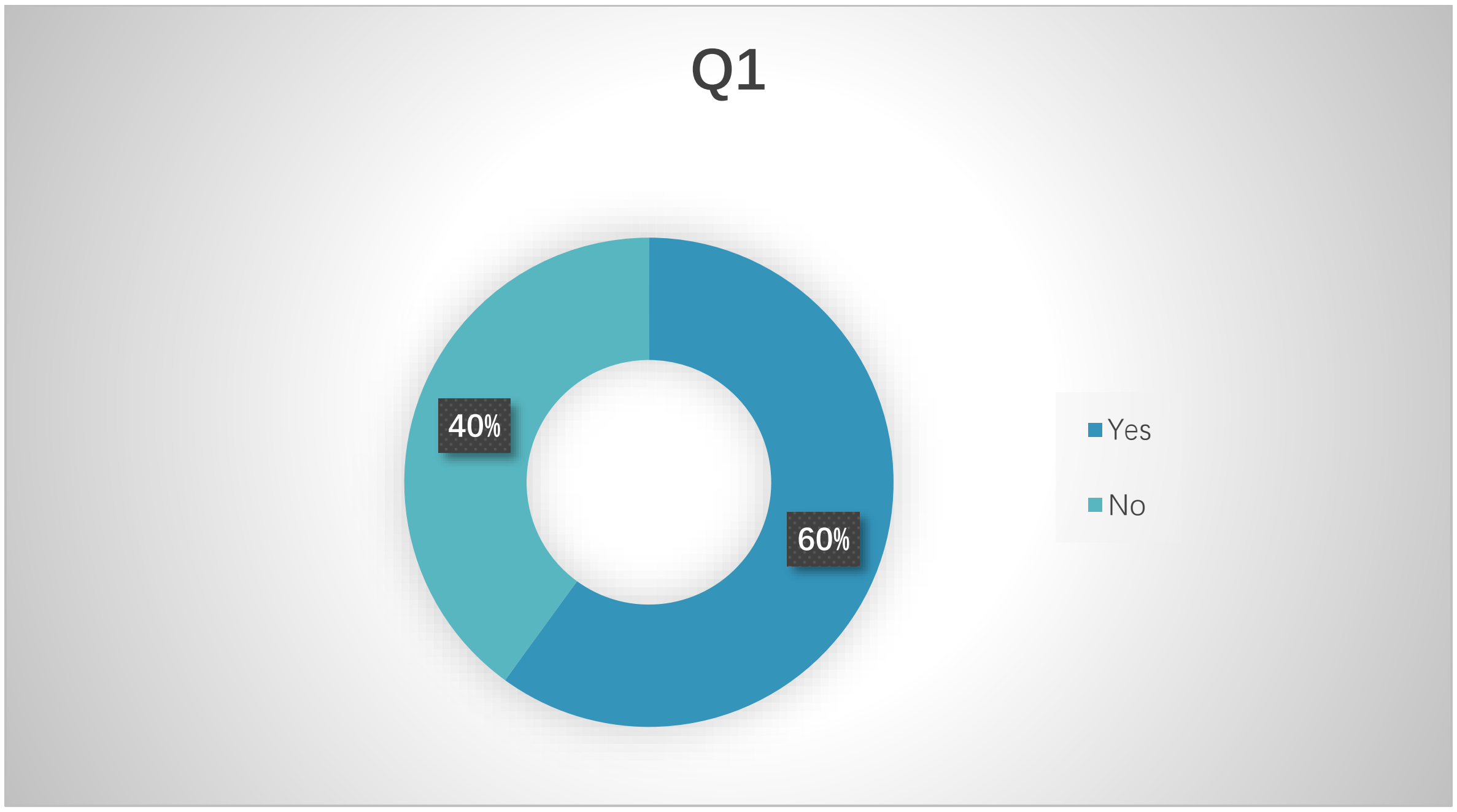

2. What is the size of your enterprise? Which industry is your enterprise in?
   All 10 companies we interviewed has 100+ employees

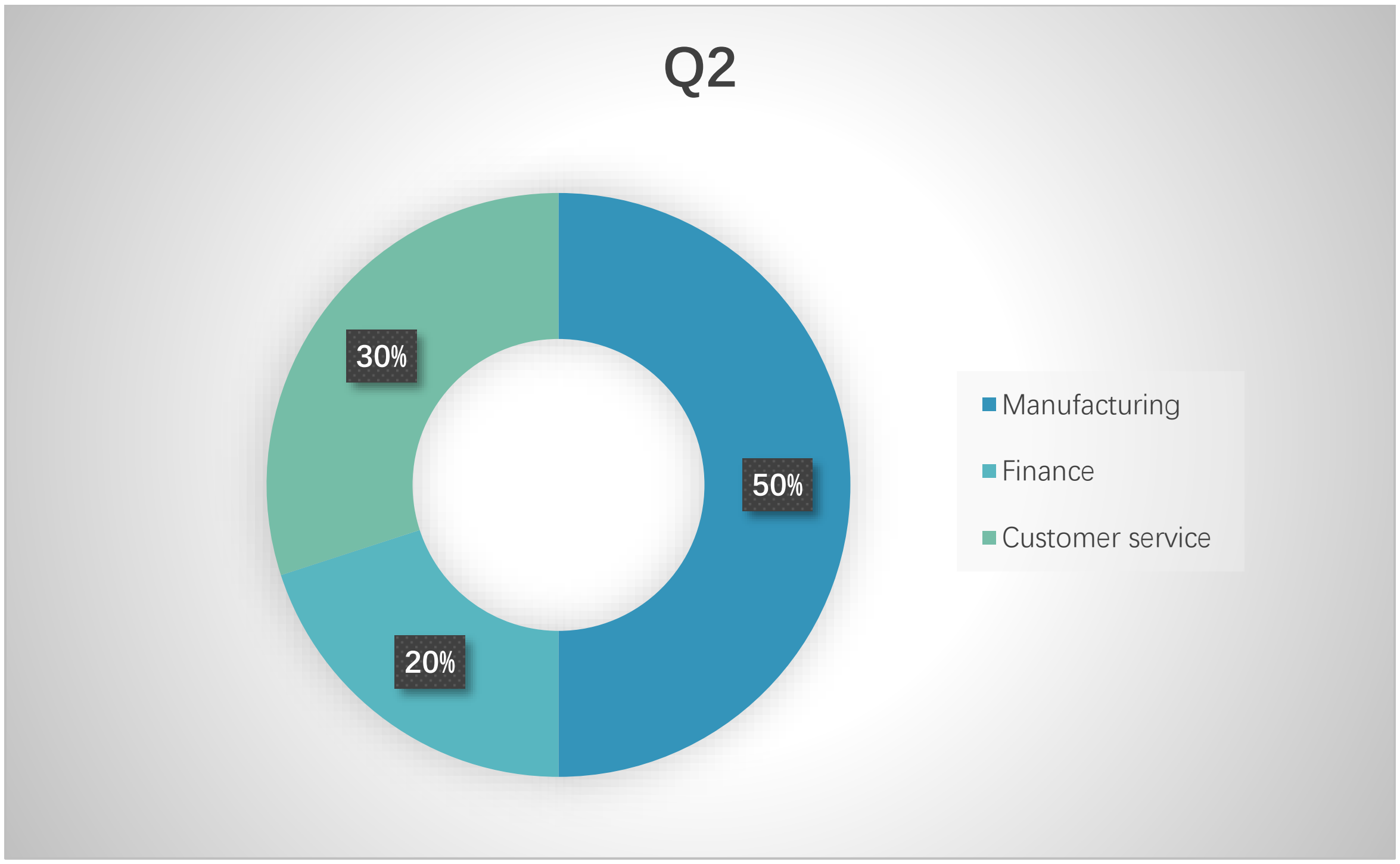

3. Is there any situation that the noise caused by your cooperation has affects to the employee, the cooperate or the environment?

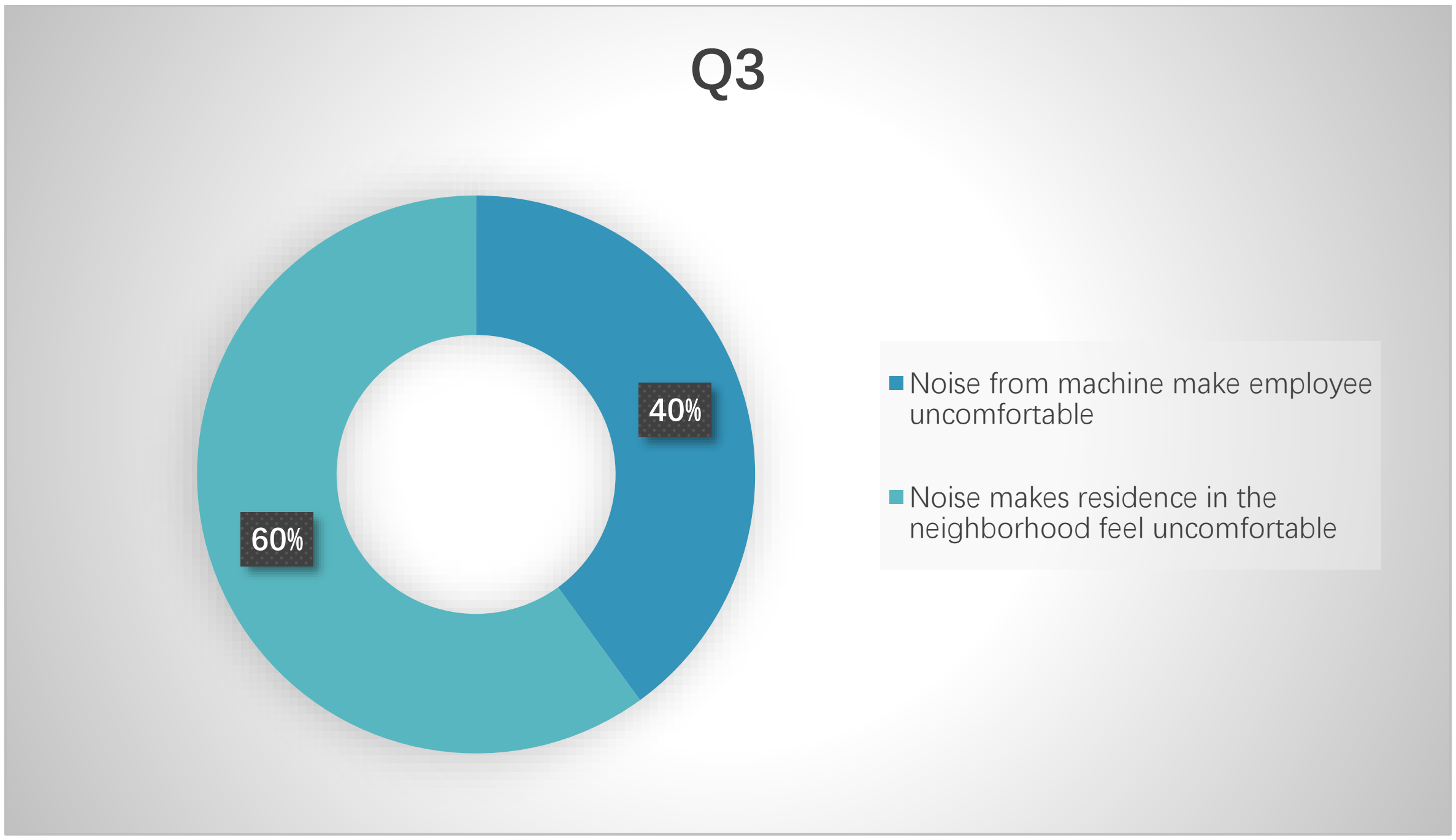

4. Are you willing to let us monitor and analyze the noise data collected by our device?

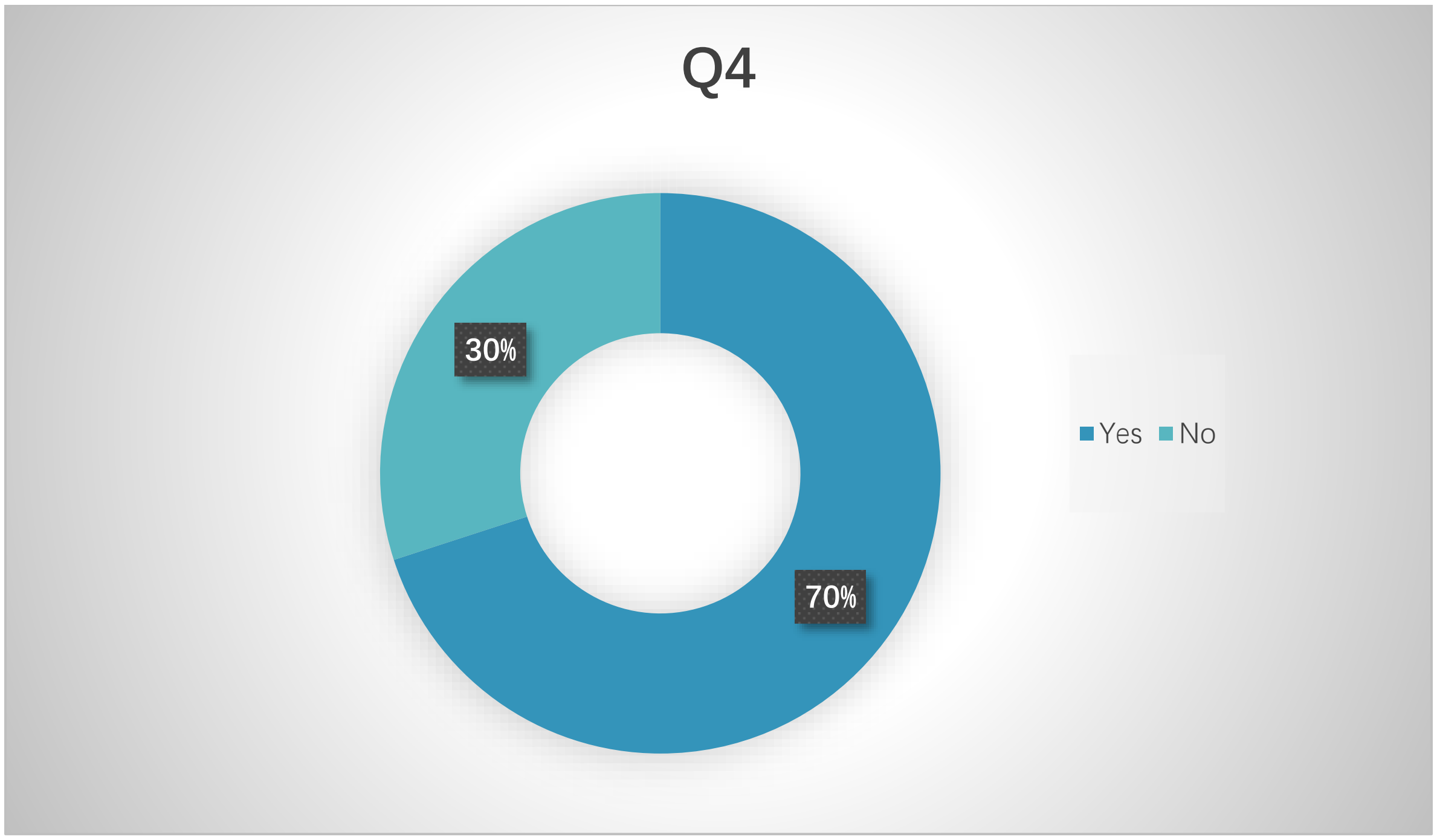

5. Any other functionality that you think is necessary as a cooperate user?

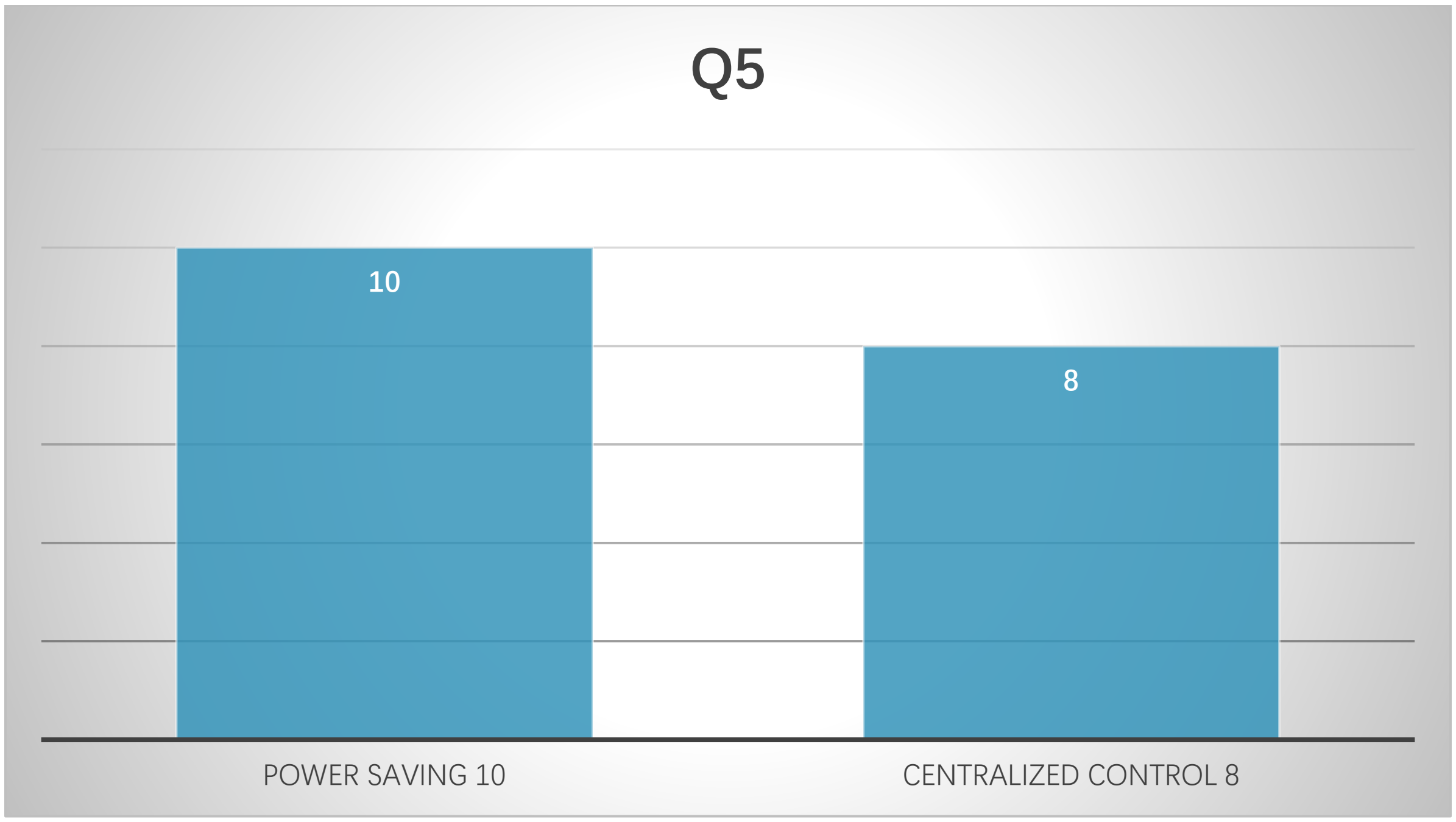

## 3.2 Interpreting Raw Data in Terms of Customer Needs

| No. | Target Audience | Question | Customer's Statement | Interpreted Need(s) |
|---|---|---|---|---|
| 1 | Customers | Do you think noise pollution related problems are very serious? | Yes, I think it is very serious. | Noise pollution problem is very serious. The smart noise reduction device using the high technology is developing to resolve the problem. |
| 2 | Customers | Do you have any trouble caused by noise? | Yes, I have a lot of trouble caused by it. | Most people have been troubled by noise. The functions of the smart noise reduction device able eliminate the noise encountered. |
| 3 | Customers | Normally, when will the noise appear? | Scattered time for whole day | Noise can be anywhere and happened anytime. The smart noise reduction device able to self-adjust to suit different situation. |
| 4 | Customers | Usually, how often will you be affected by the noise appeared? | Quite often affected by the noise appeared. | Most people think they are influenced by noise usually. The smart noise reduction device able to monitor and record the noise level accordingly. |
| 5 | Customers | Normally, the noise happens in which location? | Work place 12.9%<br>School 17.8%<br>Home 26.6%<br>And other place like MRT, Library, Restaurants 39.5% | Noise can happen everywhere including workplace, school, park, MRT train, library, restaurants, and even at home. The smart noise reduction device can use at different kind of environment to reduce the noise. |
| 6 | Customers | Usually, what is the source of the noise? | Construction site 26.1%<br>Transport traffic 20.0%<br>Neighbors 19.6%<br>Others load talking 27.3% | Sources of noise can come from construction, transport, neighbors, load music, others' talking, etc. The smart noise reduction device able to auto adjusts to different source of noise. |
| 7 | Customers | Is the source you identified from Q6 a permanent or temporary source? | Permanent 57.67%<br>Temporary 42.33% | The sources in Q6 are permanent for most people. |
| 8 | Customers | Currently, what is your solution to deal with the noise? | Ear plug 76%<br>Calling respective government agency to control 24%<br>Others 0% | Most people used ear plug to shield noise, but with the smart noise reduction device, it can replace the function of Earplug and provide additional functions to the people. |
| 9 | Customers | Do you know the effects that can be caused by long-term noise? | Yes, I know the effects caused by it. | Most people know noise is harmful to their health in the long term. The smart noise reduction device will provide people with a more comfort and relax living environment. |
| 10 | Customers | Has the current solution applied by the government, that is to reduce the noise cause by construction sites, MRT or airport, meet your requirement? | No, it not meets requirement. | Most people are not satisfied with the current solutions government provided for reducing noise. The smart noise reduction device able to provide an alternative solution to resolve the noise problem. |
| 11 | Customers | What kind of the following sound do you think is noise? | Mostly are Train crossing track, Airplane flying over and Any sound that is too loud. | Most people think train crossing track, airplane flying over or other load voices are noises. The smart noise reduction devices able to auto adjust and eliminate different source of noise. |
| 12 | Product Functionality | If there is a product that can help you to reduce noise, what are the functions that you think is good to have on this product? | Noise reduction 21.4%<br><br>Sleep quality monitor and data record 15.5%<br><br>Sound (e.g. phone ringing tone) amplification and speaker 10.6%<br><br>Alert customer based on the sound detected (e.g. fire alarm) 9.7%<br><br>Record noise data and analysis of data that can help government take action 17.7%<br><br>Integrated with current smart home system 11.6%<br><br>Other (please specify)<br><br>can play white noise 11.3% | People hope there is a product which can reduce noise, as well as monitor sleep quality and record data, amplify music and integrate with smart home system, like electronic furniture and fire alarm.<br><br>The smart noise reduction device has multi functions for noise reduction, sleep quality monitor and data record, amplification and alarm customer Integrated with current smart home system. |
| 13 | Product Functionality | If it is possible for government to monitor and control noise source based on the noise reduction product, do you think it is necessary and useful? | Yes, it is possible for government to monitor and control noise source based on the noise reduction product. | Most people think it’s necessary and useful for the government to monitor and control noise source. The smart noise reduction device has these functions and able to help government to monitor and control noise source. |
| 14 | Product Functionality | If this product can also help you to monitor your sleep quality, do you think the function is necessary? | Yes. I think it is necessary. | Most people think it's necessary and useful to monitor sleep quality by a product.<br>The smart noise reduction device can monitor and record our sleep cycle quality. |
| 15 | Product Functionality | If the device will be using green energy as part of the power supply, which do you think is more suitable? | Solar Energy 76.67%<br>Wind Energy 23.33% | 77% of people think solar energy can be used as power supply. 23% of people think wind energy can be used.<br>The smart noise reduction device designed with solar energy panel as one of its power source. |
| 16 | Product Functionality | Do you prefer the product has the wireless remote control function? Which do you prefer? | Wi-Fi 31%<br>Bluetooth 36.67%<br>Both 19.67%<br>Not important 12.67% | Most people prefer Bluetooth and WIFI as remote control function.<br>The smart noise reduction device is equipped with Bluetooth and wireless remote control function. |
| 17 | Product Functionality | If can do remote control, do you want it be controlled on your mobile device like your smart phone using an app? | Yes, I want it to be controlled on the mobile. | Most people prefer the device can be controlled with our mobile device.<br>The smart noise reduction device can be controlled with mobile software application. |
| 18 | Other information | If the product with these functions is priced at S$ 300, will you purchase it? If No, please specify the percentage the price should be reduced. | Yes 49.33%<br>May consider 17.33%<br>No (with specify reduction)<br>Total:33.33% | 50% of people will purchase the product if the price is S$300, 59% of the remaining people will buy if the price is reduced by 25%.<br>With the multi-function of the smart noise reduction device, it is considering affordable to people. |
| 19 | Other information | Will you make the noise into consideration when you making the following decision? | Looking for job 6.7%<br>Purchasing house 60.9%<br>Applying for school 32.4%<br>Other (please Specify) 0% | Most people will consider the factor of noise when purchasing a house.<br>The smart noise reduction device will make decision easier. |
| 20 | Investors | Are you willing to invest in a product that can do the noise reduction for domestic and business usage? | Yes. I will invest this product. | 70% of investors will invest a product that can reduce the noise.<br>The smart noise reduction device is multifunction device suitable for both domestic and business usage. |
| 21 | Investors | If the predicted market share is 1 billion, return rate for the project is 10%. We need S$ 500,000 to start up the business. Are you willing to be an investor? If no, please specify the suitable amount that you are willing to invest in. | Yes 70%<br>No 30% | 70% of investors are willing to fund S$ 500,000 to start up the business, 67% of the remaining can provide S$ 250K. |
| 22 | Investors | In which form are you willing to invest? | Cash 50%<br>Provide office 20%<br>Capital 30% | 50% of investors like to Invest by cash. |
| 23 | Investors | Which type of return do you prefer? | Share 60%<br>Cash 40% | 60% of investors prefer the return as a share. The smart noise reduction device |
| 24 | Enterprise customers | If the product can help you to reduce the noise in your working environment, do you want to purchase? | Yes 60%<br>No 40% | 60% of people want to buy this product which can reduce the noise in your working environment. The smart noise reduction device helps to reduce the noise in working environment. |
| 25 | Enterprise customers | What is the size of your enterprise? Which industry is your enterprise in? | Mostly are 100+ 0%<br>Manufacturing 50%<br>Finance 20%<br>Customer service 30% | 50% of the enterprises are in the manufacturing field, 20% is in Finance and 30% is in customer service. |
| 26 | Enterprise customers | Is there any situation that the noise caused by your cooperation has affects to the employee, the cooperate or the environment? | Noise from machine make employee uncomfortable and Noise makes residence in the neighborhood feel uncomfortable. | 60% of the noise makes residence in the neighborhood feel uncomfortable, and 40% of the noise makes the employee uncomfortable. The smart noise reduction device can deal these situations efficiently. |
| 27 | Enterprise customers | Are you willing to let us monitor and analyze the noise data collected by our device? | Yes 70%<br>No 30%<br>Yes, I am willing to let us monitor and analyze the noise data collected by our device | Most people are willing to share the data for monitoring and analyzing. The smart noise reduction device can monitor and analyze the noise data for enterprise customers. |
| 28 | Enterprise customers | Any other functionality that you think is necessary as a cooperate user? | Power saving 55.56%<br>Centralized control 44.44% | Some other functions, like power saving or centralized control are preferred by some enterprises. The smart noise reduction device has Power saving and Centralized control functions. |

## 3.3 Organize Needs into Hierarchy

The needs are organized in hierarchy with primary and secondary customer needs. The voting reflects the collective ideology of team that reflects important functions for noise reducing system. Important rating for the needs are indicated by the number of *'s.

| Hierarchy | Customer Needs |
|---|---|
| ***** | The system can deal with different level of noise |
| ** | The noise reducing system has memory and recording function |
| *** | The noise reducing system monitors noise level and send information to user |
| *** | The noise reducing system may install solar panel for power supply |
| ** | The noise reducing system can distinguish the particular sounds |
| **** | The noise reducing device needs to be user friendly |
| *** | The noise reducing system can help tracking our sleeping quality |
| ** | The noise reducing system wireless speaker function |
| * | The noise reducing system has reminding function for specific sound |
| *** | Amplify specific sound（fire alarm, alarm, ring tune etc.） |
| **** | Record noise data and analysis of data that can help government take action |
| ** | Integrated with current smart home system |
| *** | The size of noise reducing device need to be considered |
| * | The appearance design of noise reducing device need to be considered |
| * | The noise reducing device should have different colors |
| *** | The noise reducing system can suit for different environment |
| *** | The noise reducing device should have longer battery life |
| **** | The noise reducing system need to be affordable |
| **** | The noise reducing device need to be durable |
| **** | The noise reducing device can vary the reduction level |
| *** | The noise reducing device is easy to maintain |
| *** | The noise reducing device allow multiple user access |
| ** | The noise reducing device should be power saving |

## 3.4 Establish Relative Importance of the Needs

We established the relative importance of customer's needs through the response we obtained from our survey, including online survey for domestic users and face-to-face interview questionnaire. Team unanimously accepted the following 3 types surveys with indicated ranking as following.

For each of the following the team has indicated on scale of 1 to 5 as follows.

1. Feature is undesirable, I would not consider product with this feature.
2. Feature is not important, but I would not mind having it.
3. Feature would be nice to have, but is not necessary.
4. Feature is highly desirable, but I would consider a product without it.
5. Feature is critical, I would not consider product without this feature.

| Importance | Customer Needs | Feature is Unique |
|---|---|---|
| 5 | The system can deal with different level of noise | Yes |
| 3 | The noise reducing system has memory and recording function | Yes |
| 4 | The noise reducing system monitors noise level and send information to user | Yes |
| 4 | The noise reducing system may install solar panel for power supply | No |
| 3 | The noise reducing system can distinguish the sounds | Yes |
| 5 | The noise reducing device needs to be user friendly | No |
| 4 | The noise reducing system can help tracking our sleeping quality | Yes |
| 3 | The noise reducing system wireless speaker function | No |
| 2 | The noise reducing system has reminding function for specific sound | Yes |
| 4 | Amplify specific sound（fire alarm, alarm, ring tune etc.） | Yes |
| 5 | Record noise data and analysis of data that can help government take action | Yes |
| 3 | Integrated with current smart home system | No |
| 4 | The size of noise reducing device need to be considered | No |
| 2 | The appearance design of noise reducing device need to be considered | No |
| 2 | The noise reducing device should have different colors | No |
| 4 | The noise reducing system can suit for different environment | No |
| 4 | The noise reducing device should have longer battery life | No |
| 5 | The noise reducing system need to be affordable | No |
| 5 | The noise reducing device need to be durable | No |
| 5 | The noise reducing device can vary the reduction level | Yes |
| 4 | The noise reducing device is easy to maintain | No |
| 4 | The noise reducing device allow multiple user access | No |
| 3 | The noise reducing device should be power saving | No |

# 4 Product Specification

## 4.1 Metrics and Units

The Metrics table below shown the translation from relative importance of customer indicated in Table 4.1 to form product attributes in various measurable units.

| Metrics | Units |
|---|---|
| temperature measurement | °C |
| Noise level | dB |
| WI-FI speed | Mb/s |
| LCD image resolution | ppi |
| Life Limits | Years |
| Power consumption | W |
| Solar panel output power | kWh |
| S/N ratio | dB |
| Notes: Accuracy of readings | ±0.1dB (noise) ±0.2°C (temperature) |

Table 1 metrics and units for the product attributes

## 4.2 Customer Needs and Its Benchmark

Based on the data compilation and analysis, we could determine the various needs of the customers and the relative importance. In Table 3.2 shows the benchmarking of the customer's needs and its importance level. (1-least important, 5-most important)

| No. | Needs | | Imp Level |
|---|---|---|---|
| 1 | Eliminoise™ | user interface on the application is simple to use | 5 |
| 2 | Eliminoise™ | is able to communicate with other product based on different user requirement | 5 |
| 3 | Eliminoise™ | sends the information through either software or SMS | 4 |
| 4 | Eliminoise™ | is user friendly for apps and device operation | 4 |
| 5 | Eliminoise™ | is light weight | 5 |
| 6 | Eliminoise™ | is easy to install | 5 |
| 7 | Eliminoise™ | is small and compact | 5 |
| 8 | Eliminoise™ | runs on solar polar battery | 4 |
| 9 | Eliminoise™ | has long battery life span | 5 |
| 10 | Eliminoise™ | is easy to operate | 5 |
| 11 | Eliminoise™ | has a LCD display | 4 |
| 12 | Eliminoise™ | is affordable | 5 |
| 13 | Eliminoise™ | is easy to maintain | 4 |
| 14 | Eliminoise™ | able to remote control through software application | 4 |
| 15 | Eliminoise™ | able to auto detect and reduce the noise level | 5 |
| 16 | Eliminoise™ | able to reduce noise effectively | 5 |
| 17 | Eliminoise™ | able to sleep tracking | 3 |
| 18 | Eliminoise™ | provide data on sleep cycle, room temperature, weather, time & noise level, | 3 |
| 19 | Eliminoise™ | provide smart sound detection e.g. fire alarm, doorbell, phone call etc. | 5 |
| 20 | Eliminoise™ | provide different types of music or white noise | 4 |
| 21 | Eliminoise™ | Uses Bluetooth for wireless application data transfer | 4 |
| 22 | Eliminoise™ | can be used for long durations | 5 |
| 23 | Eliminoise™ | can recharge quickly | 5 |
| 24 | Eliminoise™ | has removable batteries | 4 |
| 25 | Eliminoise™ | can operate independent (without companion software) | 5 |
| 26 | Eliminoise™ | can function under extreme temperature | 5 |
| 27 | Eliminoise™ | can work stable | 4 |

Table 2 Customer Needs and its benchmark

## 4.3 Establish Metrics & Units and Benchmark on Customers' Needs

In this product, there are specifications that are aligned towards the needs of consumer. In the next Table, these establish Metrics & Units and Benchmark on Customers' Needs specifications are worked out to form the basis of our product during prototyping. (1-least important, 5-most important).

| No. | Need No. | Metric (Attribute) | Imp Level | Units |
|---|---|---|---|---|
| 1 | 5, 7 | Weight & Size | 4 | kg & m |
| 2 | 3, 21 | Data transfer Speed | 5 | bps |
| 3 | 15, 16, 25 | Noise Reduction Effectiveness & Accuracy | 5 | dB |
| 4 | 1, 2, 14, 17, 18, 19, 20 | Data collection and analysis | 5 | Nil |
| 5 | 12 | Price | 5 | $ |
| 6 | 9, 22, 25 | Durability | 5 | year |
| 7 | 8, 9, 23, 24 | Battery capacity | 5 | mAh |
| 8 | 4, 6, 10, 11, 13,27 | Product function & Accessibility | 4 | Nil |
| 9 | 26 | Temperature | 5 | °C |

Table 3 Metrics & Units and Benchmark on Customers' Needs

## 4.4 Shows Benchmark on Consumers' Needs against Competitors

We benchmarked our product Eliminoise$^{TM}$ against our closest competitors to make improvement on our design and features as shown in the below table Benchmark on Consumers' Needs against Competitors. (1-least important, 5-most important)

| No | Need No. | Metric (Attribute) | Imp Level | Units | Competitor #1: Marsona 1288A Sound Conditioner 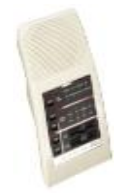 | Competitor #2: Pyle PLGI35T 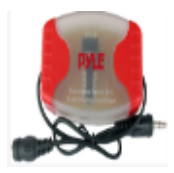 | Competitor #3: iSonicavct AEC-2020 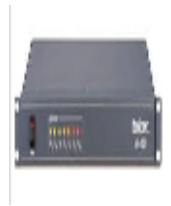 |
|---|---|---|---|---|---|---|---|
| 1 | 5 | Weight | 4 | kg | 1kg | 0.8kg | 2.4 kg |
| 2 | 7 | Size | 4 | m | (230x150x80) mm | (200x150x80) mm | (440x44x180) mm |
| 3 | 3 | Bluetooth | 5 | bps | No | No | No |
| 4 | 15, 16, 26 | Noise reduction level | 5 | dB | 18 dB - 35 dB | 20 dB - 35 dB | 12 dB - 36 dB |
| 5 | 1,2, 14,17, 18,19, 20,27 | Mobile application usage | 5 | Nil | No | No | No |
| 6 | 12 | Price | 5 | $ | $179 | $79 | $279 |
| 7 | 9, 22 | Durability | 5 | year | 4 | 3 | 4 |
| 8 | 21 | Data transfer through Bluetooth | 4 | Nil | No | No | No |
| 9 | 8, 9, 23, 24 | Battery usage | 5 | kAh | 1 day | Power Plug | Power Plug |
| 10 | 11 | LCD display | 4 | Nil | No | No | No |
| 11 | 28 | Temperature | 5 | °C | No | No | No |
| 12 | 29 | Music & white noise | 4 | Nil | White noise only | White noise only | White noise only |

Table 4 Benchmark on Consumers' Needs against Competitors

## 4.5 Concept Generation

After keeping into consideration the various needs which from customer and investor as well as relative priority which we determined in previous Section. A function diagram and concept combination chart is drawing up to express our concept generation process. Those two diagrams and chart help us to better understanding the basic function for our product.

### 4.5.1 Problem Decomposition: Function Diagram

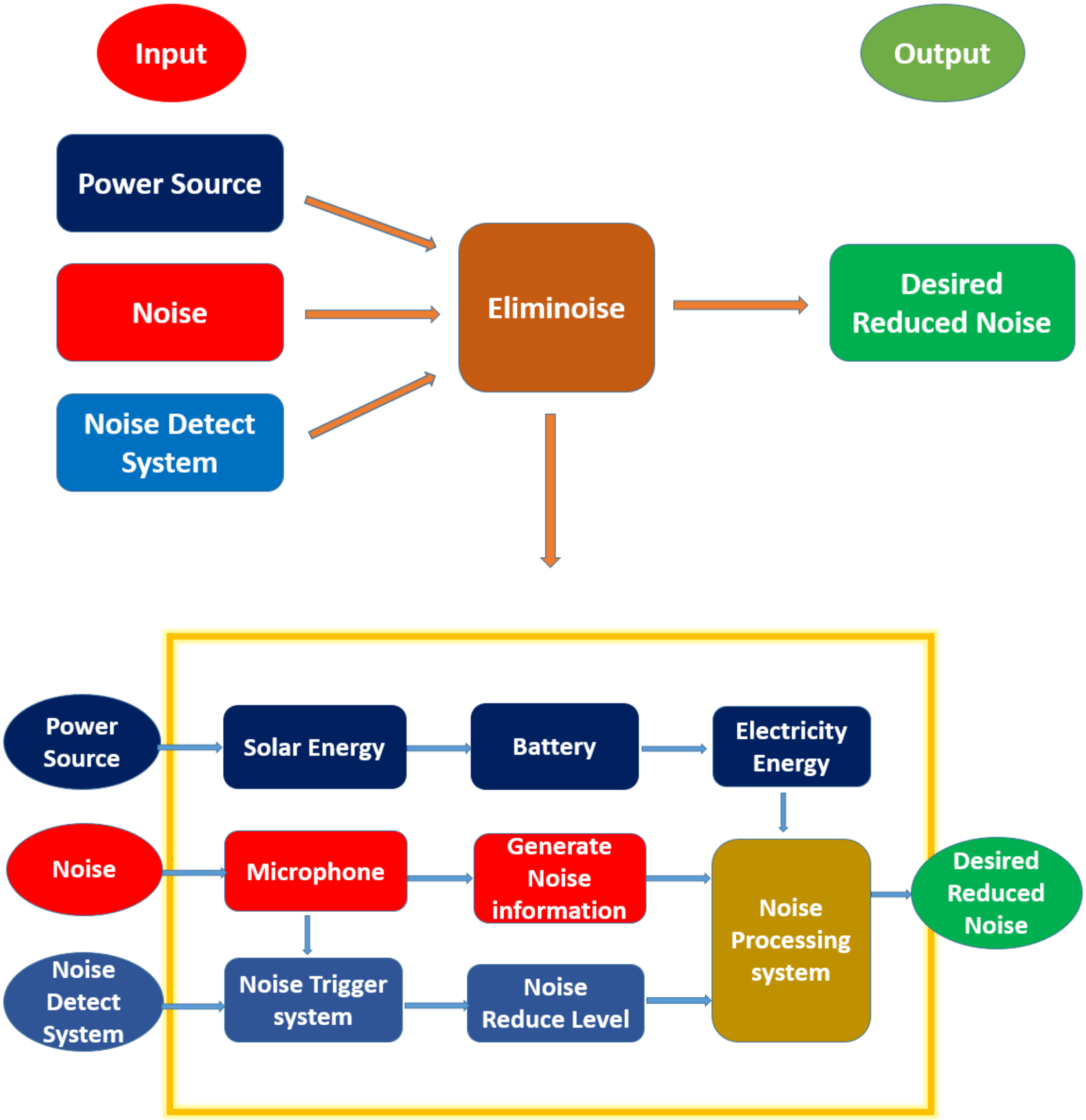

**Figure 9: Eliminoise™ Function Diagram**

### 4.5.2 System Exploration: Concept Combination Chart

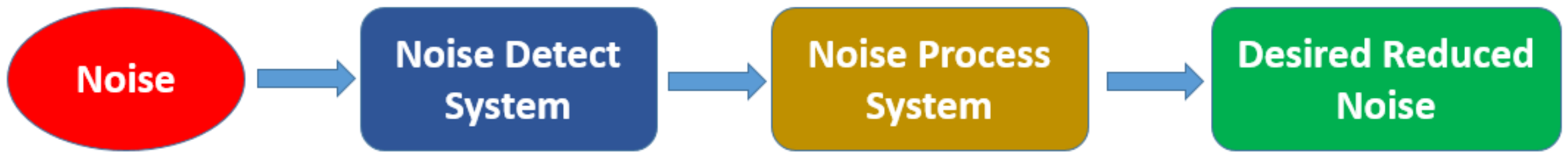

**Figure 10: Concept Combination Chart**

### 4.5.3 Concept Selection

To better fulfill our customer's needs and address the investors' concern of making profit, three concepts were come out after we take into consideration few factors which like power supply type, device cover material, connection type, power consumption and some other factors.

**Concept A:**

Using solar power and electric power as the primary and secondary power supply. There is a wire connection between speaker and silencer to transfer the noise data if the speaker detects any noise from outside. For the device cover material, we consider to choose plastic.

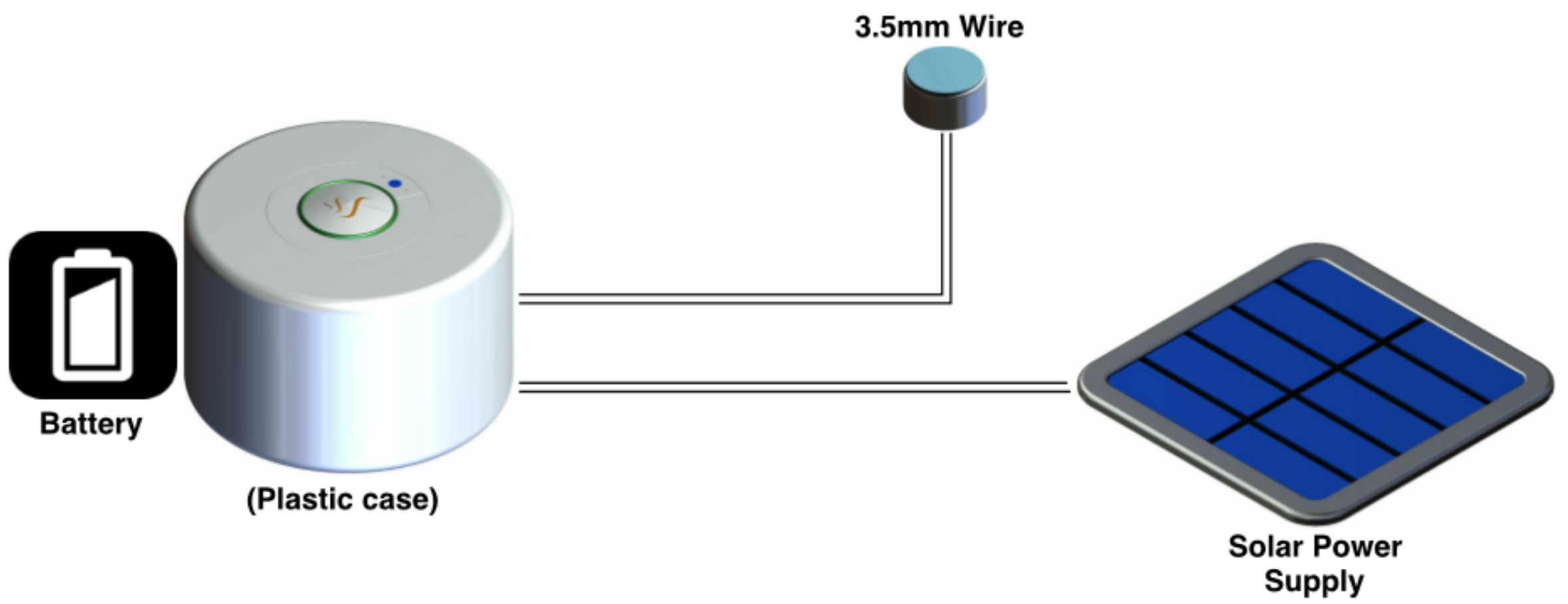

**Figure 11: Concept A Design**

## Concept B:

Using wind energy and electric power as the primary and secondary power supply. There is a Bluetooth connection between speaker and silencer to transfer the noise data after the speaker detects noise. For the device cover material, we consider to choose wood.

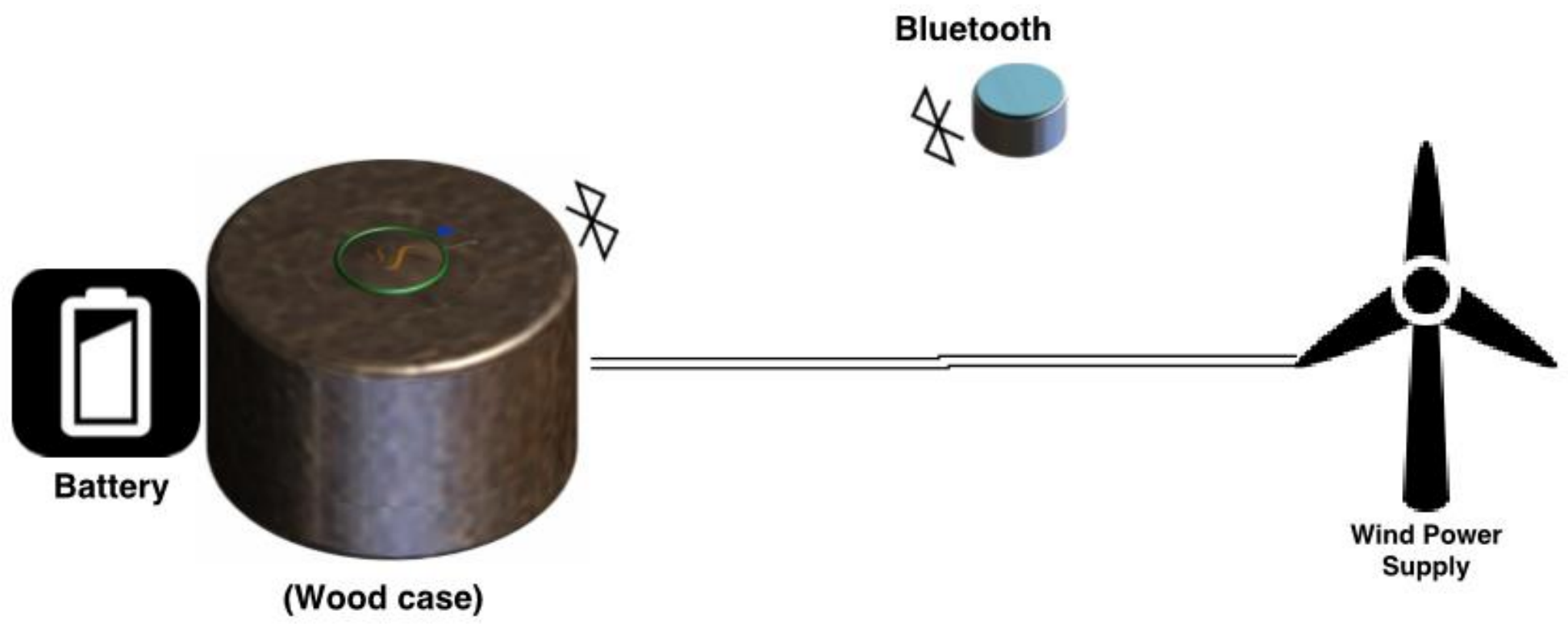

**Figure 12: Concept B Design**

## Concept C:

Using electric power as the only power supply. There is a Wi-Fi connection between speaker and silencer to transfer the noise date once the speaker detects noise. For the device cover material, we consider to choose Stainless steel.

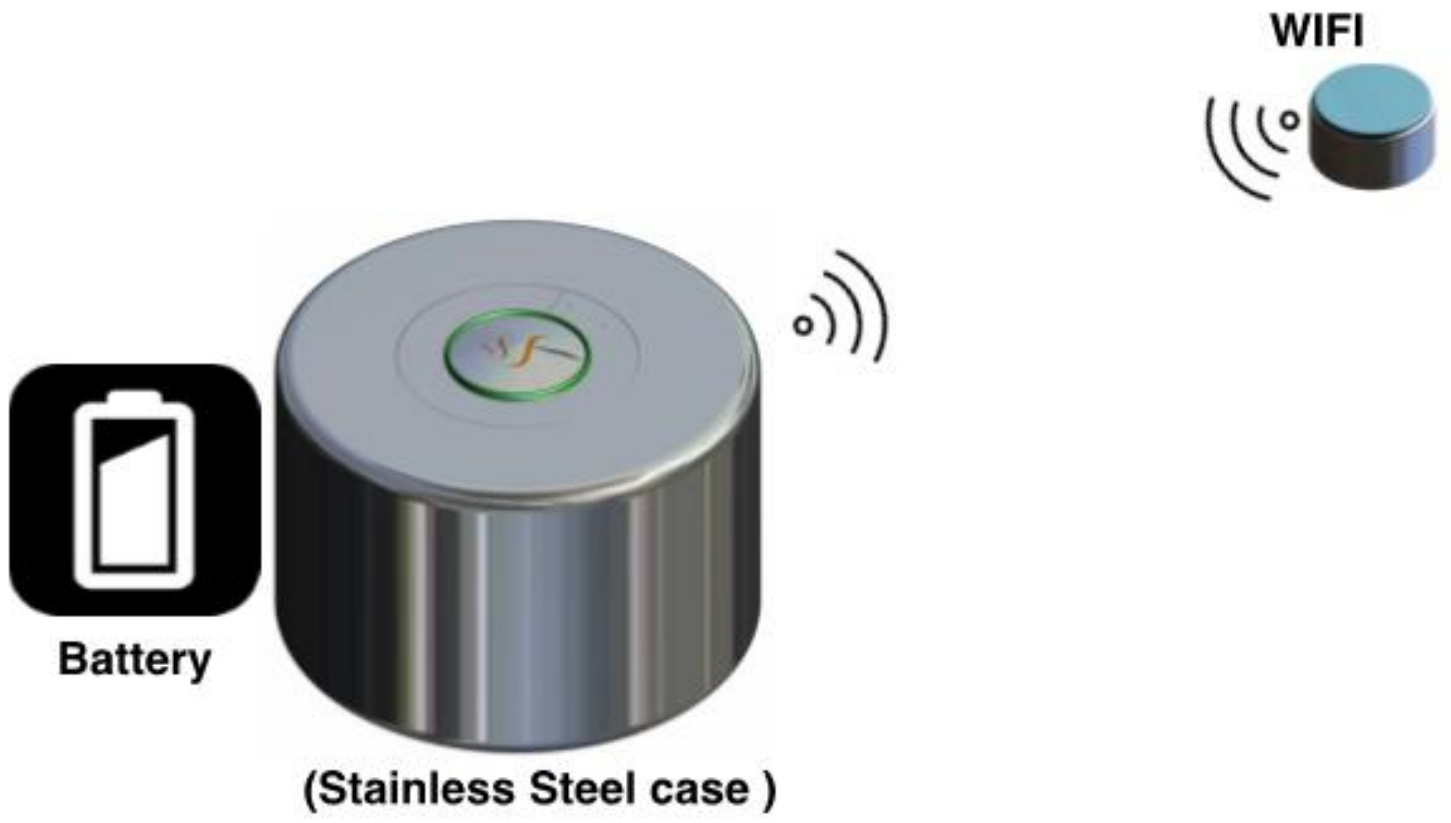

**Figure 13: Concept C Design**

## 4.6 Concept Screening

The following table indicated the ranking of the criteria for the three concepts. We will use six different factors to screen and test our three concepts to find the final one which could be further developed.

| Criteria | Concept A | Concept B | Concept C |
|---|---|---|---|
| Appearance Design | Very Good | Good | Very Good |
| Maintenance | Easy | Not Easy | Very Easy |
| Power Supply | Stable | Stable | Very Stable |
| Operation | Easy | Easy | Easy |
| Connection between Speaker and Silencer | Low delay | High delay | High delay |
| Noise reduce level | 0 – 40db | 0 – 40db | 0 – 40db |
| Manufacturing Cost | 20 USD | 22 USD | 16 USD |
| Material cost | 32 USD | 43 USD | 21 USD |
| Power consumption* | 0.1Watt/Day | 0.1Watt/Day | 2 Watt/Day |
| Noise data Delay time | 0.02s | 1.2s | 1.5s |
| Collect & Analyze Customer Data | Yes | Yes | Yes |
| Physical Size (Eliminoise™, Speaker) | 100mm X 68mm(DxH) 120mmx120mmx5mm (LXWXH) | 100mm X 68mm(DxH) 100X100(DXH) | 100mm X 68mm(DxH) 50CMX50CM |
| Battery Capacity | 3000mAh | 3000mAh | 3000mAh |
| Wi-Fi Reception Signal Strength | Decrease 3% | Decrease 21% | Decrease 6% |
| Touch Screen | Yes | Yes | Yes |
| Multi-User access | Yes | Yes | Yes |

*Power consumption on electricity power.

Table 5: Concept Screening

All three concepts were evaluated and tested with our potential customers and investor. As it shown in the survey, these concepts are well accepted and attractive to our potential customers, and addressed our investors' concern of making profit. This is a strong indication on customer purchase-intent, as we solicited from the survey.

## 4.7 Concept Scoring

The following table has been created and ranked from these three concepts. The weightage for the criteria are determined based on customer's desired and technical concerns and scored accordingly. It is obviously that Concept A has the highest score product. Hence, we will use concept A as our prototype to further develop.

## Concept Scoring

| Selection Criteria | Weight | Concept A | | Concept B | | Concept C | |
|---|---|---|---|---|---|---|---|
| | | Rating* | Weighted Score | Rating* | Weighted Score | Rating* | Weighted Score |
| Appearance Design | 8% | 3 | 0.24 | 2 | 0.16 | 1 | 0.08 |
| Power Supply Type | 7% | 3 | 0.21 | 2 | 0.14 | 1 | 0.07 |
| Connection Between Speaker and Silencer | 6% | 3 | 0.18 | 2 | 0.12 | 1 | 0.06 |
| User Friendly | 6% | 2 | 0.12 | 2 | 0.12 | 2 | 0.12 |
| Maintenance | 6% | 2 | 0.12 | 1 | 0.06 | 3 | 0.18 |
| Noise Reduce Level | 6% | 3 | 0.18 | 3 | 0.18 | 3 | 0.18 |
| Manufacturing Cost | 10% | 2 | 0.2 | 1 | 0.1 | 3 | 0.3 |
| Material Cost | 10% | 2 | 0.2 | 1 | 0.1 | 3 | 0.3 |
| Power Consumption | 12% | 3 | 0.36 | 3 | 0.36 | 1 | 0.12 |
| Noise Data Delay Time | 8% | 3 | 0.24 | 2 | 0.16 | 1 | 0.08 |
| Collect & Analyze Customer Data | 4% | 3 | 0.12 | 3 | 0.12 | 3 | 0.12 |
| Physical Size | 5% | 3 | 0.15 | 2 | 0.1 | 3 | 0.15 |
| Battery Capacity | 3% | 3 | 0.09 | 3 | 0.09 | 3 | 0.09 |
| Wi-Fi Reception Signal Strength | 5% | 3 | 0.15 | 1 | 0.05 | 2 | 0.1 |
| Touch Screen | 2% | 3 | 0.06 | 3 | 0.06 | 3 | 0.06 |
| Multi-User Access | 2% | 3 | 0.06 | 3 | 0.06 | 3 | 0.06 |
| Total Score | | 2.68 | | 1.98 | | 2.07 | |

| | | | | |
|---|---|---|---|---|
| Ranking | | 1 | 3 | 2 |
| Continue | | Develop | No | No |

* Highest score is 3, lowest score is 1

Table 6: Concept Scoring

# 4.8 Concept Testing

## 4.8.1 Purpose of Concept Test

The purpose of the concept test is to obtain feedback on which of the 3 concepts designed is the most desirable to meet the customer's needs, and provides additional considerations for areas of improvement.

## 4.8.2 Sample population

We managed to sample about 500 peoples of our target market – Domestic consumer, office consumer, potential investor and manager of MRT, construction site, Airport.

## 4.8.3 Survey Results

Refer to Appendix B for Testing Concept Survey Data.

**Testing Concept Survey**

1) Which design do you prefer based on concept descriptions?

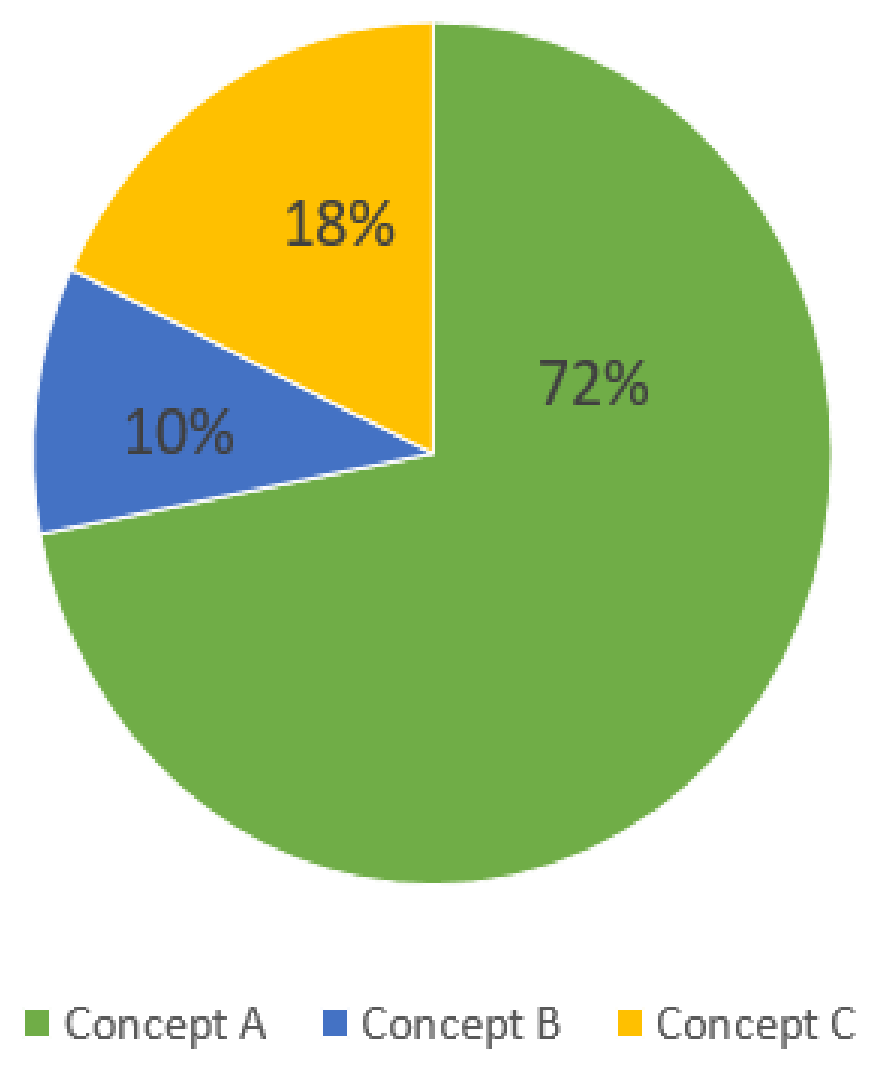

2) Which feature of our product is your favorite?

a) User Friendly
b) Touch Screen
c) Low Maintenance Cost
d) Solar Power Supply
e) Wind Power Supply
f) Wi-Fi Remote Control
g) Work on all Mobile Platforms
h) Wide Noise Reduced Level
i) Improve user sleeping quality
j) Battery Capacity
k) Dual power Supply
l) Physical Size
m) Easy to Maintenance
n) Energy Saving
o) Appearance Design
p) Multi-user access at the same time

3) What additional features would you like to have?

a) Able to Share the sleep data on apps and compare to others.
b) There are some regular updates on apps about how to sleep well, when should to sleep, how to reduce insomnia, etc.
c) Sleeping and waking up time auto reminder according to user sleep data analyses.

### 4.8.4 Interpret Results

We did an online survey to know more about view on the three concepts' aesthetics and feedback through our concept testing survey, we could get an understanding of what our target market needs are and which part we can improve for the next phase of our product.

| Selection Criteria | Concepts Variants | | |
|---|---|---|---|
| | A | B | C |
| **Prefer** | 72% | 10 | 18% |
| **Rank** | 3 | 1 | 2 |

Table7: Concept Survey Results

| Feature | Percentage | Needs | Importance Level |
|---|---|---|---|
| **Noise Reduced Level** | 89% | Eliminator provide 0-40db wide noise reduce range | 5 |
| **User Friendly** | 80% | Eliminator provide a touch screen for user to operate and mobile apps to remote control via Wi-Fi | 5 |
| **Energy Saving** | 70% | Eliminator provider Solar Energy for main power source | 4 |
| **Low Maintenance Cost** | 56% | 5-year warranty | 3 |
| **Improve user sleeping quality.** | 86% | Eliminator provider user sleep data collect and analyses function | 5 |
| **Dual power Supply** | 51% | Eliminator provider dual power supply function | 3 |
| **Sleep and wake up auto reminder** | 67% | Eliminator provider smarter reminder function based on big data that collect from user | 4 |
| **Knowledge sharing on how to sleep well on apps** | 70% | We will add this feature at next phase. | 3 |
| **Share sleep data on apps and compare to others user** | 54% | We will add this feature at next phase. | 3 |

Table 8: Concept Test Survey Analysis

## 4.9 Final Product Specifications

According to the results archived from the survey carried out for the customers' needs, we summarized the customers' needs, technical feasibility and implementation cost to conclude our product specification and final design for Eliminoise™. As shown below:

## Eliminoise™ component details:

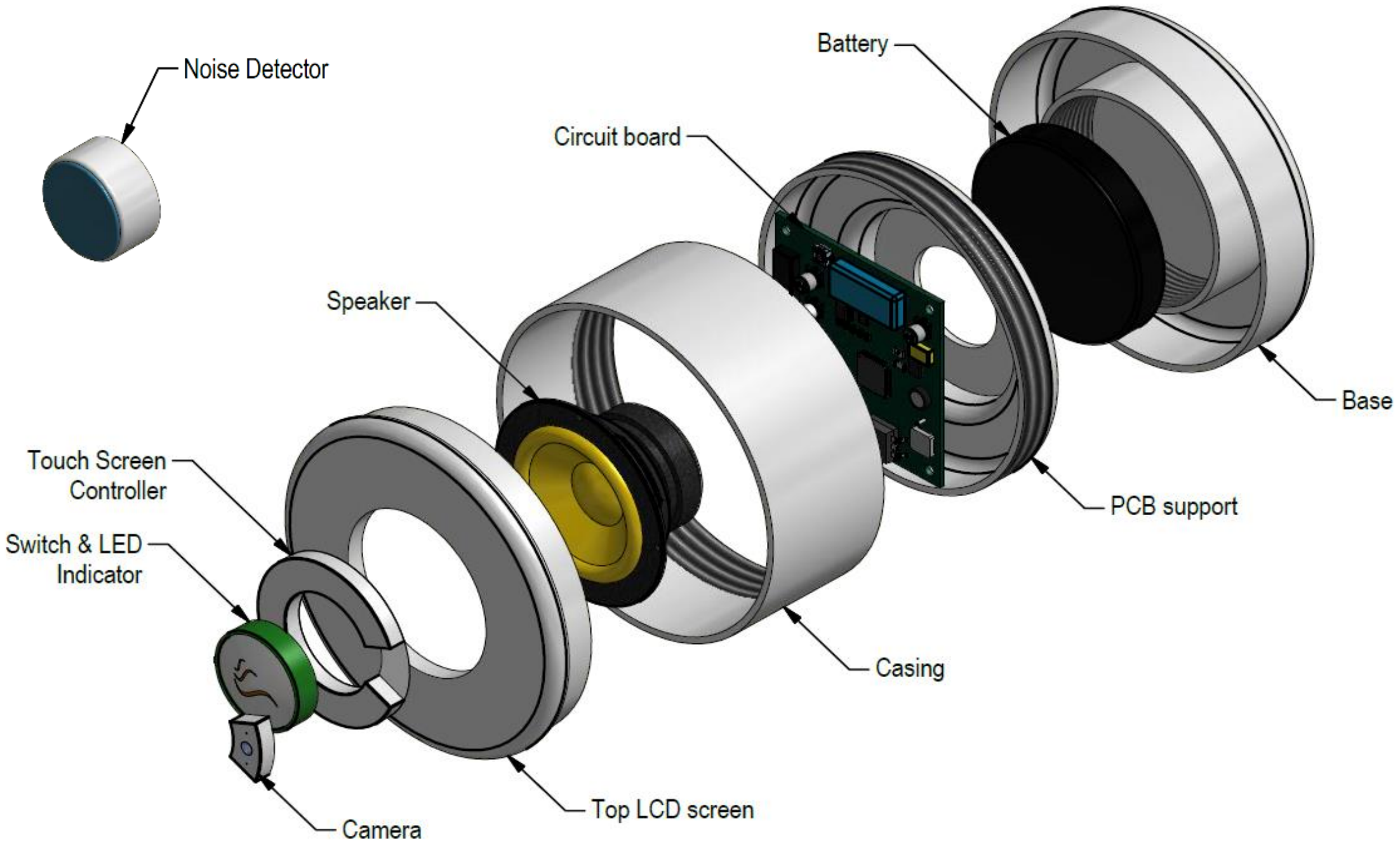

**Figure 14: Component Detail**

## Product specification details:

| Product Specification | |
|---|---|
| **Physical:** | |
| Dimensions | Ø100mm x 68mm |
| Weight | 300g |
| Colors | white, black, red, blue |
| Materials | Plastic |
| Working Temperature | [-20 ~ 60°C] |
| Life Span | 5 years |
| Solar Panel | 92mm x 61mm x 3mm (LxWxD) |
| | Sundance Solar |
| | Part Number: 700-10850-19-wires |
| Display Size | ID 92 mm x OD 100mm |
| Water and Dust Resistant | IP67 under IEC standard 60529 |
| | |
| **Mobile APP:** | |
| System Requirement | Above IOS 9 & Android 7.0 |
| APP Size | 50MB |
| WI-FI | 802.11a/b/g/n/ac Wi-Fi with MIMO |
| Bluetooth | Bluetooth 4.2 wireless technology |
| | |
| **Electrical:** | |
| Battery | Built-in rechargeable lithium-ion battery |
| | 3000mAh, 300hrs |
| | Cameron Sino |
| | Part number:CS-GRS210SH |
| Memory | 16GB， Intel |
| CPU | 17.0 mm × 17.0 mm (LxW) |
| | Texas Instruments |
| | Part Number: AM4372ZDN |
| Temperature Sensor | SMD NTC |
| | Part number: SNC101B12700X0402E |
| Bluetooth | 28.0mm x 12.7mm x 2.6mm (LxWxD) |
| | KEDSUM |
| | Part Number: KDF001-F73A |
| Camera | 12-megapixel camera |
| | PI KIDS Publications Intl |
| | Part Number: 100003 |
| Microphone | LM393 Sound Detection Sensor |

Table 9. Product specification details

# 5. Project Planning and Architecture

## 5.1 Product Development Planning

Product planning is the most important process to identify and develop the new product based on the customer requirements or needs. It is also a process that can create a product idea and work on it until the product is implemented.

Product planning also manages the product throughout its life using various marketing strategies, including product extensions or improvements, increased distribution, price changes and promotions. The planning should be done before the product development process, and it should be considered from the following several points:

- Identify opportunities
- Evaluate and prioritize projects
- Market segmentation
- Technological trajectory
- Fosters derivative products
- Technology roadmap
- Process matrix

### 5.1.1 Identify Opportunities

Based on our survey and customer needs identification, we find that eliminoise$^{TM}$ is a very good product in today's market. With the life cost increasing, more people need a good sleep or quiet environment for rest, study and work.  The eliminoise$^{TM}$ targets the people in the whole world who are suffering the noise problem during the rest or work.

With the active noise control technology development, we can use this technology to implement noise deducting function, to satisfy the people's requirement and give them a good environment. We can improve the noise reducing efficiency of our product to 40dB and integrate with many other useful functions, such as temperature indicator, sleep tracking system, home alarm system and so on.

All those functions will increase our competitiveness in current market.

Currently, we plan to produce three versions of the product to satisfy the different groups of customers and conduct some upgrading with the technology development, the details are shown in the below figure:

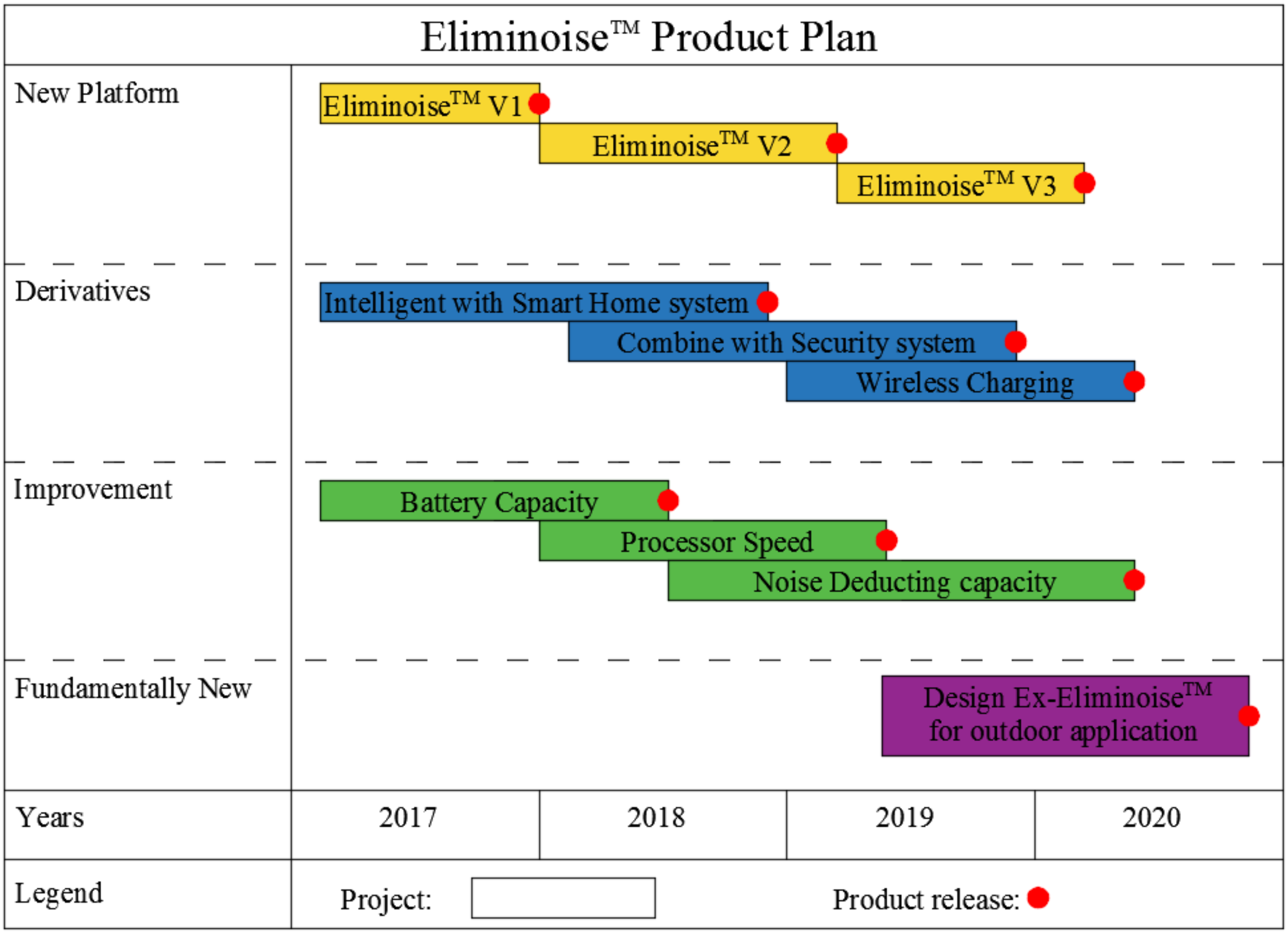

Table 10: Eliminoise product plan

### 5.1.2 Evaluate & Prioritize Projects

In order to be competitive in the current market, we need to evaluate and prioritize our product design from below points:

1. We need to improve significantly and break through the current noise deducting product and integrate with more functions.

2. We need to reduce the cost in both manufacturing and service.

3. Focus on customers' needs.

4. Quickly launch products that seize the market.

### 5.1.3 Market Segmentation: Product Segment Map

Based on the previous report analysis, we have identified the market segment will be changed based on the noise reduction level and cost. We plan to set the noise reduction value from 0 dB to 40 dB and the price we set to SGD 350. We plan to complete our product design from 01-Feb-2017 to 20-May-2017. After the testing and calibration, we hope that we can finish the product before 30-Aug-2017, so that we can launch eliminoise™ to the market on 30-Dec-2017 on time.

And we also decided to develop three types of our products which will be needed. As well as considering the development and performances increase, we determined the different prices for the different group requirements. The following figure shows the market segmentation and the price versus noise reduction performance.

| Eliminoise™ Product Segment Map | | | | |
|---|---|---|---|---|
| Personal | Eliminoise™ V1<br>For indoor use only<br>Price: $350<br>Reducing: 40dB | | | |
| Work Group | | Eliminoise™ V2<br>For indoor use only<br>Price: $350<br>Reducing: 50dB | EX-Eliminoise™<br>For Outdoor use<br>Price: $450<br>Reducing: 30dB | |
| Governments | | | Eliminoise™ V3<br>For indoor use only<br>Price: $380<br>Reducing: 55dB | |
| Years | 2017 | 2018 | 2019 | 2020 |
| Legend | Version vs Price: | | | |

Table 11: Market segmentation and the price VS noise reduction performance

### 5.1.4 Technological Trajectory

For the eliminoise™, we desire to develop it to reduce the noise and transform the data back to the mobile device to be analyzed, so the most important points we need to consider are the noise deducting efficiency and data transform performance.

To implement the noise reducing system, there are two types of noise control technology. One is passive noise control, and the other is active noise control. After careful comparison between the two, we decide to choose the active noise control technology, because we find that passive noise control technology needs to control the material or transport media for the noise source, which is more difficult to be implemented and the cost will be very high.

The following figure is technology S-curve, showing that the eliminoise™ performance changes versus time. Besides, the technologies, such as activate noise control, passive noise control, Bluetooth data transform and Wi-Fi data transform may help our product increase the marketing share. By applying the new technology developed, we believe that we can find a way to solve the outdoor noise, which will help the factory or government to monitor the noise and record the noise data to improve the environment in the future.

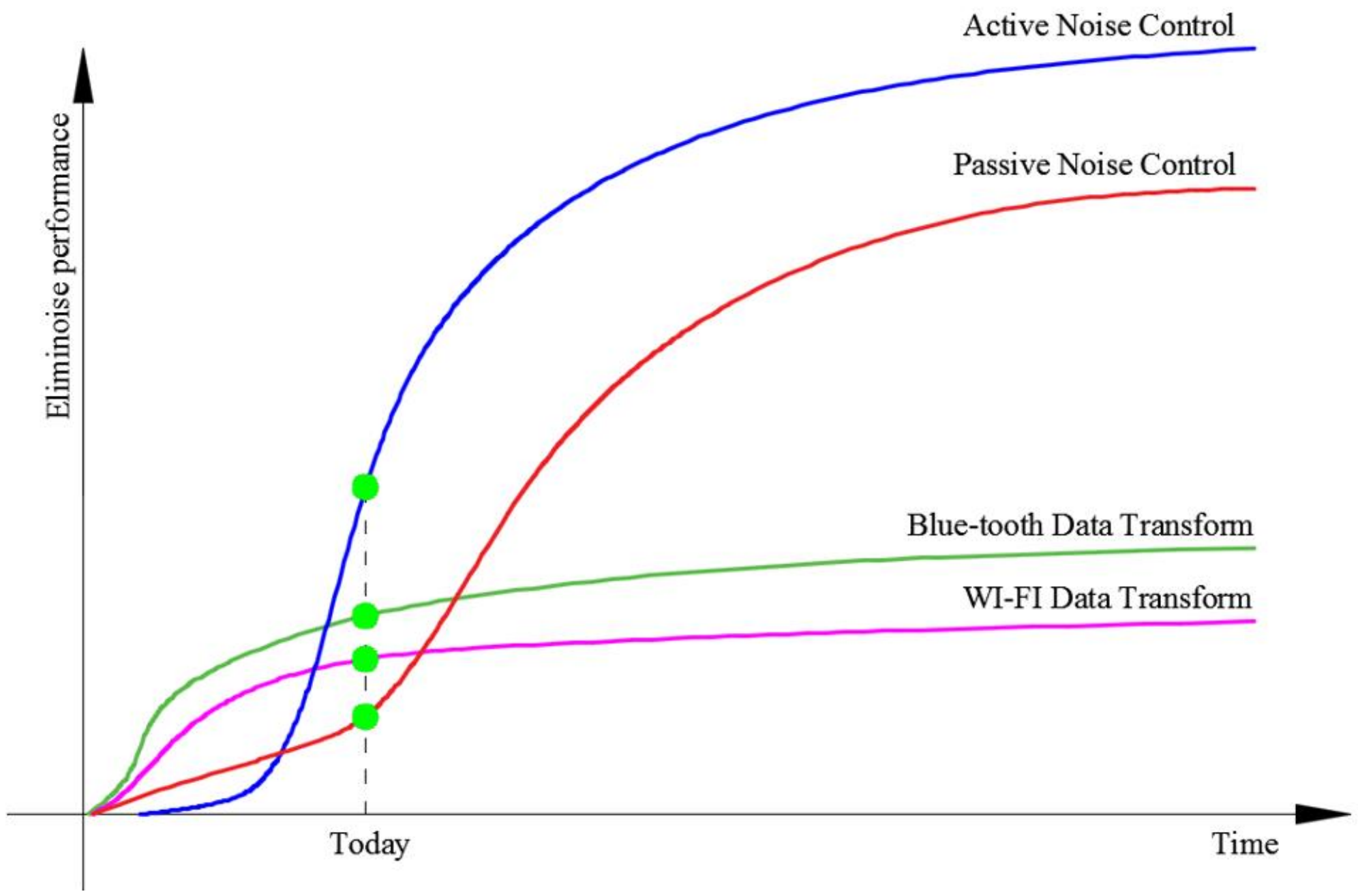

**Figure15: Technology S-curve**

### 5.1.5 Product Platform Planning: Fosters Derivative Products

In addition to the mentioned functions, we must consider the Derivative Products, which are also the important sections for the new product development. It will clearly cut the timeline of our product development and improvement. It can also help us decide whether we can afford the time cost and understand when we need to prepare the money for our new technology development and when we should have the net incoming before we do that:

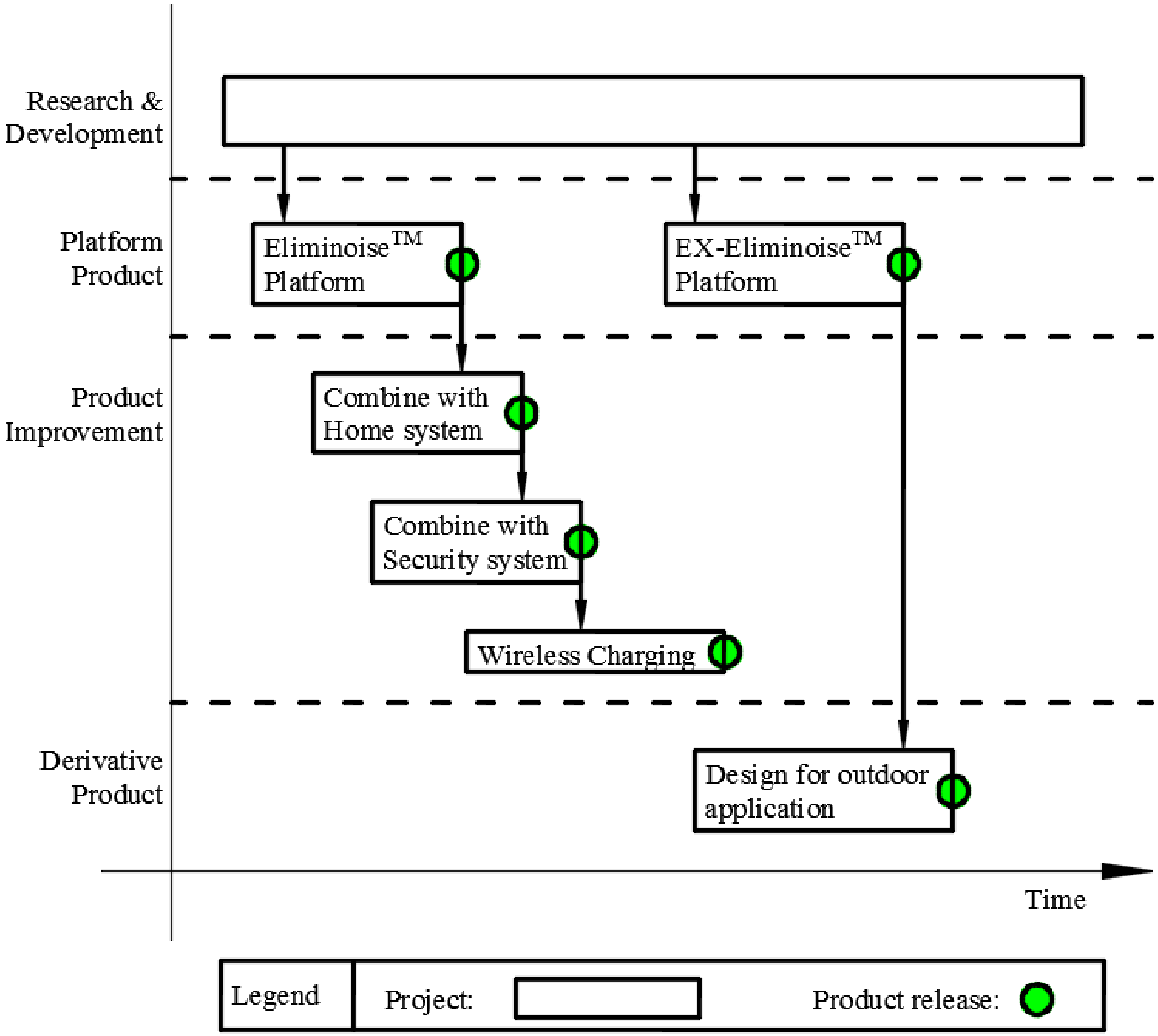

**Figure 16: Fosters Derivative Products**

### 5.1.6 Product Planning: technology roadmap

Based on the previous product design, we have analyzed the importance of developing the common base platform and continue to improve the functions and performance. In the initial design, we focus on the product implementation and stability of our product. For the version 1, we set our noise deduction level to 40dB and the working environment should be inside the house or office. And we also use IR camera to track and analyze the sleeping quality. We also set the initial working hours as 5 years. But with the business increase and based on the marketing requirement, we must upgrade the performance of eliminoise™. For example, increase the noise reduction capability, improve the battery life, and increase the accuracy of sleeping tracking system.

The technology development road map is shown in the below figure, the incremental variation with various technologies employed in variant of eliminoise™.

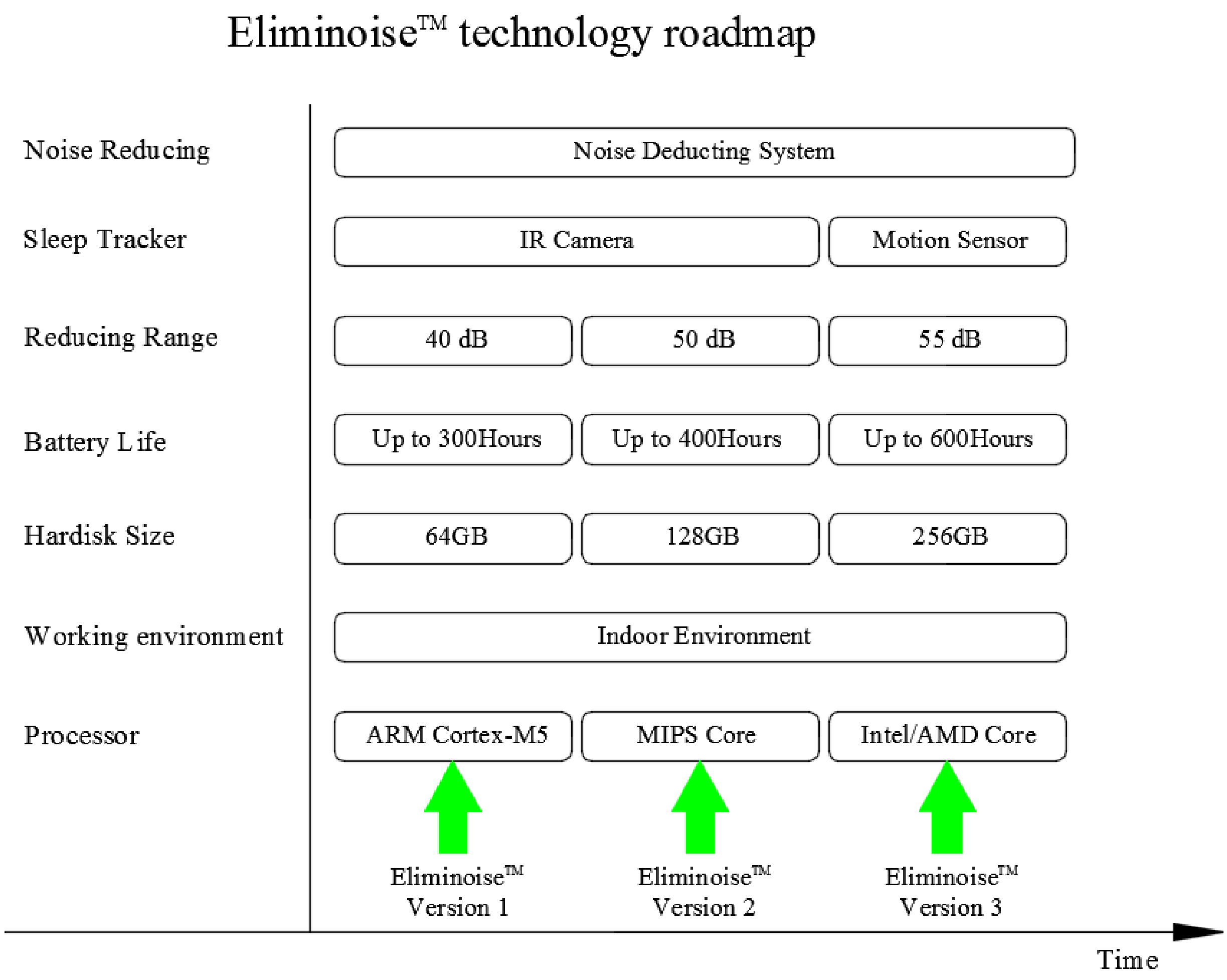

**Figure 17: Technology roadmap**

### 5.1.7 Eliminoise™ Product –Process Matrix

According to the previous analysis, it is very important to balance the development of new project and research cost of the company. In order to make our product have a big marketing share and analyze an overall cost of our project in the future development, we generate the following matrix:

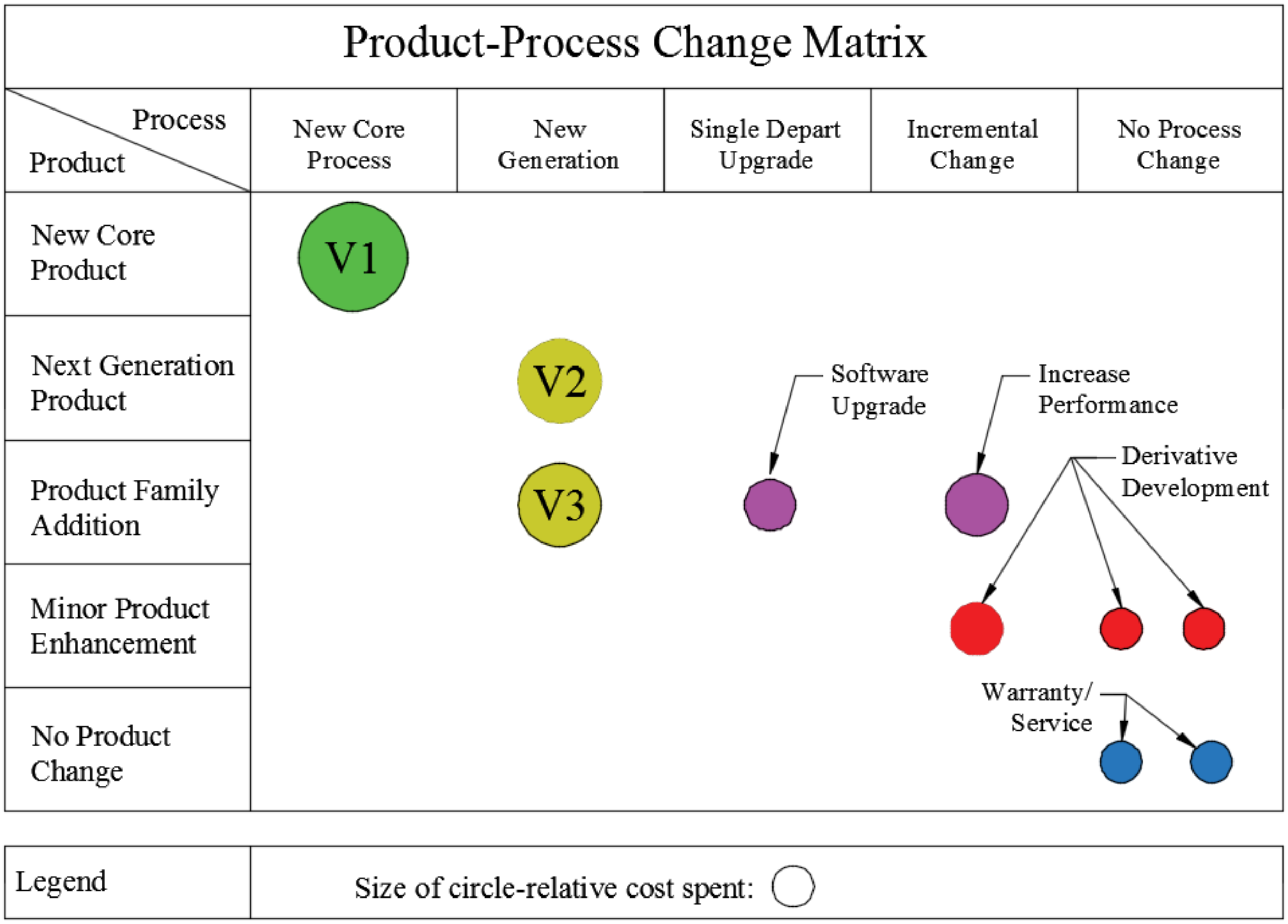

**Figure 18: Process change matrix**

## 5.2 Product architecture

Product architecture is to arrange the different functional elements of a product to the physical building blocks of the product. And the functional elements include the individual operations and transformations. The physical elements of a product contain the parts, components, and subassemblies, and they are typically organized into several major physical building blocks, called chunks. The purpose of the product architecture is to define the basic chunks in terms of what they do and what their interfaces are.

### 5.2.1 Importance of Product architecture

Product architecture decisions affect product change, product variety, component standardization, product performance, manufacturability, and PD management. It should be decided early and drives the design.
Because the different designs need different technologies and different materials and so on, so it also impacts the manufacturing cost and product evolution.
A key characteristic of product architecture is whether it is modular or integral

### 5.2.2 Modular Architecture

As mentioned earlier, the overall function of the product is reviewed and several sub-functions are identified, which need to take place in order to achieve the overall function. Likewise, assemblies and parts must be assigned to carry out these sub-functions and in-turn the whole function. There are principally two types of product development architecture – integral design and modular design.

For our product, we choose the modular architecture. We separated each function which can be achieved by a separate element, which is easy to track the problem and diagnose the system part by part. It will be convenient to integrate with other system in the future development to improve the user experience. Modular design identifies functions (or individual operations) necessary to achieve the overall product purpose. Standard assemblies are then developed to undertake these individual operations. Subsequently, the assemblies or modules are brought together to form the complete product, which can then perform its complete function. The assemblies have standard interfaces with each other. With modular design, assemblies are treated like individual components. Below is the modular architecture for the eliminoise™:

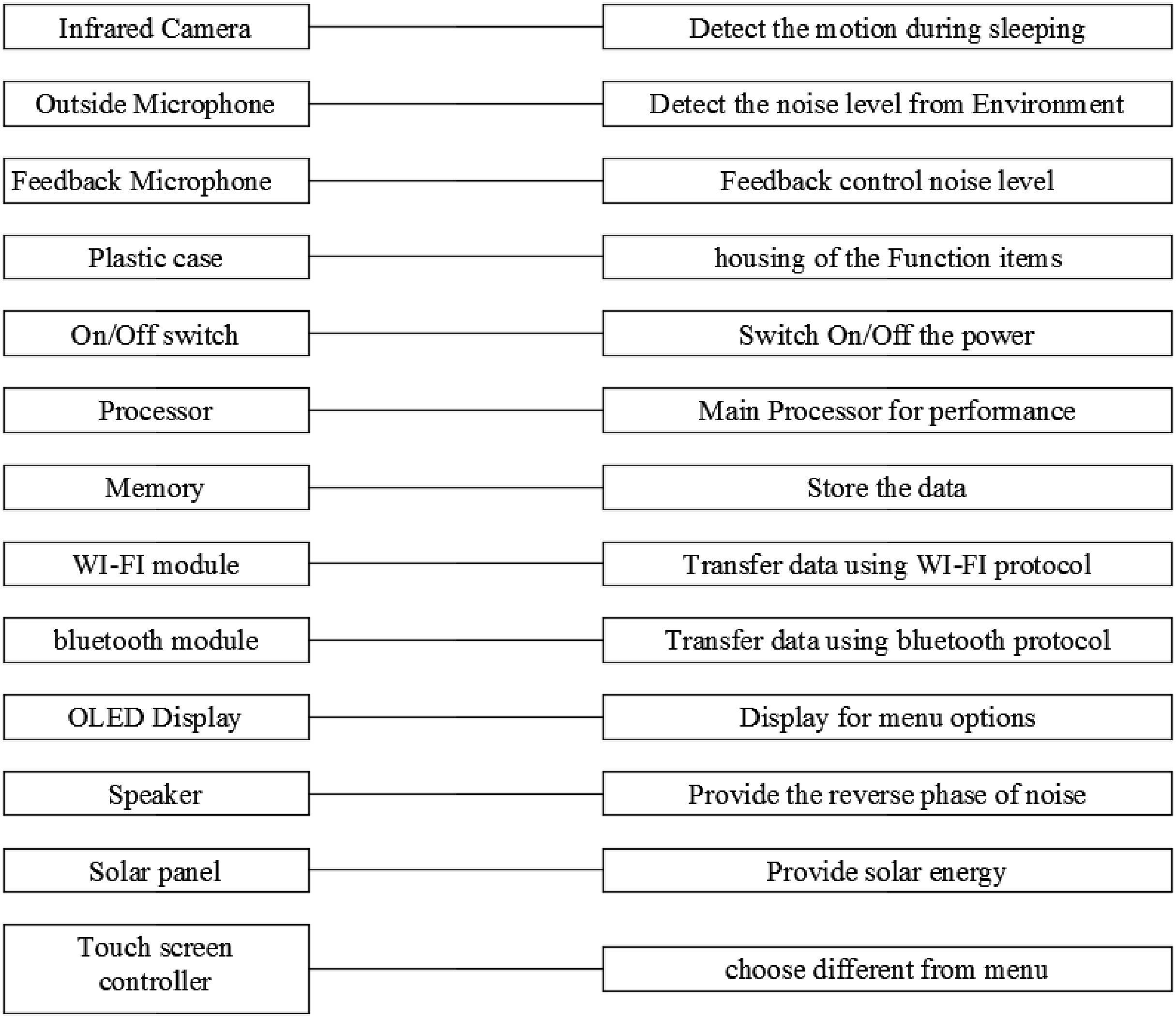

**Figure 19: Modular Architecture for the eliminoise™**

### 5.2.3 The type of modularity

There are some types of modularity for the product architecture, and in order to make our product more intelligent and functional to work for a long period, we decide to use mix-modular architecture to construct the eliminoise™.

This type of modular design to allow components to be removed, upgraded or replaced. Compared with other types, the mix modular architecture can allow to separate the most complicated section and easy for us to do the service and can avoid waste the space to place the components. Below are some additional advantages for Mix Modularity module:

- Easy to do the maintenance and service

- Reduce the assembly process than others
- Minimum the size of elminoise™
- Easy to add the solar panel power supply for the customers
- Relying on modular designs for ease of disassembly of dissimilar recyclable materials

The following figure shows Mix Modularity module:

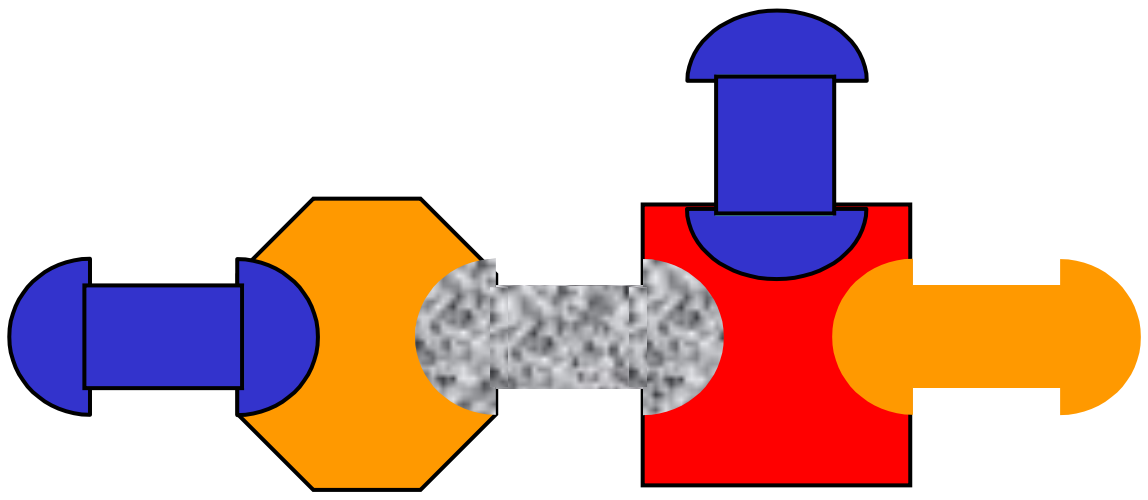

**Figure 20: Modular Architecture for the eliminoise™**

## 5.2.4 Establishing the Architecture

After we have selected the module of eliminoise™ and identified the main functions of the product, we should think about more details and start to establish the architecture using the following steps:

1) Create schematic (illustrating product architecture)
2) Cluster elements
3) Create rough geometric layout
4) Identify fundamental and incidental interactions

**Create schematic**

Functional elements are indeed the functional requirements of the product. And the organization of functional elements can be called as functional structure.

The functional elements mainly include the exchange of signals, materials, force and energy. In a few examples, some elements may not have relationship with other functional elements. There should be less than 30 elements recommended to be used to determine the initial product architecture.

In order to be very clear about our product design functions, we generate the below schematic drawings for the functional arrangement of eliminoise™.

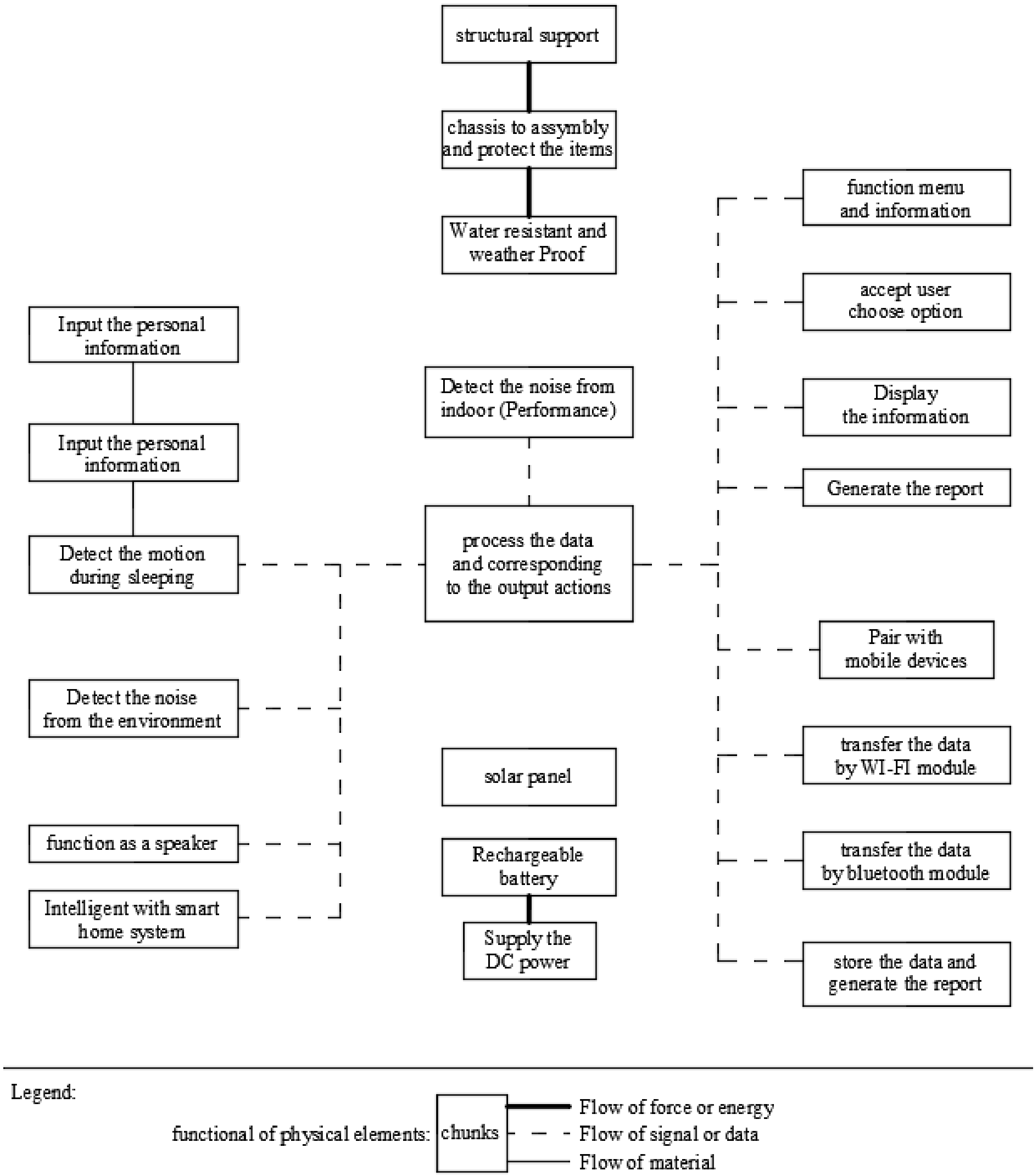

**Figure21: functional arrangement of eliminoise™**

**Cluster elements**

After we completed the schematic drawing, we know the functions and relationship between them. The next step is we need to cluster the elements based on the similar functions, which is very helpful for our new product development. We can understand the geometric integration and precision. We can just combine some similar function to reduce the product size. And we can also enable standardization-portability of the interface between different systems.

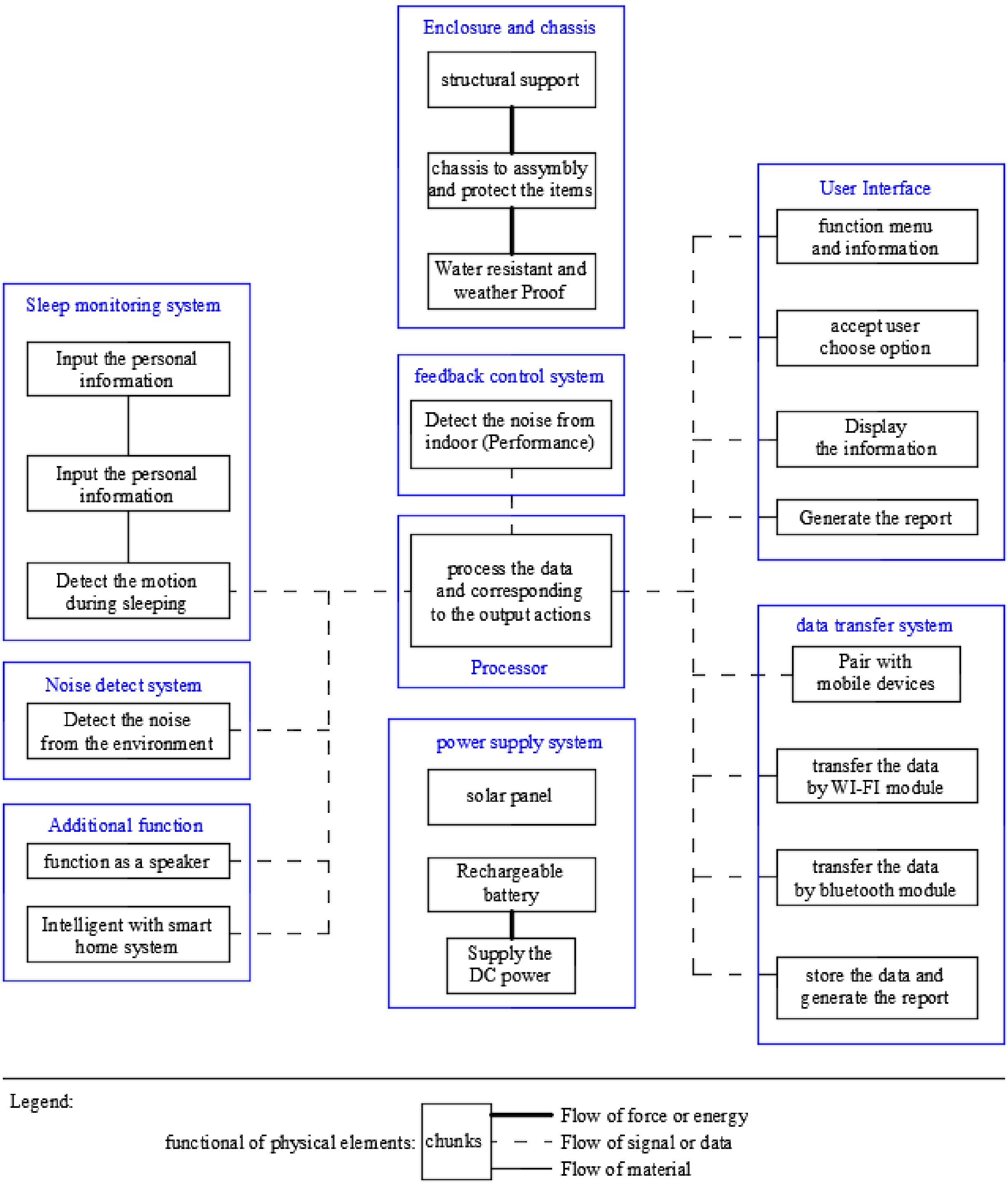

**Figure 22: Cluster elements of eliminoise™**

**Create rough geometric layout**

This procedure is to determine if there is any possibility of geometrical, thermal and / or electrical interfaces between any two components. Although a two-dimensional drawing is enough in many cases, a three-dimensional model may often be preferred. Drawing a geometric layout can force us to determine if the geometric interfaces between the components are feasible or not.

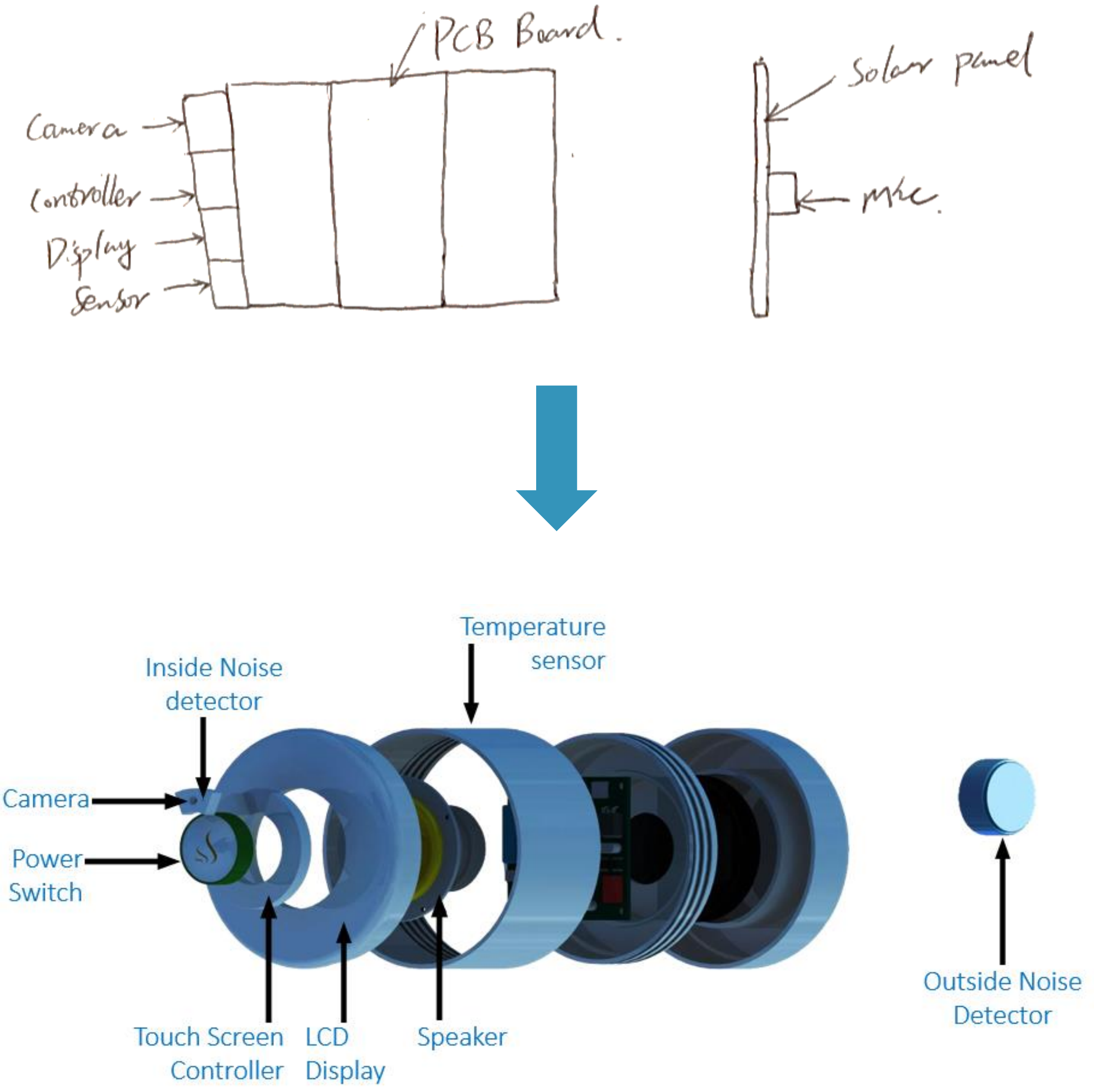

**Figure 23: Geometric layout of Eliminoise™**

**Identify fundamental and incidental interactions**

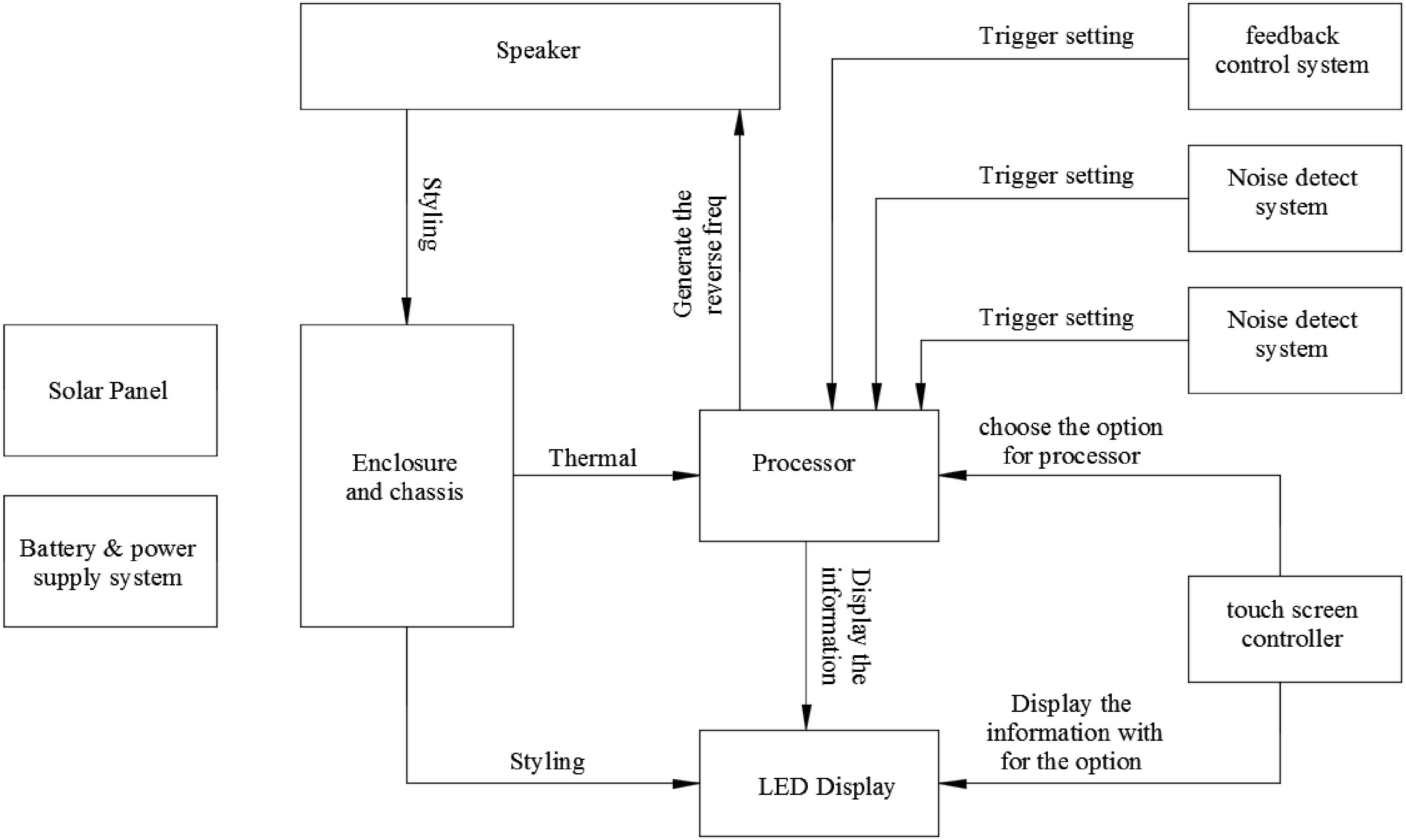

**Figure 24: fundamental and incidental interactions**

# 6. Product Development Economics

## 6.1 Design for Manufacturing (DFM) and Design for Assembly

All potential problems must be identified and handled in the early stages before launching the product into the market. Factors such as the tolerance of each components, dimension of product, and so on, need to be considered in the design stage.

DFM is the design methodology for the ease of manufacturing of the parts collection that will form after assembly the part or system. Ssuccessful implementation of DFM will lower the production cost without having to sacrifice product quality, altering what our customer’s expectation and rely on. Furthermore, the benefits of lower production cost are high profit margins, lower unit cost, and higher sale volumes and increased of customers.

The objective is to integrate both the product design and process planning are to have one common activity. The importance of designing for assembly/manufacturing is being highlighted by the fact that about 70% of assembly/manufacturing costs of a part or system are determined by design decisions, with production decisions responsible for only 20%.

DFM system is a design principles or guidelines that are structured to help the designer reduce the cost and minimized the difficulty of manufacturing an item. Our DFM method consists of five steps plus iteration:

- Estimate the manufacturing costs
- Reduce the Costs of Components and reduce the total number of parts
- Reduce the costs of assembly
- Reduce the costs of supporting production
- Consider the impact of DFM decisions on other factors

To determine DFM, we need to estimate the manufacturing cost (the costs of components, costs of assembly and costs of supporting production). After reviewing the estimated the manufacturing costs, we need to find ways to reduce all these costs. After reducing all the costs, we have to consider the impact of DFM on other factors and recalculate the manufacturing costs. Additionally we also have to determine the recalculating cost is good enough to accept the design.

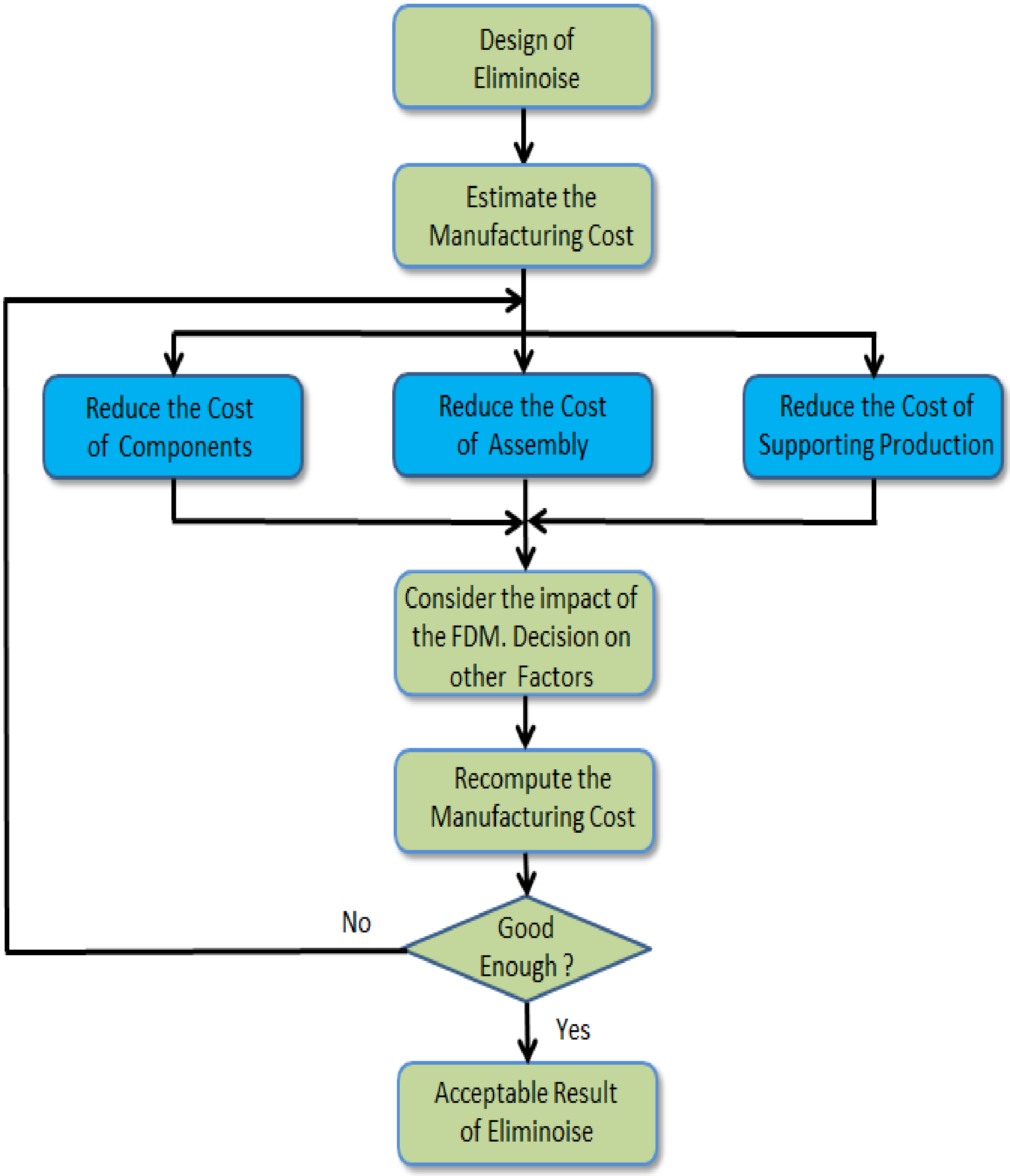

**Figure 25: DFM approach flow chart**

## 6.2 Estimating the Manufacturing Costs

Manufacturing costs can categories into 3 broad:

1) Direct materials cost.

2) Direct labor cost.

3) Manufacturing overhead cost.

In order to start DFM decisions, first we need to have Bill of Materials (BOM) developed to estimate the manufacturing costs.

Below figure is an example of a simple input-output model of a manufacturing system. All input are raw materials, purchased components, employees' efforts, energy, and equipment. All output is finished goods and waste.

Manufacturing cost is accounted for all expenditures for the inputs of the system and disposal of the wastes produced by the system. Companies generally use unit manufacturing cost as the metric of cost for a product and computed by dividing the total manufacturing costs for some period by the number of units of the product manufactured during that period.

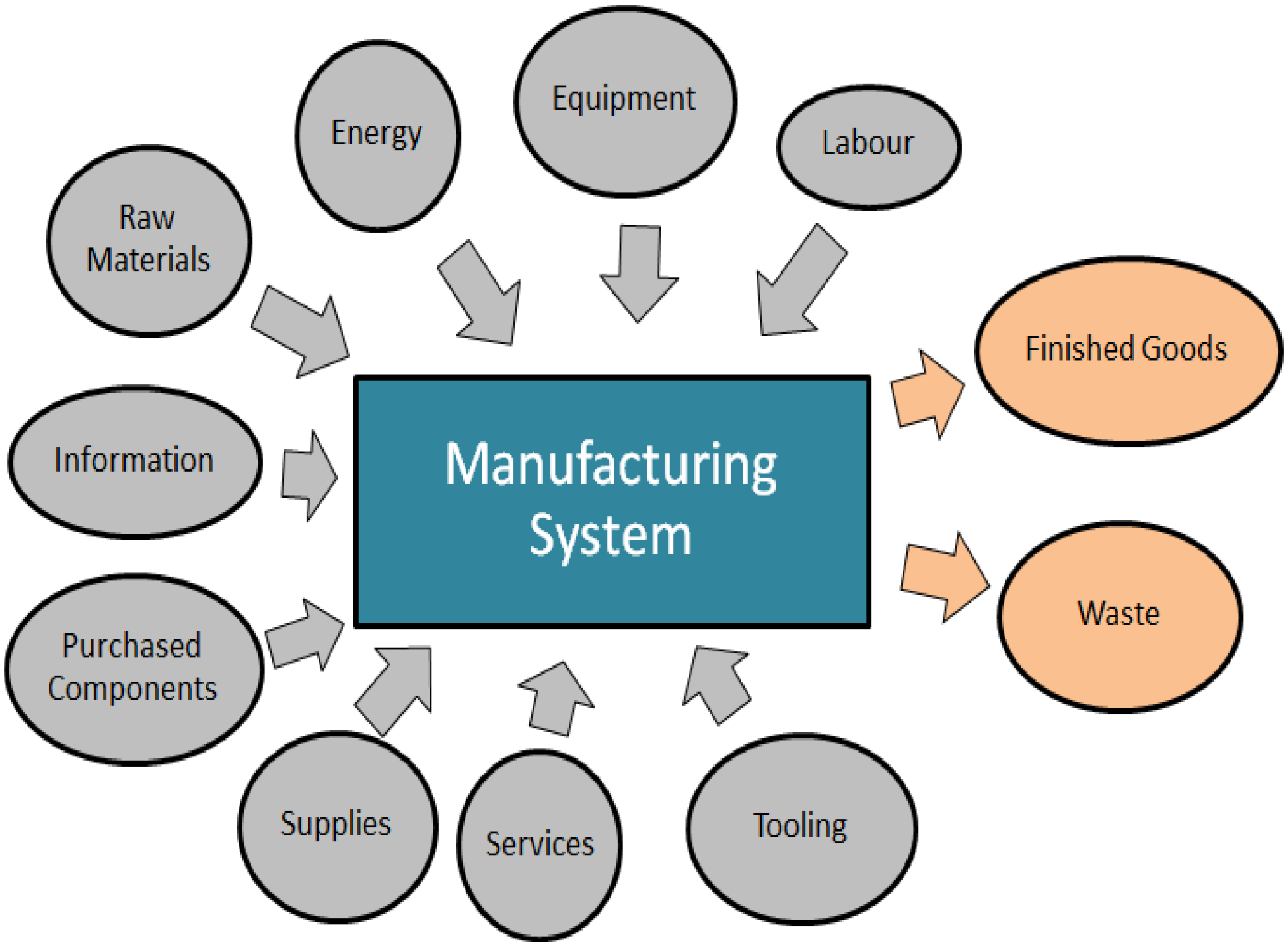

**Figure 26: A simple input-output model of a manufacturing system**

However, the transportation costs are not included in the manufacturing cost. But this cost is an important factor because the transportation of parts to the factory to be assembled and the shipment of product for retail would represent a significant amount in our expenses. This cost will have to be accounted for after the DFM process.

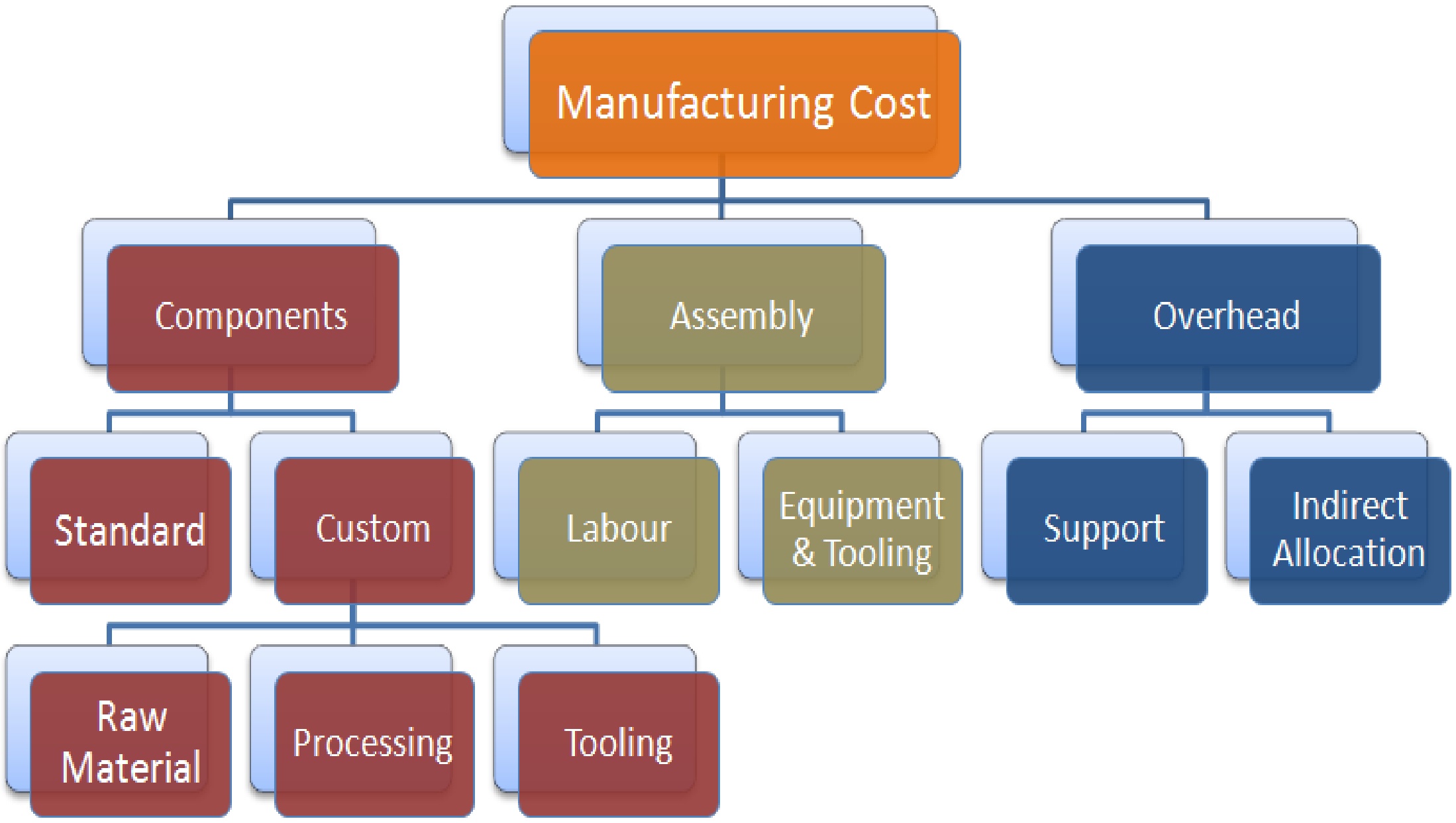

**Figure 27: Elements of the manufacturing cost of a plant**

### 6.2.1 Estimated Component Costs

The component cost determines how the results of material costing are updated and differentiate costs for each material according to cost component such as material costs, internal activities, external activities, and overhead.

The product's components may include as the standard parts purchased from the suppliers. Parts can be electronic chips, straps and screws. Other customizes parts which are made according to the manufacturer’s design from raw materials such as casing, PCB board, etc. These customize parts are made in the manufacturer’s own plant while others may be produced by suppliers according to the manufacturer’s design specifications. All these parts will be e included in the initial Bill of Materials.

During the starting phase, due to lack or insufficient capital, spare parts, machineries and equipment, the process of acquired these components have to be contracted to various external companies. Hence, it will increase additional estimated cost of components by adding up the costs of raw materials, processing and tooling during the manufacturing process. The table below shows their estimated cost.

| Item No | Item Description | Purchase Price for Different Volumes (SGD) | | | Suppliers |
|---|---|---|---|---|---|
| | | 1 | 100 | 1000 | |
| 1 | Wi-Fi RF Transceiver Module | 4.04 | 3.75 | 3 | Advance Technology |
| 2 | Bluetooth RF Transceiver Module | 3.75 | 3.2 | 2.8 | Advance Technology |
| 3 | Lithium-Ion Battery | 13.48 | 10 | 8 | Shenzhen MIKY technology |
| 4 | Flash memory (32G) | 12.5 | 11.5 | 10.5 | LeiZhan |
| 5 | USB port | 5.66 | 4.6 | 3.8 | Solomo |
| 6 | Sleep Tracker sensor (Motion Sensor) | 3.82 | 3.4 | 3 | Teamdewhole Technology Co., Ltd. |
| 7 | Temperature sensor | 2.91 | 2.4 | 2 | A+++ Electronics Maker |
| 8 | 5MP Camera Board Module | 12.12 | 11.3 | 10.5 | Reland Tech |
| 9 | LM393 Sound Detection Sensor | 0.71 | 0.62 | 0.58 | SemiDek Store |
| 10 | Audio Amplifier Module | 1.17 | 1.03 | 0.92 | Feiyang electronics |
| 11 | Microphone Amplifier Sound Sensor Module | 1.35 | 1.2 | 1.08 | STIME Electronics Technology Co., Ltd. |
| 12 | Small Solar Panel | 3.52 | 3.1 | 2.8 | Magic Office |
| 13 | Noise Detector | 15.59 | 13.8 | 12.9 | Real Support Electronic Co.,Ltd. |
| 14 | Speaker | 1.52 | 1.45 | 1.3 | High Fashion Electronics Store |
| 15 | PCB Board | 3.68 | 3.3 | 3 | Teamdewhole Technology Co., Ltd. |
| 16 | Plastic Casing | 6 | 5.5 | 5 | Eletronic Connector & Enclosure World |
| 17 | Push Button Switch | 0.7 | 0.65 | 0.6 | Shenzhen king-Dash Technology Co., LTD |
| 18 | LCD display | 2.53 | 2.4 | 2.25 | E-visiontek Store |
| 19 | Bolts | 0.05 | 0.04 | 0.03 | Summer's Hardware Store |
| 20 | Nuts | 0.05 | 0.04 | 0.03 | Summer's Hardware Store |
| 21 | LEDs | | | | |
| | LED,Red,0603 | 0.14 | 0.1 | 0.07 | Professional semiconductor suppliers |
| | LED,Green,0603 | 0.14 | 0.1 | 0.07 | Professional semiconductor suppliers |
| | LED,Blue,0603, | 0.14 | 0.1 | 0.07 | Professional semiconductor suppliers |
| | LED,Orange,0603 | 0.14 | 0.1 | 0.07 | Professional semiconductor suppliers |
| 22 | ICs | | | | |
| | IC,LD1117STR,SOT-223,ST,(X3 Code 9176138) | 0.35 | 0.28 | 0.21 | Vertical.com |
| | IC,LD1086D2M33TR,D2PAK/A,ST,(X3 Code 9176205) | 0.35 | 0.28 | 0.21 | Vertical.com |
| | IC,ST3241EBPR,SSOP28,ST,(X3 Code 9176312) | 0.35 | 0.28 | 0.21 | Vertical.com |
| | IC,EMIF02-USB03F2,flip ship,ST,(X3 Code 9179037) | 0.35 | 0.28 | 0.21 | Vertical.com |
| | IC,USBULC6-2F3,4FLIPCHIP,ST,(X3 Code9151090) | 0.32 | 0.26 | 0.21 | Vertical.com |
| | IC,STM32F103C8T6(2-2),48-LQFP,ST,(X3 Code9175981) | 0.35 | 0.28 | 0.21 | Vertical.com |
| | IC,LD1117S18TR,SOT-223,ST,(X3 Code9176180) | 0.35 | 0.28 | 0.21 | Vertical.com |
| | IC,ESDA14V2-2BF3,SMT,ST,(X3 Code 9179040) | 0.35 | 0.28 | 0.21 | Vertical.com |
| | IC,MP34DT01TR,HCLGA-4LD,STM | 0.35 | 0.28 | 0.21 | Vertical.com |
| | IC,STMPE1600QTR,QFN24(4X4),STM | 0.35 | 0.28 | 0.21 | Vertical.com |
| | IC,STMPS2151STR@54,SOT-23-5L,STM | 0.35 | 0.28 | 0.21 | Vertical.com |
| 23 | Connectors | | | | |
| | Micro SD card connector,SMT,PJS008-2003-1,MB672,Yamaichi | 4.07 | 2.94 | 2.46 | Mouser Electronics |
| 24 | Resistors | | | | |
| | resistor,0R,0603,5%,YAGEO | 0.15 | 0.1 | 0.05 | Mouser Electronics |
| | resistor,150R,0603,1% | 0.15 | 0.1 | 0.05 | Mouser Electronics |
| | resistor,1.5K,0603,1%,YAGEO | 0.15 | 0.1 | 0.05 | Mouser Electronics |
| | resistor,10K,0603,1% | 0.15 | 0.1 | 0.05 | Mouser Electronics |
| | resistor,100R,0603,1% | 0.15 | 0.1 | 0.05 | Mouser Electronics |
| | resistor,100K,0603,1% | 0.15 | 0.1 | 0.05 | Mouser Electronics |
| | resistor,680R,0603,1% | 0.15 | 0.1 | 0.05 | Mouser Electronics |
| | resistor,510R,0603,1% | 0.15 | 0.1 | 0.05 | Mouser Electronics |
| | resistor,3.3K,0603,1% | 0.15 | 0.1 | 0.05 | Mouser Electronics |
| | resistor,330R,0603,1% | 0.15 | 0.1 | 0.05 | Mouser Electronics |
| | resistor,820R,0603,1% | 0.15 | 0.1 | 0.05 | Mouser Electronics |
| 25 | Capacitors | | | | |
| | capacitor,10nF,0603,50V,10%,X7R,YAGEO; | 0.67 | 0.22 | 0.19 | Mouser Electronics |
| | capacitor,1uF,0603,16V,10%,X5R,YAGEO | 0.67 | 0.22 | 0.19 | Mouser Electronics |
| | capacitor,10pF,5%,50V,NPO,0603,YAGEO | 0.67 | 0.22 | 0.19 | Mouser Electronics |
| | capacitor,22pF,5%,25V,NPO,0603,YAGEO | 0.67 | 0.22 | 0.19 | Mouser Electronics |
| | capacitor,20pF,5%,50V,NPO,0603,YAGEO | 0.67 | 0.22 | 0.19 | Mouser Electronics |
| | capacitor,100nF,10%,16V,X7R,0402,YAGEO | 0.67 | 0.22 | 0.19 | Mouser Electronics |
| | capacitor,100nF,10%,16V,X7R,0402,YAGEO | 0.67 | 0.22 | 0.19 | Mouser Electronics |
| | capacitor,2.2uF,10%,16V,X7R,1206 | 0.67 | 0.22 | 0.19 | Mouser Electronics |
| | capacitor,100nF,10%,16V,X7R,0603,YAGEO | 0.67 | 0.22 | 0.19 | Mouser Electronics |
| 26 | Amplifier LM380N-8/NOPB,Audio Amp Speaker Mono 2.5W | 1.2 | 0.9 | 0.72 | Texas Instruments |
| **27** | **ADXL202JE-Dual-AxisAccelerometer with duty cycle ouptut** | **9.9** | **8** | **5.5** | **Analog Devices** |
| **28** | **2SC732-SILICON NPN TRANSISTOR-date code** | **0.25** | **0.2** | **0.1** | **Toshiba** |
| **29** | **2SC734-UHF BAND, NPN, RF SMALL SIGNAL TRANSISTOR, TO-92** | **1.5** | **0.8** | **0.33** | **Toshiba Electronic Components** |
| **30** | **MCP6024 - low pass filter** | **0.95** | **0.75** | **0.3** | **shen zhen yan xin Technology** |
| **31** | **TSC2046E-Wire Touch Screen Controller with low volt digital I/O** | **1.5** | **1.2** | **0.8** | **Texas Instruments** |

**Table 11: Estimated Costs of the Components**

### 6.2.2 Estimated Assembly Costs

A product is generally assembled from different parts. The process of assembling a product always incurs labor costs, costs for equipment and tooling.

Our product, Elimimoise$^{TM}$ requires various components such as Casing, PCB boards, sensors, amplifiers, LCD, etc to assemble. Assembling of parts required manual labour and the total assembly time are the sum of handling and insertion time.

The estimated assembly cost of the whole process is calculated by multiplying the time of each assembly operation with the cost per hour of the labour required. Using a software tool, Design for Manufacture and Assembly (DFMA®) by Boothroyd Dewhurst Inc. ,we can accurately model the assembly process so that the money and time spent in the assembly process can be optimized in the long run. The table below shows the estimated cost of assembly:

| Parts | Quantity/ Eliminoise™ | Handling Time (s) | Insertion Time (s) | Total Time (s) |
|---|---|---|---|---|
| | | | | |
| LED display panel | 1 | 20 | 25 | 45 |
| PCB board | 1 | 20 | 25 | 45 |
| Amplifiers | 5 | 30 | 120 | 150 |
| Sensors | 5 | 30 | 120 | 150 |
| Plastic casing | 1 | 10 | 60 | 70 |
| Controllers | 5 | 30 | 120 | 150 |
| Miscellaneous (ICs, Resistors, Capacitors, LEDs, Connectors, RAM, Solar Panel & USB port | 38 | 550 | 680 | 1230 |
| Total Time (s) | | | | 1840 |
| Assembly Cost at $10/hour | | | | $ 5.15 |

**Table 12: Estimated Cost of Assembly**

### 6.2.3 Estimated Overhead Costs

Manufacturing a product required power, supplies and employees whose functions are essential to the operation even though they aren't part of the manufacturing process itself. These are the manufacturing overhead costs. All these support costs used in materials handling, quality assurance, purchasing, shipping, receiving, facilities, and equipment/tooling maintenance are required to manufacture the product, and the product design will directly influence these costs. Indirect costs are costs shared during the manufacturing process; they include any expenses that are not directly tied to the specific operation. For example, the salary of the cleaner and the maintenance cost of the building and grounds are indirect costs. These activities are shared among several different products and difficult to allocate directly to a specific product. Although indirect cost contribute to the cost of the product, but it is not relevant in DFM

The table below shows the estimated overhead costs:

| Cost Drivers | | Overhead Rates | Overhead Costs |
|---|---|---|---|
| Purchased materials | $94.21 | 10% | $9.42 |
| Assembly labor | $5.15 | 80% | $4.12 |
| Total Overhead Cost | | | $13.54 |

**Table 13: Estimated Overhead Costs**

### 6.2 .4 Transportation and Logistic Costs

The operation of transportation determines the efficiency of moving products. The progress in techniques and management principles improves the moving load, delivery speed, service quality, operation costs, the usage of facilities and energy saving. Transportation takes a crucial part in the manipulation of logistic. Reviewing the current condition, a strong system needs a clear frame of logistics and a proper transport implements and techniques to link the producing procedures. Although the table above does not include the logistics and transportation cost, it is still part of the COST minimization process. Based on the rates on volume and weight, the product design team can easily include transportation costs in its analysis, and by doing so it may be warranted when the team faces design decisions involving the physical volume or weight of the product.

### 6.2 .5 Fixed Costs versus Variable Costs

All manufacturing costs can be broken into two main categories:

1) Fixed costs - Fixed costs are costs that are independent of output. These remain constant throughout the relevant range and are usually considered sunk for the relevant range (not relevant to output decisions). Fixed costs often include rent, buildings, machinery, etc.

2) Variable costs - Variable costs are costs that vary with output. Generally variable costs increase at a constant rate relative to labor and capital. Variable costs may include wages, utilities, materials used in production, etc. When comparing fixed costs to variable costs, or when trying to determine whether a cost is fixed or variable, simply ask whether or not the particular cost would change if the company stopped its production or primary business activities. If the company would continue to incur the cost, it is a fixed cost. If the company would no longer incur the cost, then it is most likely a variable cost.

## 6.2.6 Estimating Manufacturing Costs - Bill of Materials (BOM)

| Component | Qty required | Purchased Costs | Processing | Assembly (labor) | Total Unit Variable | Suppliers |
|---|---|---|---|---|---|---|
| **Component** | | | | | | |
| Wi-Fi RF Transceiver Module | 1 | 3 | 0.3 | 0.2 | 3.5 | Advance Technology |
| Bluetooth RF Transceiver Module | 1 | 2.8 | 0.3 | 0.2 | 3.3 | Advance Technology |
| Lithium-Ion Battery | 1 | 8 | 0.2 | 0.1 | 8.3 | Shenzhen MIKY technology |
| Flash memory (32G) | 1 | 10.5 | 0.5 | 0.3 | 11.3 | LeiZhan |
| USB port | 1 | 3.8 | 0.2 | 0.1 | 4.1 | Solomo |
| Sleep Tracker sensor (Motion Sensor) | 1 | 3 | 0.2 | 0.1 | 3.3 | Teamdewhole Technology Co., Ltd. |
| Temperature sensor | 1 | 2 | 0.2 | 0.1 | 2.3 | A+++ Electronics Maker |
| 5MP Camera Board Module | 1 | 10.5 | 0.2 | 0.1 | 10.8 | Reland Tech |
| LM393 Sound Detection Sensor | 1 | 0.58 | 0.2 | 0.1 | 0.88 | SemiDek Store |
| Audio Amplifier Module | 1 | 0.92 | 0.25 | 0.15 | 1.32 | Feiyang electronics |
| Microphone Amplifier Sound Sensor Module | 1 | 1.08 | 0.25 | 0.15 | 1.48 | STIME Electronics Technology Co., Ltd. |
| Small Solar Panel | 1 | 2.8 | 0.3 | 0.2 | 3.3 | Magic Office |
| Noise Detector | 1 | 12.9 | 0.5 | 0.4 | 13.8 | Real Support Electronic Co.,Ltd. |
| Speaker | 1 | 1.3 | 0.3 | 0.2 | 1.8 | High Fashion Electronics Store |
| PCB Board | 1 | 3 | 0.3 | 0.2 | 3.5 | Teamdewhole Technology Co., Ltd. |
| Plastic Casing | 1 | 5 | 0.4 | 0.3 | 5.7 | Eletronic Connector & Enclosure World |
| Push Button  Switch | 1 | 0.6 | 0.2 | 0.1 | 0.9 | Shenzhen king-Dash Technology Co., LTD |
| LCD display | 1 | 2.25 | 0.4 | 0.3 | 2.95 | E-visiontek Store |
| Bolts | 8 | 0.24 | 0.1 | 0.05 | 0.39 | Summer's Hardware Store |
| Nuts | 8 | 0.24 | 0.1 | 0.05 | 0.39 | Summer's Hardware Store |
| LEDs | 6 | 0.42 | 0.2 | 0.1 | 0.72 | Professional semiconductor suppliers |
| Amplifier -LM380N-8/NOPB,Audio Amp Speaker 1-CH Mono 2.5W | 1 | 0.72 | 0.1 | 0.1 | 0.92 | Texas Instruments |
| ADXL202JE-Dual-AxisAccelerometer with duty cycle output | 1 | 5.5 | 0.3 | 0.3 | 6.1 | Analog Devices |
| 2SC732-SILICON NPN TRANSISTOR-date code | 1 | 0.1 | 0.1 | 0.1 | 0.3 | Toshiba |
| MCP6024 - low pass filter | 2 | 0.3 | 0.1 | 0.1 | 0.5 | Toshiba Electronic Components |
| 2SC734-UHF BAND, Si, NPN, RF SMALL SIGNAL TRANSISTOR, TO-92 | 1 | 0.33 | 0.1 | 0.15 | 0.58 | shen zhen yan xin Technology |
| LM380N-8/NOPB-Audio Amp Speaker | 2 | 0.72 | 0.1 | 0.1 | 0.92 | Texas Instruments |
| ICs | 10 | 2.1 | 0.3 | 0.2 | 2.6 | Vertical.com |
| Connectors | 1 | 2.46 | 0.4 | 0.3 | 3.16 | Mouser Electronics |
| Resistors | 30 | 1.5 | 0.2 | 0.1 | 1.8 | Mouser Electronics |
| Capacitors | 25 | 4.75 | 0.2 | 0.1 | 5.05 | Mouser Electronics |
| TSC2046E-Wire Touch Screen Controller with low voltage digital I/O | 1 | 0.8 | 0.1 | 0.1 | 1 | Texas Instruments |
| Transport | | | | | | |
| Shipment | | | | | $ 0.20 | |
| Total Direct Cost | | 94.21 | 7.6 | 5.15 | $ 107.16 | |
| Overhead  Charges | | | | | $13.54 | |
| Warranty Cost | | | | | $ 0.32 | |
| Total Manufacturing Cost | | | | | $ 121.02 | |

**Table 14: Table Bill of Material (BOM)**

## 6.3 Reduce the Costs of Manufacturing

### 6.3.1 Reduce the Costs of Components and the total number of parts

The cost of the materials is probably one of the largest expenses, directly affecting profitability. There are a few ways to reduce the cost of components such as lesser components used in the product, purchase components in bulk, eliminate unnecessary product features, and also reduced by outsourcing rather than manufactured by own factory.

Probably the best way to reduce manufacturing costs is by reduce the number of parts in a product. Fewer parts will have lesser purchase, inventory, handling, processing time, development time, equipment, engineering time, assembly difficulty, service inspection, testing and manpower.

In general, it will reduce the level of intensity of all of the activity that is related to the product. One of the main focus of part reduction are based on the using one-piece structures and selection of manufacturing processes such as injection molding, extrusion, precision castings, and powder metallurgy.

#### 6.3.1.1 Develop a modular design

Modular product design simplifies manufacturing activities such as inspection, testing, assembly, purchasing, redesign, maintenance, service. The advantages of modular design are numerous. Some examples are listed below:

- Minimizing cost, by reducing the diversity of parts in a product range
- Savings in design time, as assemblies/modules are simply selected like bought out parts, as their reliability, cost and quality are documented and easily available.
- Modules can be modified or replaced without changing anything else on the product.
- Modular design simplifies the information processing in a design project.
- Easy and quick installation of products
- Easy and quick servicing and maintenance of products

Modular design add versatility to product update in the redesign process, help run tests before final assembly is put together, and allow the use of standard components to minimize product variations.

#### 6.3.1.2 Use of standard components

A standard component is a **pre-prepared part** that is used in the production of the product. The advantage of using a standard components are it can ensures consistency, saves time and effort, reduce costs, and also the components can be bought in bulk. Standard components are less expensive than custom-made items. The high availability and bulk manufactured of these components reduces product lead times and cost. Furthermore, the use of standard components refers to the production pressure to the supplier, relieving in part the manufacture’s concern of meeting production schedules.

#### 6.3.1.3 Design parts for multi-use

In a manufacturing firm, in order to increase repetitiveness, different products can share parts that have been designed for multi-use. In order to do this, parts that have the same or different functions when used in different products is identified as the parts that are suitable for multi-use. Then, parts are categories to similar parts in each group. The goal is to minimize the number of categories, the variations within the categories, and the number of design features within each variation After categorizes all the parts into different part families, the manufacturing processes are standardized for each part family. Specific part will be assigned to a given part family that has been setup for its family, skipping the operations that are not required for it. Furthermore, in design changes to existing products and especially in new product designs, the standard multi-use components should be used.

#### 6.3.1.4 Design for ease of fabrication

By selecting the processes compatible with the materials and fabrication process combination will minimize the overall manufacturing cost while meeting functional requirements. Final operations such as painting, polishing, finishing machining can be avoided to save cost. Excessive tolerance, surface-finish requirement are commonly found problems that will surface and result in higher than necessary production cost.

#### 6.3.1.5 Avoid separate fasteners

Due to the handling and feeding operations, the use of fasteners increases the cost of manufacturing. Besides the high cost of the equipment required for them, the fastening processes are not 100% successful thus the efficiency will drop. Avoid screws that are too long, too short, separate washes, tapped holes, and road heads and flatheads. Self-tapping and chamfered screws are preferred because they are instant and improve placement success.

### 6.3.2 Reduce the Costs of Assembly

#### 6.3.2.1 Keeping Score

Boothroyd Dewhurst Method involves systematically looking at the part geometry and how they're fastened to estimate a time to put the different parts together and also maintaining an ongoing estimate of the cost of assembly. This is measured as an index that is the ratio of the theoretical minimum assembly time to an estimate of the actual assembly time for the product. This concept is useful in driving down the cost of assembly.

Below is the expression for the DFA index is:

$$\text{DFA index} = \frac{\text{(Theoretical minimum number of parts) X (3 seconds)}}{\text{Estimated total assembly time}}$$

#### 6.3.2.2 Minimize assembly directions

It is the designer's role to ensure that assembly can be completed in the most efficient way. All parts should be assembled from one direction. If possible, the best way to add parts is from above, in a vertical direction, parallel to the gravitational direction. Using the effect of gravity, it can help to minimize energy used by the labor workers, maximize work done and increase in efficiency.

### 6.3.3 Reduce the Costs of Supporting Production

#### 6.3.3.1 Maximize compliance

Maximize compliance is sometimes best achieved through the use of a variety of tools that are interdependent and that also complement investigations and systemic actions. The aim is clearly to prevent non-compliance and to facilitate compliance. In many situations, due to variations in part dimensions or on the accuracy of the positioning device used, errors can occur during insertion operations. This error or faulty behavior can cause damage to the part and to the equipment. Therefore, it is necessary to include compliance in the part design and in the assembly process. A simple solution is to use high-quality parts with designed-in-compliance, a rigid-base part, and selective compliance in the assembly tool.

#### 6.3.3.2 Minimize handling

Handling comprises on positioning, orienting, and fixing a part or components. To minimize non-value-added manual effort and ambiguity in orienting and merging parts, use external guiding features to help the orientation of a part. During parts designing, minimize the flow of material waste and also select appropriate and safe packaging for the product. Parts must be designed to consistently orient themselves when fed into a process. Avoid using flexible parts, use slave circuit boards instead, and also avoid heavy parts that will increase worker fatigue, increase risk of worker injury, and slow the assembly process.

### 6.3.4 The impact on DFM on Development Time

Product development time in any manufacturing firm is directly linked to the overall performance and profits of the organization. DFM plays a very important role for their impact on development time as well as the impact on manufacturing cost. Manufacturers will have better insights about their designed product and modifications can be done without compromising with the time-to-market. Thus, Companies can gain better position in the market and initiate their marketing and sales activities early, allowing to realize their revenue faster.

#### 6.3.4.1 The impact on DFP on Development Cost

At the beginning of product development, the requirement has many unknown factors and the product design has uncertainty problem. That led to the uncertainty of product's cost. Design for production (DFP) evaluates manufacturing system performance as a function of product design variables. In general, by aggressively pushing for low manufacturing costs as an integral part of the development process, it will able to develop the products in at the same time and with about the same budget used.

#### 6.3.4.2 The impact on DFP on Product Quality

Design for production (DFP) can advise a product development team to consider changing the product design to avoid problems or improve profitability. In addition, DFP can provoke suggestions to improve the existing manufacturing system, this will help DFM to focus  primarily on manufacturing cost reduction to also result in improved serviceability, ease of disassembly, and recycling, hence, this will also improve the product quality.

## 6.4 Revised Bill of Materials

### 6.4.1 Reduction Costs of Components

By reducing the cost of the some expensive components, we manage to bring down the cost of the product from $121.02 to $92.50. A cost saving is $28.52, which is 23.56%. No doubt, we have replaced some components with cheaper price, but we have done our research on the quality these components to ensure it will not affect the performance of our product.

<table>
<tr><th>Item Description</th><th>Cost reduction method</th><th>Previous cost</th><th>New cost</th><th>Savings</th></tr>
<tr><td>Wifi RF Transceiver Module</td><td rowspan="2">WiFi + Bluetooth RF Transceiver Module</td><td>3</td><td rowspan="2">2.5</td><td rowspan="2">3.3</td></tr>
<tr><td>Bluetooth RF Transceiver Module</td><td>2.8</td></tr>
<tr><td>LCD display</td><td>Outsource to different supplier</td><td>2.25</td><td>1.8</td><td>0.45</td></tr>
<tr><td>Flash RAM (32MB)</td><td>Outsource to different supplier</td><td>10.5</td><td>4.8</td><td>5.7</td></tr>
<tr><td>Lithium-Ion Battery</td><td>Outsource to different supplier</td><td>8</td><td>3.2</td><td>4.8</td></tr>
<tr><td>5MP Camera Board Module</td><td>Change to 2MP Camera Board Module</td><td>10.8</td><td>2.8</td><td>8</td></tr>
<tr><td>Plastic Casing</td><td>Outsource to different supplier</td><td>5</td><td>2.3</td><td>2.7</td></tr>
<tr><td>Connectors</td><td>Outsource to different supplier</td><td>2.46</td><td>1.21</td><td>1</td></tr>
</table>

| | | | | |
|---|---|---|---|---|
| Noise Detector | Outsource to different supplier | 12.9 | 10.33 | 2.57 |
| Total | | | | $ 28.52 |

**Table 15: Reduction of cost of components**

### 6.4.2 Finalize Bill of Materials (BOM)

Below are the finalize Bill of Materials (BOM) after reduction cost of Components.

| Component | Qty required | Purchased Costs | Processing | Assembly (labor) | Total Unit Variable | Suppliers |
|---|---|---|---|---|---|---|
| Component | | | | | | |
| Wi-Fi - Bluetooth RF Transceiver Module | 1 | 2.5 | 0.3 | 0.2 | 3 | Shenzhen CAIZHIXING Electronic |
| Lithium-Ion Battery | 1 | 3.2 | 0.2 | 0.1 | 3.5 | Shenzhen MIKY technology |
| Flash memory (32G) | 1 | 4.8 | 0.5 | 0.3 | 5.6 | ViLonHK Store |
| USB port | 1 | 3.8 | 0.2 | 0.1 | 4.1 | Solomo |
| Sleep Tracker sensor (Motion Sensor) | 1 | 3 | 0.2 | 0.1 | 3.3 | Teamdewhole Technology Co., Ltd. |
| Temperature sensor | 1 | 2 | 0.2 | 0.1 | 2.3 | A+++ Electronics Maker |
| 2MP Camera Board Module | 1 | 2.8 | 0.2 | 0.1 | 3.1 | HKESCCTV STORE |
| LM393 Sound Detection Sensor | 1 | 0.58 | 0.2 | 0.1 | 0.88 | SemiDek Store |
| Audio Amplifier Module | 1 | 0.92 | 0.25 | 0.15 | 1.32 | Feiyang electronics |
| Microphone Amplifier Sound Sensor Module | 1 | 1.08 | 0.25 | 0.15 | 1.48 | STIME Electronics Technology Co., Ltd. |
| Small Solar Panel | 1 | 2.8 | 0.3 | 0.2 | 3.3 | Magic Office |
| Noise Detector | 1 | 10.33 | 0.5 | 0.4 | 11.23 | STIME Electronics Technology Co., Ltd. |
| Speaker | 1 | 1.3 | 0.3 | 0.2 | 1.8 | High Fashion Electronics Store |
| PCB Board | 1 | 3 | 0.3 | 0.2 | 3.5 | Teamdewhole Technology Co., Ltd. |
| Plastic Casing | 1 | 2.3 | 0.4 | 0.3 | 3 | The Plastic Forming Company |
| Push Button Switch | 1 | 0.6 | 0.2 | 0.1 | 0.9 | Shenzhen king-Dash Technology Co., LTD |
| LCD display | 1 | 2.25 | 0.4 | 0.3 | 2.95 | E-visiontek Store |
| Bolts | 8 | 0.24 | 0.1 | 0.05 | 0.39 | Summer's Hardware Store |
| Nuts | 8 | 0.24 | 0.1 | 0.05 | 0.39 | Summer's Hardware Store |
| LEDs | 6 | 0.42 | 0.2 | 0.1 | 0.72 | Professional semiconductor suppliers |
| Amplifier -LM380N-8/NOPB,Audio Amp Speaker 1-CH Mono 2.5W | 1 | 0.72 | 0.1 | 0.1 | 0.92 | Texas Instruments |
| ADXL202JE-Dual-AxisAccelerometer with duty cycle output | 1 | 5.5 | 0.3 | 0.3 | 6.1 | Analog Devices |
| 2SC732-SILICON NPN TRANSISTOR-date code | 1 | 0.1 | 0.1 | 0.1 | 0.3 | Toshiba |
| MCP6024 - low pass filter | 2 | 0.3 | 0.1 | 0.1 | 0.5 | Toshiba Electronic Components |
| 2SC734-UHF BAND, Si, NPN, RF SMALL SIGNAL TRANSISTOR, TO-92 | 1 | 0.33 | 0.1 | 0.15 | 0.58 | shen Zhen yan xin Technology |
| LM380N-8/NOPB-Audio Amp Speaker | 2 | 0.72 | 0.1 | 0.1 | 0.92 | Texas Instruments |
| ICs | 10 | 2.1 | 0.3 | 0.2 | 2.6 | Vertical.com |
| Connectors | 1 | 1.21 | 0.4 | 0.3 | 1.91 | Digi-Key.com |
| Resistors | 30 | 1.5 | 0.2 | 0.1 | 1.8 | Mouser Electronics |
| Capacitors | 25 | 4.75 | 0.2 | 0.1 | 5.05 | Mouser Electronics |
| TSC2046E-Wire Touch Screen Controller with low voltage digital I/O | 1 | 0.8 | 0.1 | 0.1 | 1 | Texas Instruments |
| Transport | | | | | | |
| Shipment | | | | | $ 0.20 | |
| Total Direct Cost | | 66.46 | 7.3 | 4.95 | $ 78.91 | |
| Overhead Charges | | | | | $13.54 | |
| Warranty Cost | | | | | $ 0.32 | |
| Total Manufacturing Cost | | | | | $ 92.50 | |

**Table 16: Finalize Bill of Materials**

## 6.5 Product Development Economics

Industry competitiveness always look out for the best possible quality–price combination, hence, cost analysis have been used during the early stages of product design. At the product and process design stages off–line quality control methods are conducted to improve manufacturing costs and quality of the product. There are several basic cash inflows (revenues) and cash outflows (costs) in the life cycle of a successful new product. Basic cash inflows come from product sales. Cash outflows consists of spending on product and process development; costs of production ramp-up such as equipment purchases and tooling; costs of marketing and supporting the product; and ongoing production costs such as raw materials, components, and labor. The cumulative cash inflows and outflows over the life cycle of a typical successful product are shown schematically as below.

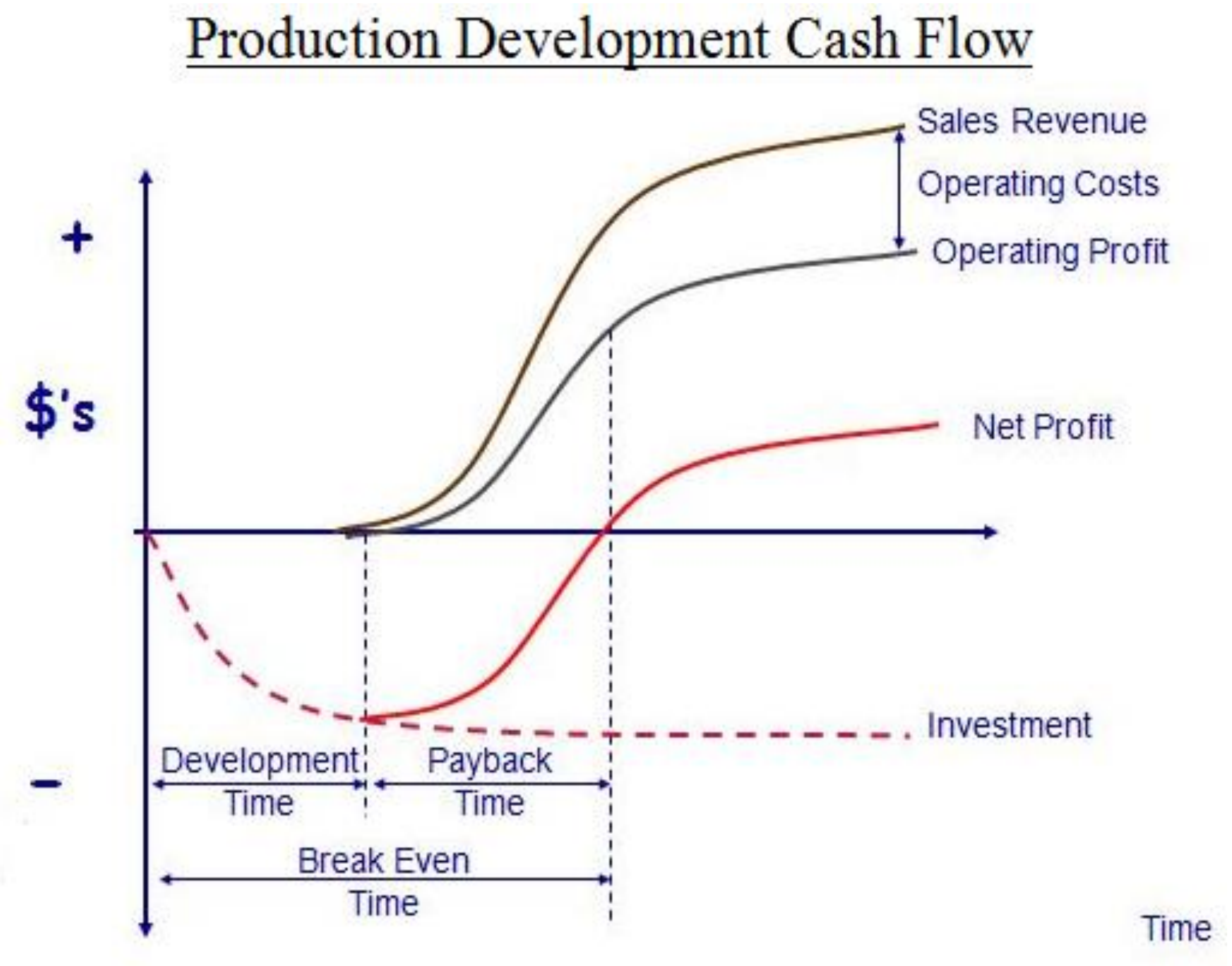

**Figure 28: Typical cash flows for new product**

When the product generate more cumulative inflows than cumulative outflows, it presents the products are profitable. This economic analysis method uses Net Present Value (NPV) techniques because NPV is one of the most used techniques. Net Present Value (NPV), defined as the present value of the future net cash flows from an investment project, is one of the main ways to evaluate an investment. The quantitative part of the economic analysis method is the estimated NPV of the project's expected cash flows. The value of quantitative analysis often reduces and restructures a complex problem to a limited numbers of variables and also in bringing a more reliable and objective of product development projects.

Internal rate of return (IRR) is a metric used in capital budgeting measuring the profitability of potential investments. Internal rate of return is a discount rate that makes the net present value (NPV) of all cash flows from a particular project equal to zero. IRR calculations rely on the same formula as NPV does. Internal rate of return able to provides a different view as the internal discount rate of the Company can be miscalculated, so the IRR (higher the IRR, the better) will be able to give insight as to how good a project really is.

## 6.5.1 Element of Economic Analysis and Qualitative Analysis

### 6.5.1.1 Qualitative Analysis

Qualitative research gathers information that is not in numerical form. Qualitative data is typically descriptive data and as such is harder to analyze than quantitative data. Qualitative research is useful for studies at the individual level, and to find out, in depth, the ways in which people think or feel. Analysis of qualitative data is difficult and requires an accurate description of participant responses, for example, sorting responses to open questions and interviews into broad themes. A good example of a qualitative research method would be unstructured and group interviews which generate qualitative data through the use of open questions. This allows the respondent to talk in some depth, choosing their own words. This helps the researcher develop a real sense of a person's understanding of a situation. However, it can be time consuming to conduct the unstructured interview and analyze the qualitative data.

### 6.5.1.2 Economic Analysis Process

The following four-step method for the economic analysis of a product development project:

1) Build a base-case financial model
2) Perform a sensitivity analysis to understand the relationships between financial success and the key assumptions and variables of the model
3) Use the sensitivity analysis to understand project trade-offs
4) Consider the influence of the qualitative factors on project success

## 6.5.2 Build a Base-Case Financial Model

### 6.5.2.1 Estimating the Timing and Magnitude of Future Cash Inflows and Outflows

There are several basic categories of cash flow for a typical new product development project as shown below:

- Development cost
- Testing cost
- Tooling cost
- Ramp-up cost
- Marketing instruction cost
- Ongoing Marketing Costs
- Production cost

Depending on the types of decisions the model will support, greater levels of details for one or more areas may be required. More detailed modelling may consider these same five cash flows in greater detail, or it may consider other flows. Typical refinements include:

- Breakdown of production costs into direct costs and indirect costs. (i.e., overhead)
- Breakdown of marketing and support costs into launch costs, promotion costs, direct sales costs, and service costs
- Inclusion of tax effects, including depreciation and investment tax credits
- Inclusion of such miscellaneous inflows and outflows as working capital requirements, cannibalization, salvage costs, and opportunity costs

### 6.5.2.2 Compute the Net Present Value of the Cash Flows and Internal Rate of Return

To determine the value of a project is challenging because there are many several of method to measure the value of future cash flows available in the market. We're choosing to use Net Present Value (NPV) and Internal Rate of Return (IRR) as our financial analysis method in this project.

I ) Net present value (NPV) is a method of determining the current value of all future cash flows generated by a project after accounting for the initial capital investment. It is widely used in capital budgeting to establish which projects are likely to turn the greatest profit.

Below are the inputs for NPV Base Case:

- Development cost and timing
- Testing cost and timing
- Tooling investment and timing
- Ramp-up cost and timing
- Marketing and support cost and timing
- Sales volume and lifetime
- Unit production cost
- Unit revenue
- Discount rate

The following is the formula for calculating NPV:

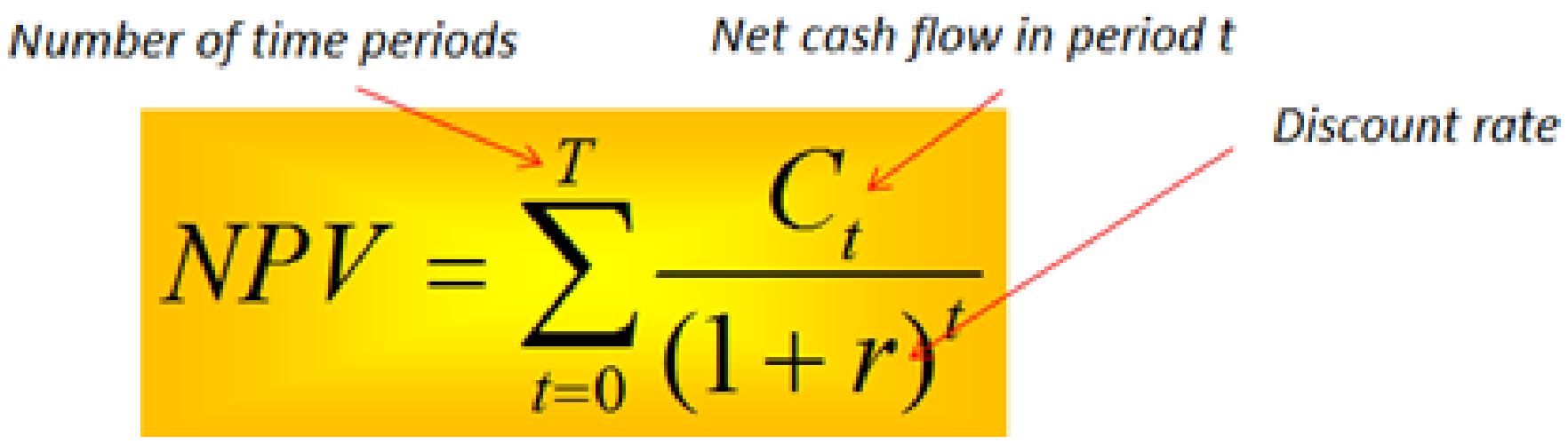

$$NPV = \sum_{t=0}^{T} \frac{C_t}{(1+r)^t}$$

A positive net present value indicates that the projected earnings generated by a project or investment (in present dollars) exceed the anticipated costs (also in present dollars). Generally, an investment with a positive NPV will be a profitable one and one with a negative NPV will result in a net loss. This concept is the basis for the Net Present Value Rule, which dictates that the only investments that should be made are those with positive NPV values.

II) The IRR is the discount rate at which the NPV of the project is zero. The IRR provides a different insight as sometimes the internal discount rate of the Company can be miscalculated, so the IRR (higher the IRR, the better) will be able to give insight as to how good a project really is.

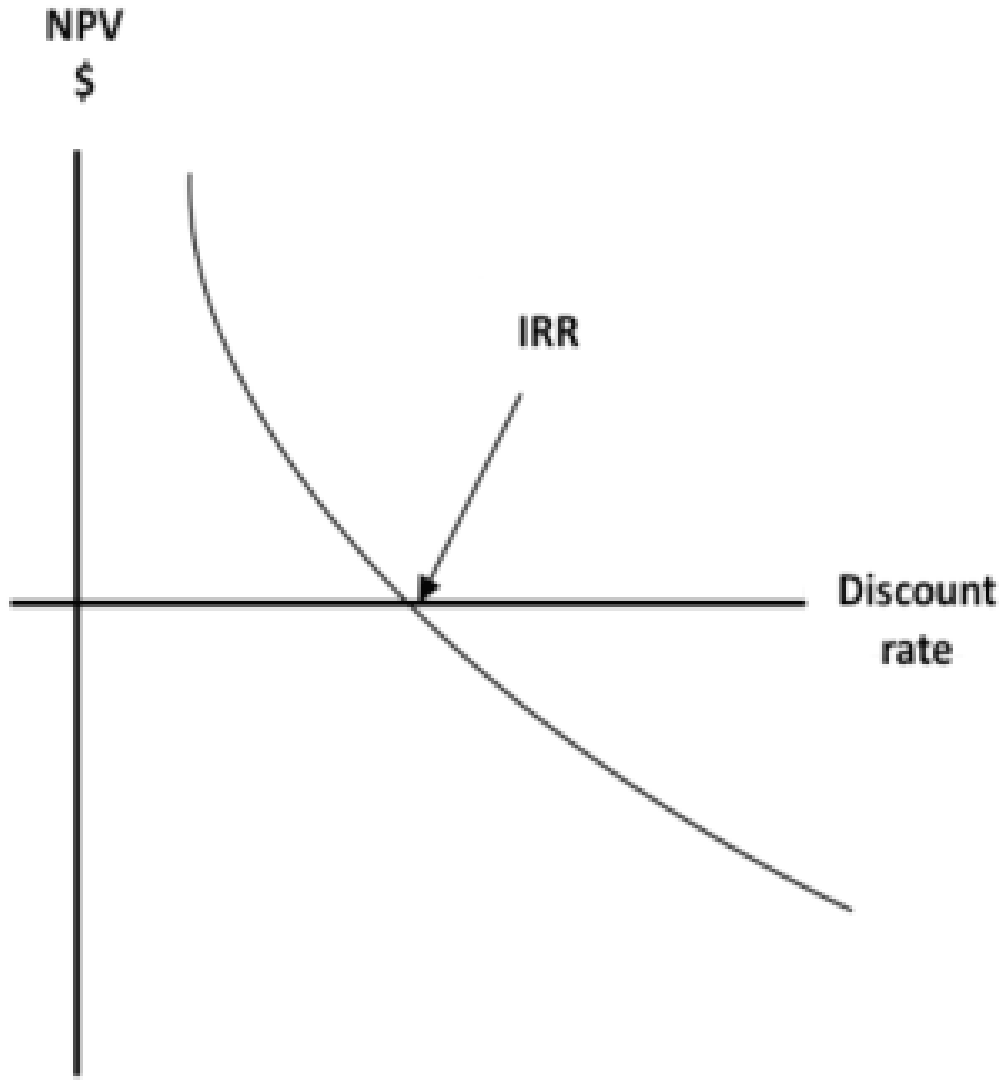

The following is the formula for calculating IRR:

$$IRR = r_a + \frac{NPV_a \, (r_b - r_a)}{(NPV_a - NPV_b)}$$

$r_a$ = lower discount rate
$r_a$ = higher discount rate
$NPV_a$ = NPV using the lower discount rate
$NPV_b$ = NPV using the higher discount rate

### 6.5.3 Perform Sensitivity Analysis

Sensitivity analysis is an analysis that finds out how sensitive an output is to any change in an input while keeping other inputs constant. Both internal and external factors influence project value. Internal factors are those over which the development team has a large degree of influence, including development program expense, development speed, production cost, and product performance. External factors are those that the team cannot arbitrarily change, including the competitive environment, sales volume, and product price. While external factors are not directly controlled by product development teams, they are often influenced by the internal factors. Use Sensitivity Analysis to Understand Project Trade-Offs

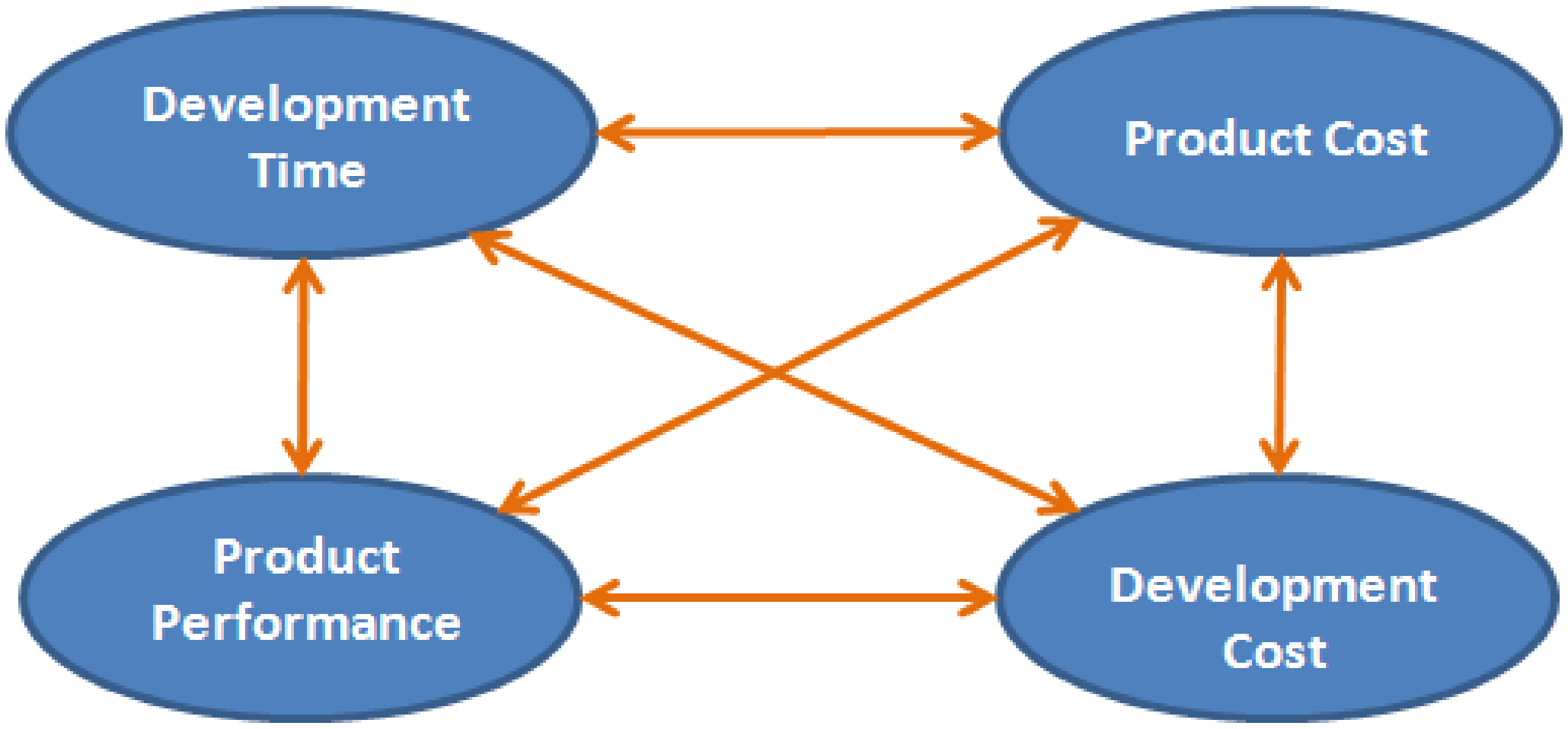

A trade-off (or tradeoff) is a situation that involves losing one quality or aspect of something in return for gaining another quality or aspect. More colloquially, if one thing increases, some other thing must decrease. Development teams attempt to manage six potential interactions between the internally driven factors. This potential interaction between any two internal factors depends on the characteristics of the specific product context. In many cases the interactions are trade-offs. However, some of these interactions are more complex than a simple trade-off.

In general, these interactions are important because of the linkage between the internal factors and the external factors. For example, increasing development cost or time may enhance product performance and therefore increase sales volume or allow higher prices. Decreasing development time may allow the product to reach the market sooner and thus increase sales volume.

### 6.5.4 Consider the Influence of the Qualitative Factors on Project Success

There are many factors influencing development projects. Due to complex or uncertain, they are difficult to quantify, these factors are qualitative factors. The quantitative model implicitly accounts for and many other issues with several board assumptions. The model assumes that decisions made by the project team do not affect actions of groups external to the project, or alternatively that the external forces do not change the team's actions. It is particularly crucial in the study of many other financial models such that other relevant factors influencing these are assumed to be constant by the assumption of Ceteris Paribus.

## 6.6 ELIMINOSIE™ Financial Models

To measure these capital budgeting measures, we make use of financial models to study the parameters of each model and assigned with specific periods and burn rates. We used 24 quarters to do financial projection for our product.

### 6.6.1 Base case Model

The first step of analysis is the construction of a base case financial model. After resetting the percentage derivation to 0%, the base NPV is $4,050,146 and IRR is 51%. The Figure below shows that the break-even time is exactly 5 periods, which is around 1 year 3 month.

**IRR = 51 %**

PROJECT NPV $ 4,050,146

Set Base

| Base NPV |
|---|
| 4,050,146 |

Changes from Base NPV

| % of NPV | $ change |
|---|---|
| 0.0% | 0 |

MODEL VALUES

| | first | last | base burn rate | adjusted burn rate | %∆ from base value | $∆ from base value |
|---|---|---|---|---|---|---|
| Development | 1 | 3 | -50000 | -50000 | 0.0% | 0 |
| Testing | 1 | 4 | -20000 | -20000 | 0.0% | 0 |
| Tooling and Ramp-Up Costs | 4 | 5 | -15000 | -15000 | 0.0% | 0 |
| Market Introduction | 4 | 5 | -20000 | -20000 | 0.0% | 0 |
| Ongoing Marketing Costs | 5 | 12 | -10000 | -10000 | 0.0% | 0 |
| Unit Sales | 5 | 24 | 1500 | 1500 | 0.0% | 0 |
| Unit Price | 5 | 24 | 300.000 | 300.000 | 0.0% | 0.00 |
| Unit Production Cost | 5 | 24 | -92.500 | -92.500 | 0.0% | 0.00 |
| Discount Rate (per time period) | 2.50% | | | | | |

**Table 17: TaNPV analysis for base case model**

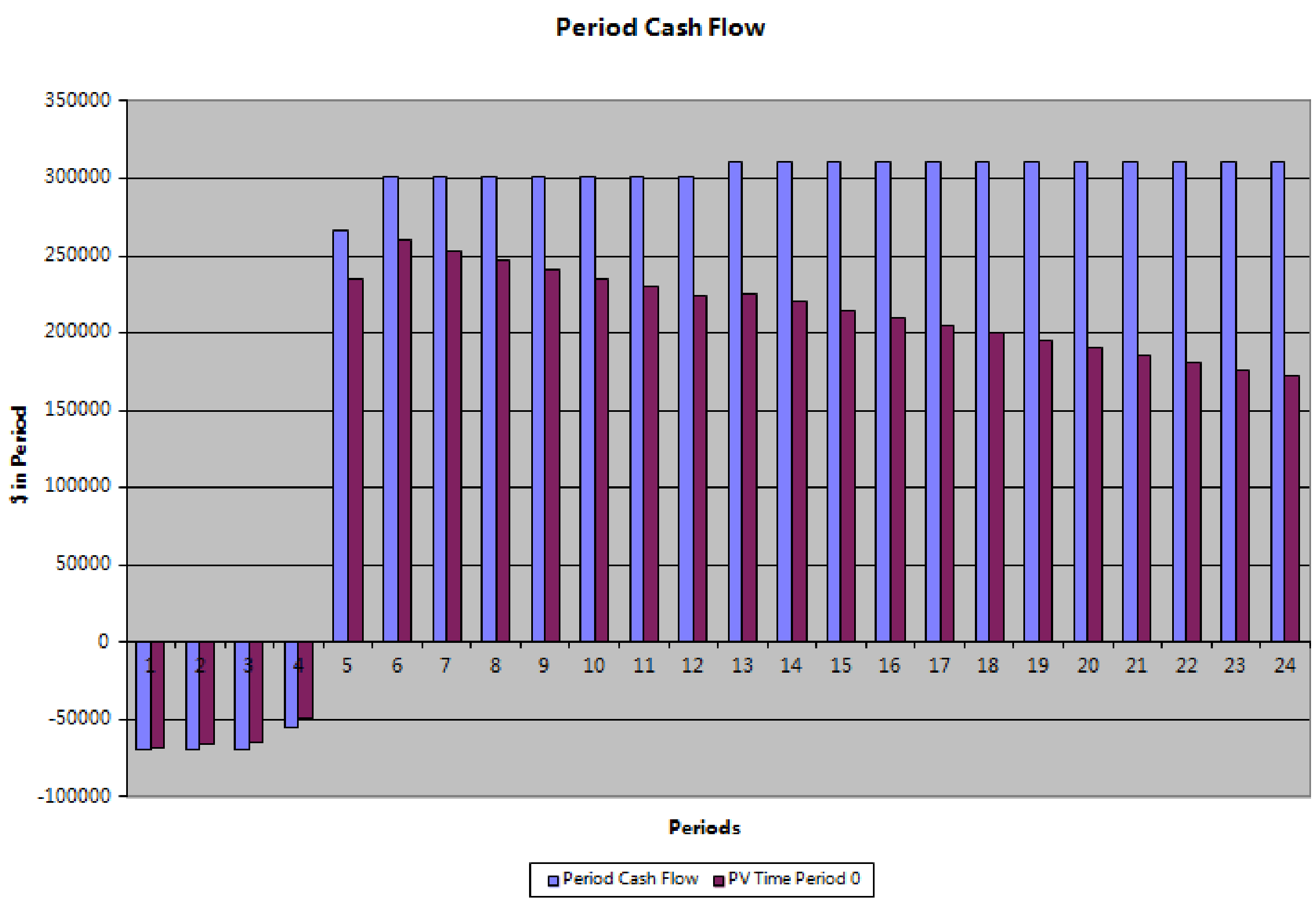

**Figure 29: Cash Flow Charts for every period in base case model**

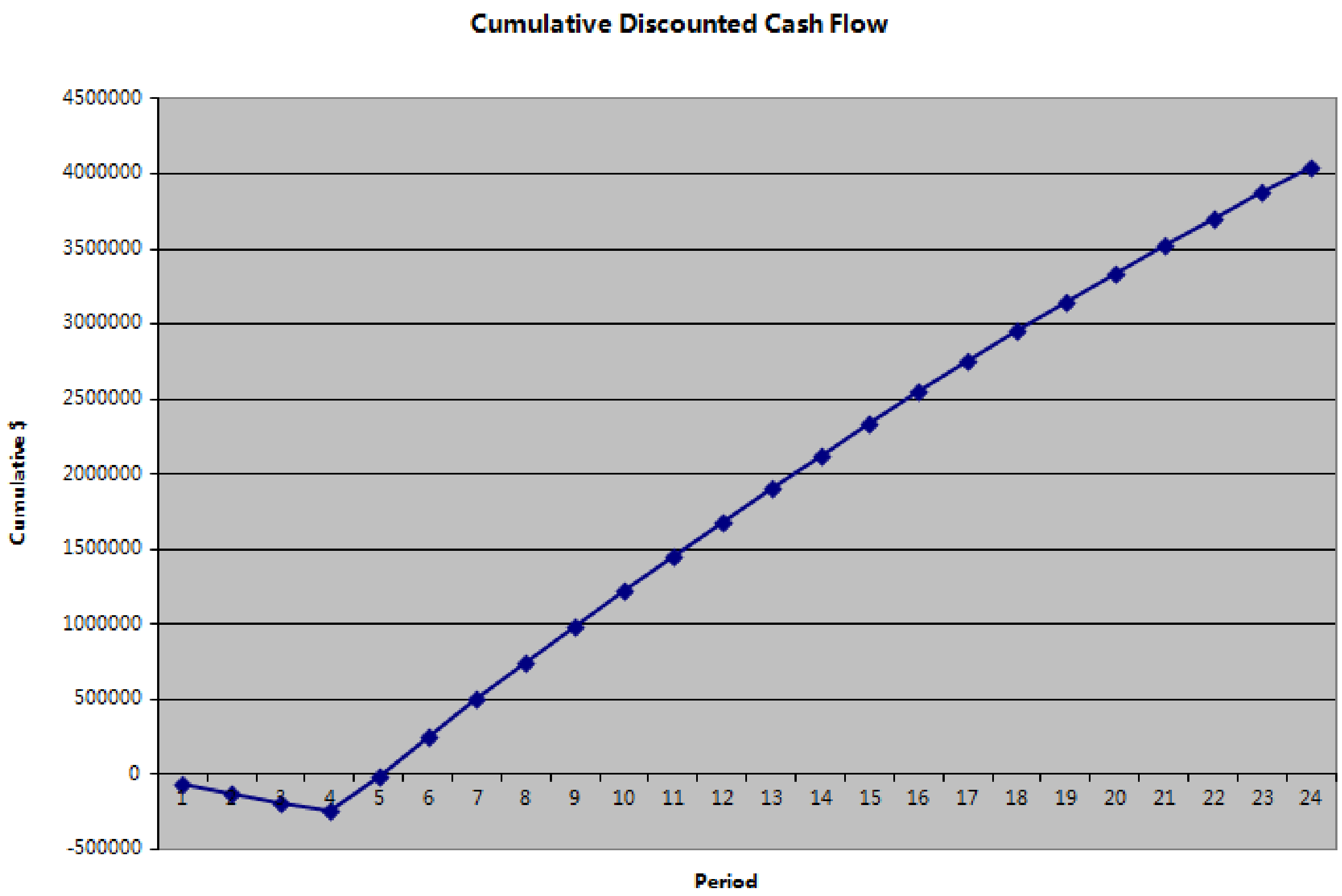

**Figure 30: Cumulative Discounted Cash Flow for the base case model showing the break-even point**

### 6.6.2 Best Case Model

The base case financial model is constructed at a discount rate at 2.5%. The 2.5% discount rate implies that the cash flows are discounted back at a rate of 2.5% per annum. For consistency for the rest of the models, we will continue to use 2.5% as the Discount Rate is an internal rate for the Company and should the NPV should be computed by discounting back by the same rate.

The product lifespan is estimated to be five years (20 periods) excluding the development phase of 3 periods. From below Figure, it is observed that there is a deviation in each parameter.

These deviations such as the increment in unit price and sales are essential in order to achieve the best case NPV of $10,031,183. The 147.7% change from the base NPV implies the amount of profit after six years. For this case, the profit is increased by $5,981,037. The best case IRR is 97%, which is staggering as this means that we have obtained a 97% return on initial capital outlay, nearly doubling all initial capital outlay. In this case, it is noticed that the initial cash outflows are much smaller than the ones in the base case. It is also noticed that the break-even time for this period is only 4 periods, which is 1 year.

**IRR = 97 %**

PROJECT NPV $ **10,031,183**

Set Base

| Base NPV |
|---|
| **4,050,146** |

Changes from Base NPV

| % of NPV | $ change |
|---|---|
| 147.7% | 5981037 |

**MODEL VALUES**

| | first | last | base burn rate | adjusted burn rate | %∆ from base value | $∆ from base value |
|---|---|---|---|---|---|---|
| Development | 1 | 3 | -50000 | -35000 | -30.0% | 15000 |
| Testing | 1 | 4 | -20000 | -16000 | -20.0% | 4000 |
| Tooling and Ramp-Up Costs | 4 | 5 | -15000 | -12000 | -20.0% | 3000 |
| Market Introduction | 4 | 5 | -20000 | -16000 | -20.0% | 4000 |
| Ongoing Marketing Costs | 5 | 12 | -10000 | -9000 | -10.0% | 1000 |
| Unit Sales | 5 | 24 | 1500 | 2550 | 70.0% | 1050 |
| Unit Price | 5 | 24 | 300.000 | 360.000 | 20.0% | 60.00 |
| Unit Production Cost | 5 | 24 | -92.500 | -74.000 | -20.0% | 18.50 |
| Discount Rate (per time period) | 2.50% | | | | | |

**Table 18: NPV for Best Case Model**

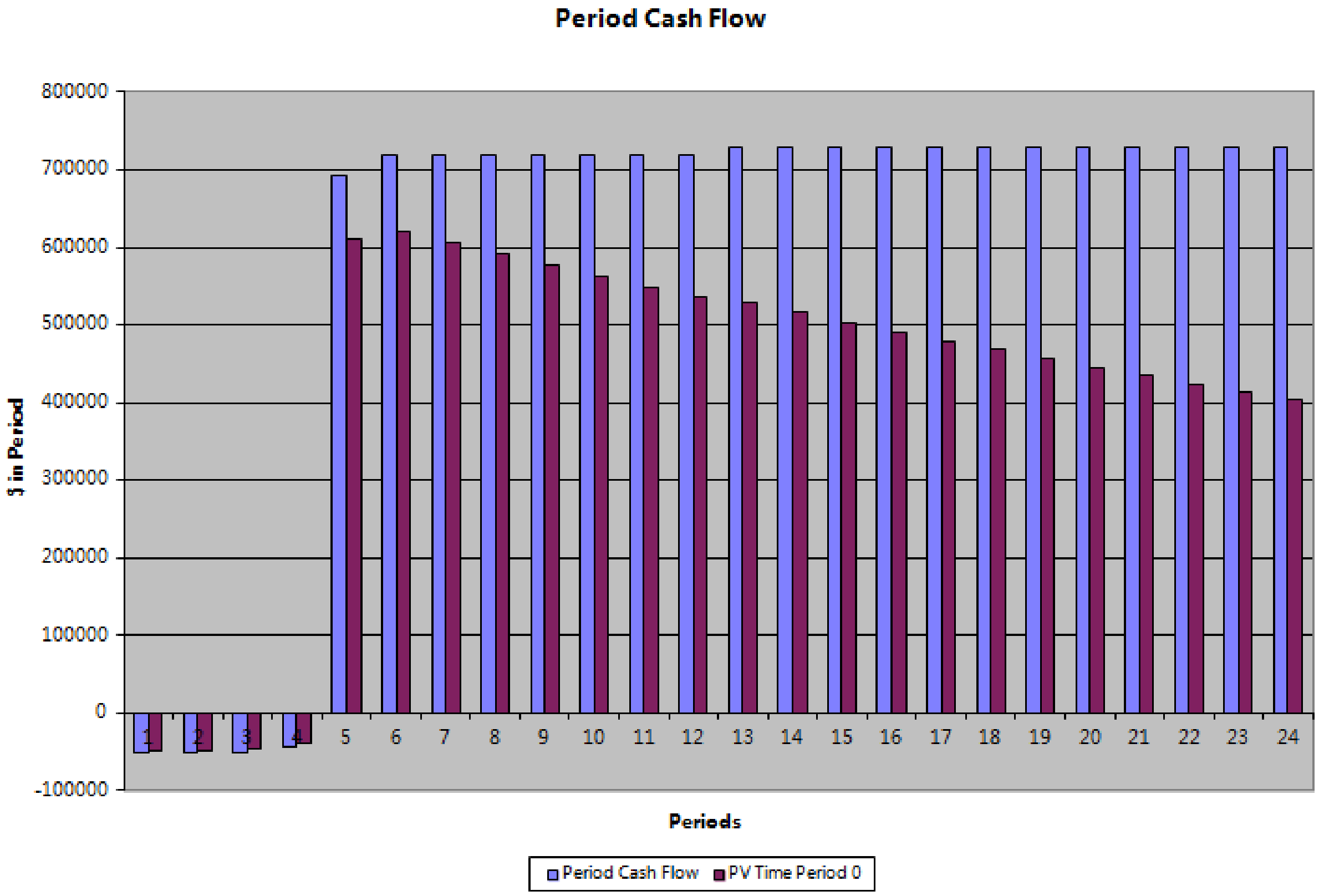

**Figure 31: Cash Flow Charts for every period in best case model**

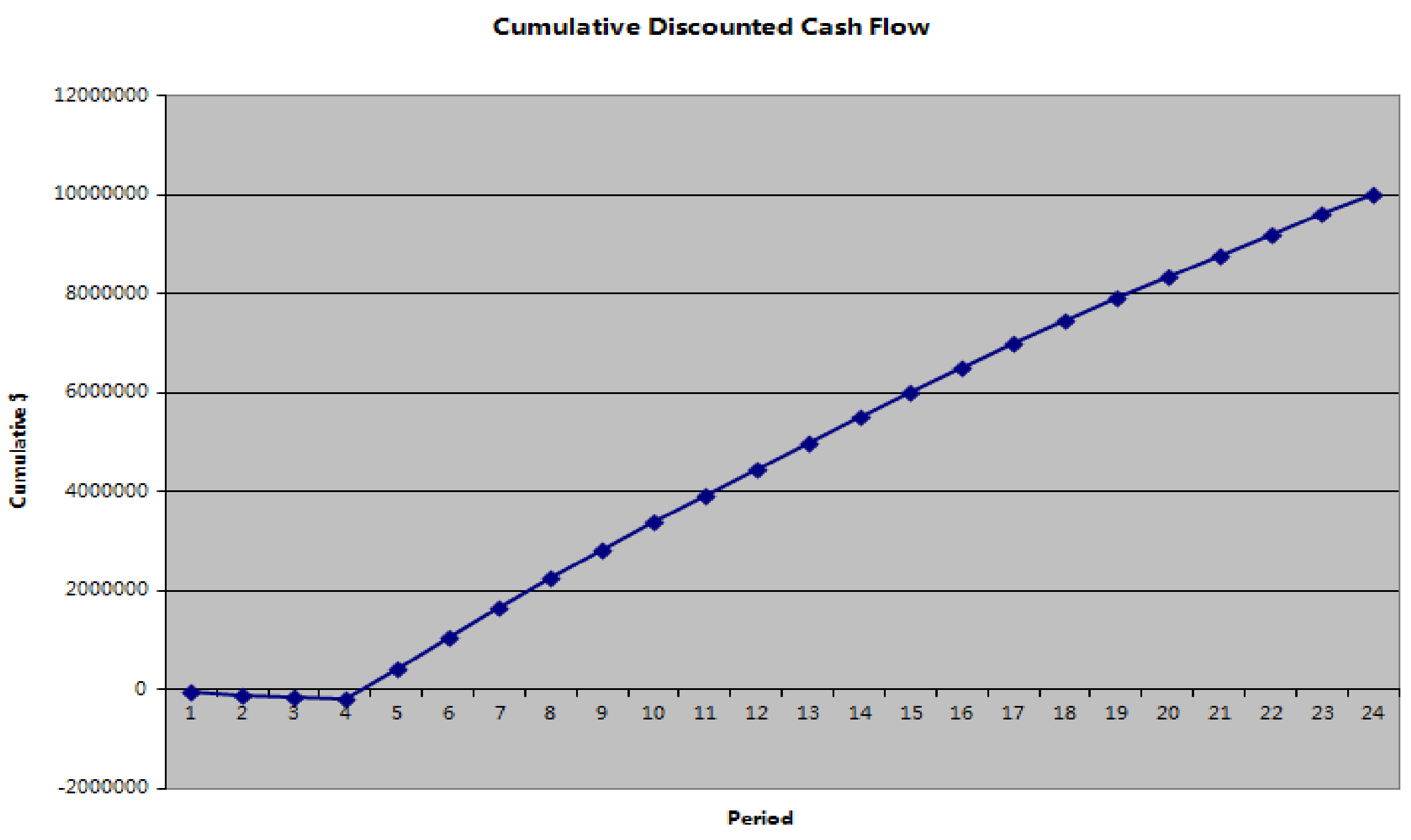

**Figure 32: Cumulative Cash Flow for best case model showing the breakeven point**

### 6.6.3 Bare Minimum Model

The next is the bare minimum model which is NPV equates to zero. At this rate, the discounted rate is consistent at 2.5% and thus it would still be acceptable as a capital budgeting decision to take up the project. For this model, the IRR almost is exactly discount rate as expected, and the NPV of the project is non-negative, so this would still be acceptable as a capital budgeting decision to take up the project. For this model, the break-even would be exactly at the end of the 24 periods (6 years). This is not surprising as the discounted cash flows sum to zero, meaning that the break-even is at the end of the project. This model is seen as a suitable measure to win the market share. At this point, we would be indifferent whether to take up the project or not.

**IRR = 3 %**

| | |
|---|---|
| PROJECT NPV $ | 0 |
| Set Base | Base NPV 4,050,146 |

Changes from Base NPV

| % of NPV | $ change |
|---|---|
| -100.0% | -4050146 |

MODEL VALUES

| | first | last | base burn rate | adjusted burn rate | %Δ from base value | $Δ from base value |
|---|---|---|---|---|---|---|
| Development | 1 | 3 | -50000 | -75500 | 51.0% | -25500 |
| Testing | 1 | 4 | -20000 | -30200 | 51.0% | -10200 |
| Tooling and Ramp-Up Costs | 4 | 5 | -15000 | -24033 | 60.2% | -9033 |
| Market Introduction | 4 | 5 | -20000 | -30520 | 52.6% | -10520 |
| Ongoing Marketing Costs | 5 | 12 | -10000 | -15000 | 50.0% | -5000 |
| Unit Sales | 5 | 24 | 1500 | 900 | -40.0% | -600 |
| Unit Price | 5 | 24 | 300.000 | 180.000 | -40.0% | -120.00 |
| Unit Production Cost | 5 | 24 | -92.500 | -138.750 | 50.0% | -46.25 |
| Discount Rate (per time period) | 2.50% | | | | | |

**Table 19: NPV for Bare Minimum model**

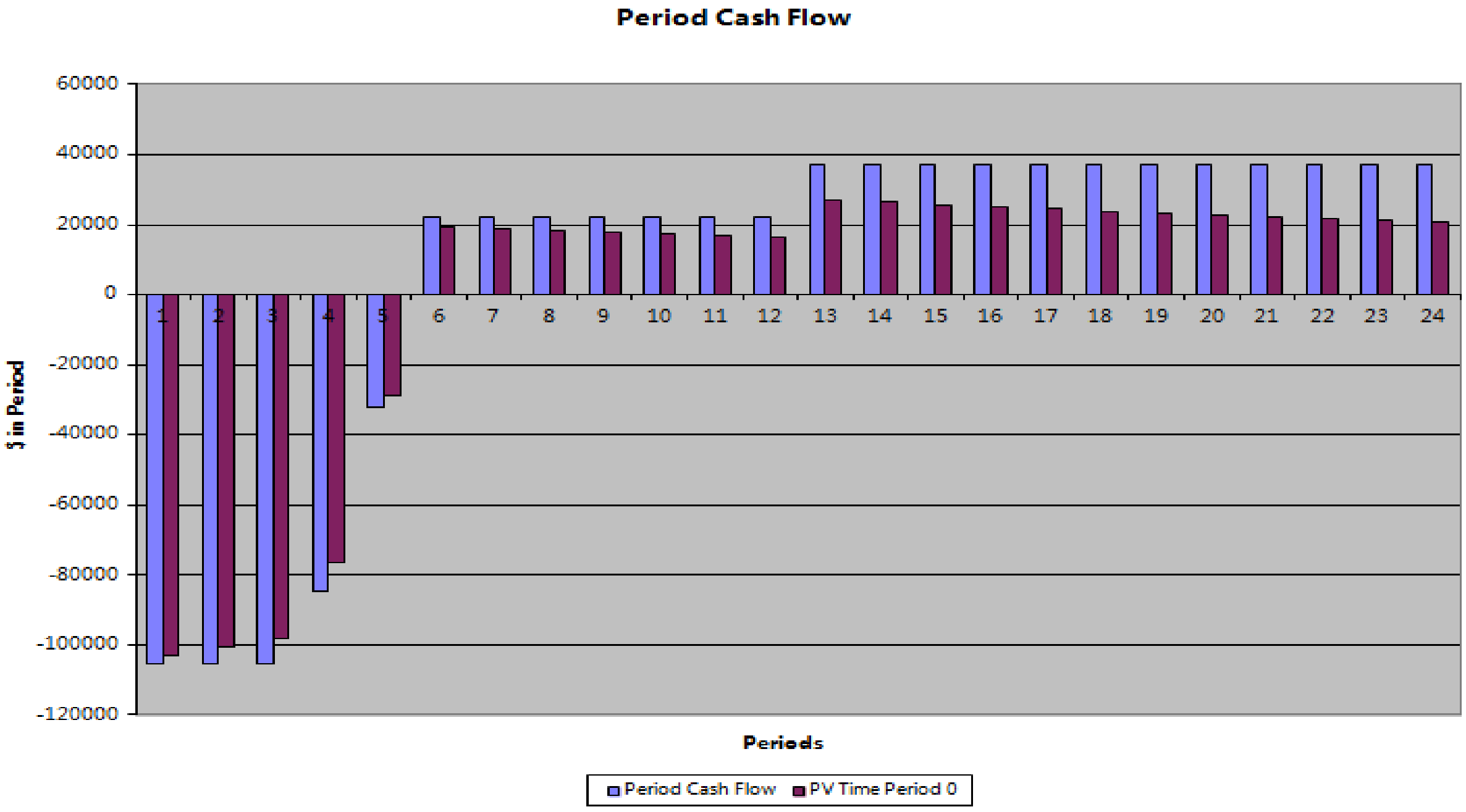

**Figure 33: Cash Flow Charts for every period in bare minimum model**

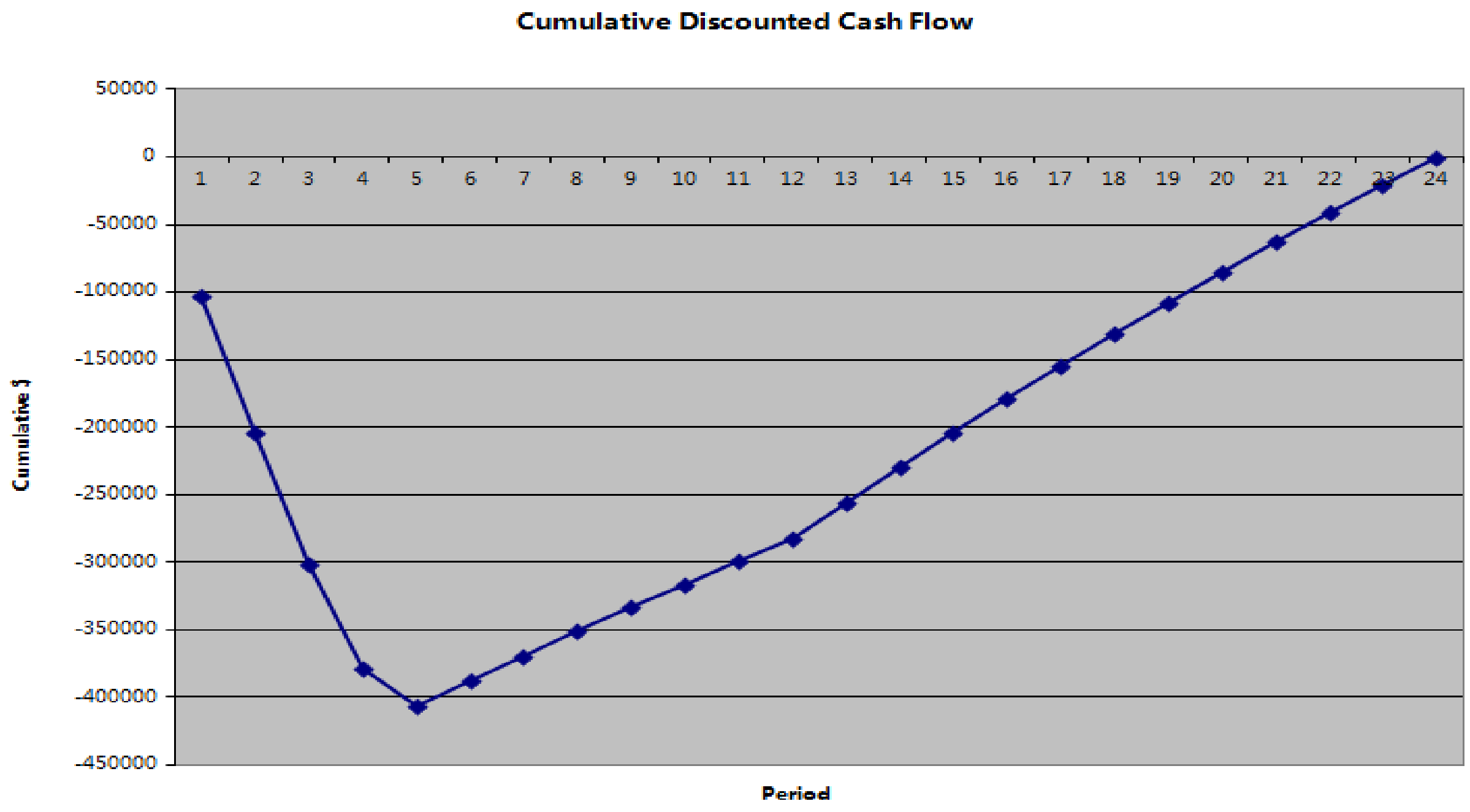

**Figure 34: Cumulative Discounted Cash Flow for bare minimum model showing the break-even point**

### 6.6.4 Worst Case Model

Using the best case model to derive the NPV and IRR, the financial department used this model to construct the worst case model. The worst case model would show as guidance for the situation under the worst of economic cases. From below figure, it can be observed that there is a large deviation as compared to the best case. Due to economy recession, poor sales, overstocking, prices factors and strong competitors, it may force the selling prices to drop to $135 while bringing production costs to $138.75, ultimately causing NPV and IRR to be negative. In this case, we would not take up this the project as NPV is negative, as we noticed that cash flows never become positive, and therefore there will never be a break-even.

**PROJECT NPV $** -542,295

**IRR = #DIV/0!**

Set Base

| Base NPV |
|---|
| 4,050,146 |

Changes from Base NPV

| % of NPV | $ change |
|---|---|
| -113.4% | -4592441 |

**MODEL VALUES**

| | first | last | base burn rate | adjusted burn rate | %Δ from base value | $Δ from base value |
|---|---|---|---|---|---|---|
| Development | 1 | 3 | -50000 | -75000 | 50.0% | -25000 |
| Testing | 1 | 4 | -20000 | -30000 | 50.0% | -10000 |
| Tooling and Ramp-Up Costs | 4 | 5 | -15000 | -22500 | 50.0% | -7500 |
| Market Introduction | 4 | 5 | -20000 | -30000 | 50.0% | -10000 |
| Ongoing Marketing Costs | 5 | 12 | -10000 | -15000 | 50.0% | -5000 |
| Unit Sales | 5 | 24 | 1500 | 450 | -70.0% | -1050 |
| Unit Price | 5 | 24 | 300.000 | 135.000 | -55.0% | -165.00 |
| Unit Production Cost | 5 | 24 | -92.500 | -138.750 | 50.0% | -46.25 |
| Discount Rate (per time period) | 2.50% | | | | | |

**Table 20: NPV analysis for Worst Case model**

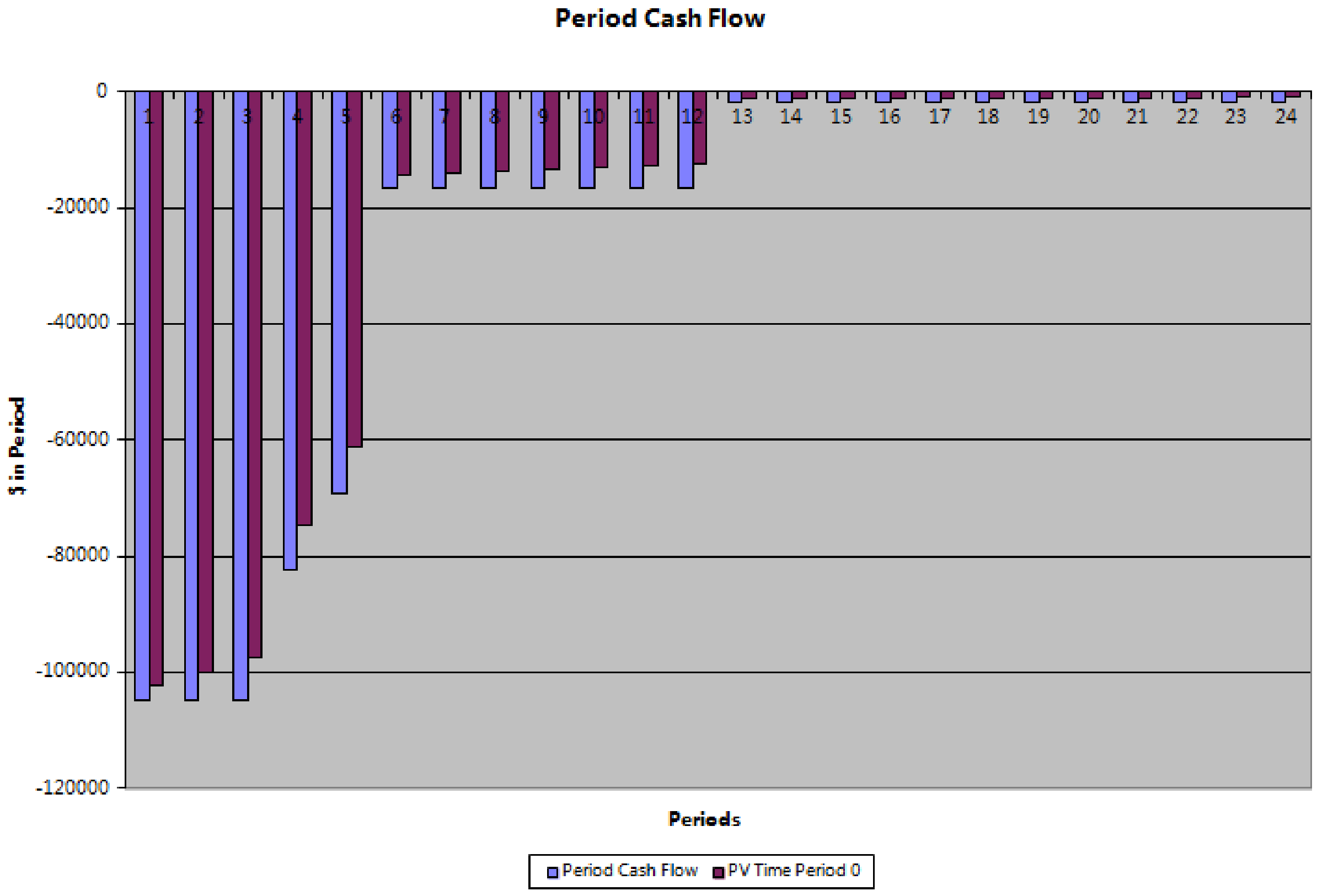

**<u>Figure 35: Cash Flow Charts for every period in worst case model</u>**

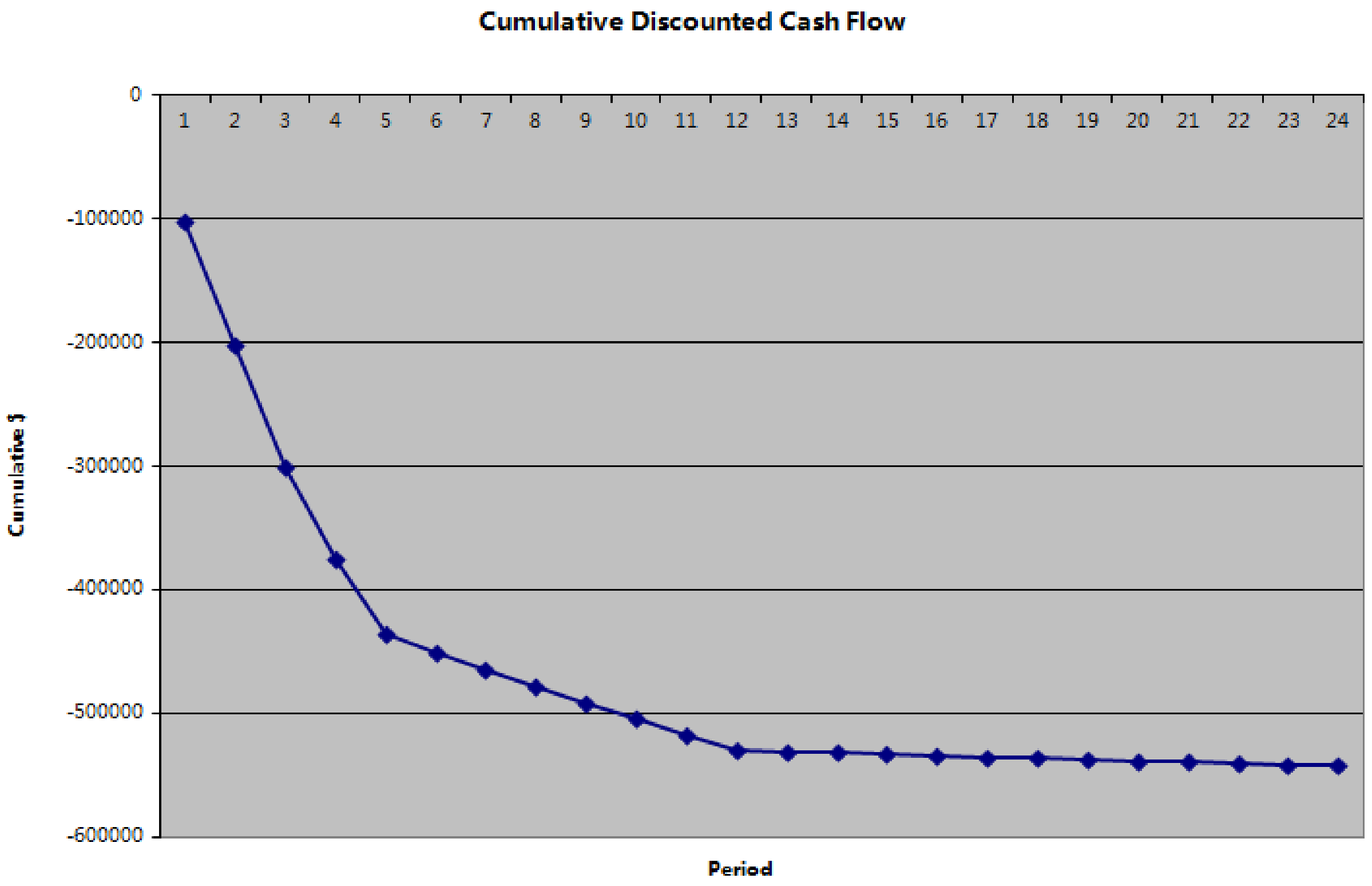

**Figure 36: Cumulative discount Cash Flow for worst case model showing NO break-even point**

## 6.7 Development Cost and Timing

In today's technology driven world, the important of innovation and product development is growing. We are expecting to minimize our development cost as much as possible. The planning is to use lesser amount of money to invest during the development process, the shorter the time to cover back the cost of it.

We are keeping the duration of the development timing to a period of 3. Knowing the fact that a single period refers to 3 months, therefore the total development timing would be 9 months. We have realized that our product would take no more than 8 months of the time to complete but we are giving ourselves a month of extra time to improve on the safety aspect of the product and better alternatives of materials and also in case something happen during the process of development such as time delays or staff turnover.

### 6.7.1 Sensitivity Analysis

To analysis the fluctuation of the development cost on how it will affect the overall NPV. Therefore, we decided to set other settings all to 0% to see purely the effect of it. We are simulating the account into 4 scenarios for comparison.

For the 1st scenario whereby the fluctuation will be 0%, which means that we are not going allocate any extra funds during the development periods. This would mean that investing $50,000 per period with no extra funds in it. The project NPV is $4,050,146.

**IRR = 51 %**

**PROJECT NPV $ 4,050,146**

Set Base

| Base NPV |
|---|
| **4,050,146** |

Changes from Base NPV

| % of NPV | $ change |
|---|---|
| 0.0% | 0 |

**MODEL VALUES**

| | first | last | base burn rate | adjusted burn rate | %∆ from base value | $∆ from base value |
|---|---|---|---|---|---|---|
| Development | 1 | 3 | -50000 | -50000 | 0.0% | 0 |
| Testing | 1 | 4 | -20000 | -20000 | 0.0% | 0 |
| Tooling and Ramp-Up Costs | 4 | 5 | -15000 | -15000 | 0.0% | 0 |
| Market Introduction | 4 | 5 | -20000 | -20000 | 0.0% | 0 |
| Ongoing Marketing Costs | 5 | 12 | -10000 | -10000 | 0.0% | 0 |
| Unit Sales | 5 | 24 | 1500 | 1500 | 0.0% | 0 |
| Unit Price | 5 | 24 | 300.000 | 300.000 | 0.0% | 0.00 |
| Unit Production Cost | 5 | 24 | -92.500 | -92.500 | 0.0% | 0.00 |
| Discount Rate (per time period) | 2.50% | | | | | |

**Table 21: NPV analysis for 1st scenario**

For the 2nd scenario, we are setting the fluctuation percentage of the development cost to -30%. A reduce fund of $15000 from the development cost of $50000. The percentage of NPV raised by 1.1%, with projection NPV increased by $42,840.

**IRR = 58 %**

| PROJECT NPV $ | 4,092,986 |
|---|---|

Set Base

| Base NPV |
|---|
| 4,050,146 |

Changes from Base NPV

| % of NPV | $ change |
|---|---|
| 1.1% | 42840 |

MODEL VALUES

| | first | last | base burn rate | adjusted burn rate | %Δ from base value | $Δ from base value |
|---|---|---|---|---|---|---|
| Development | 1 | 3 | -50000 | -35000 | -30.0% | 15000 |
| Testing | 1 | 4 | -20000 | -20000 | 0.0% | 0 |
| Tooling and Ramp-Up Costs | 4 | 5 | -15000 | -15000 | 0.0% | 0 |
| Market Introduction | 4 | 5 | -20000 | -20000 | 0.0% | 0 |
| Ongoing Marketing Costs | 5 | 12 | -10000 | -10000 | 0.0% | 0 |
| Unit Sales | 5 | 24 | 1500 | 1500 | 0.0% | 0 |
| Unit Price | 5 | 24 | 300.000 | 300.000 | 0.0% | 0.00 |
| Unit Production Cost | 5 | 24 | -92.500 | -92.500 | 0.0% | 0.00 |
| Discount Rate (per time period) | 2.50% | | | | | |

**Table 22: NPV analysis for 2nd scenario**

For the 3rd scenario, we are setting the fluctuation percentage of the development cost to 10%. An increased fund of $5000 from the development cost of $50000. The percentage of NPV reduced by 0.4%, with projection NPV decreased to $14,280

**IRR = 50 %**

| PROJECT NPV $ | 4,035,866 |
|---|---|

Set Base

| Base NPV |
|---|
| 4,050,146 |

Changes from Base NPV

| % of NPV | $ change |
|---|---|
| -0.4% | -14280 |

MODEL VALUES

| | first | last | base burn rate | adjusted burn rate | %Δ from base value | $Δ from base value |
|---|---|---|---|---|---|---|
| Development | 1 | 3 | -50000 | -55000 | 10.0% | -5000 |
| Testing | 1 | 4 | -20000 | -20000 | 0.0% | 0 |
| Tooling and Ramp-Up Costs | 4 | 5 | -15000 | -15000 | 0.0% | 0 |
| Market Introduction | 4 | 5 | -20000 | -20000 | 0.0% | 0 |
| Ongoing Marketing Costs | 5 | 12 | -10000 | -10000 | 0.0% | 0 |
| Unit Sales | 5 | 24 | 1500 | 1500 | 0.0% | 0 |
| Unit Price | 5 | 24 | 300.000 | 300.000 | 0.0% | 0.00 |
| Unit Production Cost | 5 | 24 | -92.500 | -92.500 | 0.0% | 0.00 |
| Discount Rate (per time period) | 2.50% | | | | | |

**Table 23: NPV analysis for 3rd scenario**

For the 4th scenario, we are setting the fluctuation percentage of the development cost to 40%. This would mean that we are allocating extra $20,000 funds to the development cost. The percentage of NPV reduced by 1.4% with the project NPV decreased by $57121.

**IRR = 45 %**

**PROJECT NPV $** 3,993,025

Set Base

| Base NPV |
|---|
| 4,050,146 |

Changes from Base NPV

| % of NPV | $ change |
|---|---|
| -1.4% | -57121 |

**MODEL VALUES**

| | first | last | base burn rate | adjusted burn rate | %Δ from base value | $Δ from base value |
|---|---|---|---|---|---|---|
| Development | 1 | 3 | -50000 | -70000 | 40.0% | -20000 |
| Testing | 1 | 4 | -20000 | -20000 | 0.0% | 0 |
| Tooling and Ramp-Up Costs | 4 | 5 | -15000 | -15000 | 0.0% | 0 |
| Market Introduction | 4 | 5 | -20000 | -20000 | 0.0% | 0 |
| Ongoing Marketing Costs | 5 | 12 | -10000 | -10000 | 0.0% | 0 |
| Unit Sales | 5 | 24 | 1500 | 1500 | 0.0% | 0 |
| Unit Price | 5 | 24 | 300.000 | 300.000 | 0.0% | 0.00 |
| Unit Production Cost | 5 | 24 | -92.500 | -92.500 | 0.0% | 0.00 |
| Discount Rate (per time period) | 2.50% | | | | | |

**Table 24: NPV analysis for 4th scenario**

## 6.8 Testing Cost and Timing

The overall testing process, in term of cost and timing, is consider as the essential phase before our final product is truly ready to sell to the market. Therefore in terms of cost, we are expecting to invest a relative close amount of money as our development cost.

For the duration of testing process, we are giving a slightly longer period of time. We are expecting to put our testing process and our development process concurrently as both phrases affect one another. During the phase of testing, we are taking into the consideration of various parameters such as the durability of material that we are going to choose to use, cost of our prototype and also the safety standards of our prototype. The concurrent process allow us to save time so that we can launch our product as early as possible thus increasing our net revenue.

### 6.8.1 Sensitivity Analysis

Since the testing phase and the development are closely related to each other. Therefore we are using the same method to learn how the fluctuation of the testing cost will affect the overall NPV. Therefore, we decided to set other settings all to 0% to see purely the effect of it. Similarly, we are simulating the account into 4 scenarios for comparison.

For the 1st scenario, whereby the fluctuation will be 0%, which means that we are not going allocate any extra funds during the testing periods. This would mean that investing $20,000 per period with no extra funds in it with project NPV of $4,050,146.

IRR = 51 %

PROJECT NPV $ 4,050,146

Set Base

| Base NPV |
|---|
| 4,050,146 |

Changes from Base NPV

| % of NPV | $ change |
|---|---|
| 0.0% | 0 |

MODEL VALUES

| | first | last | base burn rate | adjusted burn rate | %Δ from base value | $Δ from base value |
|---|---|---|---|---|---|---|
| Development | 1 | 3 | -50000 | -50000 | 0.0% | 0 |
| Testing | 1 | 4 | -20000 | -20000 | 0.0% | 0 |
| Tooling and Ramp-Up Costs | 4 | 5 | -15000 | -15000 | 0.0% | 0 |
| Market Introduction | 4 | 5 | -20000 | -20000 | 0.0% | 0 |
| Ongoing Marketing Costs | 5 | 12 | -10000 | -10000 | 0.0% | 0 |
| Unit Sales | 5 | 24 | 1500 | 1500 | 0.0% | 0 |
| Unit Price | 5 | 24 | 300.000 | 300.000 | 0.0% | 0.00 |
| Unit Production Cost | 5 | 24 | -92.500 | -92.500 | 0.0% | 0.00 |
| Discount Rate (per time period) | 2.50% | | | | | |

**Table 25: NPV analysis for 1st scenario**

For the 2nd scenario, we are expecting the fluctuation percentage of the testing cost to be about -30%. With -30% fluctuations, the cost reduces $6000 from $20,000 into the testing cost. The percentage of NPV raised by 0.6%, with project NPV increased by $22,571.

IRR = 54 %

PROJECT NPV $ 4,072,717

Set Base

| Base NPV |
|---|
| 4,050,146 |

Changes from Base NPV

| % of NPV | $ change |
|---|---|
| 0.6% | 22571 |

MODEL VALUES

| | first | last | base burn rate | adjusted burn rate | %Δ from base value | $Δ from base value |
|---|---|---|---|---|---|---|
| Development | 1 | 3 | -50000 | -50000 | 0.0% | 0 |
| Testing | 1 | 4 | -20000 | -14000 | -30.0% | 6000 |
| Tooling and Ramp-Up Costs | 4 | 5 | -15000 | -15000 | 0.0% | 0 |
| Market Introduction | 4 | 5 | -20000 | -20000 | 0.0% | 0 |
| Ongoing Marketing Costs | 5 | 12 | -10000 | -10000 | 0.0% | 0 |
| Unit Sales | 5 | 24 | 1500 | 1500 | 0.0% | 0 |
| Unit Price | 5 | 24 | 300.000 | 300.000 | 0.0% | 0.00 |
| Unit Production Cost | 5 | 24 | -92.500 | -92.500 | 0.0% | 0.00 |
| Discount Rate (per time period) | 2.50% | | | | | |

**Table 26: NPV analysis for 2nd scenario**

For the 3rd scenario, we are setting the fluctuation percentage of the testing cost to 10%. An increased fund of $2000 from the testing cost of $20000. The percentage of NPV reduced by 0.2%, with projection NPV decreased by $7,524.

IRR = 50 %

PROJECT NPV $ 4,042,622

Set Base

| Base NPV |
|---|
| 4,050,146 |

Changes from Base NPV

| % of NPV | $ change |
|---|---|
| -0.2% | -7524 |

**MODEL VALUES**

| | first | last | base burn rate | adjusted burn rate | %Δ from base value | $Δ from base value |
|---|---|---|---|---|---|---|
| Development | 1 | 3 | -50000 | -50000 | 0.0% | 0 |
| Testing | 1 | 4 | -20000 | -22000 | 10.0% | -2000 |
| Tooling and Ramp-Up Costs | 4 | 5 | -15000 | -15000 | 0.0% | 0 |
| Market Introduction | 4 | 5 | -20000 | -20000 | 0.0% | 0 |
| Ongoing Marketing Costs | 5 | 12 | -10000 | -10000 | 0.0% | 0 |
| Unit Sales | 5 | 24 | 1500 | 1500 | 0.0% | 0 |
| Unit Price | 5 | 24 | 300.000 | 300.000 | 0.0% | 0.00 |
| Unit Production Cost | 5 | 24 | -92.500 | -92.500 | 0.0% | 0.00 |
| Discount Rate (per time period) | 2.50% | | | | | |

**Table 27: NPV analysis for 3rd scenario**

For the 4th scenario, we are setting the fluctuation percentage of the testing cost to 40%. This would mean that we are allocating extra $8000 funds to the testing cost. The percentage of NPV reduced by 0.6% with the project NPV decreased by $30,096.

IRR = 48 %

PROJECT NPV $ 4,020,050

Set Base

| Base NPV |
|---|
| 4,050,146 |

Changes from Base NPV

| % of NPV | $ change |
|---|---|
| -0.7% | -30096 |

**MODEL VALUES**

| | first | last | base burn rate | adjusted burn rate | %Δ from base value | $Δ from base value |
|---|---|---|---|---|---|---|
| Development | 1 | 3 | -50000 | -50000 | 0.0% | 0 |
| Testing | 1 | 4 | -20000 | -28000 | 40.0% | -8000 |
| Tooling and Ramp-Up Costs | 4 | 5 | -15000 | -15000 | 0.0% | 0 |
| Market Introduction | 4 | 5 | -20000 | -20000 | 0.0% | 0 |
| Ongoing Marketing Costs | 5 | 12 | -10000 | -10000 | 0.0% | 0 |
| Unit Sales | 5 | 24 | 1500 | 1500 | 0.0% | 0 |
| Unit Price | 5 | 24 | 300.000 | 300.000 | 0.0% | 0.00 |
| Unit Production Cost | 5 | 24 | -92.500 | -92.500 | 0.0% | 0.00 |
| Discount Rate (per time period) | 2.50% | | | | | |

**Table 28: NPV analysis for 4th scenario**

## 6.9 Tooling Investment and Ramp-up Cost and Timing

At the initial stage, we are expecting to have a relatively low ramp-up cost due to our product demand may not be high because there will be a low firm production. However, as our product is introduced to the market after some time, we are expecting to see a high increase in firm production as there will be increase in product demand. The amount of money set for the tooling and ramp-cost may also vary accordingly, therefore strategy adjustment for the cost is also needed in the future and the phase is needed to be adjusted as well. The total amount for both of our tooling investment and ramp-up cost are critically plan to find the right balance because we perceive that more money we add in during the tooling and ramp-up period, we are expected to take a longer time to recoup the cost.

Taking into this consideration, we are putting about $15,000 per period for both tooling investment and ramp-up cost as it can serve as a part of the reserve for the company just in case there is a need in the future.

### 6.9.1 Sensitivity Analysis

We want to know how the fluctuation of the tooling investment and ramp-up cost will affect the overall NPV. Therefore, we decided to set other settings all to 0% to see purely the effect of it. Similarly, we are simulating the account into 4 scenarios for comparison.

For the 1st scenario, whereby the fluctuation will be 0%, which means that we are not going allocate any extra funds during the tooling investment and ramp-up periods. This would mean that investing $15,000 per period with no extra funds in it with project NPV of $4,050,146.

PROJECT NPV $ **4,050,146**

**IRR = 51 %**

Set Base

| Base NPV |
|---|
| 4,050,146 |

Changes from Base NPV

| % of NPV | $ change |
|---|---|
| 0.0% | 0 |

**MODEL VALUES**

| | first | last | base burn rate | adjusted burn rate | %Δ from base value | $Δ from base value |
|---|---|---|---|---|---|---|
| Development | 1 | 3 | -50000 | -50000 | 0.0% | 0 |
| Testing | 1 | 4 | -20000 | -20000 | 0.0% | 0 |
| Tooling and Ramp-Up Costs | 4 | 5 | -15000 | -15000 | 0.0% | 0 |
| Market Introduction | 4 | 5 | -20000 | -20000 | 0.0% | 0 |
| Ongoing Marketing Costs | 5 | 12 | -10000 | -10000 | 0.0% | 0 |
| Unit Sales | 5 | 24 | 1500 | 1500 | 0.0% | 0 |
| Unit Price | 5 | 24 | 300.000 | 300.000 | 0.0% | 0.00 |
| Unit Production Cost | 5 | 24 | -92.500 | -92.500 | 0.0% | 0.00 |
| Discount Rate (per time period) | 2.50% | | | | | |

**Table 29: NPV analysis for 1st scenario**

For the 2nd scenario, we are expecting that the fluctuation would be around -30% of fluctuation. With -30% of fluctuation, it would mean will be reduce $4,500. The percentage of NPV raised by 0.2% with project NPV increased by $8,054.

**IRR = 52 %**

**PROJECT NPV $** 4,058,200

Set Base

| Base NPV |
|---|
| 4,050,146 |

Changes from Base NPV

| % of NPV | $ change |
|---|---|
| 0.2% | 8054 |

**MODEL VALUES**

| | first | last | base burn rate | adjusted burn rate | %Δ from base value | $Δ from base value |
|---|---|---|---|---|---|---|
| Development | 1 | 3 | -50000 | -50000 | 0.0% | 0 |
| Testing | 1 | 4 | -20000 | -20000 | 0.0% | 0 |
| Tooling and Ramp-Up Costs | 4 | 5 | -15000 | -10500 | -30.0% | 4500 |
| Market Introduction | 4 | 5 | -20000 | -20000 | 0.0% | 0 |
| Ongoing Marketing Costs | 5 | 12 | -10000 | -10000 | 0.0% | 0 |
| Unit Sales | 5 | 24 | 1500 | 1500 | 0.0% | 0 |
| Unit Price | 5 | 24 | 300.000 | 300.000 | 0.0% | 0.00 |
| Unit Production Cost | 5 | 24 | -92.500 | -92.500 | 0.0% | 0.00 |
| Discount Rate (per time period) | 2.50% | | | | | |

**Tale 30: NPV analysis for 2nd scenario**

For the 3rd scenario, we are setting the fluctuation percentage of the tooling cost to 10%. An increased fund of $1,500 from the tooling cost of $15,000. The percentage of NPV reduced by 0.1% with projection NPV decreased by $2,685.

**IRR = 51 %**

**PROJECT NPV $** 4,047,461

Set Base

| Base NPV |
|---|
| 4,050,146 |

Changes from Base NPV

| % of NPV | $ change |
|---|---|
| -0.1% | -2685 |

**MODEL VALUES**

| | first | last | base burn rate | adjusted burn rate | %Δ from base value | $Δ from base value |
|---|---|---|---|---|---|---|
| Development | 1 | 3 | -50000 | -50000 | 0.0% | 0 |
| Testing | 1 | 4 | -20000 | -20000 | 0.0% | 0 |
| Tooling and Ramp-Up Costs | 4 | 5 | -15000 | -16500 | 10.0% | -1500 |
| Market Introduction | 4 | 5 | -20000 | -20000 | 0.0% | 0 |
| Ongoing Marketing Costs | 5 | 12 | -10000 | -10000 | 0.0% | 0 |
| Unit Sales | 5 | 24 | 1500 | 1500 | 0.0% | 0 |
| Unit Price | 5 | 24 | 300.000 | 300.000 | 0.0% | 0.00 |
| Unit Production Cost | 5 | 24 | -92.500 | -92.500 | 0.0% | 0.00 |
| Discount Rate (per time period) | 2.50% | | | | | |

**Table 31: NPV analysis for 3rd scenario**

For the 4th scenario, we are setting the fluctuation to about 40% which is an addition of $6000 is added to it. The percentage of NPV reduced by 0.3% with project NPV decreased by $10,739.

| | | | | | | | |
|---|---|---|---|---|---|---|---|
| IRR = 51 % | | PROJECT NPV $ | 4,039,407 | | Changes from Base NPV | | |
| | | Set Base | Base NPV | | % of NPV | $ change | |
| | | | 4,050,146 | | -0.3% | -10739 | |
| MODEL VALUES | | | | | | | |
| | first | last | base burn rate | adjusted burn rate | %Δ from base value | $Δ from base value | |
| Development | 1 | 3 | -50000 | -50000 | 0.0% | 0 | |
| Testing | 1 | 4 | -20000 | -20000 | 0.0% | 0 | |
| Tooling and Ramp-Up Costs | 4 | 5 | -15000 | -21000 | 40.0% | -6000 | |
| Market Introduction | 4 | 5 | -20000 | -20000 | 0.0% | 0 | |
| Ongoing Marketing Costs | 5 | 12 | -10000 | -10000 | 0.0% | 0 | |
| Unit Sales | 5 | 24 | 1500 | 1500 | 0.0% | 0 | |
| Unit Price | 5 | 24 | 300.000 | 300.000 | 0.0% | 0.00 | |
| Unit Production Cost | 5 | 24 | -92.500 | -92.500 | 0.0% | 0.00 | |
| Discount Rate (per time period) | 2.50% | | | | | | |

**Table 31: NPV analysis for 4th scenario**

## 6.10 Market Introduction and Timing

In today's product development environment, products and the technologies they're based on change rapidly, as do the number of competitors for market share. This means that time to market and finding ways of optimizing it are critical components that directly affect revenue. The better control you have over your product development processes, the better you'll be able to control and predict your time to market and get new technology out while it's still new. Effective market introduction and timing are critical to attracting customers and optimizing revenue and profit on a new product.

### 6.10.1 Sensitivity Analysis

We want to know how the fluctuation of the market introduction will affect the overall NPV. Therefore, we decided to set other settings all to 0% to see purely the effect of it. Similarly, we are simulating the account into 4 scenarios for comparison.

For the 1st scenario, whereby the fluctuation will be 0%, this would mean that investing $20,000 per period with project NPV at $4,050,146.

**IRR = 51 %**

PROJECT NPV $ **4,050,146**

Set Base

| Base NPV |
|---|
| **4,050,146** |

Changes from Base NPV

| % of NPV | $ change |
|---|---|
| 0.0% | 0 |

**MODEL VALUES**

| | first | last | base burn rate | adjusted burn rate | %Δ from base value | $Δ from base value |
|---|---|---|---|---|---|---|
| Development | 1 | 3 | -50000 | -50000 | 0.0% | 0 |
| Testing | 1 | 4 | -20000 | -20000 | 0.0% | 0 |
| Tooling and Ramp-Up Costs | 4 | 5 | -15000 | -15000 | 0.0% | 0 |
| Market Introduction | 4 | 5 | -20000 | -20000 | 0.0% | 0 |
| Ongoing Marketing Costs | 5 | 12 | -10000 | -10000 | 0.0% | 0 |
| Unit Sales | 5 | 24 | 1500 | 1500 | 0.0% | 0 |
| Unit Price | 5 | 24 | 300.000 | 300.000 | 0.0% | 0.00 |
| Unit Production Cost | 5 | 24 | -92.500 | -92.500 | 0.0% | 0.00 |
| Discount Rate (per time period) | 2.50% | | | | | |

**Table 32: NPV analysis for 1st scenario**

For the 2nd scenario, we are expecting that the fluctuation would be around -30%. With -30% of fluctuation, it would mean will be reduce $6,000 is allocated to it. The percentage of NPV raised by 0.3% with project NPV increased by $10,738.

**IRR = 52 %**

PROJECT NPV $ **4,060,884**

Set Base

| Base NPV |
|---|
| **4,050,146** |

Changes from Base NPV

| % of NPV | $ change |
|---|---|
| 0.3% | 10738 |

**MODEL VALUES**

| | first | last | base burn rate | adjusted burn rate | %Δ from base value | $Δ from base value |
|---|---|---|---|---|---|---|
| Development | 1 | 3 | -50000 | -50000 | 0.0% | 0 |
| Testing | 1 | 4 | -20000 | -20000 | 0.0% | 0 |
| Tooling and Ramp-Up Costs | 4 | 5 | -15000 | -15000 | 0.0% | 0 |
| Market Introduction | 4 | 5 | -20000 | -14000 | -30.0% | 6000 |
| Ongoing Marketing Costs | 5 | 12 | -10000 | -10000 | 0.0% | 0 |
| Unit Sales | 5 | 24 | 1500 | 1500 | 0.0% | 0 |
| Unit Price | 5 | 24 | 300.000 | 300.000 | 0.0% | 0.00 |
| Unit Production Cost | 5 | 24 | -92.500 | -92.500 | 0.0% | 0.00 |
| Discount Rate (per time period) | 2.50% | | | | | |

**Table 33: NPV analysis for 2nd scenario**

For the 3rd scenario, we are setting the fluctuation percentage to 10%. An increased fund of $2,000 from the market introduction cost of $15,000. The percentage of NPV reduced by 0.1%, with projection NPV decreased by $3,580.

| | | | | | | |
|---|---|---|---|---|---|---|
| IRR = 51 % | | PROJECT NPV $ | 4,046,566 | | Changes from Base NPV | |
| | | Set Base | Base NPV | | % of NPV | $ change |
| | | | 4,050,146 | | -0.1% | -3580 |

MODEL VALUES

| | first | last | base burn rate | adjusted burn rate | %Δ from base value | $Δ from base value |
|---|---|---|---|---|---|---|
| Development | 1 | 3 | -50000 | -50000 | 0.0% | 0 |
| Testing | 1 | 4 | -20000 | -20000 | 0.0% | 0 |
| Tooling and Ramp-Up Costs | 4 | 5 | -15000 | -15000 | 0.0% | 0 |
| Market Introduction | 4 | 5 | -20000 | -22000 | 10.0% | -2000 |
| Ongoing Marketing Costs | 5 | 12 | -10000 | -10000 | 0.0% | 0 |
| Unit Sales | 5 | 24 | 1500 | 1500 | 0.0% | 0 |
| Unit Price | 5 | 24 | 300.000 | 300.000 | 0.0% | 0.00 |
| Unit Production Cost | 5 | 24 | -92.500 | -92.500 | 0.0% | 0.00 |
| Discount Rate (per time period) | 2.50% | | | | | |

**Table 34: NPV analysis for 3rd scenario**

For the 4th scenario, we are setting the fluctuation to about 40% which is an addition of $6,000 is added to it. The percentage of NPV reduced by 0.4% with project NPV decreased by $14,319.

| | | | | | | |
|---|---|---|---|---|---|---|
| IRR = 51 % | | PROJECT NPV $ | 4,035,827 | | Changes from Base NPV | |
| | | Set Base | Base NPV | | % of NPV | $ change |
| | | | 4,050,146 | | -0.4% | -14319 |

MODEL VALUES

| | first | last | base burn rate | adjusted burn rate | %Δ from base value | $Δ from base value |
|---|---|---|---|---|---|---|
| Development | 1 | 3 | -50000 | -50000 | 0.0% | 0 |
| Testing | 1 | 4 | -20000 | -20000 | 0.0% | 0 |
| Tooling and Ramp-Up Costs | 4 | 5 | -15000 | -15000 | 0.0% | 0 |
| Market Introduction | 4 | 5 | -20000 | -28000 | 40.0% | -8000 |
| Ongoing Marketing Costs | 5 | 12 | -10000 | -10000 | 0.0% | 0 |
| Unit Sales | 5 | 24 | 1500 | 1500 | 0.0% | 0 |
| Unit Price | 5 | 24 | 300.000 | 300.000 | 0.0% | 0.00 |
| Unit Production Cost | 5 | 24 | -92.500 | -92.500 | 0.0% | 0.00 |
| Discount Rate (per time period) | 2.50% | | | | | |

**Table 35: NPV analysis for 4th scenario**

## 6.11 On-going Marketing Costs

Marketing cost identifies the risks and potential gains in a marketing campaign. The 'cost' refers to money, energy and/or time spent on a campaign. At this point, we understand the costs associated with developing a new marketing plan. Examining the cost associated with each individual marketing activity to assess the profitability of each. On-going marketing should not be overlook as well, this is because with proper on-going marketing, it can increase the familiarity of our product and it can also increase or maintain our current state of market share.

### 6.11.1 Sensitivity Analysis

We want to know how the fluctuation of the on-going marketing will affect the overall NPV. Therefore, we decided to set other settings all to 0% to see purely the effect of it. Similarly, we are simulating the account into 4 scenarios for comparison.

For the 1st scenario, whereby the fluctuation will be 0, this would mean that investing $10,000 per period with no extra funds in it with project NPV of $4,050,146

**IRR = 51 %**

PROJECT NPV $ **4,050,146**

Set Base

| Base NPV |
|---|
| 4,050,146 |

Changes from Base NPV

| % of NPV | $ change |
|---|---|
| 0.0% | 0 |

MODEL VALUES

| | first | last | base burn rate | adjusted burn rate | %Δ from base value | $Δ from base value |
|---|---|---|---|---|---|---|
| Development | 1 | 3 | -50000 | -50000 | 0.0% | 0 |
| Testing | 1 | 4 | -20000 | -20000 | 0.0% | 0 |
| Tooling and Ramp-Up Costs | 4 | 5 | -15000 | -15000 | 0.0% | 0 |
| Market Introduction | 4 | 5 | -20000 | -20000 | 0.0% | 0 |
| Ongoing Marketing Costs | 5 | 12 | -10000 | -10000 | 0.0% | 0 |
| Unit Sales | 5 | 24 | 1500 | 1500 | 0.0% | 0 |
| Unit Price | 5 | 24 | 300.000 | 300.000 | 0.0% | 0.00 |
| Unit Production Cost | 5 | 24 | -92.500 | -92.500 | 0.0% | 0.00 |
| Discount Rate (per time period) | 2.50% | | | | | |

**Table 36: NPV analysis for 1st scenario**

For the 2nd scenario, we are expecting that the fluctuation would be around -30% of fluctuation. With -30% of fluctuation, it would mean will be reduce $3,000 is allocated to it. The percentage of NPV raised by 0.5% with project NPV increased by $19,487.

IRR = 52 %

| PROJECT NPV $ | 4,069,633 |
|---|---|
| Base NPV | 4,050,146 |

Changes from Base NPV

| % of NPV | $ change |
|---|---|
| 0.5% | 19487 |

MODEL VALUES

| | first | last | base burn rate | adjusted burn rate | %∆ from base value | $∆ from base value |
|---|---|---|---|---|---|---|
| Development | 1 | 3 | -50000 | -50000 | 0.0% | 0 |
| Testing | 1 | 4 | -20000 | -20000 | 0.0% | 0 |
| Tooling and Ramp-Up Costs | 4 | 5 | -15000 | -15000 | 0.0% | 0 |
| Market Introduction | 4 | 5 | -20000 | -20000 | 0.0% | 0 |
| Ongoing Marketing Costs | 5 | 12 | -10000 | -7000 | -30.0% | 3000 |
| Unit Sales | 5 | 24 | 1500 | 1500 | 0.0% | 0 |
| Unit Price | 5 | 24 | 300.000 | 300.000 | 0.0% | 0.00 |
| Unit Production Cost | 5 | 24 | -92.500 | -92.500 | 0.0% | 0.00 |
| Discount Rate (per time period) | 2.50% | | | | | |

**Table 37: NPV analysis for 2nd scenario**

For the 3rd scenario, we are setting the fluctuation percentage to 10%. An increased fund of $1,000 from the marketing cost of $10,000. The percentage of NPV reduced by 0.2%, with projection NPV decreased by $6,496.

IRR = 51 %

| PROJECT NPV $ | 4,043,650 |
|---|---|
| Base NPV | 4,050,146 |

Changes from Base NPV

| % of NPV | $ change |
|---|---|
| -0.2% | -6496 |

MODEL VALUES

| | first | last | base burn rate | adjusted burn rate | %∆ from base value | $∆ from base value |
|---|---|---|---|---|---|---|
| Development | 1 | 3 | -50000 | -50000 | 0.0% | 0 |
| Testing | 1 | 4 | -20000 | -20000 | 0.0% | 0 |
| Tooling and Ramp-Up Costs | 4 | 5 | -15000 | -15000 | 0.0% | 0 |
| Market Introduction | 4 | 5 | -20000 | -20000 | 0.0% | 0 |
| Ongoing Marketing Costs | 5 | 12 | -10000 | -11000 | 10.0% | -1000 |
| Unit Sales | 5 | 24 | 1500 | 1500 | 0.0% | 0 |
| Unit Price | 5 | 24 | 300.000 | 300.000 | 0.0% | 0.00 |
| Unit Production Cost | 5 | 24 | -92.500 | -92.500 | 0.0% | 0.00 |
| Discount Rate (per time period) | 2.50% | | | | | |

**Table 38: NPV analysis for 3rd scenario**

For the 4th scenario, we are setting the fluctuation to about 40% which is an addition of $4,000 is added to it. The percentage of NPV reduced by 1.6% with project NPV decreased by $65,427.

IRR = 45 %

PROJECT NPV $ 3,984,719

Set Base

| Base NPV |
|---|
| 4,050,146 |

Changes from Base NPV

| % of NPV | $ change |
|---|---|
| -1.6% | -65427 |

MODEL VALUES

| | first | last | base burn rate | adjusted burn rate | %Δ from base value | $Δ from base value |
|---|---|---|---|---|---|---|
| Development | 1 | 3 | -50000 | -50000 | 0.0% | 0 |
| Testing | 1 | 4 | -20000 | -20000 | 0.0% | 0 |
| Tooling and Ramp-Up Costs | 4 | 5 | -15000 | -15000 | 0.0% | 0 |
| Market Introduction | 1 | 4 | -20000 | -20000 | 0.0% | 0 |
| Ongoing Marketing Costs | 5 | 12 | -10000 | -14000 | 40.0% | -4000 |
| Unit Sales | 5 | 24 | 1500 | 1500 | 0.0% | 0 |
| Unit Price | 5 | 24 | 300.000 | 300.000 | 0.0% | 0.00 |
| Unit Production Cost | 5 | 24 | -92.500 | -92.500 | 0.0% | 0.00 |
| Discount Rate (per time period) | 2.50% | | | | | |

**Table 39: NPV analysis for 4th scenario**

## 6.12 Unit Sales

Unit sales is a way of quantifying the total revenue that a business earns from selling products over a particular accounting period in terms of the earnings per product item sold. Companies use unit sales to determine how profitable a certain product is and, if necessary, to make improvements, expand, or cancel the product.

### 6.12.1 Sensitivity Analysis

We want to know how the fluctuation of the on-going marketing will affect the overall NPV. Therefore, we decided to set other settings all to 0% to see purely the effect of it. Similarly, we are simulating the account into 4 scenarios for comparison.

For the 1st scenario, whereby the fluctuation will be 0%, which is means no increased or reduce in sales. This would mean that no fluctuation in sales with project NPV at $4,050,146.

**IRR = 51 %**

PROJECT NPV $ **4,050,146**

Set Base

| Base NPV |
|---|
| 4,050,146 |

Changes from Base NPV

| % of NPV | $ change |
|---|---|
| 0.0% | 0 |

**MODEL VALUES**

| | first | last | base burn rate | adjusted burn rate | %Δ from base value | $Δ from base value |
|---|---|---|---|---|---|---|
| Development | 1 | 3 | -50000 | -50000 | 0.0% | 0 |
| Testing | 1 | 4 | -20000 | -20000 | 0.0% | 0 |
| Tooling and Ramp-Up Costs | 4 | 5 | -15000 | -15000 | 0.0% | 0 |
| Market Introduction | 4 | 5 | -20000 | -20000 | 0.0% | 0 |
| Ongoing Marketing Costs | 5 | 12 | -10000 | -10000 | 0.0% | 0 |
| Unit Sales | 5 | 24 | 1500 | 1500 | 0.0% | 0 |
| Unit Price | 5 | 24 | 300.000 | 300.000 | 0.0% | 0.00 |
| Unit Production Cost | 5 | 24 | -92.500 | -92.500 | 0.0% | 0.00 |
| Discount Rate (per time period) | 2.50% | | | | | |

**Table 40: NPV analysis for 1st scenario**

For the 2nd scenario, we are expecting that the around -30% of fluctuation. With -30% of fluctuation, it would mean will be reduce sale of 450 units. The percentage of NPV reduced by 32.6% with project NPV  decreased by $1,318,737.

**IRR = 41 %**

PROJECT NPV $ **2,731,409**

Set Base

| Base NPV |
|---|
| 4,050,146 |

Changes from Base NPV

| % of NPV | $ change |
|---|---|
| -32.6% | -1318737 |

**MODEL VALUES**

| | first | last | base burn rate | adjusted burn rate | %Δ from base value | $Δ from base value |
|---|---|---|---|---|---|---|
| Development | 1 | 3 | -50000 | -50000 | 0.0% | 0 |
| Testing | 1 | 4 | -20000 | -20000 | 0.0% | 0 |
| Tooling and Ramp-Up Costs | 4 | 5 | -15000 | -15000 | 0.0% | 0 |
| Market Introduction | 4 | 5 | -20000 | -20000 | 0.0% | 0 |
| Ongoing Marketing Costs | 5 | 12 | -10000 | -10000 | 0.0% | 0 |
| Unit Sales | 5 | 24 | 1500 | 1050 | -30.0% | -450 |
| Unit Price | 5 | 24 | 300.000 | 300.000 | 0.0% | 0.00 |
| Unit Production Cost | 5 | 24 | -92.500 | -92.500 | 0.0% | 0.00 |
| Discount Rate (per time period) | 2.50% | | | | | |

**Table 41: NPV analysis for 2nd scenario**

For the 3rd scenario, we are setting the fluctuation percentage of the unit sales to 10%. Sales increase 150 units. The percentage of NPV raised by 10%, with projection NPV increased by $439,578.

IRR = 54 %

PROJECT NPV $ 4,489,724

Set Base

| Base NPV |
|---|
| 4,050,146 |

Changes from Base NPV

| % of NPV | $ change |
|---|---|
| 10.9% | 439578 |

MODEL VALUES

| | first | last | base burn rate | adjusted burn rate | %Δ from base value | $Δ from base value |
|---|---|---|---|---|---|---|
| Development | 1 | 3 | -50000 | -50000 | 0.0% | 0 |
| Testing | 1 | 4 | -20000 | -20000 | 0.0% | 0 |
| Tooling and Ramp-Up Costs | 4 | 5 | -15000 | -15000 | 0.0% | 0 |
| Market Introduction | 4 | 5 | -20000 | -20000 | 0.0% | 0 |
| Ongoing Marketing Costs | 5 | 12 | -10000 | -10000 | 0.0% | 0 |
| Unit Sales | 5 | 24 | 1500 | 1650 | 10.0% | 150 |
| Unit Price | 5 | 24 | 300.000 | 300.000 | 0.0% | 0.00 |
| Unit Production Cost | 5 | 24 | -92.500 | -92.500 | 0.0% | 0.00 |
| Discount Rate (per time period) | 2.50% | | | | | |

**Table 42: NPV analysis for 3rd scenario**

For the 4th scenario, we are setting the fluctuation to about 40% which is an increased sales of 600 units. The percentage of NPV raised by 41.8% with project NPV increased by $1,692,888.

IRR = 55%

PROJECT NPV $ 5,743,034

Set Base

| Base NPV |
|---|
| 4,050,146 |

Changes from Base NPV

| % of NPV | $ change |
|---|---|
| 41.8% | 1692888 |

MODEL VALUES

| | first | last | base burn rate | adjusted burn rate | %Δ from base value | $Δ from base value |
|---|---|---|---|---|---|---|
| Development | 1 | 3 | -50000 | -50000 | 0.0% | 0 |
| Testing | 1 | 4 | -20000 | -20000 | 0.0% | 0 |
| Tooling and Ramp-Up Costs | 4 | 5 | -15000 | -15000 | 0.0% | 0 |
| Market Introduction | 1 | 4 | -20000 | -20000 | 0.0% | 0 |
| Ongoing Marketing Costs | 5 | 12 | -10000 | -14000 | 40.0% | -4000 |
| Unit Sales | 5 | 24 | 1500 | 2100 | 40.0% | 600 |
| Unit Price | 5 | 24 | 300.000 | 300.000 | 0.0% | 0.00 |
| Unit Production Cost | 5 | 24 | -92.500 | -92.500 | 0.0% | 0.00 |
| Discount Rate (per time period) | 2.50% | | | | | |

**Table 43: NPV analysis for 4th scenario**

## 6.13 Unit Prices

Each unit sold has a marketing cost associated with it, enabling us to measure how much it cost to market each unit. It is important to find the right balance on the price of product to ensure maximum profits. Therefore, by finding the right balance while dominated the market share is much more critical.

### 6.13.1 Sensitivity Analysis

We want to know how the fluctuation of the production revenue cost will affect the overall NPV. Therefore, we decided to set other settings all to 0% to see purely the effect of it. Similarly, we are simulating the account into 4 scenarios for comparison.

For the 1st scenario, whereby the fluctuation will be 0%, which means that the unit price will be remain as $300 with project NPV at $4,050,146.

**IRR = 51 %**

**PROJECT NPV $** 4,050,146

| Set Base | Base NPV |
|---|---|
| | 4,050,146 |

Changes from Base NPV

| % of NPV | $ change |
|---|---|
| 0.0% | 0 |

**MODEL VALUES**

| | first | last | base burn rate | adjusted burn rate | %Δ from base value | $Δ from base value |
|---|---|---|---|---|---|---|
| Development | 1 | 3 | -50000 | -50000 | 0.0% | 0 |
| Testing | 1 | 4 | -20000 | -20000 | 0.0% | 0 |
| Tooling and Ramp-Up Costs | 4 | 5 | -15000 | -15000 | 0.0% | 0 |
| Market Introduction | 4 | 5 | -20000 | -20000 | 0.0% | 0 |
| Ongoing Marketing Costs | 5 | 12 | -10000 | -10000 | 0.0% | 0 |
| Unit Sales | 5 | 24 | 1500 | 1500 | 0.0% | 0 |
| Unit Price | 5 | 24 | 300.000 | 300.000 | 0.0% | 0.00 |
| Unit Production Cost | 5 | 24 | -92.500 | -92.500 | 0.0% | 0.00 |
| Discount Rate (per time period) | | 2.50% | | | | |

**Table 44: NPV analysis for 1st scenario**

For the 2nd scenario, we are expecting -30% of fluctuation. With -30% of fluctuation, it would mean will be reduce sale price of $90. The percentage of NPV reduced by 47.1% with project NPV decreased by $1,906,607.

**IRR = 35 %**

**PROJECT NPV $** 2,143,539

| Set Base | Base NPV |
|---|---|
| | 4,050,146 |

Changes from Base NPV

| % of NPV | $ change |
|---|---|
| -47.1% | -1906607 |

**MODEL VALUES**

| | first | last | base burn rate | adjusted burn rate | %Δ from base value | $Δ from base value |
|---|---|---|---|---|---|---|
| Development | 1 | 3 | -50000 | -50000 | 0.0% | 0 |
| Testing | 1 | 4 | -20000 | -20000 | 0.0% | 0 |
| Tooling and Ramp-Up Costs | 4 | 5 | -15000 | -15000 | 0.0% | 0 |
| Market Introduction | 4 | 5 | -20000 | -20000 | 0.0% | 0 |
| Ongoing Marketing Costs | 5 | 12 | -10000 | -10000 | 0.0% | 0 |
| Unit Sales | 5 | 24 | 1500 | 1500 | 0.0% | 0 |
| Unit Price | 5 | 24 | 300.000 | 210.000 | -30.0% | -90.00 |
| Unit Production Cost | 5 | 24 | -92.500 | -92.500 | 0.0% | 0.00 |
| Discount Rate (per time period) | | 2.50% | | | | |

**Table 45: NPV analysis for 2nd scenario**

For the 3rd scenario, we are setting the fluctuation percentage to 10%. An increased selling price of $30. The percentage of NPV raised by 15.7%, with projection NPV increased by $635,535.

**IRR = 56 %**

PROJECT NPV $ 4,685,681

Set Base

| Base NPV |
|---|
| 4,050,146 |

Changes from Base NPV

| % of NPV | $ change |
|---|---|
| 15.7% | 635535 |

**MODEL VALUES**

| | first | last | base burn rate | adjusted burn rate | %∆ from base value | $∆ from base value |
|---|---|---|---|---|---|---|
| Development | 1 | 3 | -50000 | -50000 | 0.0% | 0 |
| Testing | 1 | 4 | -20000 | -20000 | 0.0% | 0 |
| Tooling and Ramp-Up Costs | 4 | 5 | -15000 | -15000 | 0.0% | 0 |
| Market Introduction | 4 | 5 | -20000 | -20000 | 0.0% | 0 |
| Ongoing Marketing Costs | 5 | 12 | -10000 | -10000 | 0.0% | 0 |
| Unit Sales | 5 | 24 | 1500 | 1500 | 0.0% | 0 |
| Unit Price | 5 | 24 | 300.000 | 330.000 | 10.0% | 30.00 |
| Unit Production Cost | 5 | 24 | -92.500 | -92.500 | 0.0% | 0.00 |
| Discount Rate (per time period) | 2.50% | | | | | |

**Table 46: NPV analysis for 3rd scenario**

For the 4th scenario, we are setting the fluctuation percentage to 40%. Selling price is increasing to $120. The percentage of NPV raised by 129.7%, with projection NPV increased by $5,251,887.

**IRR = 72%**

PROJECT NPV $ 9,302,033

Set Base

| Base NPV |
|---|
| 4,050,146 |

Changes from Base NPV

| % of NPV | $ change |
|---|---|
| 129.7% | 5251887 |

**MODEL VALUES**

| | first | last | base burn rate | adjusted burn rate | %∆ from base value | $∆ from base value |
|---|---|---|---|---|---|---|
| Development | 1 | 3 | -50000 | -50000 | 0.0% | 0 |
| Testing | 1 | 4 | -20000 | -20000 | 0.0% | 0 |
| Tooling and Ramp-Up Costs | 4 | 5 | -15000 | -15000 | 0.0% | 0 |
| Market Introduction | 1 | 4 | -20000 | -20000 | 0.0% | 0 |
| Ongoing Marketing Costs | 5 | 12 | -10000 | -14000 | 40.0% | -4000 |
| Unit Sales | 5 | 24 | 1500 | 2100 | 40.0% | 600 |
| Unit Price | 5 | 24 | 300.000 | 420.000 | 40.0% | 120.00 |
| Unit Production Cost | 5 | 24 | -92.500 | -92.500 | 0.0% | 0.00 |
| Discount Rate (per time period) | 2.50% | | | | | |

**Table 47: NPV analysis for 4th scenario**

## 6.14 Unit production cost

Unit production cost is the total expenditure incurred by a company to produce, store and sell one unit of a particular product or service. Unit costs include all fixed costs, or overhead costs, and all variable costs, or direct material costs and direct labor costs, involved in production. Determining the unit cost is a quick way to check if companies are efficient in producing their products.

### 6.14.1 Sensitivity Analysis

We want to know how the fluctuation of the unit production cost will affect the overall NPV. Therefore, we decided to set other settings all to 0% to see purely the effect of it. Similarly, we are simulating the account into 4 scenarios for comparison.

For the 1st scenario, whereby the fluctuation remains at 0%, this would mean the cost per unit for our product remains at $92.5. The percentage of NPV is at 0% with project NPV of $4,050,146.

IRR = 51 %

PROJECT NPV $ 4,050,146

Set Base

| Base NPV |
|---|
| 4,050,146 |

Changes from Base NPV

| % of NPV | $ change |
|---|---|
| 0.0% | 0 |

MODEL VALUES

| | first | last | base burn rate | adjusted burn rate | %Δ from base value | $Δ from base value |
|---|---|---|---|---|---|---|
| Development | 1 | 3 | -50000 | -50000 | 0.0% | 0 |
| Testing | 1 | 4 | -20000 | -20000 | 0.0% | 0 |
| Tooling and Ramp-Up Costs | 4 | 5 | -15000 | -15000 | 0.0% | 0 |
| Market Introduction | 4 | 5 | -20000 | -20000 | 0.0% | 0 |
| Ongoing Marketing Costs | 5 | 12 | -10000 | -10000 | 0.0% | 0 |
| Unit Sales | 5 | 24 | 1500 | 1500 | 0.0% | 0 |
| Unit Price | 5 | 24 | 300.000 | 300.000 | 0.0% | 0.00 |
| Unit Production Cost | 5 | 24 | -92.500 | -92.500 | 0.0% | 0.00 |
| Discount Rate (per time period) | 2.50% | | | | | |

**Table 48: NPV analysis for 1st scenario**

For the 2nd scenario, we are going to allocate about -30% fluctuation which is about reduce of $27.75 per unit production. The percentage of NPV raised by 14.5% with project NPV increased by $587,870.

**IRR = 55 %**

| | |
|---|---|
| PROJECT NPV $ | 4,638,016 |
| Set Base | Base NPV 4,050,146 |

Changes from Base NPV

| % of NPV | $ change |
|---|---|
| 14.5% | 587870 |

MODEL VALUES

| | first | last | base burn rate | adjusted burn rate | %Δ from base value | $Δ from base value |
|---|---|---|---|---|---|---|
| Development | 1 | 3 | -50000 | -50000 | 0.0% | 0 |
| Testing | 1 | 4 | -20000 | -20000 | 0.0% | 0 |
| Tooling and Ramp-Up Costs | 4 | 5 | -15000 | -15000 | 0.0% | 0 |
| Market Introduction | 4 | 5 | -20000 | -20000 | 0.0% | 0 |
| Ongoing Marketing Costs | 5 | 12 | -10000 | -10000 | 0.0% | 0 |
| Unit Sales | 5 | 24 | 1500 | 1500 | 0.0% | 0 |
| Unit Price | 5 | 24 | 300.000 | 300.000 | 0.0% | 0.00 |
| Unit Production Cost | 5 | 24 | -92.500 | -64.750 | -30.0% | 27.75 |
| Discount Rate (per time period) | 2.50% | | | | | |

**Table 49: NPV analysis for 2nd scenario**

For the 3rd scenario, we are setting the fluctuation percentage to 10%. An increased of unit production cost of $9.25 per unit. The percentage of NPV reduced by 4.8%, with projection NPV decreased by $195,957.

**IRR = 50 %**

| | |
|---|---|
| PROJECT NPV $ | 3,854,189 |
| Set Base | Base NPV 4,050,146 |

Changes from Base NPV

| % of NPV | $ change |
|---|---|
| -4.8% | -195957 |

MODEL VALUES

| | first | last | base burn rate | adjusted burn rate | %Δ from base value | $Δ from base value |
|---|---|---|---|---|---|---|
| Development | 1 | 3 | -50000 | -50000 | 0.0% | 0 |
| Testing | 1 | 4 | -20000 | -20000 | 0.0% | 0 |
| Tooling and Ramp-Up Costs | 4 | 5 | -15000 | -15000 | 0.0% | 0 |
| Market Introduction | 4 | 5 | -20000 | -20000 | 0.0% | 0 |
| Ongoing Marketing Costs | 5 | 12 | -10000 | -10000 | 0.0% | 0 |
| Unit Sales | 5 | 24 | 1500 | 1500 | 0.0% | 0 |
| Unit Price | 5 | 24 | 300.000 | 300.000 | 0.0% | 0.00 |
| Unit Production Cost | 5 | 24 | -92.500 | -101.750 | 10.0% | -9.25 |
| Discount Rate (per time period) | 2.50% | | | | | |

**Table 50 : NPV analysis for 3rd scenario**

For the 4th scenario,   we are setting the fluctuation percentage up to 40% which is about an extra of $37 per unit production. The percentage of NPV reduced by 20.3% with project NPV of decreased by $823,271.

IRR = 67%

PROJECT NPV $ 3,226,875

Set Base

| Base NPV |
|---|
| 4,050,146 |

Changes from Base NPV

| % of NPV | $ change |
|---|---|
| -20.3% | -823271 |

MODEL VALUES

| | first | last | base burn rate | adjusted burn rate | %Δ from base value | $Δ from base value |
|---|---|---|---|---|---|---|
| Development | 1 | 3 | -50000 | -50000 | 0.0% | 0 |
| Testing | 1 | 4 | -20000 | -20000 | 0.0% | 0 |
| Tooling and Ramp-Up Costs | 4 | 5 | -15000 | -15000 | 0.0% | 0 |
| Market Introduction | 1 | 4 | -20000 | -20000 | 0.0% | 0 |
| Ongoing Marketing Costs | 5 | 12 | -10000 | -10000 | 0.0% | 0 |
| Unit Sales | 5 | 24 | 1500 | 1500 | 0.0% | 0 |
| Unit Price | 5 | 24 | 300.000 | 300.000 | 0.0% | 0.00 |
| Unit Production Cost | 5 | 24 | -92.500 | -129.500 | 40.0% | -37.00 |
| Discount Rate (per time period) | 2.50% | | | | | |

**Table 51: NPV analysis for 4th scenario**

## 6.15 Discount rate

Discount rate is the rate used to discount future cash flows to the present value. Having much consideration, we decided on the discount rate to determine the estimated returns that required capital needed for our project is invested elsewhere.

For our project, we set our discount rate of project at 2.5% and the results actually yield a profit of our NVP. This mean that we should continue with our project as we would expect this project would generate more returns than alternative investment.

## 6.16 Overall Sensitivity Analysis

Company used overall sensitivity analysis to understand the key relationship between economic return, assumptions and variables of the model. The analysis can be done by having different individual parameters by -30%, 10%, 40% and the results are shown in the Table below:

| | % Δ from base value | Periods (Quarter ) | % Change in NPV | $ Change in NPV |
|---|---|---|---|---|
| Development | -30 | 1 - 3 | 1.10% | 42840 |
| Development | 10 | 1 - 3 | -0.40% | -14280 |
| Development | 40 | 1 - 3 | -1.40% | -57121 |
| Testing | -30 | 1 - 4 | 0.60% | 22571 |
| Testing | 10 | 1 - 4 | -0.20% | -7524 |
| Testing | 40 | 1 - 4 | -0.70% | -30096 |
| Tooling and Ramp-Up Costs | -30 | 4 - 5 | 0.20% | 8054 |
| Tooling and Ramp-Up Costs | 10 | 4 - 5 | -0.10% | -2685 |
| Tooling and Ramp-Up Costs | 40 | 4 - 5 | -0.30% | -10739 |
| Marketing Introduction | -30 | 4 - 5 | 0.30% | 10738 |
| Marketing Introduction | 10 | 4 - 5 | -0.10% | -3580 |
| Marketing Introduction | 40 | 4 - 5 | -0.40% | -14319 |
| On-going Marketing Cost | -30 | 5 - 12 | 0.50% | 19487 |
| On-going Marketing Cost | 10 | 5 - 12 | -0.20% | 6496 |
| On-going Marketing Cost | 40 | 5 - 12 | -1.60% | -65427 |
| Unit Sales | -30 | 5 - 24 | -32.60% | -1318737 |
| Unit Sales | 10 | 5 - 24 | 10.90% | 439578 |
| Unit Sales | 40 | 5 - 24 | 41.80% | 1692888 |
| Unit Price | -30 | 5 - 24 | -47.10% | -1906607 |
| Unit Price | 10 | 5 - 24 | 15.70% | 635535 |
| Unit Price | 40 | 5 - 24 | 129.70% | 5251887 |
| Unit Production Cost | -30 | 5 - 24 | 14.50% | 587870 |
| Unit Production Cost | 10 | 5 - 24 | -4.80% | -195957 |
| Unit Production Cost | 40 | 5 - 24 | -20.30% | -823271 |

**Table 52: Overall Sensitivity Analysis**

The product life cycle for Eliminoise$^{TM}$ is at 24 periods. As for our development and testing phase, we will conclude at the 4th period. Subsequently, we will introduce our product into the market at the 5th period. From the sensitively analysis table, we can realised that the different percentage deviation of respective parameters will make changes in the overall NPV and revenue, which resulted the percentage deviation of some parameters to a large amount of variation in both the NPV and revenue.

The critical parameters are the unit price, unit sales and unit production cost. These parameters are very critical, and required to be monitored and adjusted closely as it affects the overall performance of the financial model. NPV is most sensitive to the change of unit price among these parameters, because any changes in unit price can be immediately reflected in the revenue.

At the later stage, when the product sales and production scale increased, production cost is unlikely to increase. It will affect the NPV negatively but it is not as prominent as the deviation in unit price and unit sales.

## 6.17 Summary of Analysis

From the base case financial models, we are able to break-even within 1 year 3 months, and have a positive NPV and IRR is higher than the Company discount rates. All these signs point towards taking up this project to develop Eliminoise™. From the sensitivity analysis, it shows that the critical parameters are the unit price, unit sales and unit production cost. These parameters will directly affect the impact of the NPV and IRR. From the competitor analysis, we believed our product, Eliminoise™, will have a great impact in the market and able to obtain the market share after introducing to the market for 1 year.

## 6.18 Risk Assessment

Businesses exist in a competitive environment, fierce competition with each other to provide the best possible value for money goods, and to offer the most suitable range of products for their customers. Under such circumstances, it is critically important to have an effective risk assessment when the worst case happened with the NPV is in negative value. It allow you understand what are the things have gone wrong and provide you alternative ways to react when facing the worst scenario. Below are the steps we can perform when we are facing an operating at a loss :

1. Reduce expenses.
    - Reduce component spending
    - Redesign the product which required lesser parts
    - Try to negotiate better deals from the suppliers
    - By out sourcing which have a lower operational and labor costs
    - Cut down or stop the production

2. Increase sales.

- Improve sales method
- Providing salespeople with consistent, repeatable sales methodology
- Create more effective marketing campaigns
- Find the best business location

3. Get advice

- Get advice from an accountant or business advisor that can help you turn it around and get your business back on track and avoid trouble ahead.

4. Market survey

- To determine whether there really is a market for the product
- Improve the product or service based on findings about what the customer's needs.
- Revised the price based on popular profit margins, competitors' prices, financing options, or the price a customer is willing to pay.
- Identify competitors

# 7. Project Management & Scheduling

## 7.1 Project Development Organization

Organization development is very important as it affects the performance of the organization and effectiveness of its people.  To maintain smooth work flow and good coordination among project Teams, it's quite important to have an effective project structure.

3S technology is a startup company with head count of approximately 30 employees. In order to ascertain the launch date of the Eliminoise™ project, it is critical to have well-structured and designed project organization. The objective of this organizational is to facilitate the coordination and implementation of project activities. By defining the responsibilities and working relationships between the different project team, it can reduce dubiousness and perplexity throughout the phase of the project.

Furthermore, by having specialized project teams, each team member can have better fixate on their respective area, and so that we can maximize the productivity of the overall project. As such, we have organized the project team into six key specializations or technology areas and show in below organization structure.

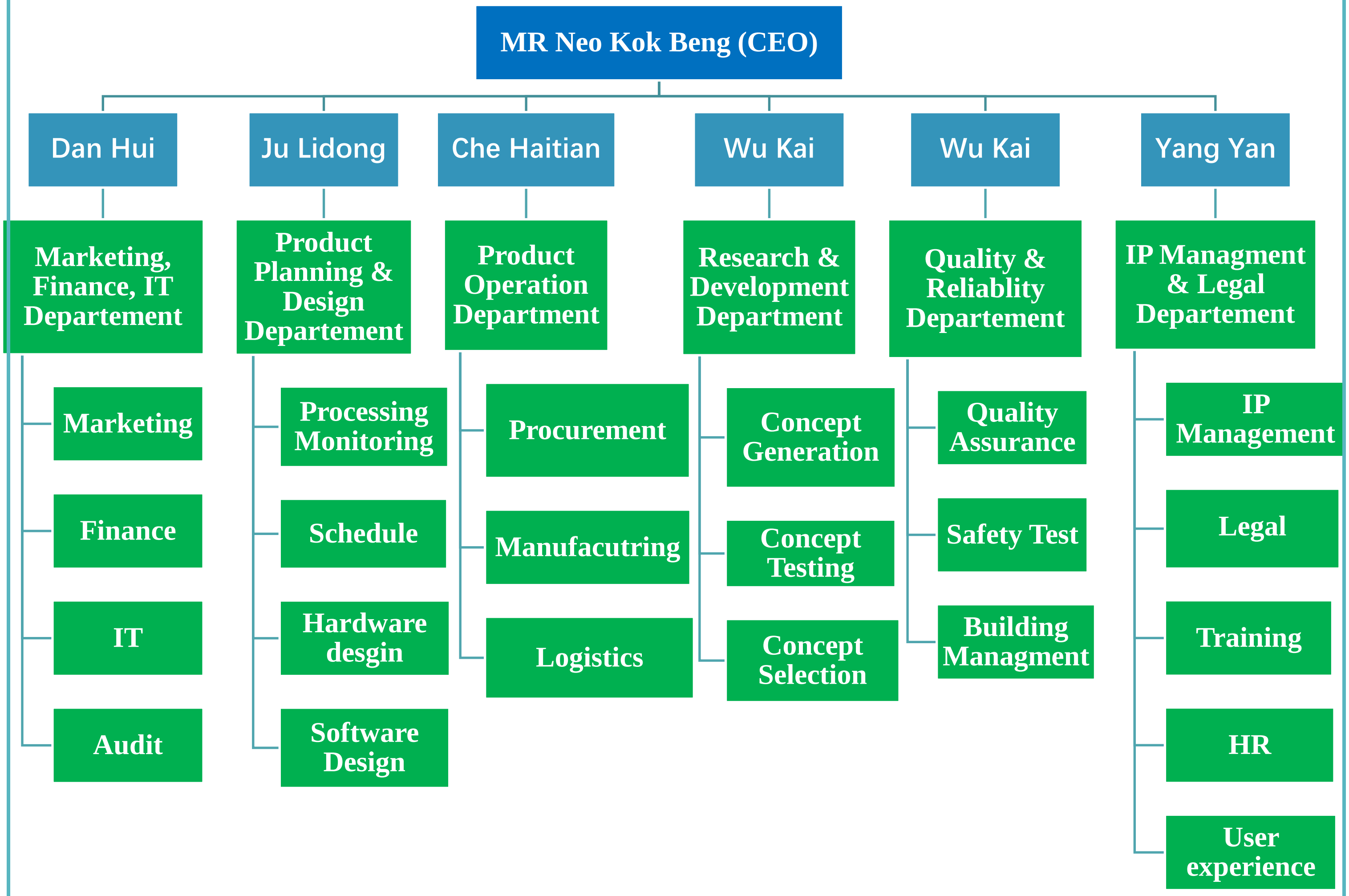

**Figure 37 : Eliminoise Organisation Chart**

- CEO(Mr Neo Kok Beng)
- Marketing , Finance, Aduit & IT Department(Lead by Dan Hui)
- Product Planning & Department(Lead by Ju Lidong)
- Product Operation Department(Lead by Che Haitian)
- Research & Development Department(Lead by Wu Kai)
- Quality & Reliablity Department(Lead by Wu Kai)
- IP Management & Legal Department(Lead by Yang Yan)

### 7.1.1 Team Members and Their Responsibilities

**Mr.Neo Kok Beng- Chief executive officer**

Executive Officer ("CEO") is responsible for leading the whole company and also the head of the corporation. He is directly in charge of all of the departments and projects. He support to all project team leader by advising and guiding them to complete the project successfully on time. In our development organization plan, the CEO is also responsible for project manager roles to make sure the final product to meet the requried quality.

**Responsibilities**

- Responsible for profit and loss of the product sales
- Maintain the good relationship with government
- Make decision in the critial crisis management
- Decide the development direction of company
- Plan for company long and short term plan with mission, vision and strategy
- Decision making in planning and organization difficulties
- Monitoring risk and internal control of the company
- Deal with stake holders effectively
- Ensure the budget of the company and monitoring the cost of organization

**Dan Hui - Marketing , Finance , IT & Aduit Department**

Marketing Team Member: Sales Executive, 1 Sales Enginner, 1 Digiter Maketing Executive

Finance Team Member:Chief Finance Officer, 2 Finance Analysts, 1 Financial Risk Manager

IT Team Member:1 Desktop Support & Infrasturure Enginner, 1 Software developer

Product Marketing team lead the company to make sure the launch of a new product become the successful one. Prepare the proposal of time to launch our product and evalute how to advertise the product and boost the sales volume. The marketing team also help other team to understand the issues that sales team face in lauching and marketing the product.

The finance team will be responsilbe for the finance planning, analysis product sales volume and finance control . The team will determine the development cost and profit for the overall product lifescyle from design ,development, manufacturing to retirement. In addition, the team will monitor supplier account, process supplier invoice and maintain the purchase order system. The team will also be responsible for monitoring cash reserves and financial records, as well as preparing financial statements,account payable (AP) and account receivable(AR).  Furthermore, the team will manage the project budget and financial risks.The team will need to identify the target market and perform primary and secondary market research.

IT Team will in charge of help desk support and maintian the company all IT relalted equipment. Make sure the internet and telephone line, printer and fax device ,server, sharepoint, website running without any major interrruption. Install all the needed software and OS for the company staff. Write the dizaster recovery (DR) plan for the data and interenet. IT team also incharge for the company internal Audit.

**Responsibilities**

- Maintenance company IT equipment and servcies to meet the company requirement
- Providing IT help desk support to all staff
- Write the company dizaster recovery (DR) plan and configure the DR related enviroment
- Internal audit and 6S(Sort,Shine,Sort,Safety,Sustain,Standardize) insepection
- Assist Product design team to develop and test  the software to meet product function requirement.
- Undesrstand current and future potential customers
- Analyzing the primary and secondary market
- Management the customer loyalty  by getting feedback from the user.
- Preparing marketing strategy and plan for local and international market
- Managing agencies and Measuring success
- Managing budgets based on current market & uncertainty
- Managing account receiveable and acccount payable

**<u>Ju Lidong – Product Planning and design Department</u>**

Product Planning and Design Department Team Member: Chief Product Analyst, 1 Marketing Manager, 2 Product Designer.

Product planning team play the critical role for pre-production to prepare for the all kinds of module of products before the stage of manufacturing. The product planning department also corporate with marketing and finance department to decide the new product price, components, specification,storage,disturbing and promotion.

The team will be responsible for collecting and analyzing survey results and data to determine customer needs. Determine the product feature and specification based on the customer needs survey result.  The team will also investigate and analyses the competitors' products or similar products to identify the suitable feature in order to distinguish our product. The product design team also corporate with IT team to design and develop the software.

**Responsibilities**

- Scheduling the current future product desgin plan
- Design the software feature and test the software after IT team develop the software
- Determine the product feature and specification
- Improve the feature based on feedback from user
- Prepare for weekly or monthly report of planning for project team
- Testing new product by building prototype
- Verification test planning and test execution result
- Analyzing verification results and reporting the analyzed data
- Document the design detail for the future reference

**<u>Che Haitian – Production and  Operation Department</u>**

Team members: Chief Operations Manager, 2 Systems Engineers, 1 Warehouse Operations Manager, 1 Supply Chain Manager

The operation team responsible for  planing the raw components and facilities to optimise the manufacturing process. The production team is responsible for producion and test the quality of the final product.

The team will manage the procurement of raw materials and components as well as the assembly equipment. The team will also be responsible for preparing the manufacturing facility through installation of production equipment and ensuring the overall operability of the manufacturing system. The team will have an important role to check the component quality and team leader should to maintain the good relationship the supplier.

**Responsibilities**

- Planning and facilitating the manufacturing components and process
- Assembling the production equipment and machines
- Organizing production schedules of manufacturing
- Managing the raw materials procument and storage
- Testing the comppnent quality before production
- Monitoring and controlling the production flow in order to avoid long queue
- Yield monitoring and analyzing the failure source
- Plan the dizaster recovery (DR) plan for the manufacutring line
- Imporve the production process to get better productivity

**Wu Kai - Research & Development Department**

Team members: 3 Product develop Engineer, 2 Research Engineer

This department will cooperate with design department to collect and research the customers' needs. Based on customer need survey to generate the concept, proceed the concept selecting and concept testing until get final product prototype. The department will cooperate with marketing department to investigate the current available product in market, primary and secondary market and competitors to improve our own product feature and user experience.

**Responsibilities**

- Collecting the survey results to organize customers' needs and wants
- Responsible for concept generating, testing and scoring
- Monitoring the current available product in market for research purpose
- Research & develop for the future production
- Building prototype product, testing and debugging software with IT department

**Wu Kai - Quality & Reliablity Department**

Team: Chief Quality Engineer, 2 Quality Engineers

The team will be responsible for the development and operation of quality control systems, as well as the development of test approaches and procedures. The team will write testing procedure to analyses the product performance with adherence to industry specifications, codes and standards to ensure quality compliance.

The quality and reliability department team monitor product quality control system after the test of the product. The department checks production defects to avoid problems before delivering the product to customers. This department also in charge of staff working places safety and production process meet the safety regulation.

**Responsibilities**

- Ensure product reliability and durability
- Testing product feature to meet the desired requirement
- Provide safety audit to all the departments
- Overall quality control
- Working places safety and environment friendly
- Update the company safety regulation based on government requirement

**<u>Yang Yan - IP Management, Legal , HR &Training Department</u>**

Team: Corporate Attorney, 1 Assistant Corporate Attorney, 1 Legal advisor, 1 HR Executive

The IP Management & Legal team will be responsible for addressing and managing legal issues concerning the intellectual property rights of 3S technology. They will be carrying out the filing of patents and applications of copyright and trademark.

This team will also in charge of staff recruitment, staff benefit, payroll, safety and production training. Deal with customer feedback and complain to improve the user experience.

The team will also be responsible for regulatory and certification reporting activities, manage product intellectual property right. The team shall provide legal advice on intellectual property infringement to the product design team. The team shall also process patents filing through the technologies of mobile application products.

**Responsibilities**

- Provide the training to internal staff
- Responsible for the user feedback and complain
- Staff benefit and payroll
- Developing guidelines for project team and customer
- Responsible for Intellectual properties
- Working with Quality department to write staff working environment safety regulation
- Any legal dispute related to product or company

## 7.2 Project Activities

| Activity | Task Name | Duration | Start | Finish | Predecessors |
|---|---|---|---|---|---|
| 1 | **Product Launch Plan** | **326 days** | **Mon 1/2/17** | **Mon 4/2/18** | |
| 2 | **Preliminary Market Analysis** | **15 days** | **Mon 1/2/17** | **Fri 1/20/17** | |
| 3 | **Phase 1 - Planning** | **19 days** | **Mon 1/23/17** | **Thu 2/16/17** | |
| 4 | **Determine sales objectives** | **2 days** | **Mon 1/23/17** | **Tue 1/24/17** | **1** |
| 5 | **Define launch goals (launch timing and publicity objectives)** | **2 days** | **Wed 1/25/17** | **Thu 1/26/17** | **3** |
| 6 | **Determine partners** | **10 days** | **Fri 1/27/17** | **Thu 2/9/17** | |
| 7 | **Identify channel partners** | **10 days** | **Fri 1/27/17** | **Thu 2/9/17** | **4** |
| 8 | **Identify retail partners** | **5 days** | **Fri 1/27/17** | **Thu 2/2/17** | **4** |
| 9 | **Identify online opportunities** | **5 days** | **Fri 1/27/17** | **Thu 2/2/17** | **4** |
| 10 | **Establish Launch Budget** | **5 days** | **Fri 2/10/17** | **Thu 2/16/17** | |
| 11 | **Identify budget requirements** | **4 days** | **Fri 2/10/17** | **Wed 2/15/17** | **6,7,8** |
| 12 | **Obtain launch budget approval from CEO** | **1 day** | **Thu 2/16/17** | **Thu 2/16/17** | **10** |
| 13 | **Phase 2 - Initiation** | **86 days** | **Fri 2/17/17** | **Fri 6/16/17** | |
| 14 | **Kickoff product launch** | **1 day** | **Fri 2/17/17** | **Fri 2/17/17** | **11** |
| 15 | **Marketing** | **85 days** | **Mon 2/20/17** | **Fri 6/16/17** | |
| 16 | **Customer needs survey** | **20 days** | **Mon 2/20/17** | **Fri 3/17/17** | **13** |
| 17 | **Interpret customer needs** | **10 days** | **Mon 3/20/17** | **Fri 3/31/17** | **15** |
| 18 | **Generate preliminary specifications** | **5 days** | **Mon 4/3/17** | **Fri 4/7/17** | **16** |
| 19 | **Benchmark on customer needs** | **4 days** | **Mon 4/3/17** | **Thu 4/6/17** | **16** |
| 20 | **Concept generation** | **10 days** | **Mon 4/10/17** | **Fri 4/21/17** | **17,18** |
| 21 | **Concept Selection** | **5 days** | **Mon 4/24/17** | **Fri 4/28/17** | **19** |
| 22 | **Concept Testing** | **10 days** | **Mon 5/1/17** | **Fri 5/12/17** | **20** |
| 23 | **Concept Scoring** | **5 days** | **Mon 5/15/17** | **Fri 5/19/17** | **21** |
| 24 | **Generate Final specifications** | **20 days** | **Mon 5/22/17** | **Fri 6/16/17** | **22** |
| 25 | **Engineering** | **5 days** | **Mon 2/20/17** | **Fri 2/24/17** | |
| 26 | **Align product release timing with sales goal** | **5 days** | **Mon 2/20/17** | **Fri 2/24/17** | **13** |
| 27 | **Manufacturing** | **10 days** | **Mon 2/20/17** | **Fri 3/3/17** | |
| 28 | **Prepare for volume manufacturing** | **10 days** | **Mon 2/20/17** | **Fri 3/3/17** | **13** |
| | **according to sales goals** | | | | |
| 29 | **Sales** | **10 days** | **Mon 2/20/17** | **Fri 3/3/17** | |
| 30 | **Plan sales staffing and training to support sales goals** | **10 days** | **Mon 2/20/17** | **Fri 3/3/17** | **13** |
| 31 | **Product Support** | **5 days** | **Mon 2/20/17** | **Fri 2/24/17** | |
| 32 | **Plan team staffing to support sales goals** | **5 days** | **Mon 2/20/17** | **Fri 2/24/17** | **13** |
| 33 | **Phase 3 - Execution** | **165 days** | **Mon 6/19/17** | **Fri 2/2/18** | |
| 34 | **Marketing** | **18 days** | **Mon 6/19/17** | **Wed 7/12/17** | |
| 35 | **Evaluate market and refine messaging** | **5 days** | **Mon 6/19/17** | **Fri 6/23/17** | **12** |
| 36 | **Create product specification literature** | **6 days** | **Mon 6/26/17** | **Mon 7/3/17** | **34** |
| 37 | **Communicate product launch details to internal organization** | **2 days** | **Tue 7/4/17** | **Wed 7/5/17** | **34,35** |
| 38 | **Create sales, field, and product support training** | **5 days** | **Thu 7/6/17** | **Wed 7/12/17** | **36** |
| 39 | **Update launch plan based on forecast** | **1 day** | **Thu 7/6/17** | **Thu 7/6/17** | **36** |
| 40 | **Engineering** | **40 days** | **Mon 6/19/17** | **Fri 8/11/17** | |
| 41 | **Architecture Design** | **30 days** | **Mon 6/19/17** | **Fri 7/28/17** | **23** |
| 42 | **Component Design** | **20 days** | **Mon 6/19/17** | **Fri 7/14/17** | **23** |
| 43 | **Software Develop** | **40 days** | **Mon 6/19/17** | **Fri 8/11/17** | **23** |
| 44 | **Manufacturing** | **125 days** | **Mon 8/14/17** | **Fri 2/2/18** | |
| 45 | **Create prototype products** | **90 days** | **Mon 8/14/17** | **Fri 12/15/17** | **42** |
| 47 | **Test prototype** | **15 days** | **Mon 12/18/17** | **Fri 1/5/18** | **44** |
| 48 | **debug prototype** | **15 days** | **Mon 1/8/18** | **Fri 1/26/18** | **45** |
| 49 | **Finalize Prototype** | **5 days** | **Mon 1/29/18** | **Fri 2/2/18** | **46** |
| 50 | **Sales** | **30 days** | **Mon 6/19/17** | **Fri 7/28/17** | |
| 51 | **Establish sales channels** | **30 days** | **Mon 6/19/17** | **Fri 7/28/17** | **12** |
| 52 | **Product Support** | **45 days** | **Mon 6/19/17** | **Fri 8/18/17** | |
| 53 | **Establish product support mechanisms** | **40 days** | **Mon 6/19/17** | **Fri 8/11/17** | **12** |
| 54 | **Patenting for hardware and software design** | **5 days** | **Mon 8/14/17** | **Fri 8/18/17** | **40,41,42** |
| 55 | **Phase 4 - Release To Manufacture** | **41 days** | **Mon 2/5/18** | **Mon 4/2/18** | |
| 56 | **Manufacturing** | **25 days** | **Mon 2/5/18** | **Fri 3/9/18** | |
| 57 | **Certify product** | **10 days** | **Mon 2/5/18** | **Fri 2/16/18** | **47** |
| 58 | **Manufacture the volume of product planned to meet sales objectives.** | **15 days** | **Mon 2/19/18** | **Fri 3/9/18** | **55** |
| 59 | **Sales** | **18 days** | **Mon 2/5/18** | **Wed 2/28/18** | |
| 60 | **Hire and train sales staff about the product specification** | **18 days** | **Mon 2/5/18** | **Wed 2/28/18** | **47** |
| 61 | **Product Support** | **30 days** | **Mon 2/5/18** | **Fri 3/16/18** | |
| 62 | **Hire and train product support sales** | **30 days** | **Mon 2/5/18** | **Fri 3/16/18** | **47** |
| 63 | **Final Quality Review** | **11 days** | **Mon 3/19/18** | **Mon 4/2/18** | |
| 64 | **Conduct final quality review** | **10 days** | **Mon 3/19/18** | **Fri 3/30/18** | **60,58,56** |
| 65 | **Launch Product Celebration** | **1 day** | **Mon 4/2/18** | **Mon 4/2/18** | **62** |

**Table 53: Project Task List and Network Analysis**

### 7.2.1 Gantt Chart

Gantt chart is developed and used in our project planning and execution in order to oversee the planning and scheduling of our project activities, it is used to help us visualize and track the timeline of our project. Through this chart, we can identify the start and end dates of project tasks, as well as the dependencies between tasks. This chart shows the entire development phase for our project up till the launch of our first product.

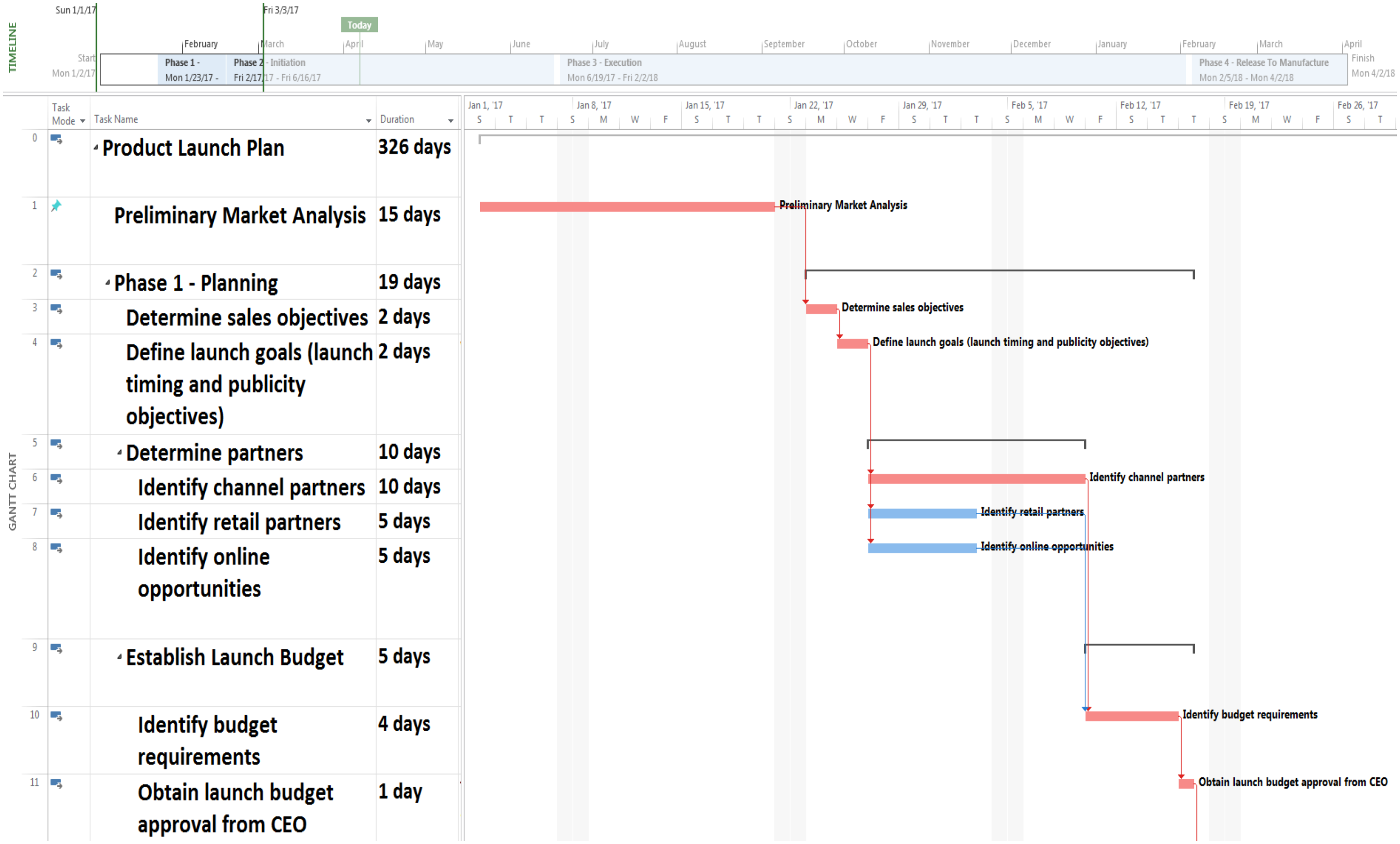

**<u>Figure 38: Gantt chart Phase 1</u>**

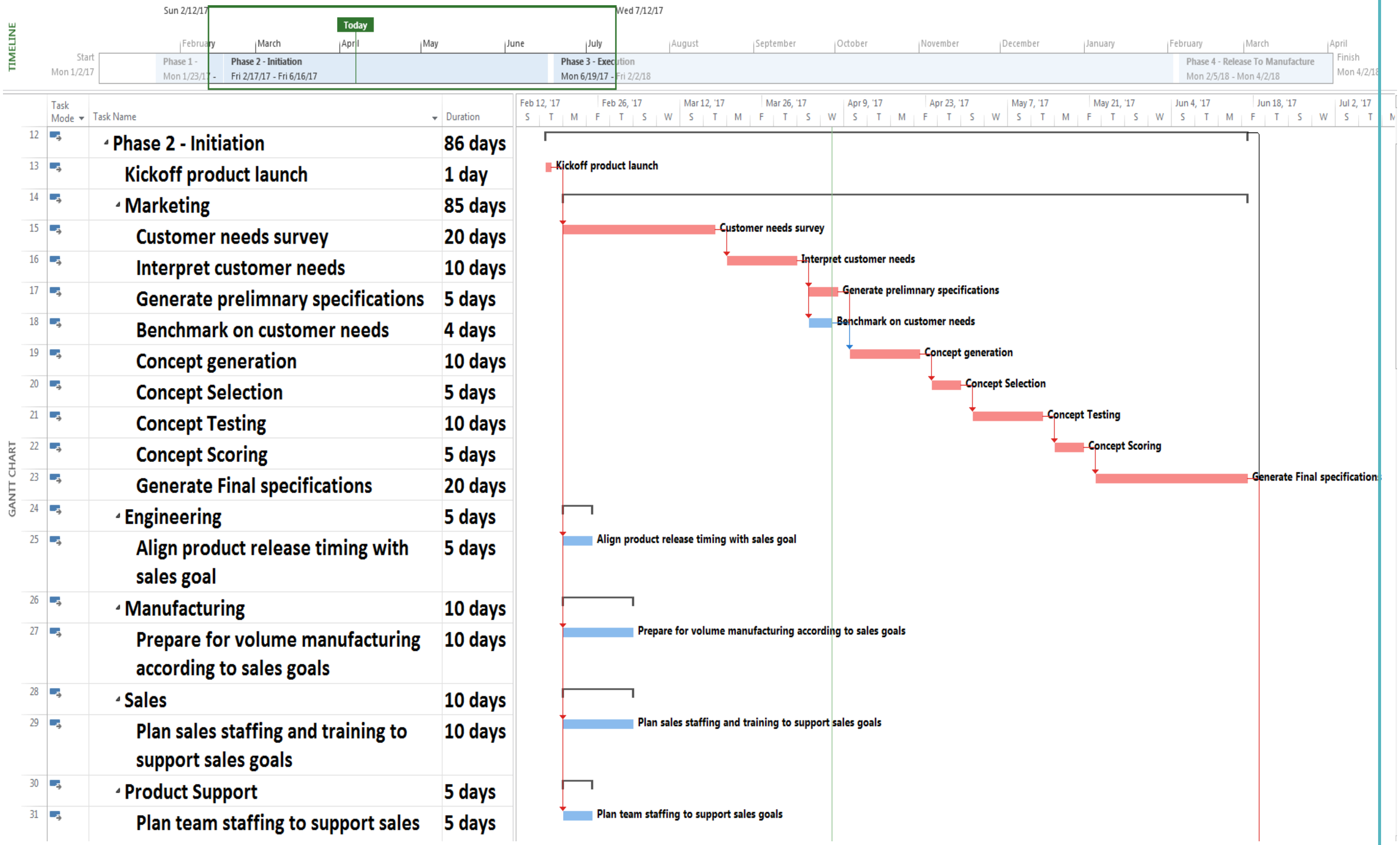

**Figure 39: Gantt chart Phase 2**

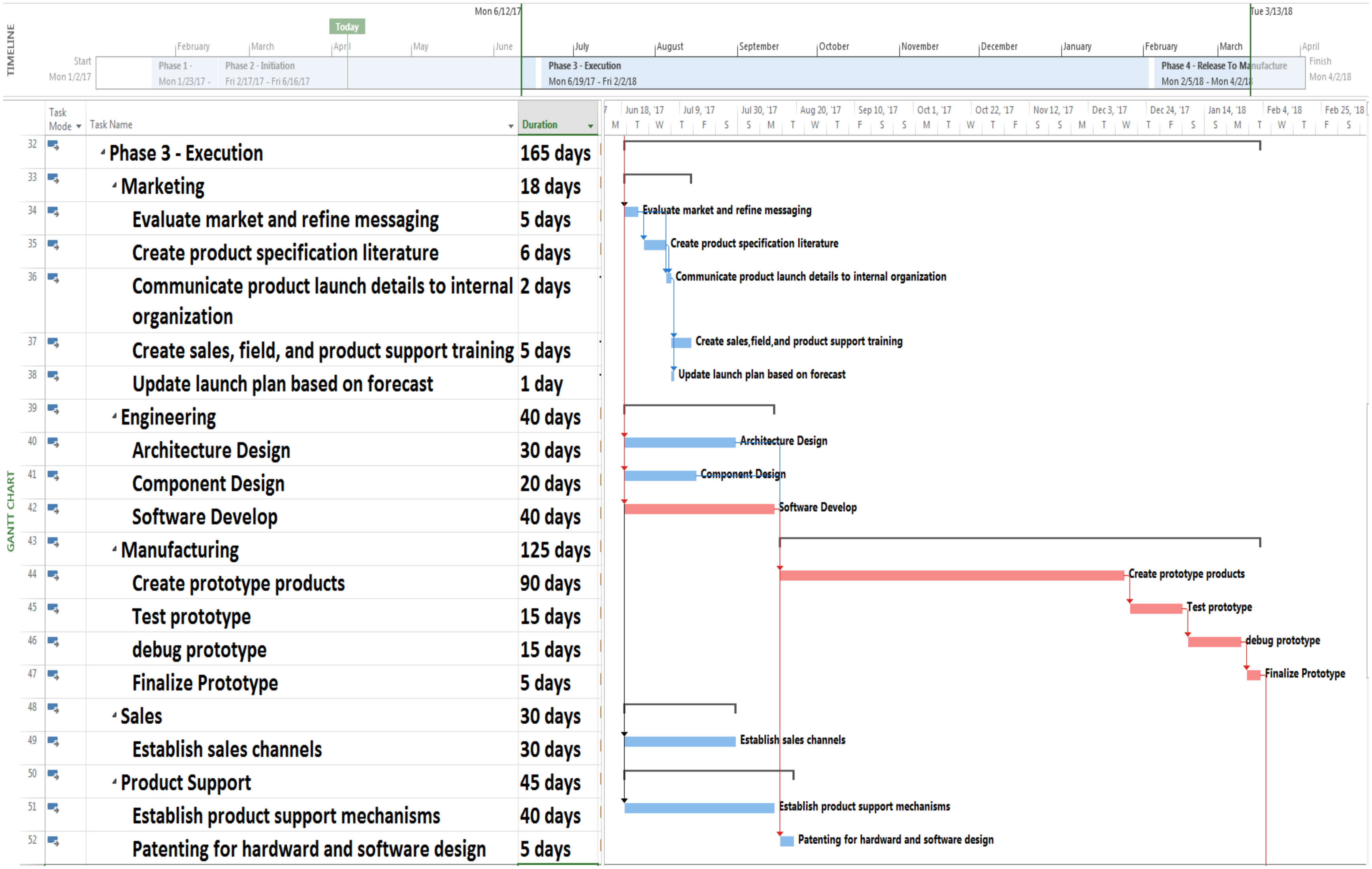

**Figure 40: Gantt chart Phase 3**

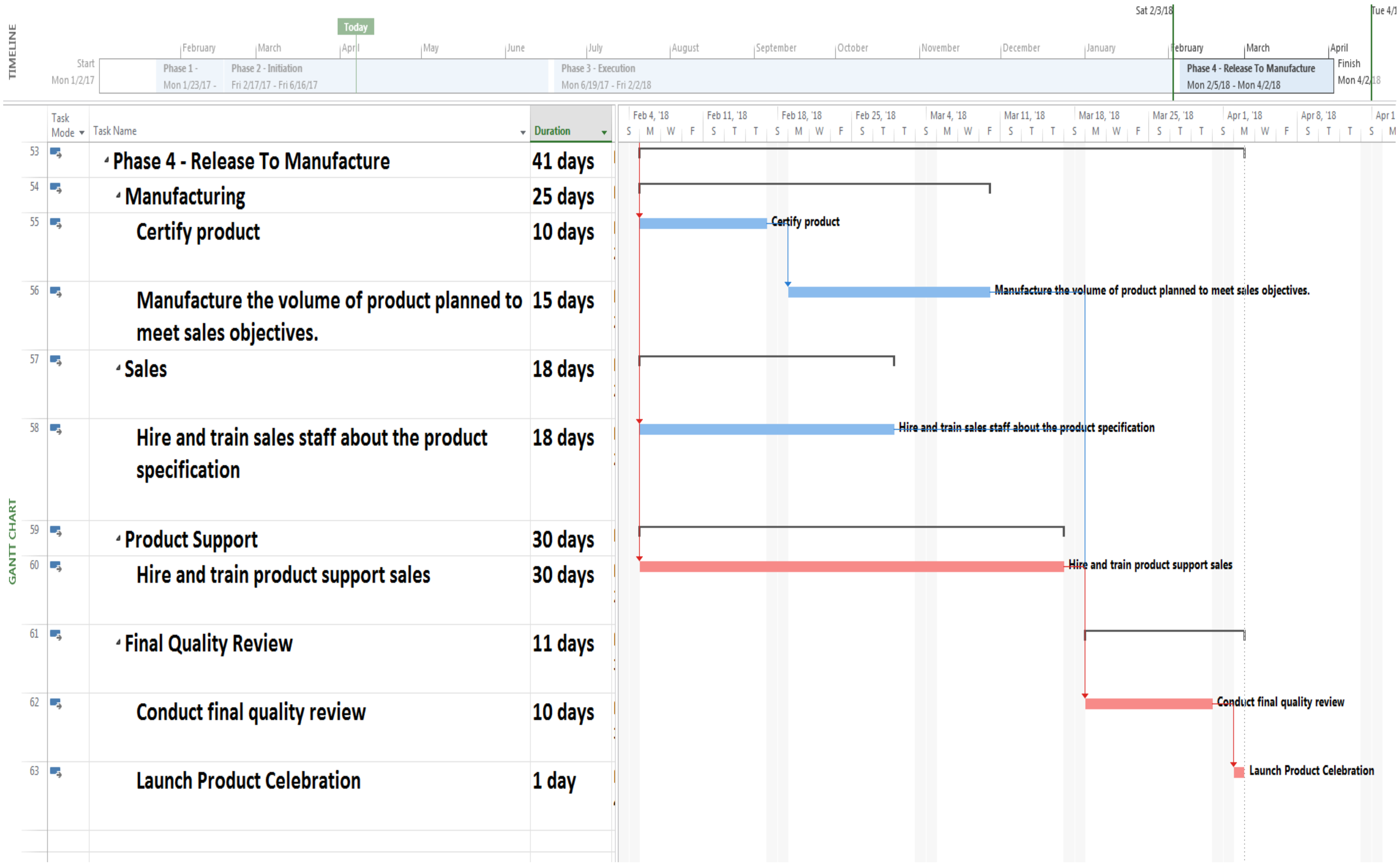

**Figure 41: Gantt chart Phase 4**

### 7.2.2 Network Diagram (PERT Chart)

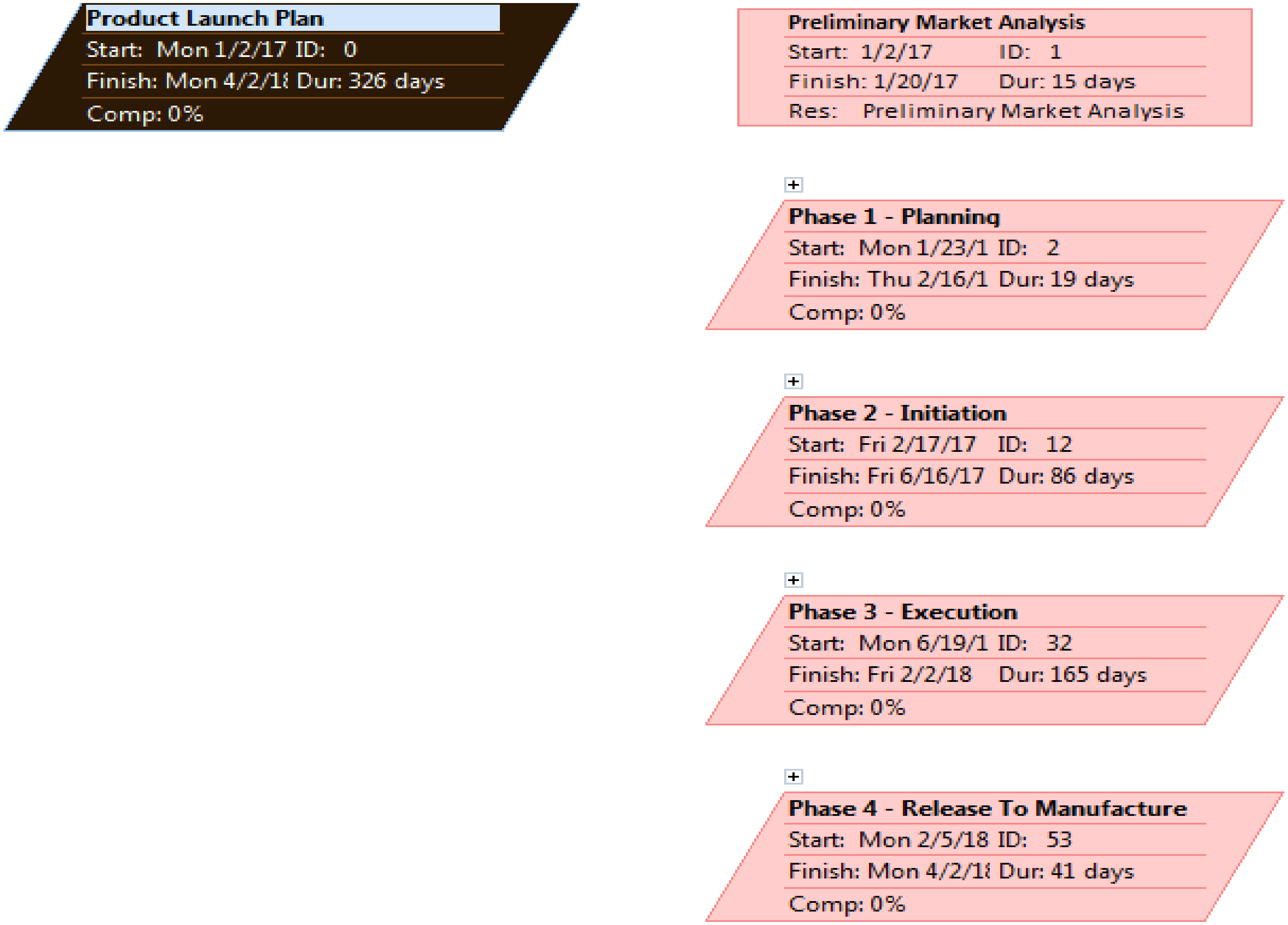

NETWORK DIAGRAM

**Figure 42: Network Diagram**

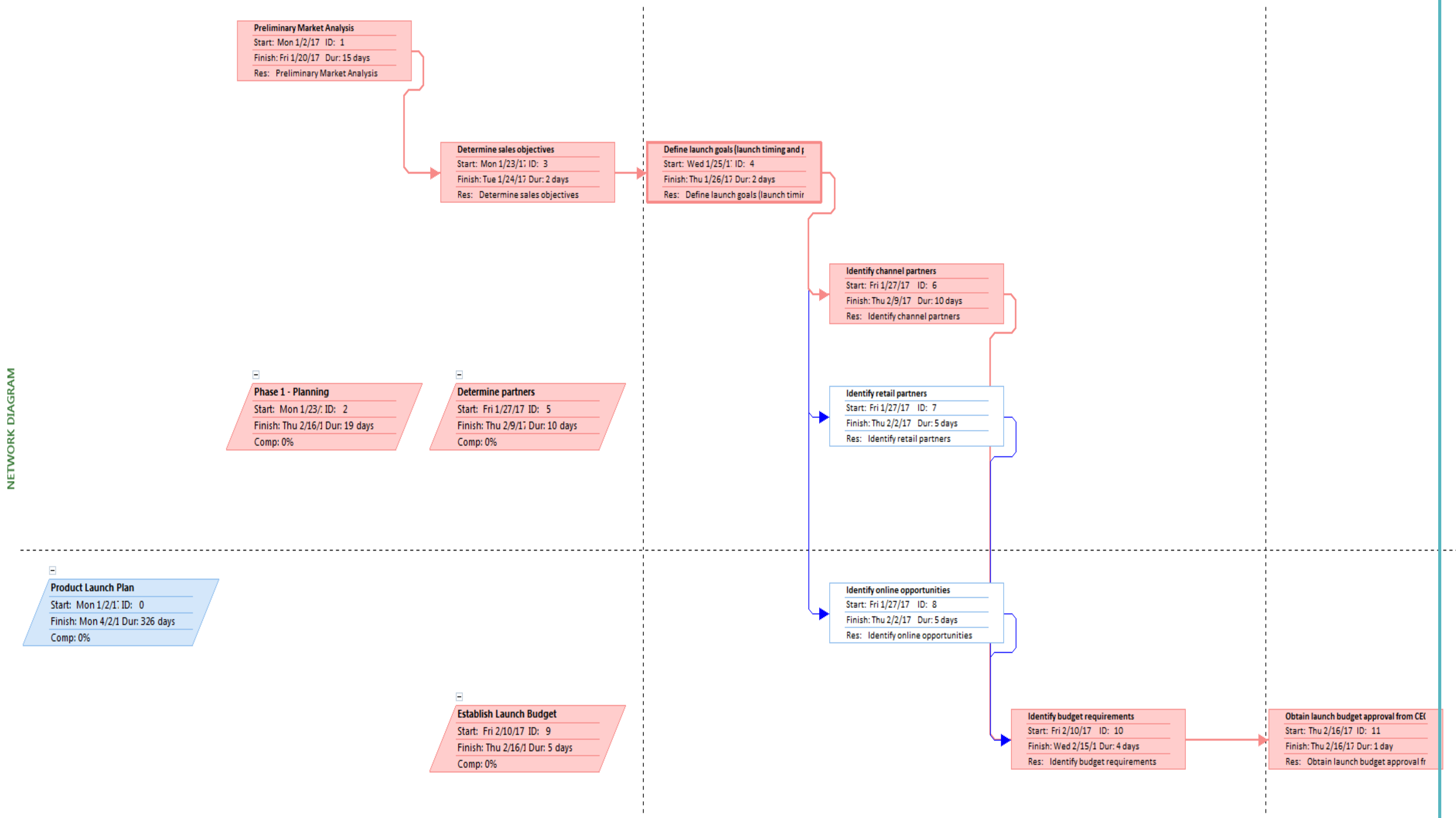

**Figure 43: Function Diagram Preliminary Market Analysis & Phase 1**

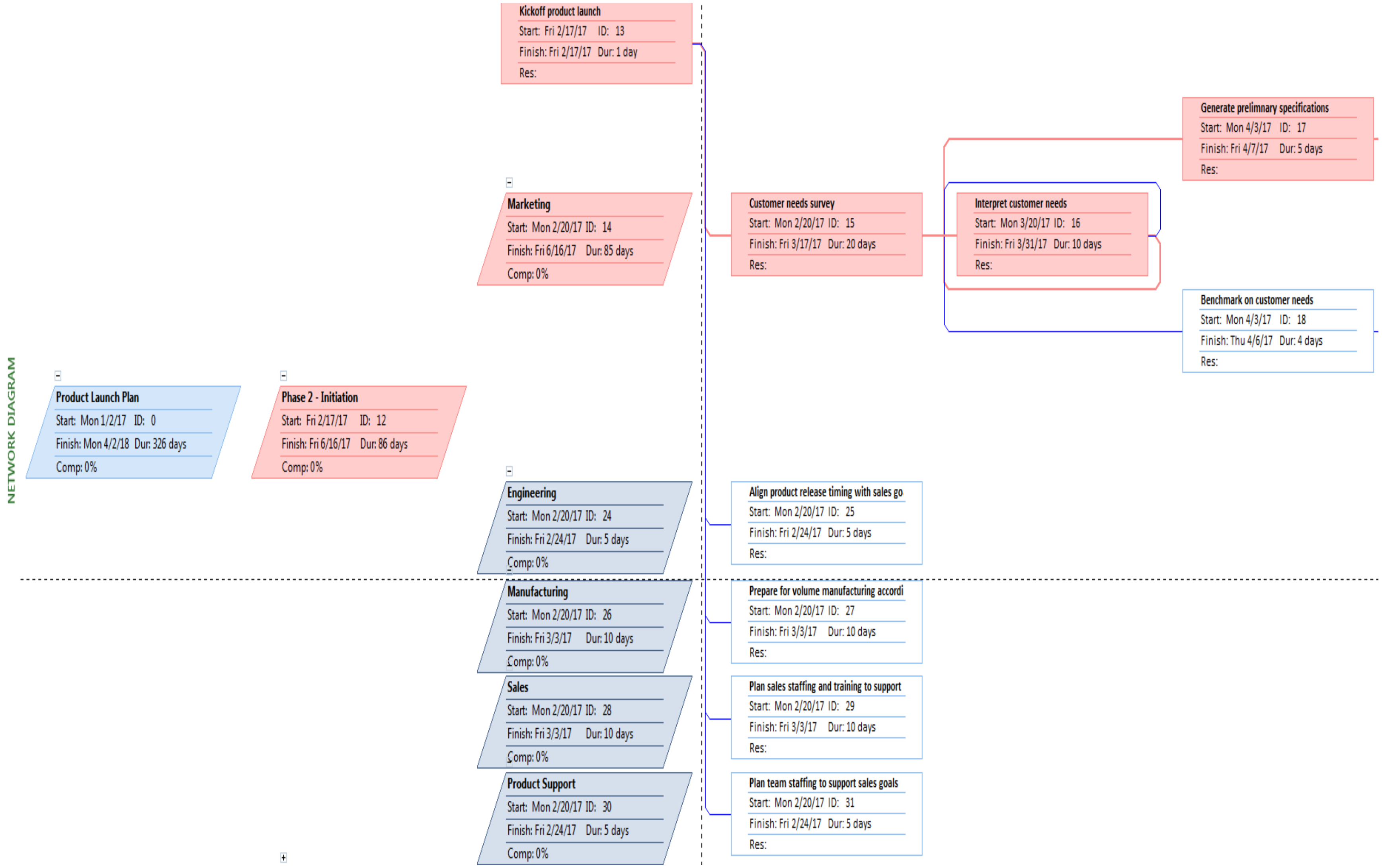

**Figure 44: Function Diagram Phase 2**

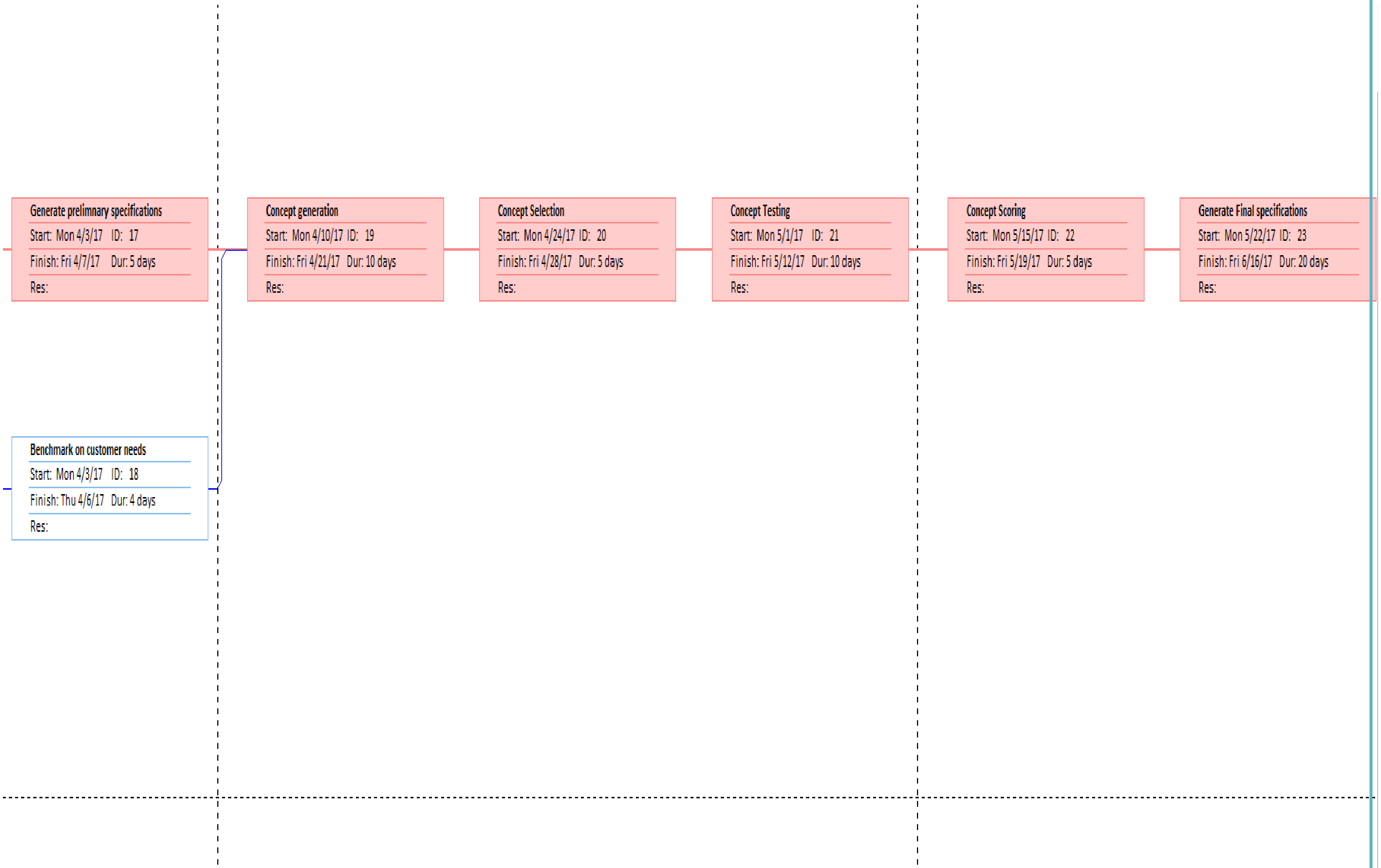

**Figure 45: Function Diagram Phase 2 Continue**

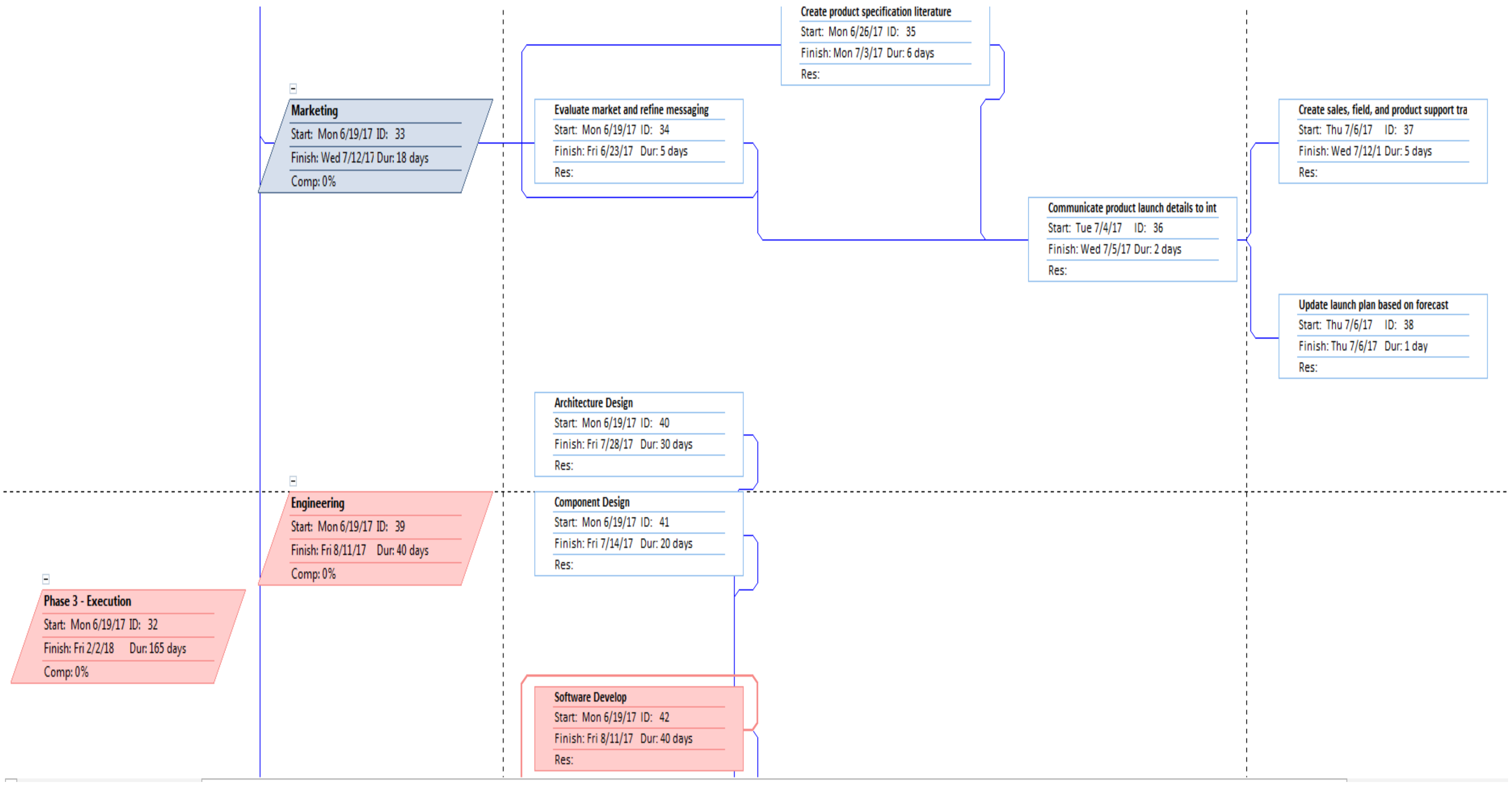

**Figure 46: Function Diagram Phase 3**

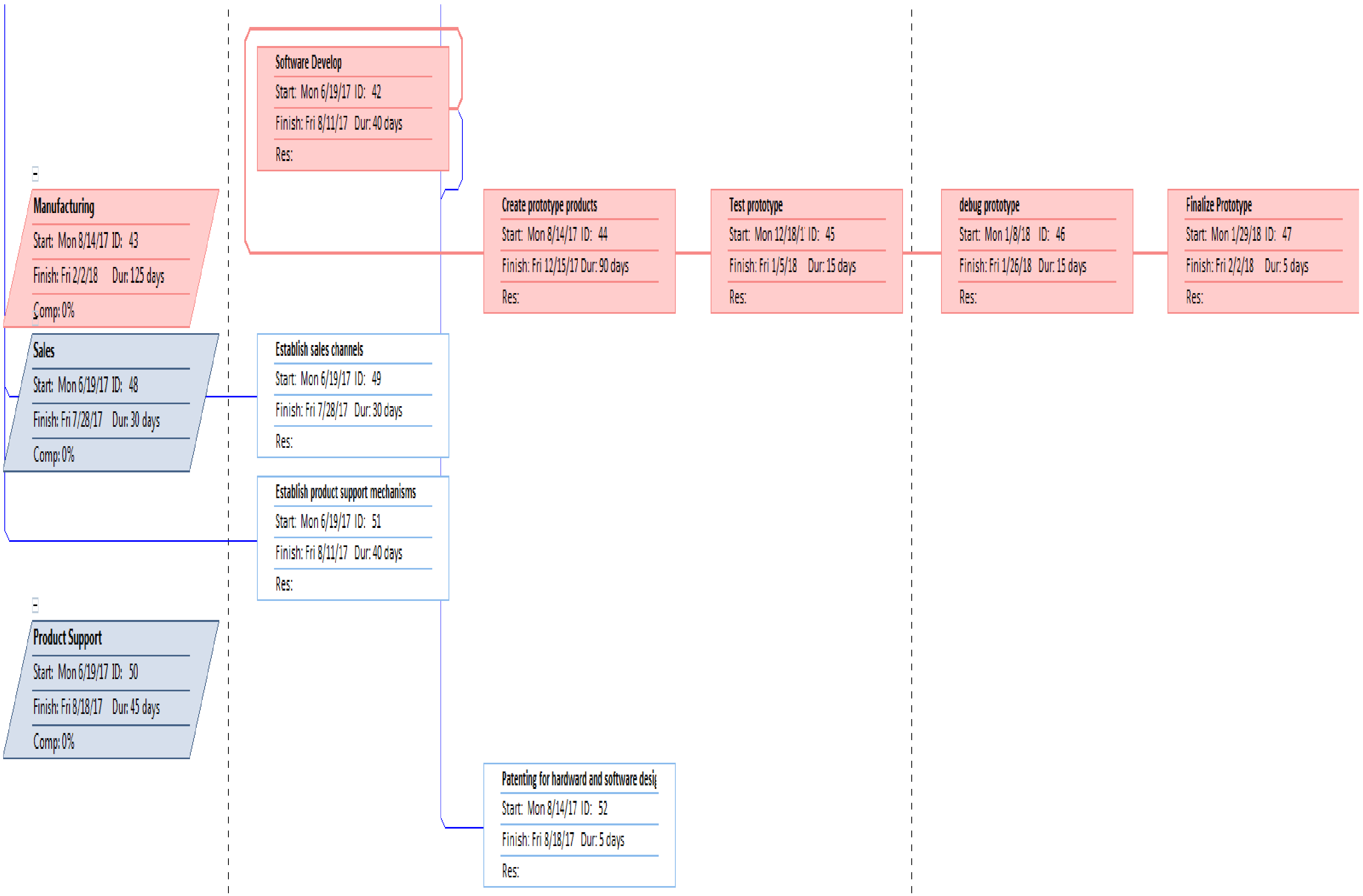

**Figure 47: Function Diagram Phase 3 Continue**

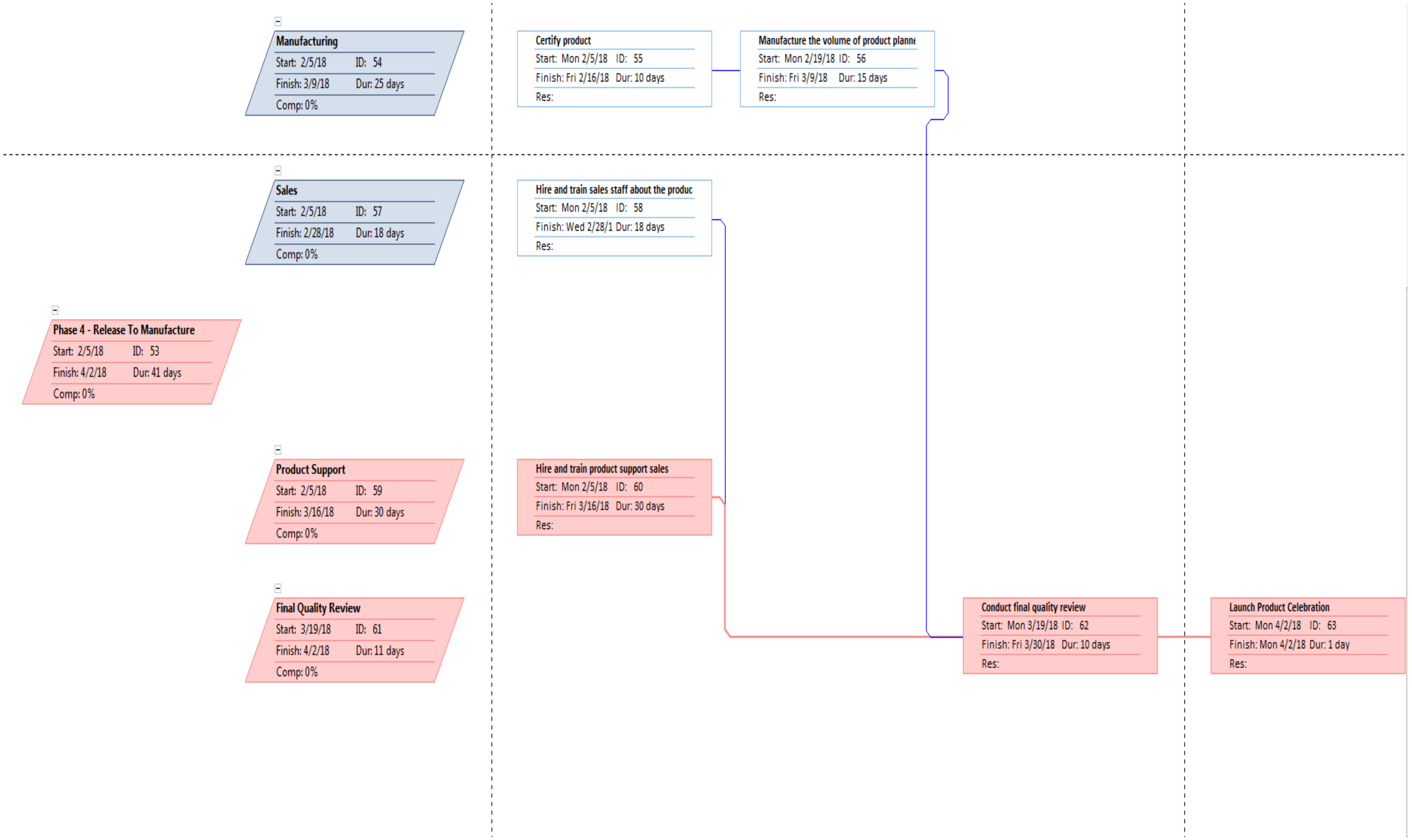

**Figure 48: Function Diagram Phase 4**

### 7.2.3 Work Breakdown Structure (WBS)

Work breakdown structure (WBS) is a key project deliverable that organizes the team's work into manageable sections. It was created by project team to identify the major functional deliverables and subdivide those deliverables into smaller systems and sub-deliverables.

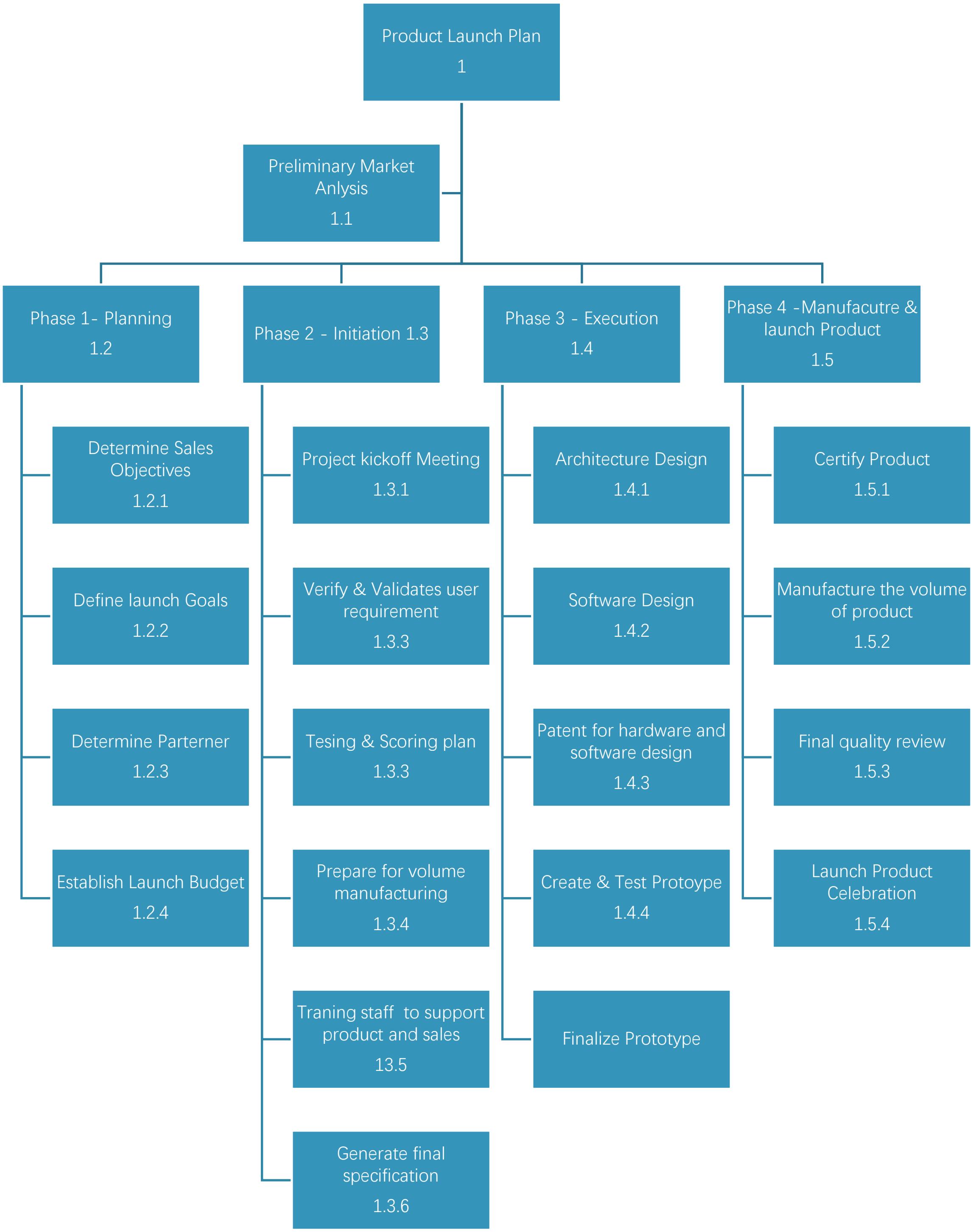

**Figure 49: Work Breakdown Structure of Eliminoise™ launch project**

## 7.3 Project Risk Analysis and Management

Risk management is the process which we can identify, analyze, control, eliminate and monitor the risk. Risk management's objective to assure uncertainty does not deflect the endeavor from the business goals. Nowadays, Project risk management is become an important aspect of project management. In most cases, Project success or fail depends on the Project manager if have proper risk management skill to control the risk. To control the risk on our project, several steps such as identify risk, perform risk analysis, monitor and control risk, plan risk solution need to be conducted.

### 7.3.1 Risk Identification

Despite the team has little experience in dealing with such uncertainties, we recognize several broad risk domains in which a very wide range of potential risks (45 peril, as for now) were identified, at least the team ken what perils will be confronting us and what might go erroneous. These perils are assembled in Table 7.3.1.1. This list is our scope of the peril assessment, in pursuit of our business objectives identified earlier.

| Risks Identified | Code |
|---|---|
| **Design-related** | |
| 1.Yield too low for the design | D1 |
| 2.Too high of the Design fee | D2 |
| 3.The product design time too long | D3 |
| 4.Product easy to damage for the design | D4 |
| 5.New industrial designs come through | D5 |
| 6.Difficult to pass the safety test | D6 |
| 7.The design not meet the customer expectation | D7 |
| **Suppliers** | |
| 1.Suppliers change component specifications | S1 |
| 2.Supplier stop supply certain component | S2 |
| 3.Current supplier increase component price | S3 |
| 4.Supplier go out of business | S4 |
| 5.Supplier change component quality | S5 |
| **Intellectual Property** | |
| 1.Certain key IP was owned by others | I1 |
| 2.Cannot get the authorization from IP owner | I2 |
| 3.Take too long to get the IP of the product | I3 |
| 4.IP was plagiarized by other company | I4 |
| **Legislation** | |
| 1.Legal issue with competitors | L1 |
| 2.Legal issue with working environment | L2 |
| 3.Legal issue with working hour | L3 |
| 4.Legal issue with building safety | L4 |
| 5.Legal issue with employee | L5 |
| **Financial** | |
| 1.High initial costs | F1 |
| 2.Low sales revenue | F2 |
| 3.Low cash flow | F3 |
| 4.Bank stop lending | F4 |
| 5.Profit too low | F5 |
| 6.Too much receivable | F6 |
| 7.Investor divestment | F7 |
| 8.Material cost over budget | F8 |
| 9.Manufacturing cost over budget | F9 |
| 10.Incorrect Pricing | F10 |
| 11.Building adequate sales | F11 |
| 12.Building/office rental fee increase | F12 |
| 13.Logistics cost too high | F13 |
| **Premises** | |
| 1.Property maintenance | P1 |
| 2.Building lease cannot extend | P2 |
| 3.Affording new premises | P3 |
| 4.Leaseholder change price | P4 |

| Technical | |
|---|---|
| 1.New technology evolves as the time | T1 |
| 2.Components easy to damage during the transit | T2 |
| 3.High wastage of material | T3 |
| 4.Some technical specifications hardly to meet | T4 |
| 5.Testing fail | T5 |
| 6.Employee don’t have enough technical skill | T6 |
| 7.Employee don’t have enough job experience | T7 |
| **Other Risk** | |
| 1.Rentension of key technical staff | O1 |
| 2.Too many similar products appear in Market | O2 |
| 3.Overstrteched management | O3 |
| 4.Rentension of key management staff | O4 |
| 5.Management personal financial issue | O5 |
| 6.Staff salary too high to afford | O6 |
| 7.Relationship with supplier and reseller | O7 |
| 8.Relationship with investor | O8 |
| 9.Key customer trust | O9 |

**Table 54: Risk identified**

## 7.3.2 Risk Rating

These perils are further assessed considering the likelihood and in cognation with its impact, as show in Table 7.3.2.1 below.

| Probability | Definition | Description |
|---|---|---|
| 1 | Unlikely | The risk is unlikely to occur within the time horizon contemplated by the objective. |
| 5 | Likely | The risk is likely to occur within the time horizon contemplated by the objective. |
| 10 | Imminent | The risk is expected to occur within the time horizon contemplated by the objective. |
| **Impact** | **Definition** | **Description** |
| 1 | Negligible | The risk will not affect the achievement of the objective, causing minimal damage to the organization's reputation. |
| 5 | Moderate | The risk will cause the objective to be delayed, causing potential damage to the organization's reputation. |
| 10 | Critical | The risk will cause the objective to not be achieved, causing damage to the organization's reputation. |

**Table 55: Risk Rating**

### 7.3.3 Risk Mapping

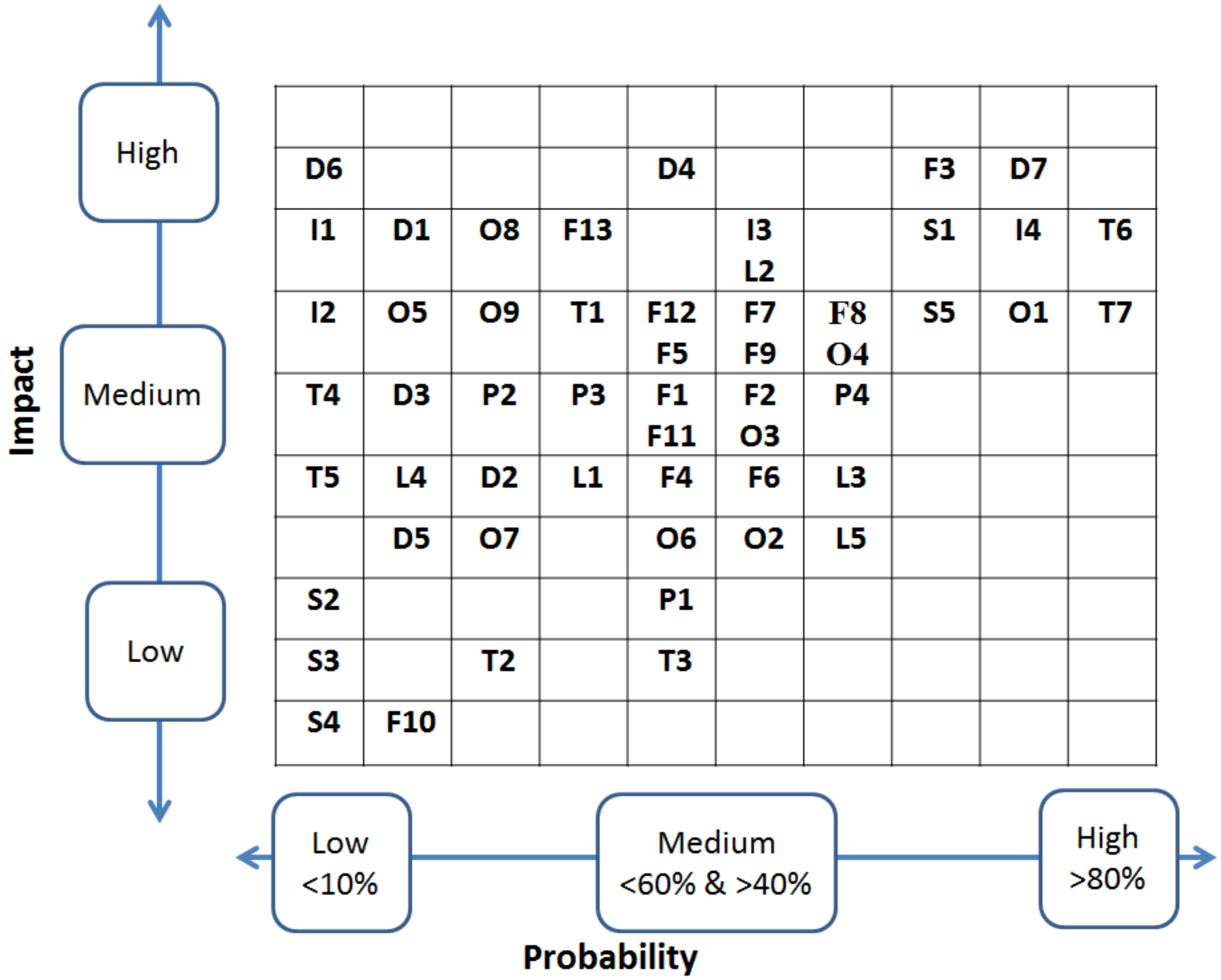

**Figure 50: Risk Mapping**

### 7.3.4 Risk Response

There are many different risks under the design, Finance, IP and other aspect. We will give a common solution on this aspect

1) Design-related risk

The company design and development team need to review the feedback and complain from the end user and survery target to understand the limitation of the product, solve the software bug and hardware issue,improve the end user product experience, understand the customer need and current market technology, future technology trend.

2) Suppliers risk

Regular meeting with supplier and sub-contractor to have a better understanding their faced issue so that can build a good relationship to avoid any miscommunication problem. Regular check and test component quality from suppliers and sub-contractor after component delivery. Write a clear legal contract with supplier to minimize the effect of the component delay and quality issue.

Carefully consider all the delay situation that maybe faced. Order the important components a bit earlier than exact plan schedule but calculation of pricing for early order must not be costly or heavy burden on organization revenue.

3) Intellectual Property risk

Learn the competitor patent in the product design and appearance to avoid legal issue. Apply appearance patent to protect our own product to be duplicated by others competitor. The patenting of specific features (Solar power supply) must prepare details writing of patent to avoid replica products.

4) Legislation risk

Careful read the safety and environment regulation and requirement for the manufactory process in China (The manufacturing places will consider based in China). Understand the labor law in Singapore and China to avoid any legal issue which related to employee working hours and salary.

5) Financial risk

The Finance team should balance the cost, time taken, target quality and constrain the resource. We need set the price based on the market demanding, competitor price, our product feature and others factor. There should be some tactics if company face the situation which have loss in certain period.

The Finance team should set the lowest limitation of cash flow in order to meet the company cash turnover requirement. The cost of the product and production must be estimated periodically and update if have any changes in component.

6) Technical risk

The technical team must proposal standard of procedure and instructions for each process in manufacturing, provide employee trainings to improve productivity and avoid technical risk.

The technical team should request the leader of technical team to conduct product preliminary discussion with the team member and share the experience to improve productivity.
Provide some reward to employee who provide some good suggestion to improve the production efficiency.

7) Premises risk

Sign a proper contract lease with the landlord to minimize the risk of the landlord increase the price. Regular check the building safety to avoid any disaster like fire.

Write the disaster recovery (DR) proposal for the IT infrastructure.

8) Other Risks

Our company don't have much cash to award to key staff and management team. We can provide some share-based compensation to encourage and reward.

The management team and design leader should to proposal new ideas and design in future. Moreover, the management need to keep up with upcoming industry's technology by learning and discovering from the industrial conference and technical magazine.

# 8. Quality & Reliability

When customer purchase a product, what they will be focusing on are the brand, function, product design and the quality of the product. Better the quality, longer the usage time.

At 3S Technology Pet Ltd, we always focus on providing customer with our best quality product and services. To achieve the goal, we decided to follow certain standards in quality control and management.

In the market, there are a number of standards for industrial and products. With certification from these organizations, customer will be more trustful in our product and our brand.

## 8.1 Necessary for Quality & Reliability

Overwhelmed products always have good hardware quality and stable system reliability. Together with low failure rate, these three are the key factors that can help us gain the reputation in the market. However, good quality also means higher material cost for the product. We need to find a balance between cost and quality for our product.

To aim to get great balance for the product cost and product quality, we decided to imply ISO 9001:2015 quality management standard and get certified.

International Organization for Standardization (ISO) is an organization that set the standards to ensure the quality reliability and safety of the products or services are aligning with the same high standards.

By implying ISO 9001:2015 standard, we aim to provide a quality and customer oriented quality management system. For customer oriented side, we would encourage customer to agree to send product log files back to us for us to better fix the bugs and develop our product. This will help us to monitor the development and usage of your product. With the development and improvement on our product, it can give us a good quality and reputation in the market. With the continuous monitor on the product performance and development, it will also help us develop a process on project improvement. This will ensure the just right amount man-hour and material is used for product improvement and development. Cost saving is counting on this. When successfully certified, customer will be assured that our company and product is set to meet high standards. They can set a free mind on purchasing our product.

Besides focusing on the product, we also want to contribute to improve our environment. As we all know, global warming is a sever issue for us and our future generation. We will get ISO 14001:2015 certification to contribute to the environment. We want our staff and customer to know that we are aware the environmental issue and will follow the standards to make some contribution.

The ISO 14000 family of standards provides useful tools for companies and organizations to manage the environmental responsibility. ISO 14001:2015 sets the criteria for an environmental management system and how it can be certified to. It gives a framework to the company or organization to follow to set up the effective environmental management system.

Besides that, we also want to make not only our product but also the packaging and other parts be environmental friendly. We will also follow the standards in different countries to make the material be recyclable and help to reduce the burden to the environment.

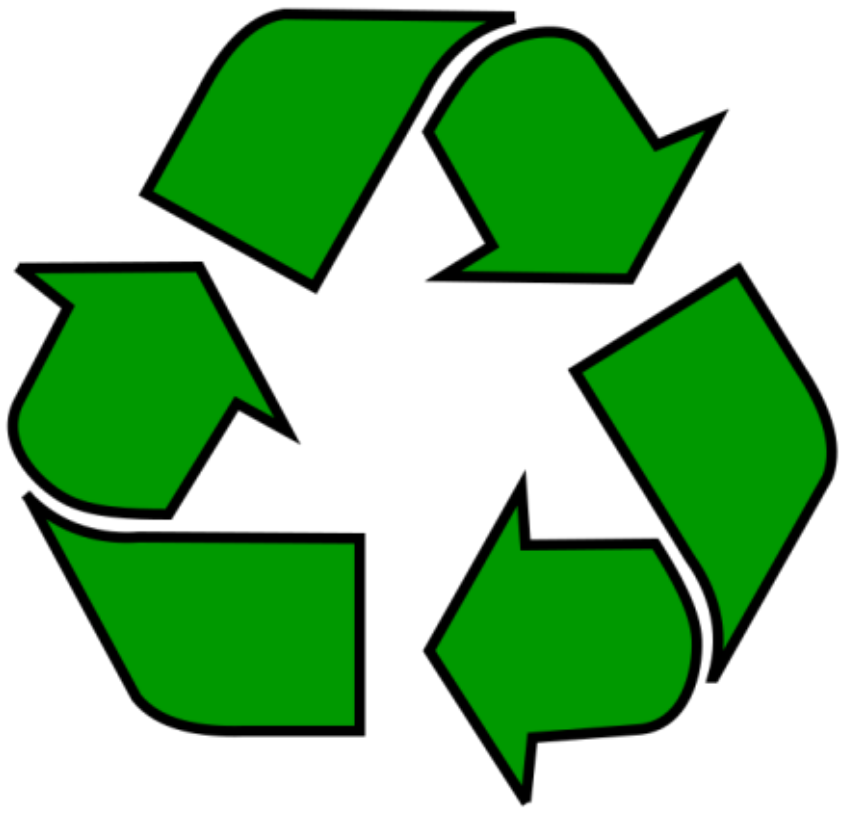

Maintaining a high standard can demonstrate that we have the ability to consistently provide high-standard products and service to customer and meet their expectation. This will also give the customer high satisfaction through our reliable products and high standard service.

## 8.2 Robust Design's Influence on Quality

Although we are designing the product in an ideal situation, the actual working environment may not be so friendly to our product. We need a robust design to make it suitable for the use for different environment and also can be resist to sudden changes for the environment or any emergency situation. We have considered the following parameters or factors to make the design more robust:

- Material – able to give a strong casing for the device
- Power supply –able to continuously provide power to device
- Electromagnetic Compatibility – able to resist to the EM interference
- Camera sensitivity – able to detect detail of customer movement
- Sensor accuracy – able to provide accurate environmental data
- Manufacturing process – able to provide high quality parts and low failure rate
- Assembly process – able to assemble high quality device and low failure rate
- Programming methodology – able to provide the stable and robust software design
- Operating manual – able to provide customer easy and accurate operating steps to follow.

Detailed design process and specification you can refer to respective section in this report.

By following the design process, we can be sure that we can build a reliable and high-quality product.

## 8.3 Supplier Quality Management – Certification & Standards

A product is built from different parts from supplier. Whether we can build a reliable and high quality product depends on the quality and reliability of our suppliers. When we source for the suppliers, we choose the suppliers who meet the ISO 9001:2015 Standard.

When searching for suppliers, we do not just focus on the quality of the product that the suppliers can give to us. We also will consider the pricing of the product that they can provide. This will help us to reduce the product cost and increase the profit margin.

## 8.4 Product Reliability Testing Methodology

Before launch the product to the market, we will do factory acceptance test (FAT) to around 500 units of final product to see the quality. We will do three levels of tests to assure the quality:

- Component level
- Device level
- Software level

For the detailed hardware tests and the software tests, please refer to the following tables.

| Hardware Test | Requirements |
|---|---|
| Component life cycle | Last for at least one year |
| Product life cycle | Last for at least one year |
| Temperature range | -20°C – 40°C |
| Waterproof | According to IP67 under IEC standard 60529 |
| Drop test | Free fall on hard surfaces from up to 2 meters |
| Power surge tolerance | Up to 10 Volt |
| Memory storage | At least 100,000 read/write cycles |

| Software Test | Requirement |
|---|---|
| Performance | App at most responds in 100ms |
| Security | Accurate and secure encoding method should not cause user data leakage |
| Compatibility | IOS 8 and above |
| | Android 4.0 and above |
| Usability | Suitable for 5 Year+ |
| User acceptance testing | Pass all the possible situation |
| GUI test | UI is accurate respond to customer operation |

## 8.5 Product Certification

Before we can launch our product in different countries, we need to get our product certified according to the standards in different countries. Some intended certifications are listed.

Conformité Européene

This is the mark that allows certain products to be sold in European Economic Area since 1985. Besides those products, products manufactured in or designed to be sold in EEA also need to get CE marked. This mark certifies that the product meets the entire requirement in the standard.

**Safety Mark**

Safety mark is one certification under Consumer Protection Registration Scheme. This mark will be displayed on the product or its packaging after the product meets the specific safety standards and then can be sold to customer in Singapore.

**<u>CCC Mark</u>**

CCC mark is actually China Compulsory Certificate. Similar to CE mark, this mark is a compulsory safety mark for any product imported, sold or used in china market. Since China is a great potential market, we should not miss out this certification.

**<u>FCC Mark</u>**

FCC mark is a standard for any electronic products manufactured or sold in USA that certified the electromagnetic interference of the product is within the limit requirement. All the limits are set and approved by Federal Communications Commission. In order to make our product can enter USA market, we need to make sure our product pass the EMI test and get FCC mark qualified.

**TÜV SÜD**

TÜV SÜD is a service provider offering product testing and certifying services. This brand already can be a representation of reliability and high quality. We may gain more trust from our customer when we get certified by TÜV SÜD.

**IEC standards**

The International Electrotechnical Commission (IEC) is the organization that set and publishes international standards for all electrical and electronic and related products and technologies. IEC standards detailed set the standards for different type of product and introduced detailed test method for each kind of product and situation. For our product, we need to follow the following standards:

- IEC 61000 Electromagnetic compatibility (EMC)
- IEC 60488 Standard digital interface for programmable instrumentation
- IEC 60574 Audio-visual, video and television equipment and systems
- IEC 60651 Sound level meters
- IEC 61000 Electromagnetic compatibility (EMC)
- IEC 62455 Internet protocol (IP) and transport stream (TS) based service access
- IEC 60801 EMI and RFI Immunity

# 9. Intellectual Properties

What is Intellectual Property? By definition from World intellectual property organization(WIPO) that intellectual Property (IP) refers to creations of the mind, such as inventions; literary and artistic works; designs; and symbols, names and images used in commerce purpose.

In other words, Intellectual Property (IP) is currently particular important for our products to achieve the objective of a larger market and a more unique marking position. Intellectual Property (IP) refers to the protection of creations of the mind, which have both moral and commercial values. IP protection is intended to stimulate the creativity of the human mind for the benefit of all by ensuring that the advantages derived from exploiting a creation that benefit the creator. This will encourage creative activity and allow investors in research and development to receive a fair return on their investment.

Base on the Singapore law, IP typically grants the author of intellectual creation exclusive rights for exploiting and benefiting from their creation. However, these rights, also called monopoly right of exploitation, are limited in scope, duration and geographical extent.

IP confers on individuals, enterprises or other entities the right to exclude others from the use of their creations. Consequently, intellectual property rights (IPRs) may have a direct and substantial impact on industry and trade as the owner of an IPR may - through the enforcement of such a right - prevent the manufacture, use or sale of a product which incorporates the IPR.

For this reason, control over the intangible asset (IPR) connotes control of the product and markets. IP protection encourages the publication, distribution and disclosure of the creation to the public, rather than keeping it secret while at the same time encouraging commercial enterprises to select creative works for exploitation. Intellectual property legal titles relate to the acquisition and use of a range of rights covering different type of creations. These may be industrial or literary and artistic.

In order to have a better IR for our product we will give details and talk more about Copyrights, Trademarks and Potential Patent which enable people to earn recognition or financial benefit from what we invent or create product called Eliminoise$^{TM}$.

## 9.1 Copyrights

What is copyright? By definition from World intellectual property organization (WIPO) that copyright (or author's right) is a legal term used to describe the rights that creators have over their literary and artistic works. Works covered by copyright range from books, music, paintings, sculpture, and films, to computer programs, databases, advertisements, maps, and technical drawings.

A more detailed explanation is copyright protects works like novels, computer programmed, plays, sheet music and paintings. Generally, the author of a copyright work has the right to reproduce, publish, perform, communicate and adapt our work. These exclusive rights form the bundle of rights that we call copyright and enable the owner to control the commercial exploitation of our work. In total, copyrights can also financially benefits the owner, the owner could have the ability to let others use their work for a fee under copyright laws and they can also sell the copy by using copyright license. The new invention and unique features of our product are bound by copyright law and thus we can take legal measure if someone copy or use it without permission.

Therefore, keeping all relevant materials on developments and future planning as well as work made for hire for Eliminoise$^{TM}$ exclusively under 3S-Technology Pte Ltd property has to be stated and incorporated. In every document, work or any materials related to 3S-Technology Pte Ltd property, would need to include the copyright statement: © 2017 Eliminoise$^{TM}$. All rights reserved. We have provided our contact information for request for permission to reproduce, republish or reproduce 3S-Technology Pte Ltd property.

**Company Address:<u>21 Heng Mui Keng Terrace, Singapore 119613</u>**

**Contact Number:** **6516 3308**

**Email address:** **sales@3s.com.sg**

**Website:** **www.3STechnology.com.sg**

**Business Hours: From Mon to Sat: 8AM-6PM, Close on Sun & PH**

This copyright is to give us legal protection from someone else from copying our materials without our consent. This is because if a company or someone violates any of these rights without our permission, we can bring infringement suit.

Copyright is also financially beneficial to the owner, the owner can let others use their work under the copyright law charges, and you can also use the copyright license to sell the copy. The new inventions and unique features of our products are subject to copyright laws, and we may also take legal action if they are copied or used without permission.

## 9.2 Trademarks

What is a trademark? By definition from World intellectual property organization (WIPO) that a trademark is a sign capable of distinguishing the goods or services of one enterprise from those of other enterprises. Trademarks are protected by intellectual property rights.

In other words, trademarks are a symbol that you can use to distinguish between business goods or services with other traders. Trademarks may represent the form or signature of the company's logo graphics. Through a registered trademark, you can protect your brand (or "mark") by restricting others from using its name or logo. Once acquired, a trade mark can last indefinitely as long as you renew it every 10 years. Because a registered trade mark is a form of IP, you can license or assign it to others.

And also an unregistered trademark, with the letters “TM”, can be used without filing any paperwork to receive permission to use the “TM” symbol for identification. The symbol “TM” can put the competitors on notice and they will respect your intellectual property.

We have branded our product as “Eliminoise™”. It is a made up of two words “Eliminate and Noise”. By definition of Eliminate is “to remove or get rid of, especially as being in some way undesirable: to eliminate risks; to eliminate hunger.” And by definition of Noise is “sound, especially of a loud, harsh, or confused kind: deafening noises.” So we combined these two word to one word “Eliminoise™” as our product name and to show that how good our product to eliminate noise. A logo with tagline was developed as our trademark of the product. Our logo will be placed on our product to uniquely identify us from other products. It is a sign of equal with the value our product has created. Any unauthorized use is prohibited.

SILENCE IS A POEM,
THE SOUND OF SILENCE !

We use “SILENCE IS A POEM, THE SOUND OF SILENCE!” our product slogan. A wonderful poem will give a quiet feeling. Our products are applicable to the Airport, Construction site; School, hospital, office, and we hope it creates a quiet, efficient and environmentally sustainable development, make human life more and more enjoyable.

Our company named as 3S-Technology Pte Ltd. So what is 3S stand for? It is Smart Silent System. We need a smart system to create a silent environment. Eliminoise™ can achieve above purposes.

Eliminoise$^{TM}$ is an advanced noise reducing system. Our logo is designed with three "S" shape which are the three "S" curves becomes gradually smaller. It not only represents a gradual reduction in noise but also represent 3S. Eliminoise$^{TM}$ logo has been marked as a trademark. Eliminoise$^{TM}$ have filed for registered trademark and yet to be approved. Therefore, in order to identify ourselves from the competitors, Eliminoise$^{TM}$ will keep the "TM" logo as it is till we received the approval from the relevant authorities.

## 9.3 Potential Patent

### 9.3.1 Patent

What is a patent? By definition from World intellectual property organization(WIPO) that a patent is an exclusive right granted for an invention, which is a product or a process that provides, in general, a new way of doing something, or offers a new technical solution to a problem. To get a patent, technical information about the invention must be disclosed to the public in a patent application.

A patent is a right granted to the owner of an invention that prevents others from making, using, importing or selling the invention without his permission. We invented Eliminoise$^{TM}$ so we enjoy all of the above patents. A patentable invention can be a product or a process that gives a new technical solution to a problem. It can also be a new method of doing things, the composition of a new product, or a technical improvement on how certain objects work.

The inventor of the owner should pay attention to keep the secret until the invention patent application has been successful. If this idea has been talked about, commercial use, publicity or presentation, and then invented novelty may be greatly reduced. If the invention needs to be disclosed to a third party before a patent application has been made, a non-disclosure agreement should be drawn up.

The invention must be representative of a product or process that is available than any existing product. Improve the special field where people who cannot have obvious skills or knowledge are invented. If an invention is new to a person who is good at art, the invention does not meet the demand for innovation.

Once it is granted and registered, its term of a patent is 20 years from the Date of Filing, subject to the payment of annual renewal fees.it apart from using the patent to prevent others from exploiting our invention, 3S-Technology Pte Ltd can employ it to raise funds for your business, license it to third parties for commercial returns or sell the patented invention.

### 9.3.2 Patent Claim

1. Eliminoise$^{TM}$ is a smart and multiple functions noise reduction device which reduce the noise effectively even for low-frequency noise.
2. Eliminoise$^{TM}$ has three types of power source, which are solar energy, battery and electricity energy.
3. Eliminoise$^{TM}$ include the feature to monitor the sleeping quality, store data, and transfer to the smart phones.
4. Eliminoise$^{TM}$ has built in memory of 16GB.
5. Eliminoise$^{TM}$ uses rechargeable lithium‑ion battery and can last up to 300 hours.
6. Eliminoise$^{TM}$ built in CPU with Texas Instruments and the dimension is 17.0 mm × 17.0 mm (LxW).
7. Eliminoise$^{TM}$ built in SMD NTC temperature sensor which is able to monitor ambient temperature.
8. Eliminoise$^{TM}$ use Bluetooth 4.2 KEDSUM for data transfer to smart devices and the dimension is 28.0mm x 12.7mm x 2.6mm (LxWxD).
9. Eliminoise$^{TM}$ uses 12 Mega pixel infrared camera to record data and transfer the data to smart devices through Bluetooth.
10. Eliminoise$^{TM}$ has microphone built in with LM393 Sound Detection Sensor for voice recording.
11. Eliminoise$^{TM}$ compatible with at least IOS 9, Android 7.0 system.
12. Eliminoise$^{TM}$ has built in Sundance Solar panel and the dimension is 92mm x 61mm x 3mm (LxWxD).

13. Eliminoise™ ’s working temperature range is from -20°C to 60°C.
14. Eliminoise™ uses materials of plastic and life span about 5 years.
15. Eliminoise™ has built in Water and Dust Resistant which is use IP67 under IEC standard 60529.
16. Eliminoise™ has resolution ID 92 mm x OD 100mm display to show data in monitor.
17. Eliminoise™ has three available colors which are white, black, red, blue.
18. The dimension of Eliminoise™ is Ø100mm x 68mm and weight is 300g.
19. Uses one button to select modes and functions.
20. Eliminoise™ has built in speaker.
21. Eliminoise™ has microphone for voice recording.
22. Eliminoise™ has user-friendly interface.
23. Eliminoise™ has remote control system.
24. Eliminoise™ is power saving device.
25. Eliminoise™ is easy integration into smart home system
26. Eliminoise™ is able to track sleeping information.

## 9.4 Invention Disclosure

The inventive disclosure is essentially used in the literature of the present invention. These are a method of documenting the details of your invention and submit it to a patent attorney who is submitting a patent application. This is the main step of the invention. It arranges the inventor's idea of invention. It must be filled in a way so that your invention is clear to unfamiliar people. The open inventive form allows the company to avoid the invention that is not patentable. The patent preparation in the form of the present invention will expedite the process of preparing the patent agent's patent draft. Patent proceedings will become more effective if there is a good and fruitful relationship between the inventor and the patent attorney. The inventor is an expert who needs to work with a patent attorney.

The invention of an invention disclosure or disclosure report is a confidential document written by a scientist or engineer using a patent office of a company or by an external patent agent to determine whether an invention should be sought for patent protection. It may follow a standardized form based on a company.

As a part of requirement, the university requires an official filing of invention disclosure. Refer attached Disclosure form for more details.

# 10. Disclosure Form

# INVENTION DISCLOSURE FORM

**Date of Submission:** ______________________

| |
|---|
| **1. TITLE OF INVENTION** *(a short but sufficiently descriptive title to identify the general nature of the invention.)*<br>Eliminoise™ (a smart device which is invented for reducing noise even for low-frequency with high efficiency to satisfy different requirements but low energy consumption. It can also remember and store settings for the same environment. This product is also integrated with other functions, like monitor sleeping quality, music player, sensing temperature, and smart control of home furniture, fire alarm, etc. This smart system can transfer data easily by WIFI or Bluetooth and controlled by its APP.) |
| |
| **2. DESCRIPTION OF THE INVENTION**<br>*Paying particular attention to A and B, please provide as complete a description as possible. This is essential to obtain an enforceable patent.*<br>*A. The purpose of your description is to enable a person with similar skills in your field to be able to make and use the invention you describe.*<br>*B. Please do not withhold any key elements of the invention (you are obliged to describe the best way of making and using the invention known to you at the time of submission).* |
| 2.1 **Field Of The Invention:**<br>The Eliminoise™ is smart noise reduction system consist of a sensor to detect the noise in the environment. Through this smart noise reduction system, even the noise with low-frequency can be eliminated. The system is also integrated with sleep tracking and music player applications. It can also remember and store settings for the same environment, sense temperature, and smart control of home furniture, fire alarm, etc. This smart system can transfer data easily by WIFI or Bluetooth and controlled by its APP. It has high efficiency, energy saving, multi-functions and use for different environment. |
| 2.2 **Summary of the Invention:**<br>The Eliminoise™ is a smart device to reduce noise. The sensor can monitor specific environment noise level and control noise actively. This device has three types of power source, which are solar energy, battery and electricity energy. The noise comes in via the Microphone. After that it passes the signal to Signal Processing System and the generated noise with inverse phase will be sent out by speaker, and cancel the initial noise. Main functions and features of our product are:<br>1. Reduce the noise effectively even for low-frequency noise;<br>2. Customers can adjust the reduced noise levels based on different requirements, which can be adjusted easily by mobile APP;<br>3. It has low energy consumption. |

4. The product can remember the stored values and noise environment so that it will automatically control and adjust accordingly;
5. The product can monitor the sleeping quality, store data, and transfer to the smart phones by Bluetooth or WIFI conveniently, as a part of healthy data;
6. This product can sense the actual temperature and adjust wisely;
7. The product can also be used as speakers, to amplify the music and easily controlled by smart phones
8. The product can combine with the intelligent system at home, to control the electrical furniture and fire alarm, etc.

2.3 **Brief Description of The Drawings** (if any)

*Listing of the captions of each drawing or figure relevant to the invention that you have attached to this invention disclosure.*

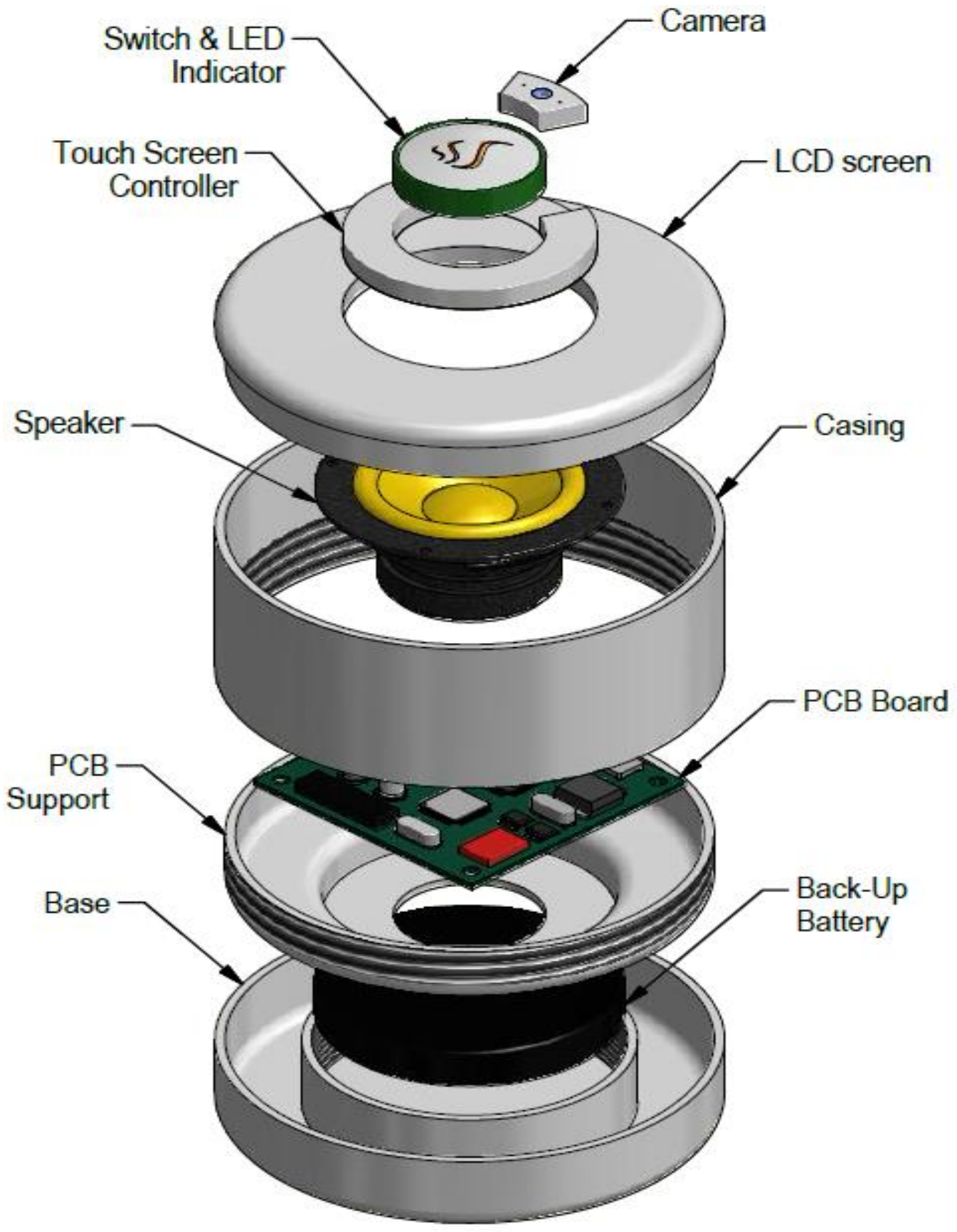

Eliminoise Details

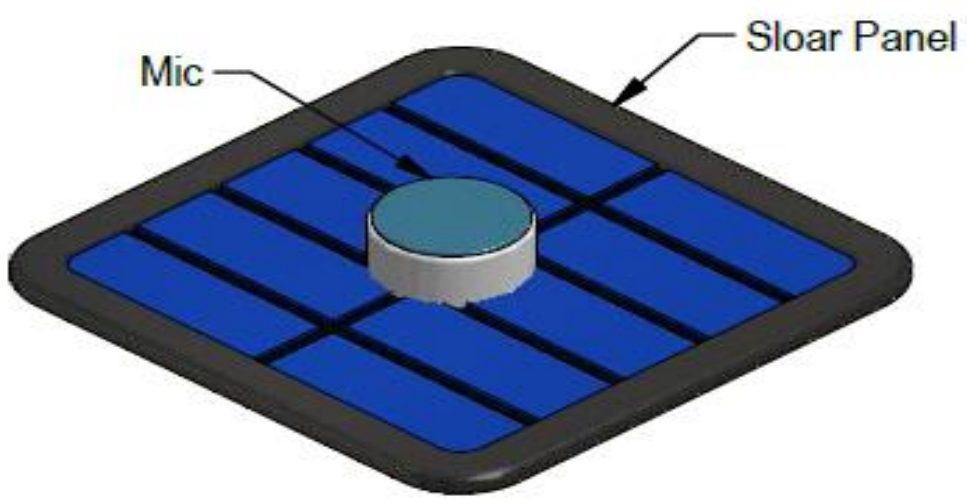

Noise Detect System

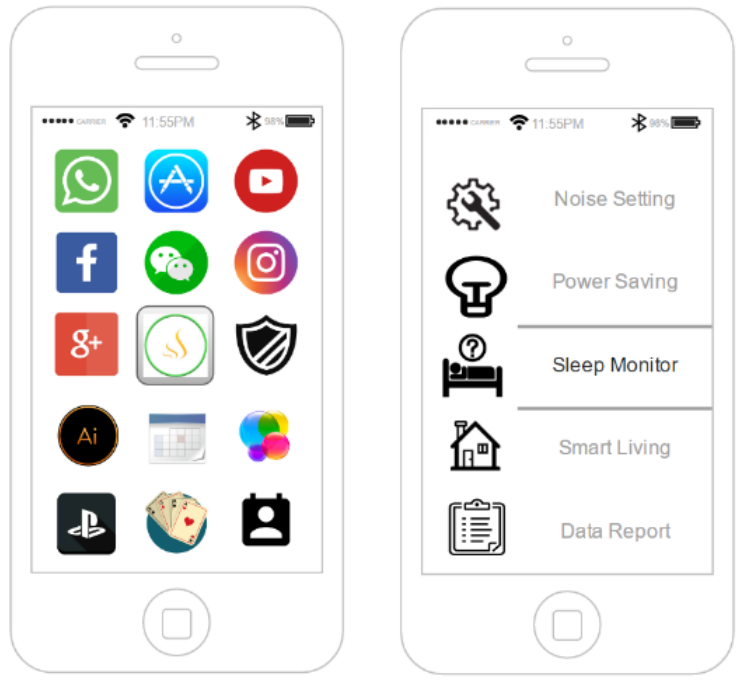

Mobile APP User Interface

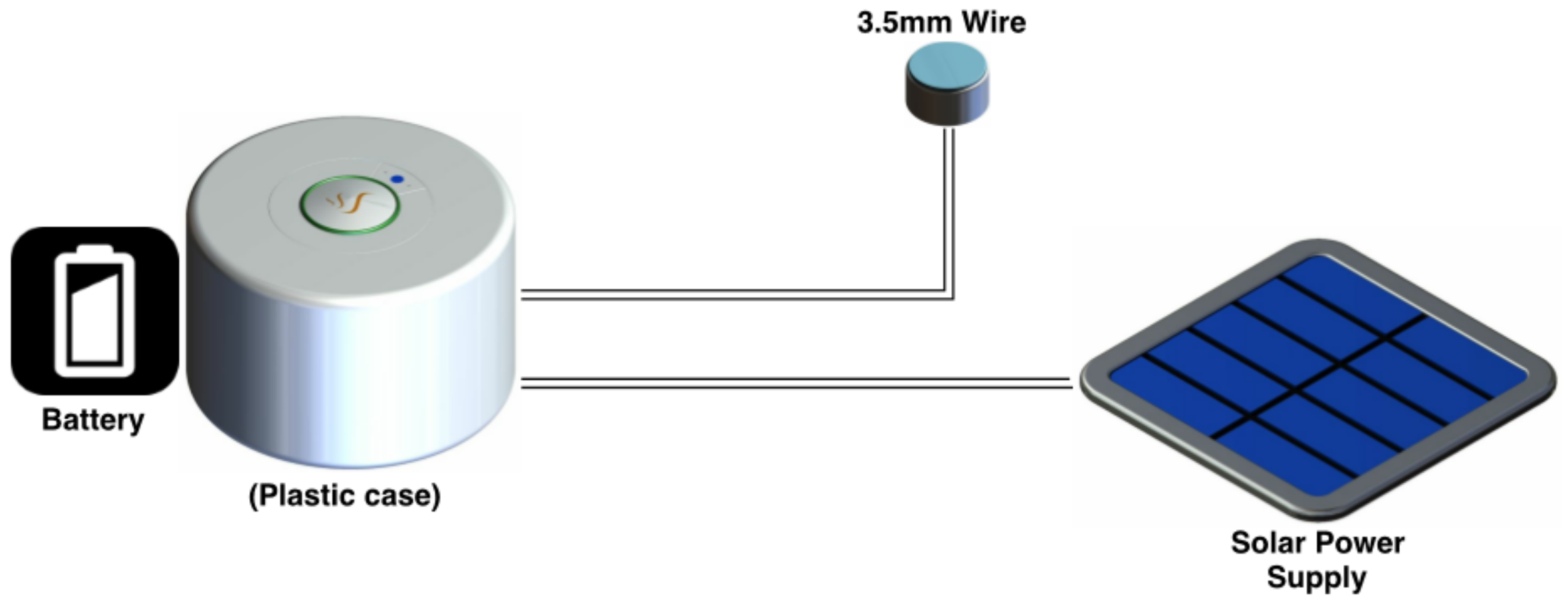

**Product Overall View**

This drawing shows the internal details for the product, and all the functions are implemented by indicated main components.

First of all, the noise will be detected by the mic which has been mounted outside of the house or office, and the detected noise will be sent to PCB board and go through the signal processing. An inverse phase frequency will be added to the initial noise frequency, so the amplitude of noise will be reduced and the noise will become smaller. After the signal processing system analysis, the output sound frequency will be generated and broadcasted by the speaker.

We also add some extra feature to the product, including tracking sleep quality by camera, playing music as a speaker by Wi-Fi block, controlling the intelligent system for the home furniture, fire alarm, and so on.

All these control behaviors can be implemented by touch screen controller and display on the LCD screen. Users can select desired level for reduction of noise. It can also be operated by the mobile app EliminoiseTM. In addition, this product can also remember the setting parameters in the same situations, record the sleep quality data, sense temperature, play music as players, and smart control the home furniture and fire alarm, and consume less energy. All stored data can be easily transferred by WIFI or Bluetooth.

2.4 **Detailed Description of The Preferred Embodiments:**

*This section should be detailed enough for a person having ordinary skill in your technical field to construct and use the invention you describe.*

*(i) A full description of the invention including background, preferred mode of practice of the invention e.g. basic nature or structure of invention, how it works with reference to relevant attached drawings etc.*

A good and quiet environment is very important. It will influence the working efficiency and the human's health. From an early report from the World Health Organization, noise exposure in long term can result in illness and even death to people in busy cities. Among various types of noise, the part only from traffic can lead to 3% of deaths due to heart attack and strokes all over the Europe. If we estimate based on the percentage, 210,000 people died of traffic noise in the world. In addition, above 600,000 potential years of healthy life were deprived by noise-related death and disabilities in Europe.

Currently, the main method for noise reducing mostly rely on the materials and transmission medium. For example, segregation board is often used to shield the noise in the construction site. These traditional ways are only effective to some extent for the high frequency noise. However, the effective reduction noise method especially for low frequency noise is very limited. To resolve this problem, we want to develop a system "active noise control" to reduce the noise by offsetting the signal frequency of noise.

Active noise control is a method to reduce the noise by adding the second sound specifically designed to cancel the initial noise. Based on Fourier theory, all the voices have different frequencies, phases and amplitude values in the frequency domain, and the voice can be regarded as a sinusoidal curve with peak amplitude. With this theory, by using the electronic circuit to generate the same frequency and amplitude value, but with -180$^{o}$C phase, it will combine to cancel out the environment noise. The details are shown as below:

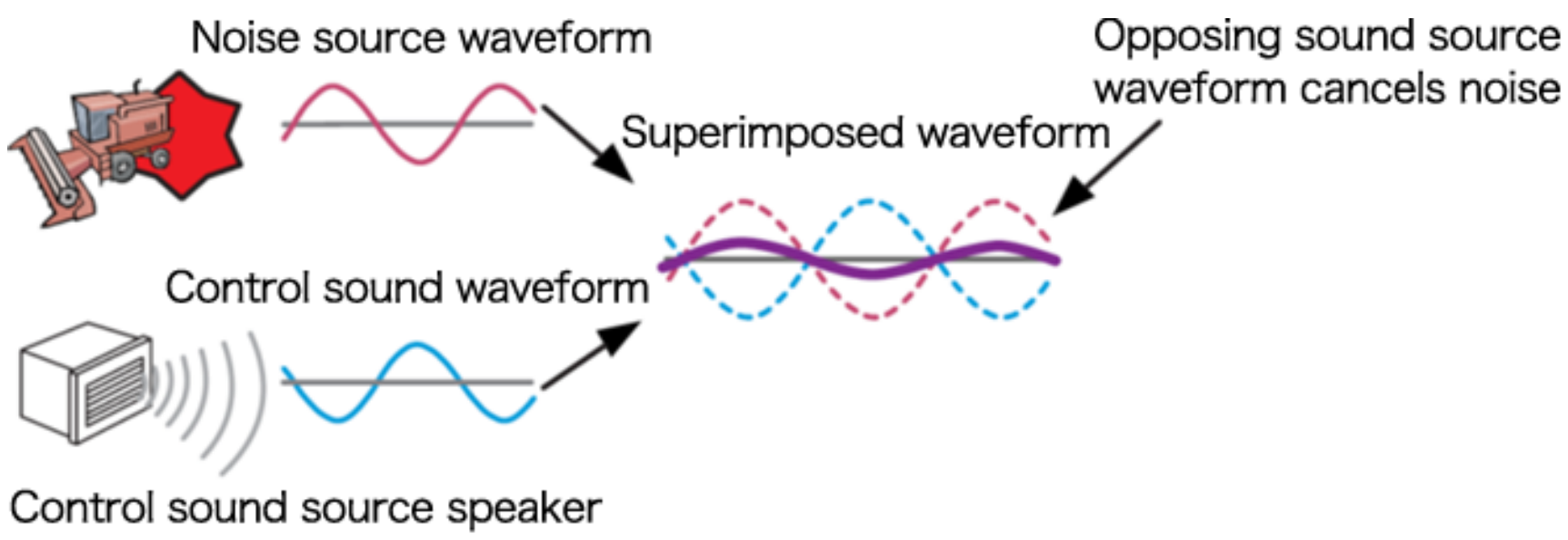

Below are the key points about the structure of the Eliminoise™, the noise reduction system:
The noise detector will detect the noise, and send to the noise reduction system.
The noise reduction system will calculate the noise frequency and generate a reverse frequency.
The noise reduction system speaker will broadcast the inverse frequency and cancel out the noise.
The system will also detect the reduction efficiency and give the system feedback to do the adjustment.

After the noise level has been reduced via the reduction system, the system is able to record the data for future analysis. The data can be transferred by Bluetooth or WIFI technology for reviewing the health record and take the action to improve the sleep quality.

Sleep tracking and music player functions are also integrated into the system. The product can be used as a speaker, and amplify the music and control the sound volume through the mobile phone. Able to choose the white noise or other voice mode in order to get a good quality of sleeping.

The product can also combine with the intelligent system at home, to control the electrical furniture and fire alarm.

*(ii) What problem(s) the invention solves and advantages over existing methods, devices or materials?*

Currently, there are not effective systems commercially to reduce the noise, including such as traffic and construction noise. Although the workers in the construction site try to build a barrier to shield noise, it’s still not effective especially for low frequency noise.

Eliminoise$^{TM}$, the noise reduction system blends comfort, user-friendliness, and advanced technology that measures and sends unwanted sounds that respond with a precise, equal and opposite signal in less than a fraction of a millisecond to counteract background noise with distinctly enhanced audio quality for a quieter environment to enjoy a moment of serenity in your living room.

Eliminoise$^{TM}$, the noise reduction system uses solar power and electric power as the primary and secondary power supply. There is a Bluetooth or Wi-Fi connection between speaker and silencer to transfer the noise data after the speaker detects noise. Using smart phone software application, it enables the system to play light music or white noise. In order to further enhance the system, it is also integrated with a sleeping tracking application and data collection function, to track the customer's sleep cycle and provide information and data for analysis.

*(iii) What are the possible specific industrial applications?*

Our main target are people who require quiet environment such as home, hospital, school, library, student center, child care center, conference hall, financial institutions, etc.

*(iv) Does your invention possess any disadvantages or limitations? Can they be overcome? What are the competing ways to solve the same problem(s)?*

The limitation would be the customers do not possess a smart phone, especially elderly and young children. In order to access its database, it will require a smart phone application. For the system to work perfectly, both smart phone & the system must be paired together through Bluetooth or Wi-Fi connection. The other limitation is wireless connection between speaker and silencer to transfer the noise data for the situations without WIFI.

To overcome these problems, the Eliminoise™, the noise reduction system has been added in a few components that can work independently without having to connect through a smartphone.

1. A touchscreen on the system, easy access of application.
2. Internal 16G, 2G RAM, for data collection and music storage
3. Built-in Bluetooth or Wi-Fi, for remote control

With the integration of multiple features as mentioned above, it might compromise with the overall size of the system. Therefore, researches could also go into designing a smaller size which comprises all the features that will be still available for the customer.

2.5 **<u>Modifications of The Preferred Embodiments:</u> (if any)**

1.Design a new model product that is small and portable so that it can be carried when the user travels.

2.Improve the product apps to have functions to share the sleeping data and compare with others, remind sleeping and waking up time automatically according to users' sleep data analyses.

2.6 References: Please list literature references that most closely describe your invention. You may if desired, conduct a patent search at http://ep.espacenet.com and http://www.uspto.gov

List of references cited in this write-up.

1. Moylan, William (2006). Understanding and crafting the mix: the art of recording. Focal Press. p. 26. ISBN 0-240-80755-3.
2. Active Noise Control, December 2005 Archived April 26, 2012, at the Wayback Machine.
3. Evaluation of an Improved Active Noise Reduction Microphone using Speech Intelligibility and Performance-Based Testing.
4. Evaluation of an Improved Active Noise Reduction Microphone using Speech Intelligibility and Performance-Based Testing.
5. Special Lay-Language Paper for the 75th Anniversary Meeting of the Acoustical Society of America, May 2004 Archived May 9, 2007, at the Wayback Machine.
6. Escape the Noise: Bose Learning Centre Archived February 19, 2012, at the Wayback Machine.
7. Anti-Noise, Quieting the Environment with Active Noise Cancellation Technology, IEEE Potentials, April 1992

**3. SOURCES OF SUPPORT AND GRANT RELATING TO INVENTION** *Please identify all outside agencies, organizations, or companies that provided funding to the research that led to the conception of the invention. Obligations of the research sponsor(s) will have to be met if patent protection and/or licensing of the technology is pursued Please also disclose any other contractual obligations entered into to come up with the invention including collaborations, research contracts material transfers etc.*

**Source(s) of Funding** *(eg. MOE ARF, A*STAR, EDB, etc. Please also include the grant no.):*

| | |
|---|---|
| **DSTA/DSO: Ref No.**<br>Title of Project Funded:<br>Collaborators (if any):<br>Was there a formal agreement signed? ☐Yes ☒No<br>Have University resources or facilities been used? ☐Yes ☒No | **NMRC: Grant No.**<br>Title of Project Funded:<br>Collaborators (if any):<br>Was there a formal agreement signed? ☐Yes ☒No<br>Have University resources or facilities been used? ☐Yes ☒No |
| **MoE: Grant No.**<br>Title of Project Funded:<br>Collaborators (if any):<br>Was there a formal agreement signed? ☐Yes ☒No<br>Have University resources or facilities been used? ☐Yes ☒No | **A*STAR(BMRC/SERC): Grant No**<br>Title of Project Funded:<br>Collaborators (if any):<br>Was there a formal agreement signed? ☐Yes ☒No<br>Have University resources or facilities been used? ☐Yes ☒No |
| **SMA/SMART: Ref No.**<br>Title of Project Funded:<br>Collaborators (if any):<br>Was there a formal agreement signed? ☐Yes ☒No<br>Have University resources or facilities been used? ☐Yes ☒No | **Other Agency: Grant No.**<br>Title of Project Funded:<br>Collaborators (if any):<br>Was there a formal agreement signed? ☐Yes ☒No<br>Have University resources or facilities been used? ☐Yes ☒No |

**4. DATES OF CONCEPTION & PUBLIC DISCLOSURE** *Please defer publication if you think that you may have patentable subject matter. Public disclosure of an invention before filing a patent application will render the invention not patentable in most countries.*

Date of documented conception of invention: 01/02/2017

Date of first public disclosure that describes invention, if any: N/A
Attach copies materials if possible.

Do you intend to disclose the invention publicly in the near future? ☐ Yes ☒ No
If yes, when and where?

Has this invention been reduced to practice? ☐ Yes ☒ No

Please indicate the status and intention for your invention. (You may ☒ more than one box)

| | | |
|---|---|---|
| ☐ Project ongoing | ☒ Looking for collaborators for further R&D | ☐ For information only |
| ☐ Project ended | ☐ Ready for Commercialization | ☐ Application to file a patent |
| ☒ Further R&D | ☐ Others (please specify) : | |

| 5. **CONTRACTUAL OBLIGATIONS:** *(Research Collaborations Agreements, Material Transfer Agreements etc.)* |
|---|
| (i) Title of Collaboration (RCA/MTA/Other):<br>N/A<br>(ii) Reference Number (NUS and/or external):<br>N/A<br>(iii) Name of Collaborator or Provider of Material:<br>N/A<br>(iv) Relevant Details of Collaboration or Material:<br>N/A |

| 6. **COMMERCIALISATION** *Please identify any potential licensees or collaborators interested in the invention.* | |
|---|---|
| List companies or organizations, if any, that could be interested in using this invention.<br><br>**SMRT Corporation**<br>**Changi Airport Group**<br>**SBS Transit**<br>**Land Transport Authority**<br>**Housing and Development Board**<br>**National Environment Agency of Singapore**<br>**City Developments Limited**<br>**Far East Organization**<br>**China Construction Development Co Pte Ltd**<br>**Straits Construction Singapore Pte Ltd**<br>**Yau Lee Constrution (Singapore) Pte Ltd**<br>**Qingdao Construction (Singapore) Pte Ltd**<br>**Koh Brother Gruop** | |
| Do you have plans to spin off a company based on your invention? | ☒Yes ☐ No |
| Would you be willing to participate in the marketing of this invention by explaining it to potential commercial partners? | ☒Yes ☐ No |

| **7. SOFTWARE DEVELOPMENT** (*If your invention involves or includes software, please answer the following questions. Else, you may skip this section*.) |
| --- |
| Is the software standalone? If not, list associated software that is required for the invention to work.<br><br>No. The application has associated with JAVA for windows OS. Object C for Apple OS. |
| What language is the software developed in and what platforms is it designed for delivery on? List the minimum hardware specifications required.<br>**<u>Computers (Windows, Mac OS):</u>**<br><br>Java for Windows OS<br>Object C for Mac Book<br><br>**<u>Smartphones:</u>**<br>Object C for iOS<br><br>C++ BlackBerry OS<br><br>Java for Android OS<br><br>**<u>Minimum hardware specifications:</u>**<br>Processor: 1 gigahertz(GHz) or faster.<br><br>Ram: 1 gigabyte(GB)(32-bit) or 2GB(64-bit)<br><br>Free hard disk Space:16 GB<br><br>Graphics Card: Microsoft DirectX9 graphics device with WDDM driver |
| Was any of the source code obtained under an open source license (e.g. BSD, GPL, Apache, etc.) or from any other source? ☐Yes ☒No<br>If yes,<br>a) Please provide a list of the sources:<br><br>N/A<br><br>b) Explain how the sources listed above have been used in the invention:<br><br>N/A |
| Are there any third party rights associated with the invention of the software? List grants or contracts if any, with third parties.<br><br>N/A |
| Is the software an improvement of existing software? Has a license been obtained on the existing software? Provide details.<br><br>N/A |
| Is the software a proof-of-concept, a demonstration, prototype or fully functional end user version?<br><br>Prototype. |

**NB: *Please attach pages as required for additional inventors**

**Kindly keep us informed of permanent/home address changes to ensure you do not miss patent-related notifications or revenue payouts**

# 11. Appendix A

## 11.1 Link to Survey

https://www.surveymonkey.com/r/DQGYL6J

## 11.2 Survey Forms

**<u>Situation investigation</u>**

1 Do you think noise pollution related problems are very serious?
- ☐Yes
- ☐No

2 Do you have any trouble caused by noise?
- ☐Yes
- ☐No

3 Normally, when will the noise appear? (Multiple choice)
- ☐In the early morning (5-7 am)
- ☐In the morning (8-11am)
- ☐In the noon (12-13 pm)
- ☐In the afternoon (14-18 pm)
- ☐In the evening (19-22 pm)
- ☐In the mid night (After 23 pm)
- ☐Whole day all possible

4 Usually, how often will you be affected by the noise appeared?
- ☐Never
- ☐hardly
- ☐seldom
- ☐occasionally
- ☐sometime
- ☐often
- ☐usually

5 Normally, the noise happens in which location? Multiple choice
- ☐Workplace
- ☐School
- ☐Home
- ☐Other (please specify)

6 Usually, what is the source of the noise?
- ☐ Construction site
- ☐ Transport traffic
- ☐ Neighbors
- ☐ Other (please specify)

7 Is the source you identified from Q6 a permanent or temporary source?
- ☐ Permanent
- ☐ Temporary

8 Are here any effects caused by the noise that affect your life?
- ☐ Difficulty to fall asleep
- ☐ Cannot focus on work and study
- ☐ High blood pressure
- ☐ Feeling stressful
- ☐ Other (please specify)

9 Currently, what is your solution to deal with the noise?
- ☐ Ear plug
- ☐ Calling respective government agency to control
- ☐ Existing noise reduction solution (please specify brand and model)
- ☐ Other (please specify)

10 Do you know the affects that can be caused by long-term noise?
- ☐ Yes
- ☐ No

11 Has the current solution applied by the government, that is to reduce the noise cause by construction sites, MRT or airport, meet your requirement?
- ☐ Yes
- ☐ No

12 What kind of the following sound do you think is noise?
- ☐ Baby cry
- ☐ Train crossing track
- ☐ Dog bark
- ☐ Airplane flying over
- ☐ Any sound that is too loud

## Product Functionality

1. If there is a product that can help you to reduce noise, what are the functions that you think is good to have on this product?
   - ☐Noise reduction
   - ☐Sleep quality monitor and data record
   - ☐Sound (e.g. phone ringing tone) amplification and speaker
   - ☐Alert customer based on the sound detected (e.g. fire alarm)
   - ☐Record noise data and analysis of data that can help government take action
   - ☐Integrated with current smart home system
   - ☐Other (please specify)

2. What are the features that you think this device should have?
   - ☐Can remote control
   - ☐Can work on its own
   - ☐User friendly
   - ☐Easy to install
   - ☐Longer battery life
   - ☐Other (please specify)

3. If it is possible for government to monitor and control noise source based on the noise reduction product, do you think it is necessary and useful?
   - ☐Yes
   - ☐No

4. If this product can also help you to monitor your sleep quality, do you think the function is necessary?
   - ☐Yes
   - ☐No

5. If the device will be using green energy as part of the power supply, which do you think is more suitable?
   - ☐Solar Energy
   - ☐Wind Energy
   - ☐Other (please specify)

6. Do you prefer the product has the wireless remote control function? Which do you prefer?
   - ☐Wi-Fi
   - ☐Bluetooth
   - ☐both
   - ☐Not important

7. If can do remote control, do you want it be controlled on your mobile device like your smart phone using an app?
   - ☐Yes
   - ☐No

## Other information

1. If the product with these function is priced at S$ 300, will you purchase it? If No, please specify the percentage the price should be reduced.
   - ☐Yes
   - ☐May consider
   - ☐No (please specify the reduction %)

2. Will you make the noise into consideration when you making the following decision?
   - ☐Looking for job
   - ☐Purchasing house
   - ☐Applying for school
   - ☐Other (please Specify)

3. Can we have your age range?
   - ☐below 20
   - ☐20-30
   - ☐30-40
   - ☐40-50
   - ☐50-60
   - ☐above

4. Gender?
   - ☐Male
   - ☐Female

5. Salary range?
   - ☐below S$2000
   - ☐S$2000 – S$3000
   - ☐S$3001- S$4000
   - ☐above S$4000

6. What is your occupation?

## 11.3 Interview Questionnaire (Sample Size 10)

### **For investors**

1. Are you willing to invest in a product that can do the noise reduction for domestic and business usage?

2. If the predicted market share is 1 billion, return rate for the project is 10%. We need S$ 500,000 to start up the business. Are you willing to be an investor? If no, please specify the suitable amount that you are willing to invest in.

3. In which form are you willing to invest?

4. Which type of return do you prefer?

### **For enterprise customer (sample size 10)**

1. If the product can help you to reduce the noise in your working environment, do you want to purchase?

2. What is the size of your enterprise? Which industry is your enterprise in?

3. Is there any situation that the noise caused by your cooperation has affects to the employee, the cooperate or the environment?

4. Are you willing to let us monitor and analyze the noise data collected by our device?

5. Any other functionality that you think is necessary as a cooperate user?

# 12. **Reference**

[1] Lam, Bhan, Woon-Seng Gan, DongYuan Shi, Masaharu Nishimura, and Stephen Elliott. "Ten questions concerning active noise control in the built environment." Building and Environment 200 (2021): 107928.

[2] Chepesiuk R. Decibel hell: the effects of living in a noisy world. Environ Health Perspect. 2005 Jan;113(1):A34-41. doi: 10.1289/ehp.113-a34. PMID: 15631958; PMCID: PMC1253729.

[3] http://www.nea.gov.sg/anti-pollution-radiation-protection/noise-pollution-control

[4] https://www.fhwa.dot.gov/publications/publicroads/03jul/06.cfm

[5] http://www.nea.gov.sg/anti-pollution-radiation-protection/noise-pollution-control/noise-pollution

[6] http://www.nea.gov.sg/anti-pollution-radiation-protection/noise-pollution-control/noise-pollution

[7] Shi, Dongyuan. "Algorithms and implementations to overcome practical issues in active noise control systems." (2020).

[8] Shi, Dongyuan, Bhan Lam, Woon-Seng Gan, and Jordan Cheer. "Active Noise Control in The New Century: The Role and Prospect of Signal Processing." arXiv preprint arXiv:2306.01425 (2023).

[9] Shen, Xiaoyi, Woon-Seng Gan, and Dongyuan Shi. "Alternative switching hybrid ANC." Applied Acoustics 173 (2021): 107712.

[10] Shen, Xiaoyi, Dongyuan Shi, and Woon-Seng Gan. "A wireless reference active noise control headphone using coherence based selection technique." In ICASSP 2021-2021 IEEE International Conference on Acoustics, Speech and Signal Processing (ICASSP), pp. 7983-7987. IEEE, 2021.

[11] Shen, Xiaoyi, Dongyuan Shi, Woon-Seng Gan, and Santi Peksi. "Implementation of coherence-based-selection multi-channel wireless active noise control in headphone." In INTER-NOISE and NOISE-CON Congress and Conference Proceedings, vol. 263, no. 4, pp. 1945-1953. Institute of Noise Control Engineering, 2021.

[12] Shen, Xiaoyi, Dongyuan Shi, Woon-Seng Gan, and Santi Peksi. "Adaptive-gain algorithm on the fixed filters applied for active noise control headphone." Mechanical Systems and Signal Processing 169 (2022): 108641.

[13] Shen, Xiaoyi, Dongyuan Shi, and Woon-Seng Gan. "A hybrid approach to combine wireless and earcup microphones for ANC headphones with error separation module." In ICASSP 2022-2022 IEEE International Conference on Acoustics, Speech and Signal Processing (ICASSP), pp. 8702-8706. IEEE, 2022.

[14] Shen, Xiaoyi, Dongyuan Shi, Santi Peksi, and Woon-Seng Gan. "A multi-channel wireless active noise control headphone with coherence-based weight determination algorithm." Journal of Signal Processing Systems 94, no. 8 (2022): 811-819.

[15] Shen, Xiaoyi, Dongyuan Shi, Santi Peksi, and Woon-Seng Gan. "Implementations of wireless active noise control in the headrest." In INTER-NOISE and NOISE-CON Congress and Conference Proceedings, vol. 265, no. 4, pp. 3445-3455. Institute of Noise Control Engineering, 2023.

[16] Shi, Chuang, Tatsuya Murao, Dongyuan Shi, Bhan Lam, and Woon-Seng Gan. "Open loop active control of noise through open windows." In Proceedings of Meetings on Acoustics, vol. 29, no. 1. AIP Publishing, 2016.

[17] Shi, Dongyuan, Jianjun He, Chuang Shi, Tatsuya Murao, and Woon-Seng Gan. "Multiple parallel branch with folding architecture for multichannel filtered-x least mean square algorithm." In 2017 IEEE International Conference on Acoustics, Speech and Signal Processing (ICASSP), pp. 1188-1192. IEEE, 2017.

[18] Shi, Chuang, Nan Jiang, Huiyong Li, Dongyuan Shi, and Woon-Seng Gan. "On algorithms and implementations of a 4-channel active noise canceling window." In 2017 International Symposium on Intelligent Signal Processing and Communication Systems (ISPACS), pp. 217-221. IEEE, 2017.

[19] Shi, Chuang, Huiyong Li, Dongyuan Shi, Bhan Lam, and Woon-Seng Gan. "Understanding multiple-input multiple-output active noise control from a perspective of sampling and reconstruction." In 2017 Asia-Pacific Signal and Information Processing Association Annual Summit and Conference (APSIPA ASC), pp. 124-129. IEEE, 2017.

[20] Lam, Bhan, Woon-Seng Gan, Dongyuan Shi, and Stephen Elliott. "HYBRID SOURCE ARRANGEMENT FOR ACTIVE CONTROL OF NOISE THROUGH APERTURES AT OBLIQUE INCI-DENCES: A PRELIMINARY INVESTIGATION." Proc. 25th Int. Congr. Sound Vib. ICSV25, Hiroshima, Japan (2018): 1-8.

[21] Lam, Bhan, Chuang Shi, Dongyuan Shi, and Woon-Seng Gan. "Active control of sound through full-sized open windows." Building and Environment 141 (2018): 16-27.

[22] Hasegawa, Rina, Dongyuan Shi, Yoshinobu Kajikawa, and Woon-Seng Gan. "Window active noise control system with virtual sensing technique." In INTER-NOISE and NOISE-CON Congress and Conference Proceedings, vol. 258, no. 1, pp. 6004-6012. Institute of Noise Control Engineering, 2018.

[23] He, Jianjun, Bhan Lam, Dongyuan Shi, and Woon Seng Gan. "Exploiting the underdetermined system in multichannel active noise control for open windows." Applied Sciences 9, no. 3 (2019): 390.

[24] Hasegawa, Rina, Dong-Yuan Shi, Yoshinobu Kajikawa, and Woon-Seng Gan. "Multi-channel ANC Window with Virtual Sensing Technique." IEICE Technical Report; IEICE Tech. Rep. 118, no. 496 (2019): 69-74.

[25] Shi, Dongyuan, Bhan Lam, and Woon-seng Gan. "Analysis of multichannel virtual sensing active noise control to overcome spatial correlation and causality constraints." In ICASSP 2019-2019 IEEE International Conference on Acoustics, Speech and Signal Processing (ICASSP), pp. 8499-8503. IEEE, 2019.

[26] Shi, Dongyuan, Woon-Seng Gan, Jianjun He, and Bhan Lam. "Practical implementation of multichannel filtered-x least mean square algorithm based on the multiple-parallel-branch with folding architecture for large-scale active noise control." IEEE Transactions on Very Large Scale Integration (VLSI) Systems 28, no. 4 (2019): 940-953.

[27] Shi, Dongyuan, Woon-Seng Gan, Bhan Lam, Rina Hasegawa, and Yoshinobu Kajikawa. "Feedforward multichannel virtual-sensing active control of noise through an aperture: Analysis on causality and sensor-actuator constraints." The Journal of the Acoustical Society of America 147, no. 1 (2020): 32-48.

[28] Shi, Dongyuan, Bhan Lam, Shulin Wen, and Woon-Seng Gan. "Multichannel active noise control with spatial derivative constraints to enlarge the quiet zone." In ICASSP 2020-2020 IEEE International Conference on Acoustics, Speech and Signal Processing (ICASSP), pp. 8419-8423. IEEE, 2020.

[29] Lam, Bhan, Dongyuan Shi, Woon-Seng Gan, Stephen J. Elliott, and Masaharu Nishimura. "Active control of broadband sound through the open aperture of a full-sized domestic window." Scientific reports 10, no. 1 (2020): 1-7.

[30] Shi, Dongyuan, Woon-Seng Gan, Bhan Lam, Shulin Wen, and Xiaoyi Shen. "Active noise control based on the momentum multichannel normalized filtered-x least mean square algorithm." In INTER-NOISE and NOISE-CON Congress and Conference Proceedings, vol. 261, no. 6, pp. 709-719. Institute of Noise Control Engineering, 2020.

[31] Lam, Bhan, Dongyuan Shi, Valiantsin Belyi, Shulin Wen, Woon-Seng Gan, Kelvin Li, and Irene Lee. "Active control of low-frequency noise through a single top-hung window in a full-sized room." Applied Sciences 10, no. 19 (2020): 6817.

[32] Shi, Dongyuan, Bhan Lam, Woon-Seng Gan, and Shulin Wen. "Block coordinate descent based algorithm for computational complexity reduction in multichannel active noise control system." Mechanical Systems and Signal Processing 151 (2021): 107346.

[33] Shi, Dongyuan, Woon-Seng Gan, Bhan Lam, and Xiaoyi Shen. "Comb-partitioned frequency-domain constraint adaptive algorithm for active noise control." Signal Processing 188 (2021): 108222.

[34] Lai, Chung Kwan, Jing Sheng Tey, Dongyuan Shi, and Woon-Seng Gan. "Robust estimation of open aperture active control systems using virtual sensing." In INTER-NOISE and NOISE-CON Congress and Conference Proceedings, vol. 265, no. 4, pp. 3397-3407. Institute of Noise Control Engineering, 2023.

[35] Shi, Dongyuan, Bhan Lam, Junwei Ji, Xiaoyi Shen, Chung Kwan Lai, and Woon-Seng Gan. "Computation-efficient solution for fully-connected active noise control window: Analysis and implementation of multichannel adjoint least mean square algorithm." Mechanical Systems and Signal Processing 199 (2023): 110444.

[36] Lam, Bhan, Kelvin Chee Quan Lim, Kenneth Ooi, Zhen-Ting Ong, Dongyuan Shi, and Woon-Seng Gan. "Anti-noise window: Subjective perception of active noise reduction and effect of informational masking." Sustainable Cities and Society 97 (2023): 104763.

[37] Shi, Dongyuan, Chuang Shi, and Woon-Seng Gan. "Effect of the audio amplifier's distortion on feedforward active noise control." In 2017 Asia-Pacific Signal and Information Processing Association Annual Summit and Conference (APSIPA ASC), pp. 469-473. IEEE, 2017.

[38] Shi, Dongyuan, Chuang Shi, and Woon-Seng Gan. "A systolic FxLMS structure for implementation of feedforward active noise control on FPGA." In 2016 Asia-Pacific Signal and Information Processing Association Annual Summit and Conference (APSIPA), pp. 1-6. IEEE, 2016.

[39] Shi, Dongyuan, Woon-Seng Gan, Bhan Lam, and Shulin Wen. "Practical consideration and implementation for avoiding saturation of large amplitude active noise control." Proc. 23rd Int. Congr. Acoust (2019): 6905-6912.

[40] Shi, DongYuan, Woon-Seng Gan, Bhan Lam, and Chuang Shi. "Two-gradient direction FXLMS: An adaptive active noise control algorithm with output constraint." Mechanical Systems and Signal Processing 116 (2019): 651-667.

[41] Shi, Dongyuan, Bhan Lam, Woon-Seng Gan, and Shulin Wen. "Optimal leak factor selection for the output-constrained leaky filtered-input least mean square algorithm." IEEE Signal Processing Letters 26, no. 5 (2019): 670-674.

[42] Wen, Shulin, Woon-Seng Gan, and Dongyuan Shi. "Convergence behavior analysis of FXLMS algorithm with different leaky term." In INTER-NOISE and NOISE-CON Congress and Conference Proceedings, vol. 261, no. 6, pp. 728-739. Institute of Noise Control Engineering, 2020.

[43] Shi, Dongyuan, Woon-Seng Gan, Bhan Lam, Shulin Wen, and Xiaoyi Shen. "Optimal output-constrained active noise control based on inverse adaptive modeling leak factor estimate." IEEE/ACM Transactions on Audio, Speech, and Language Processing 29 (2021): 1256-1269.

[44] Shi, Dongyuan, Woon-Seng Gan, Bhan Lam, and Xiaoyi Shen. "Comb-partitioned frequency-domain constraint adaptive algorithm for active noise control." Signal Processing 188 (2021): 108222.

[45] Shi, Dongyuan, Woon-Seng Gan, Bhan Lam, and Xiaoyi Shen. "Optimal penalty factor for the MOV-FxLMS algorithm in active noise control system." IEEE Signal Processing Letters 29 (2021): 85-89.

[46] Shi, Dongyuan, Woon-Seng Gan, Bhan Lam, and Xiaoyi Shen. "A Frequency-Domain Output-Constrained Active Noise Control Algorithm Based on an Intuitive Circulant Convolutional Penalty Factor." IEEE/ACM Transactions on Audio, Speech, and Language Processing 31 (2023): 1318-1332.

[47] Shi, Dongyuan, Bhan Lam, Xiaoyi Shen, and Woon-Seng Gan. "Multichannel two-gradient direction filtered reference least mean square algorithm for output-constrained multichannel active noise control." Signal Processing 207 (2023): 108938.

[48] Lai, Chung Kwan, Dongyuan Shi, Bhan Lam, and Woon-Seng Gan. "MOV-Modified-FxLMS algorithm with Variable Penalty Factor in a Practical Power Output Constrained Active Control System." IEEE Signal Processing Letters (2023).

[49] Ji, Junwei, Dongyuan Shi, Woon-Seng Gan, Xiaoyi Shen, and Zhengding Luo. "A Computation-efficient Online Secondary Path Modeling Technique for Modified FXLMS Algorithm." arXiv preprint arXiv:2306.11408 (2023).

[50] Gan, Woon-Seng, Dongyuan Shi, and Xiaoyi Shen. "Practical Active Noise Control: Restriction of Maximum Output Power." arXiv preprint arXiv:2307.10913 (2023).

[51] Shi, Dong Yuan, and Woon-Seng Gan. "Comparison of different development kits and its suitability in signal processing education." In 2016 IEEE International Conference on Acoustics, Speech and Signal Processing (ICASSP), pp. 6280-6284. IEEE, 2016.

[52] Shi, Dong Yuan, Bhan Lam, and Woon-Seng Gan. "A novel selective active noise control algorithm to overcome practical implementation issue." In 2018 IEEE International Conference on Acoustics, Speech and Signal Processing (ICASSP), pp. 1130-1134. IEEE, 2018.

[53] Wen, Shulin, Woon-Seng Gan, and Dongyuan Shi. "Using empirical wavelet transform to speed up selective filtered active noise control system." The Journal of the Acoustical Society of America 147, no. 5 (2020): 3490-3501.

[54] Wen, Shulin, Woon-Seng Gan, and Dongyuan Shi. "An improved selective active noise control algorithm based on empirical wavelet transform." In ICASSP 2020-2020 IEEE International Conference on Acoustics, Speech and Signal Processing (ICASSP), pp. 1633-1637. IEEE, 2020.

[55] Shi, Dongyuan, Woon-Seng Gan, Bhan Lam, and Shulin Wen. "Feedforward selective fixed-filter active noise control: Algorithm and implementation." IEEE/ACM Transactions on Audio, Speech, and Language Processing 28 (2020): 1479-1492.

[56] Shi, Dongyuan, Woon-Seng Gan, Bhan Lam, and Kenneth Ooi. "Fast adaptive active noise control based on modified model-agnostic meta-learning algorithm." IEEE Signal Processing Letters 28 (2021): 593-597.

[57] Shi, Dongyuan, Bhan Lam, Kenneth Ooi, Xiaoyi Shen, and Woon-Seng Gan. "Selective fixed-filter active noise control based on convolutional neural network." Signal Processing 190 (2022): 108317.

[58] Luo, Zhengding, Dongyuan Shi, and Woon-Seng Gan. "A hybrid sfanc-fxnlms algorithm for active noise control based on deep learning." IEEE Signal Processing Letters 29 (2022): 1102-1106.

[59] Luo, Zhengding, Dongyuan Shi, Junwei Ji, and Woon-seng Gan. "Implementation of multi-channel active noise control based on back-propagation mechanism." arXiv preprint arXiv:2208.08086 (2022).

[60] Luo, Zhengding, Dongyuan Shi, Woon-Seng Gan, Qirui Huang, and Libin Zhang. "Performance Evaluation of Selective Fixed-filter Active Noise Control based on Different Convolutional Neural Networks." In INTER-NOISE and NOISE-CON Congress and Conference Proceedings, vol. 265, no. 6, pp. 1615-1622. Institute of Noise Control Engineering, 2023.

[61] Shi, Dongyuan, Woon-Seng Gan, Bhan Lam, Zhengding Luo, and Xiaoyi Shen. "Transferable latent of cnn-based selective fixed-filter active noise control." IEEE/ACM Transactions on Audio, Speech, and Language Processing (2023).

[62] Luo, Zhengding, Dongyuan Shi, Xiaoyi Shen, Junwei Ji, and Woon-Seng Gan. "Deep Generative Fixed-Filter Active Noise Control." In ICASSP 2023-2023 IEEE International Conference on Acoustics, Speech and Signal Processing (ICASSP), pp. 1-5. IEEE, 2023.

[63] https://www.amazon.com/EMY-HC-SR501-Pyroelectric-Infrared-Detector/dp/B00FDPO9B8/ref=sr_1_4?s=industrial&ie=UTF8&qid=1487517101&sr=1-4&keywords=Motion+Sensor

[64] http://hypertextbook.com/facts/2003/ChrisDAmbrose.shtml

[65] https://vimeo.com/155891645

[66] https://academic.oup.com/bmb/article/68/1/243/421340/Noise-pollution-non-auditory-effects-on-health

[67] http://www.eea.europa.eu/media/infographics/noise-pollution-in-europe-1/image/image_view_fullscreen

[68] https://www.slideshare.net/imanog/remote-active-noise-control

[69] https://en.wikipedia.org/wiki/Noise_pollution

[70] https://www.amazon.com/Raspberry-5MP-Camera-Board-Module/dp/B00E1GGE40/ref=pd_sim_147_1?_encoding=UTF8&psc=1&refRID=J1VRAJW9F7P80FHRMSRM

[71] https://www.amazon.com/Small-Solar-Panel-200mA-wires/dp/B00IWZWT6S/ref=sr_1_11?s=industrial&ie=UTF8&qid=1487516695&sr=1-11&keywords=solar+panel

[72] https://www.ihi.co.jp/inc/laneng/product/product01.html

[73] https://en.wikipedia.org/wiki/Demography_of_the_United_States

[74] http://www.conserve-energy-future.com/causes-and-effects-of-noise-pollution.php

[75] http://metro.co.uk/2007/08/22/noise-kills-thousands-every-year-42454/

[76] Shi, Dongyuan, Jinhong Guo, Liang Chen, Chuncheng Xia, Zhefeng Yu, Ye Ai, Chang Ming Li, Yuejun Kang, and Zhiming Wang. "Differential microfluidic sensor on printed circuit board for biological cells analysis." Electrophoresis 36, no. 16 (2015): 1854-1858.

[77] Guo, Jinhong, Xiwei Huang, Dongyuan Shi, Hao Yu, Ye Ai, Chang Ming Li, and Yuejun Kang. "Portable resistive pulse-activated lens-free cell imaging system." RSC Advances 4, no. 99 (2014): 56342-56345.

[78] Li, Jianzhuang, Cheng Chang, Dongyuan Shi, Wei Xia, and Lin Chen. "A new firmware upgrade mechanism designed for software defined radio based system." In Proceedings of the 2012 International Conference on Information Technology and Software Engineering: Software Engineering & Digital Media Technology, pp. 277-283. Springer Berlin Heidelberg, 2013.

[79] Li, Liangliang, Wei Xia, Dongyuan Shi, and Jianzhuang Li. "Frequency estimation on power system using recursive-least-squares approach." In Proceedings of the 2012 International Conference on Information Technology and Software Engineering: Information Technology & Computing Intelligence, pp. 11-18. Springer Berlin Heidelberg, 2013.

[80] Shi, Dongyuan, Zhengping Liu, Rui Qian, Wei Xia, and Zishu He. "FPGA Implementation of DLMS Algorithm for Digital Repeater." Dianshi Jishu(Video Engineering) 36, no. 9 (2012).

[76] http://www.ti.com/product/AM4379/datasheet

[77] http://www.val.me.vt.edu/sites/default/files/publications/ActiveControlFuller_1.pdf